\documentclass[twocolumn,trackchanges]{aastex7}
\usepackage{amsfonts,amsmath,amssymb,mathrsfs}
\usepackage{comment}
\usepackage{bbm}
\usepackage{amsmath}
\usepackage{aastex}
\usepackage{graphicx}
\usepackage{subfigure}
\usepackage{epstopdf}
\usepackage{float}
\usepackage{color}
\usepackage{cancel}
\usepackage{ulem}
\usepackage{threeparttable}
\usepackage{smartref}
\usepackage{makecell}
\usepackage{multirow}
\usepackage{verbatim}
\usepackage{textcomp}
\usepackage{hyperref}
\usepackage[misc]{ifsym} 
\usepackage{changepage}
\usepackage{tabularx}
\usepackage{booktabs}

\graphicspath{{./}{figsfinal/}}

\begin{document}

\title{High Sensitivity Methodologies to Detect Radio Band Gravitational Waves}	
	
\author[0000-0001-7906-0919]{Wei Hong}
\affiliation{Institute for Frontiers in Astronomy and Astrophysics, Beijing Normal University, Beijing 102206, China}
\affiliation{Department of Astronomy, Beijing Normal University, Beijing 100875, China}	
\email{weihong@mail.bnu.edu.cn}

\author{Peng He}
\affiliation{Bureau of Frontier Science and Education, Chinese Academy of Sciences, Beijing 100190, People’s Republic of China}
\email{ }
	
\author[0000-0002-3363-9965]{Tong-Jie Zhang}
\affiliation{Institute for Frontiers in Astronomy and Astrophysics, Beijing Normal University, Beijing 102206, China}
\affiliation{Department of Astronomy, Beijing Normal University, Beijing 100875, China}	
\email[show]{tjzhang@bnu.edu.cn} 

\author[0000-0003-3193-7907]{Shi-Yu Li}
\affiliation{Beijing Planetarium, Beijing Academy of Science and Technology, Beijing 100044, China}
\email{lishiyu@bjp.org.cn}

\author[0000-0002-3386-7159]{Pei Wang}
\affiliation{National Astronomical Observatories, Chinese Academy of Sciences, Beijing 100101, China}
\affiliation{Institute for Frontiers in Astronomy and Astrophysics, Beijing Normal University, Beijing 102206, China}	
\email{wangpei@nao.cas.cn}

\received{receipt date}\revised{revision date}\accepted{acceptance date}\published{published date}

\submitjournal{ApJS}

\begin{abstract}
Gravitational waves (GWs) can resonate with magnetic fields through the Gertsenshtein-Zeldovich effect, producing electromagnetic signals at the same frequency. In pulsar magnetospheres, this conversion may yield a faint radio-band signal that could be detected. In this work, we focus on two specific pulsars, PSR J1856–3754 and PSR J0720–3125, and use numerical simulations to evaluate how well the FAST and SKA2‑MID telescopes could detect such signals. We consider transient events, including primordial-black-hole-like mergers, as well as stochastic backgrounds, including primordial GWs. To improve detection sensitivity, we propose four observational methods to lower the detectable energy-density limit of very high-frequency (VHF) GWs; the ``Multiple Pulsars with Multiple Telescopes'' (MPMT) method performs best because it allows cross-validation and rejection of false candidates. Under the assumption of nearly 6000 hours of observation at 3 GHz and a $5\sigma$ detection threshold, the minimum detectable characteristic strain is projected to be $h_c \approx 10^{-23}$ for transient events and $h_c \approx 10^{-33}$ for stochastic backgrounds. Under optimistic assumptions on integration time and conversion efficiency, these projections suggest that radio-band searches may approach the sensitivity needed to begin testing representative VHF GW scenarios. More broadly, this conversion in pulsar magnetospheres could be relevant to the origin of some repeating fast radio bursts in the our galaxy.

\end{abstract}


\section{Introduction}
As one of the most renowned predictions based on general relativity, gravitational waves (GWs) are widespread at every epoch of cosmic evolution \citep{Aggarwal:2020olq}. At the end of the Planck era, gravity was promptly decoupled from the other three fundamental interactions, making GWs important probes for studying the early universe \citep{Roshan:2024qnv}. The discovery of the first late-universe binary-merger GWs by LIGO \citep{LIGOScientific:2016aoc} marked the advent of GW astronomy. Since then, ground-based interferometers have produced a rapidly growing census of compact-binary mergers, summarized in the latest LVK Gravitational-Wave Transient Catalogs \citep{2025arXiv250818082T,2023PhRvX..13d1039A}. Complementary evidence for nanohertz GWs has been reported by multiple pulsar timing arrays \citep{NANOGrav:2023bts,EPTA:2023fyk,Reardon:2023gzh}. Motivated by the progress of radio facilities such as FAST and SKA2-MID, radio-band GW searches based on GW--photon conversion have recently attracted increasing attention \citep{Domcke:2020yzq,Domcke:2023qle,Ito:2023nkq,Ito:2023fcr,Herman:2022fau,Dandoy:2024oqg,2024PhRvD.110j3003M}.

In recent years, the largest aperture nonfully steerable single-dish radio telescope in the world, FAST, was officially inaugurated, and the SKA radio interferometer began trial operation; thus, the possibility of using radio telescopes to detect GWs in the radio band was put on the agenda \citep{Domcke:2020yzq, Domcke:2023qle, Herman:2022fau,Ito:2023nkq, Dandoy:2024oqg,2024PhRvD.110j3003M}. There are many methods and detectors, such as optically levitated sensors \citep{Arvanitaki:2012cn}, the resonant polarization rotation method \citep{Cruise:2000za}, bulk acoustic-wave devices \citep{2013NatSR...3.2132G}, superconducting rings \citep{Anandan:1982is}, laboratory microwave cavities  \citep{Bernard:2001kp,Ballantini:2003nt,Tong:2008rz, Stephenson:2009zz,Berlin:2021txa,Li:2000du,Li:2013fna, Li:2023tzw, Li:2003tv, Li:2004df, Li:2008qr, Li:2009zzy, Li:2011zzl,Li:2014bma}, and the graviton--magnon resonance \citep{Ito:2019wcb} inverse Gertsenshtein--Zeldovich (GZ) effect method \citep{Gertsenshtein:1962}, for converting GWs in the radio band into other types of detectable signals. These methods and detectors indicate that the inverse GZ effect may be a potential mechanism for observing very high-frequency (VHF, $\mathrm{MHz}-\mathrm{GHz}$) GW signals with radio telescopes. The detection sensitivity of a radio telescope to a radio signal converted via the inverse GZ effect is related to the intensity of the VHF GW source, magnetic field properties of the observed sky area, and observation method.

From the perspective of Universe evolution, VHF GW sources can be broadly grouped into (i) early-universe stochastic gravitational-wave backgrounds (SGWBs) and (ii) late-universe transient events. A broad overview of cosmological GW backgrounds and their spectral features can be found in \citet{2018CQGra..35p3001C}. Beyond standard inflationary scenarios \citep{Barnaby:2010vf}, several cosmological mechanisms can generate blue-tilted or peaked spectra at very high frequencies. Representative examples include first-order phase transitions in the early Universe \citep{1984PhRvD..30..272W,1986MNRAS.218..629H,1992PhRvD..45.4514K,1994PhRvD..49.2837K}; cosmic strings and other topological defects \citep{1985PhRvD..31.3052V,2000PhRvL..85.3761D,2007PhRvL..98k1101S,2014PhRvD..89b3512B,2020JCAP...04..034A}; and nonlinear post-inflationary dynamics such as preheating/reheating \citep{Kofman:1994rk,1997PhRvD..56..653K,2006JCAP...04..010E,2007PhRvL..98f1302G,2008PhRvD..77d3517G,Cai:2018dig}. In addition, scalar-induced second-order GWs sourced by primordial density perturbations provide another well-motivated high-frequency SGWB component \citep{2007PhRvD..75l3518A,2007PhRvD..76h4019B}. Non-standard thermal histories may further enhance the high-frequency tail of relic gravitons \citep{1999PhRvD..60l3511G}. Late-universe VHF transients include compact-binary mergers at sufficiently small separations \citep{Bauswein:2018bma} and primordial-black-hole-like merger events \citep{Bird:2016dcv,Sasaki:2016jop}. We note that additional scenarios may also source VHF GWs \citep{Andriot:2017oaz,Gasperini:2007vw,Kirzhnits:1972iw,Bisnovatyi-Kogan:2004cdg}. For a focused discussion, we concentrate on the source classes with spectra potentially lying within or near the observable GHz band of this study. Their predictions are quantitatively compared to our sensitivity range in the figures and sections that follow.

To theoretically model the expected observational signature from GW--photon conversion, which is jointly determined by the predicted GW spectrum and the magnetospheric transfer function, we employ a signal-template and correlation-statistics approach. The transfer function itself encapsulates the line-of-sight LOS conversion efficiency, along with relevant radio propagation effects (such as scintillation and polarization evolution) and the instrumental response, thereby mapping an incident GW spectrum to a predicted radio signal. Advancing beyond the geometric idealization of our earlier theoretical work, where the GW propagation was assumed to traverse the pulsar equatorial plane \citep{Hong:2024ofh}, we here employ particle-in-cell (PIC) simulations \citep{Buneman1965COMPUTEREI} to construct a realistic three-dimensional magnetospheric environment. In practice, we first use PIC simulations to generate static snapshots of the background plasma and magnetic field; the GW–photon conversion is then evaluated in post-processing by propagating GW rays through these snapshots along the observational LOS, rather than by simulating the GW dynamics directly within the PIC framework. This strategy aligns closely with actual radio observations, in which detected emission is linked to the polar-cap region and the specific LOS geometry, not to an idealized equatorial crossing. Moreover, because radio telescopes probe extended magnetospheric volumes within their wide fields of view, the more realistic geometry adopted here yields only minor adjustments for transient events but can substantially enhance the integrated response, and thus the sensitivity, for a stochastic GW background permeating the universe. On this realistic foundation, we develop physically motivated signal templates where appropriate and, for both transients and stochastic backgrounds, assess detectability via cross-correlation and consistency tests designed to suppress instrumental and systematic contaminants. In particular, we leverage the fact that independent pulsars and independent telescopes provide natural redundancy, enabling robust candidate validation through internal cross-checks in addition to raw sensitivity gains. This provides a route to inferring GW source properties in close analogy to how interstellar scintillation and polarization effects are used to diagnose pulsar radio signals \citep{1968Natur.218..326L}.

In addition, the above effects can be enhanced by selecting suitable radio observation methodologies. Tailored methods not only reduce the observation time \citep{Shao:2023agv} while maintaining the signal-to-noise ratio (S/R) but also may reveal novel and important phenomena \citep{Penzias:1965wn}. Building on the theoretical framework and particle-in-cell simulations of pulsar magnetospheres, our workflow proceeds through the calculation of graviton-photon conversion probabilities, the inclusion of propagation effects such as scintillation and polarization, telescope response modeling for FAST and SKA2-MID, and statistical validation using stability and Gaussianity tests. A custom-designed filtering scheme is then applied to extract weak signals and to quantify the detection thresholds for both transient events and stochastic GW backgrounds. Within this framework, we propose and evaluate four observational strategies: Single Pulsar with Single Telescope (SPST), Single Pulsar with Multiple Telescopes (SPMT), Multiple Pulsars with Single Telescope (MPST), and Multiple Pulsars with Multiple Telescopes (MPMT). These methods incorporate the differences in system noise across telescopes and exploit epoch-to-epoch changes in LOS geometry and propagation conditions to obtain decorrelated background-noise realizations \citep{2017LRR....20....2R}. We find that repeated observations under any of the proposed strategies lead to a marked enhancement of the post-filtering S/R. This improvement enables the detection of VHF GWs at substantially lower energy densities, thus alleviating one of the key challenges in this frequency band. Among them, the MPMT approach offers the highest sensitivity, as it combines cross-validation across pulsars and telescopes. This work is organized as follows. At the beginning, in Section \ref{Methods}, we discuss the details of obtaining and verifying the simulation results. Then, on this basis, in Section \ref{sec:results} we discuss four possible observation methods for improving detection sensitivity. Finally, in Section \ref{sec:discussion}, we briefly summarize and discuss our findings. This work employs the natural unit system where $c=\hbar=\epsilon_{0}=\mu_{0}=1$.

\section{Methods}
\label{Methods}
In this Section, we present a comprehensive overview of our observational methodology for detecting VHF GWs. The approach is developed in a step-by-step fashion, beginning with the foundational magnetosphere simulations of neutron stars, as detailed in subsection \ref{sec:PIC-simulation-NSM}. These simulations are carried out using PIC methods and provide a dynamic, self-consistent electromagnetic environment around pulsars, which forms the basis for our subsequent resonance analyses.

Then, building upon these simulated magnetic field structures, subsection \ref{sec:GP-resonance-SNSM} focuses on the calculation of the resonance response between GWs and the strong magnetic field of the pulsar. Crucially, we account for the pulsar’s actual magnetic inclination angle, which significantly influences the amplitude and angular dependence of the resonant response.

The resonance signals generated in the magnetosphere subsequently propagate outward and encounter the interstellar medium. In subsection \ref{sec:PE-RRS}, we evaluate the modulation effects induced by interstellar scintillation along the LOS, providing an essential correction to the expected observational signatures.

The modulated signal then reaches the observational end-point--the radio telescopes. In subsection \ref{sec:SR-RT}, we examine the reception of the signal, emphasizing its polarization characteristics, which are crucial for identifying potential VHF GW signatures and differentiating them from noise. This analysis provides the framework for the subsequent statistical examination of the signal's properties in subsection \ref{sec:Testing-SG-SOD}, where we assess both the stability and Gaussianity of the mock observational data. These diagnostics allow us to evaluate whether the filtered signal is consistent with theoretical expectations for a weak GW signature embedded in astrophysical noise.

Finally, subsection \ref{sec:BCKA-filter} introduces a custom-designed filtering algorithm tailored for the extraction of weak latent GW signals embedded in noisy observational data. This subsection also presents the filtering results applied to the simulated telescope outputs from subsection \ref{sec:Testing-SG-SOD}, including an evaluation of the S/R, which serves as a key metric for determining the detectability and robustness of the predicted resonance signals.

\subsection{PIC Simulation of a Neutron Star Magnetosphere.}
\label{sec:PIC-simulation-NSM}
Inspired by the collisionless Boltzmann equation, the PIC method provides a numerical solution of the kinetic equations with respect to the distribution function of many microscopic particles and is the main method for simulating the global dynamics of pulsar magnetospheres \citep{Buneman1965COMPUTEREI,Dawson:1983zz,Kane:1138693}. The PIC method can efficiently present the phenomena of partially dynamic plasmas in pulsars, model pulsar magnetospheres from first principles, and self-consistently compute the particle motion and radiation feedback between photons and electromagnetic fields. This is challenging for global 3D simulations because of the large difference between the dynamic scale $R_{\mathrm{LC}}=c/\Omega$ of the pulsar and the scale of the plasma skin effect $d_{e}=c/\omega_{p}$ and because the pulsar magnetosphere is filled with a plasma generated by pair cascades such as multiphoton Breit--Wheeler pair creation \citep{Breit:1934zz}. The energy distribution of the production rate of pairs by a hard photon can be given by the Ritus formulae and the process can be treated with a Monte-Carlo process similar to the nonlinear inverse Compton Scattering \citep{Duclous:2010zb,2016JPhCS.688a2058L}.

Considering the oblique Goldreich--Julian (GJ) model of pulsars \citep{1969ApJ...157..869G,1965JGR....70.4951H} and combining it with parameters of specific pulsars obtained from observations, we simulate the pulsar magnetosphere structure and the radiation in the polar cap region in a three-dimensional Cartesian coordinate system. Since sufficient observations of the physical phenomena in the pulsar magnetosphere are still lacking, especially in the region inside the light cylinder, we can only judge the quality of the simulation by comparing the observed pulsed signals with the simulated emission from the dispersed polar cap region. How properties such as the dispersion of the simulated pulsed signal are obtained is described in Section III. Considering the results from previous theoretical papers \citep{Hong:2024ofh}, the greater the magnetic field strength of the pulsar is and the closer the pulsar is to us, the more pronounced the observations are. Obviously, we cannot numerically simulate the magnetosphere of all eligible pulsars, so we consider two pulsars from our observation sources. These two pulsars are the primary sources marked in Table \ref{tab:pulsar-list} of our observation sources: PSR J1856-3754 \citep{2007ApJ...657L.101T} and PSR J0720-3125 \citep{1997AA...326..662H}. We take the astrometric information of these two pulsars and the observation results in the X-ray band \citep{2022MNRAS.516.4932D,2009A&A...500..861C,1997AA...326..662H,2003ApJ...590.1008K,2017A&A...601A.108H} as the initial conditions for the simulation and the test of the quality of the simulation results.

To maintain the flow of the main text, the detailed setup of the PIC simulations is presented in Appendix~\ref{app:pic-setup}, which including parameter inference from $(M,P,\dot{P},\alpha)$, choices of domain and resolution, Smilei configuration, radiation and pair-production modules, and diagnostic cadence. Below, we summarize only the key outputs that are necessary for the subsequent GW–photon conversion calculations and the development of observational strategies.

\begin{figure*}
	\centering
	\includegraphics[width=0.3\linewidth]{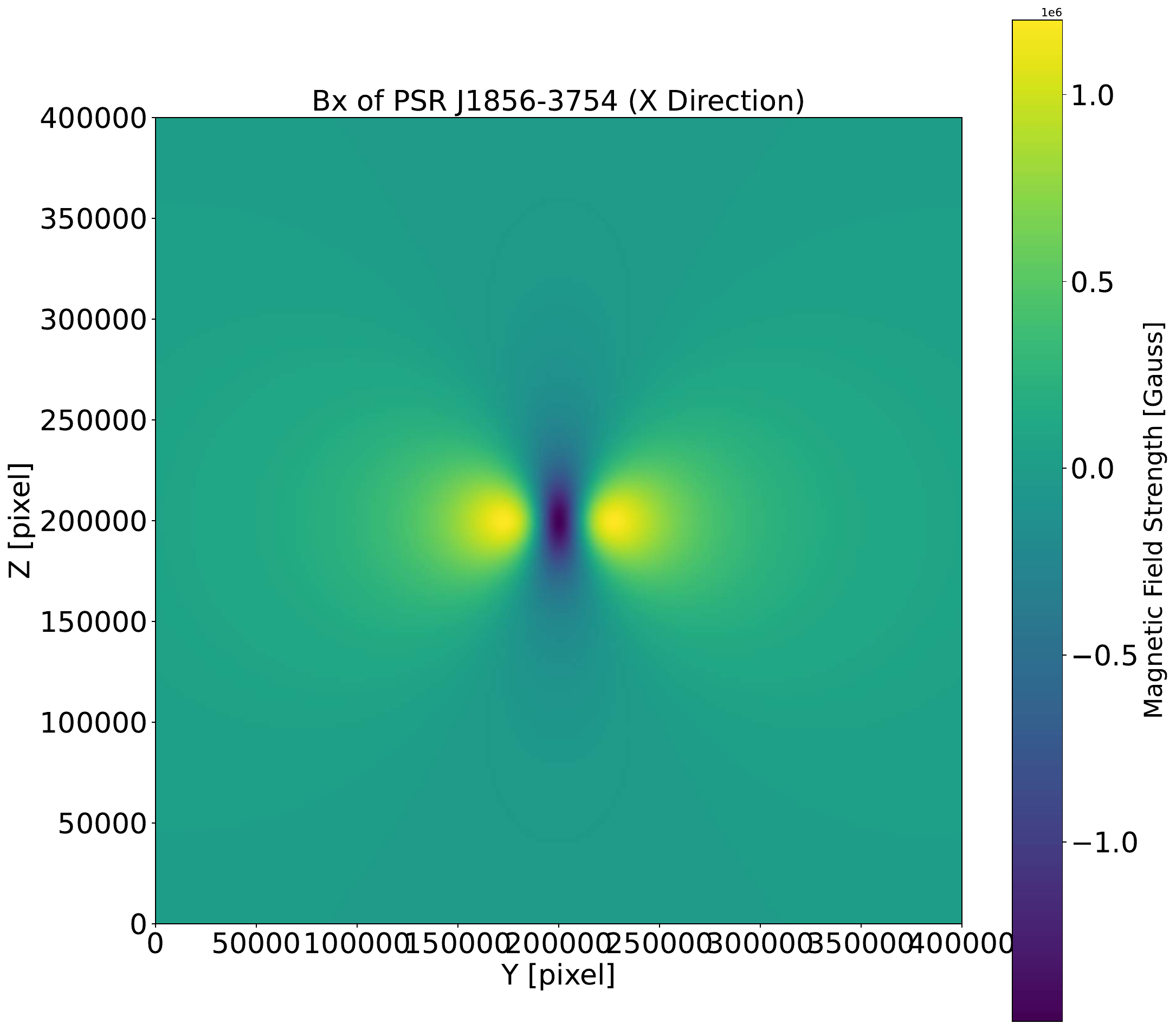}	
	\includegraphics[width=0.3\linewidth]{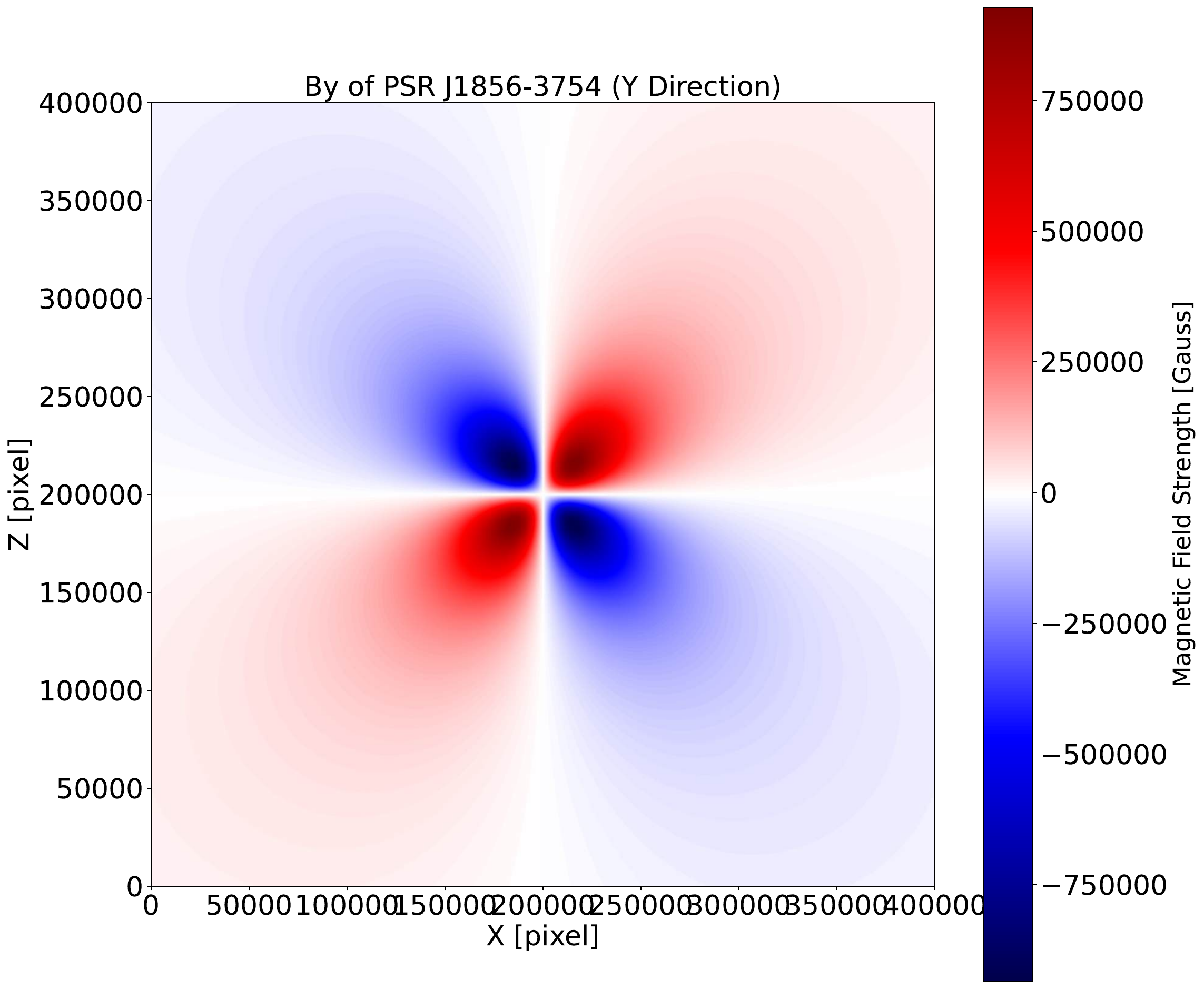}	
	\includegraphics[width=0.3\linewidth]{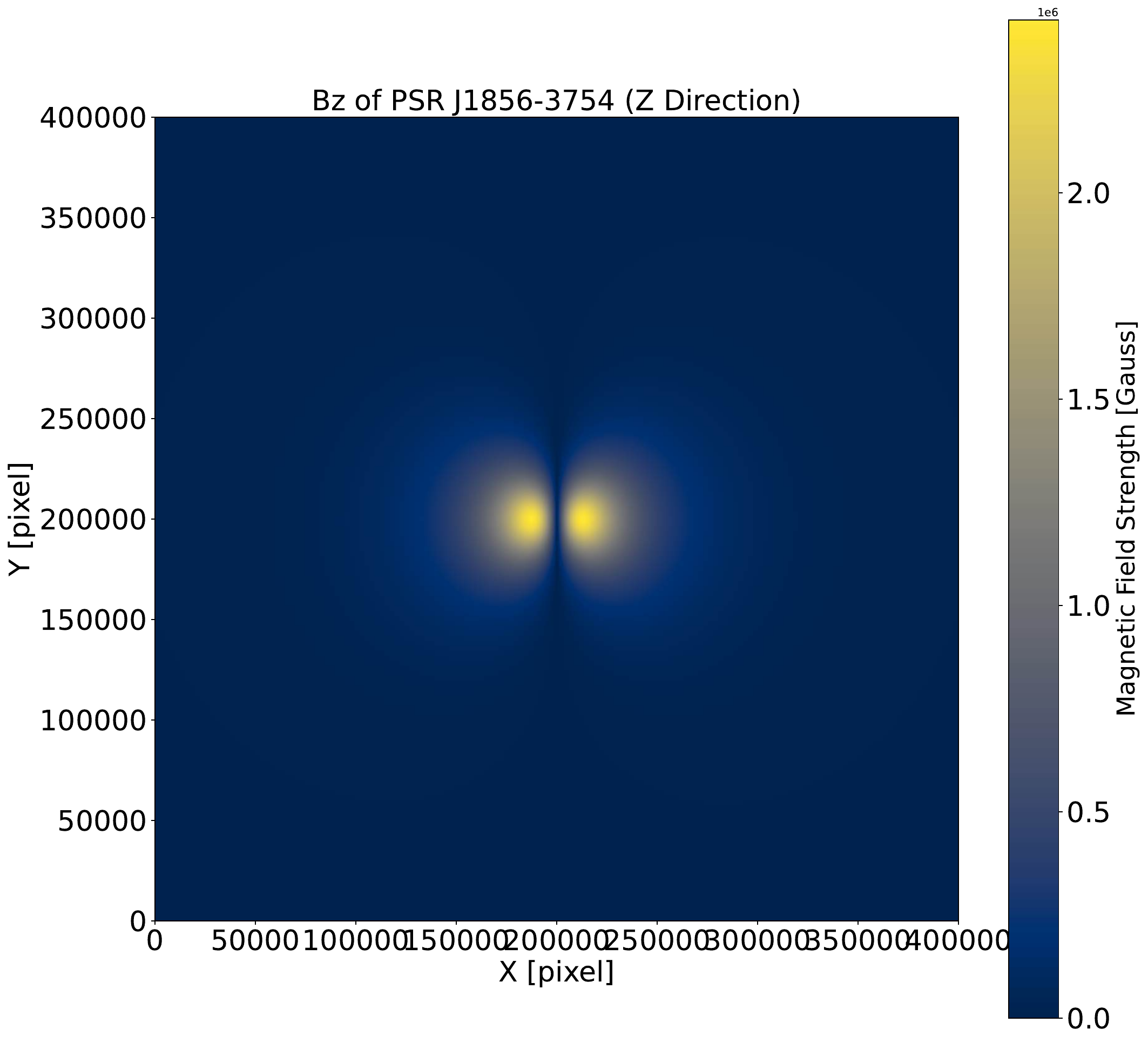}	
	\includegraphics[width=0.3\linewidth]{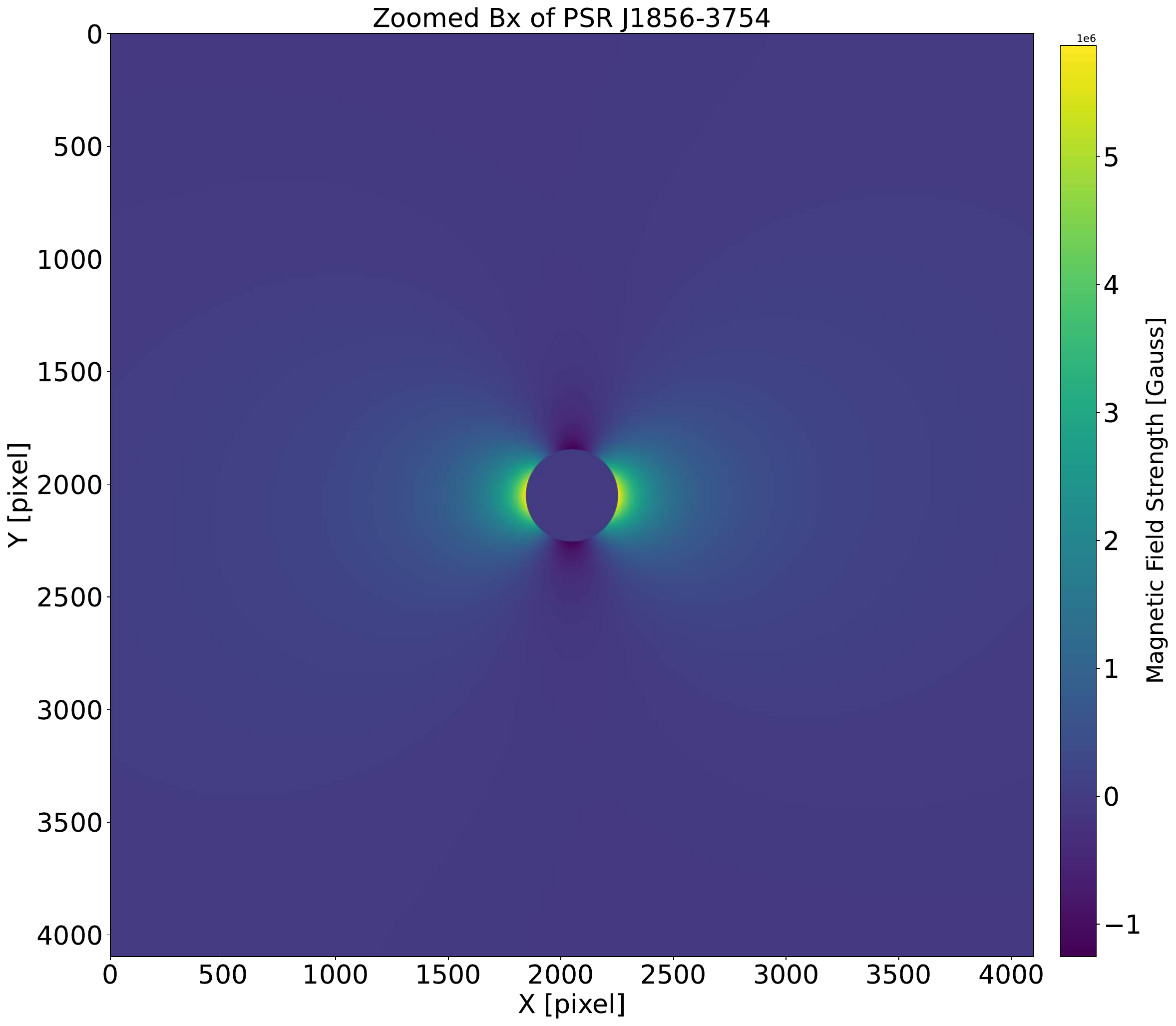}	
	\includegraphics[width=0.3\linewidth]{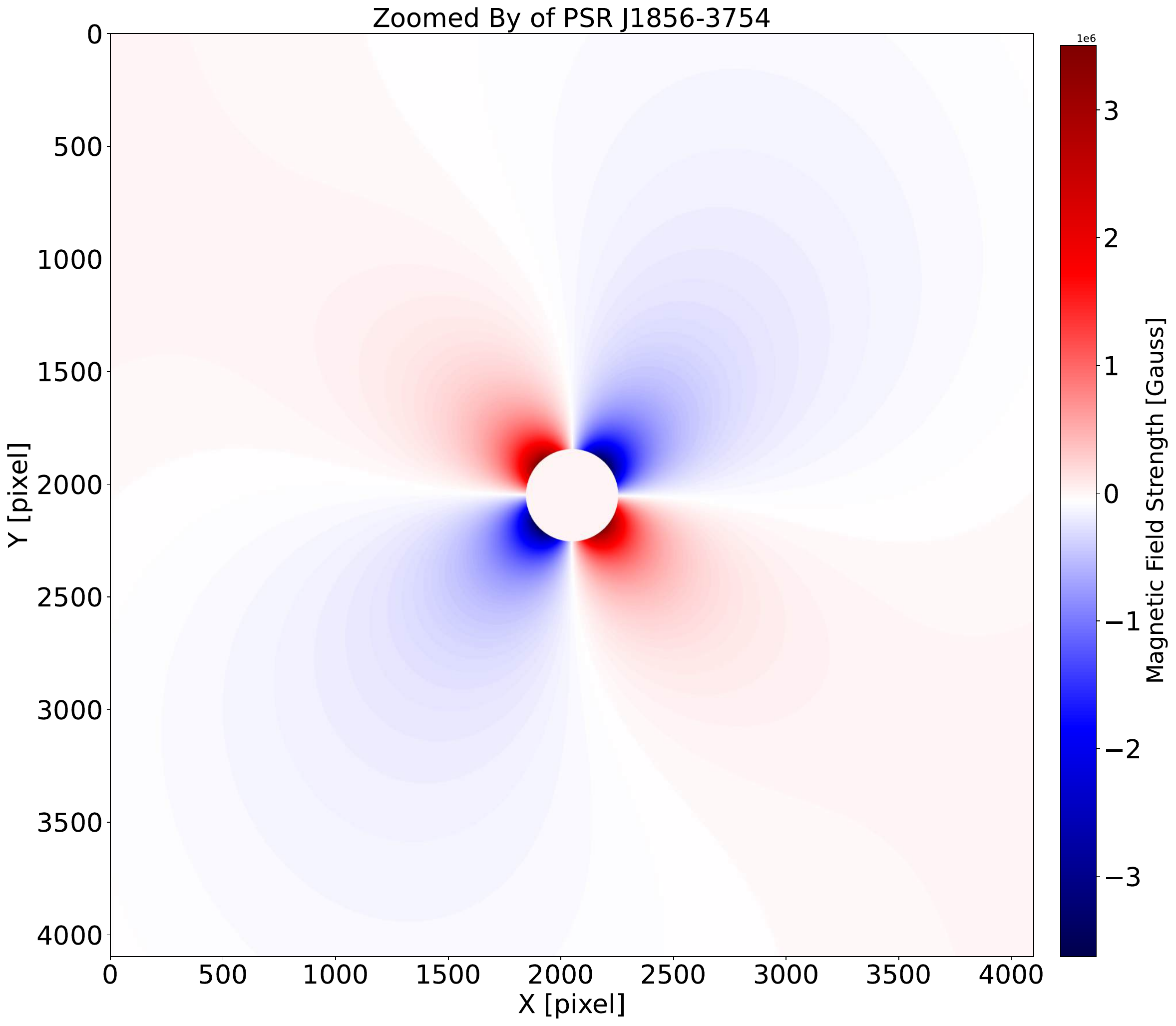}	
	\includegraphics[width=0.3\linewidth]{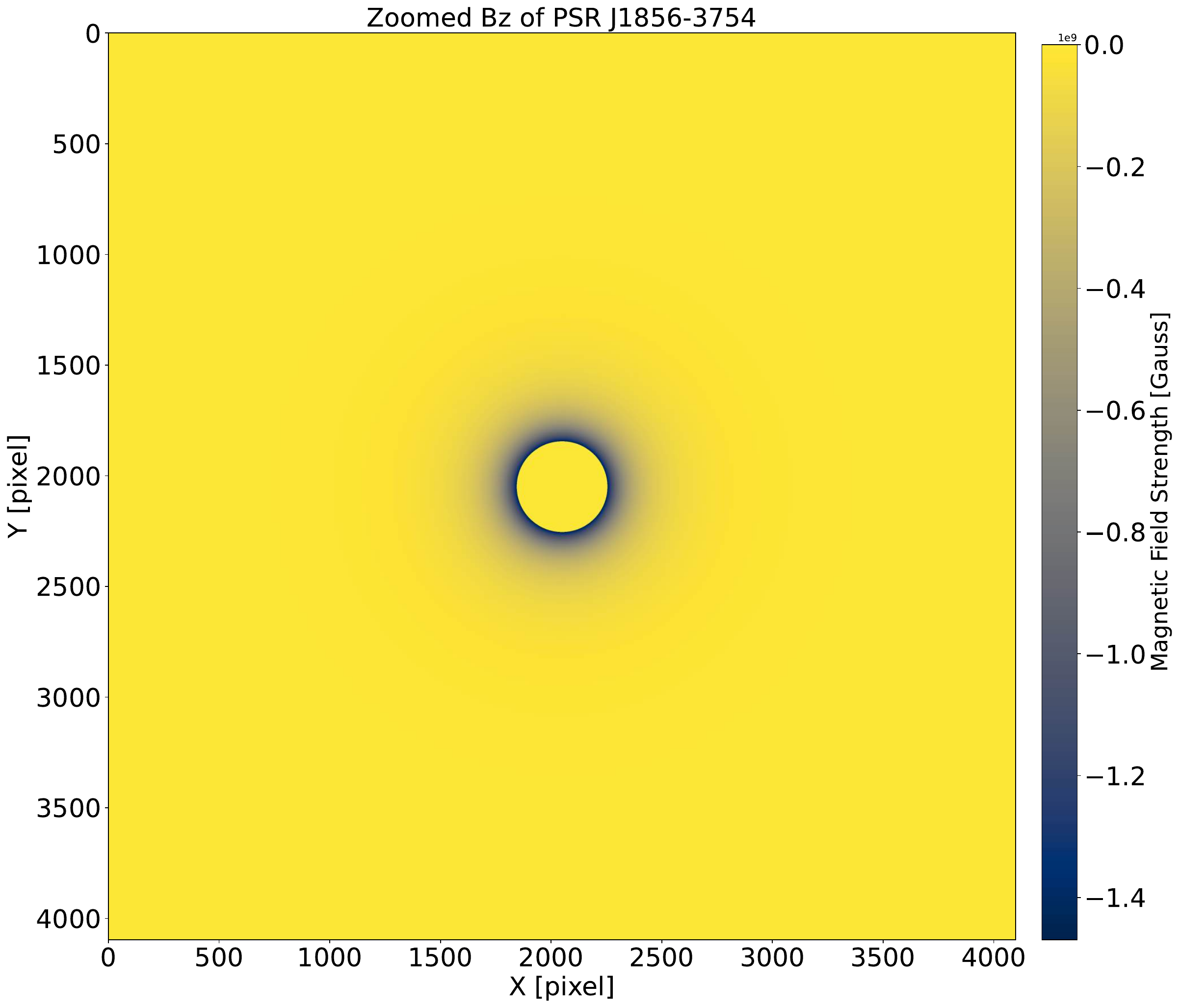}	
	\includegraphics[width=0.4\linewidth]{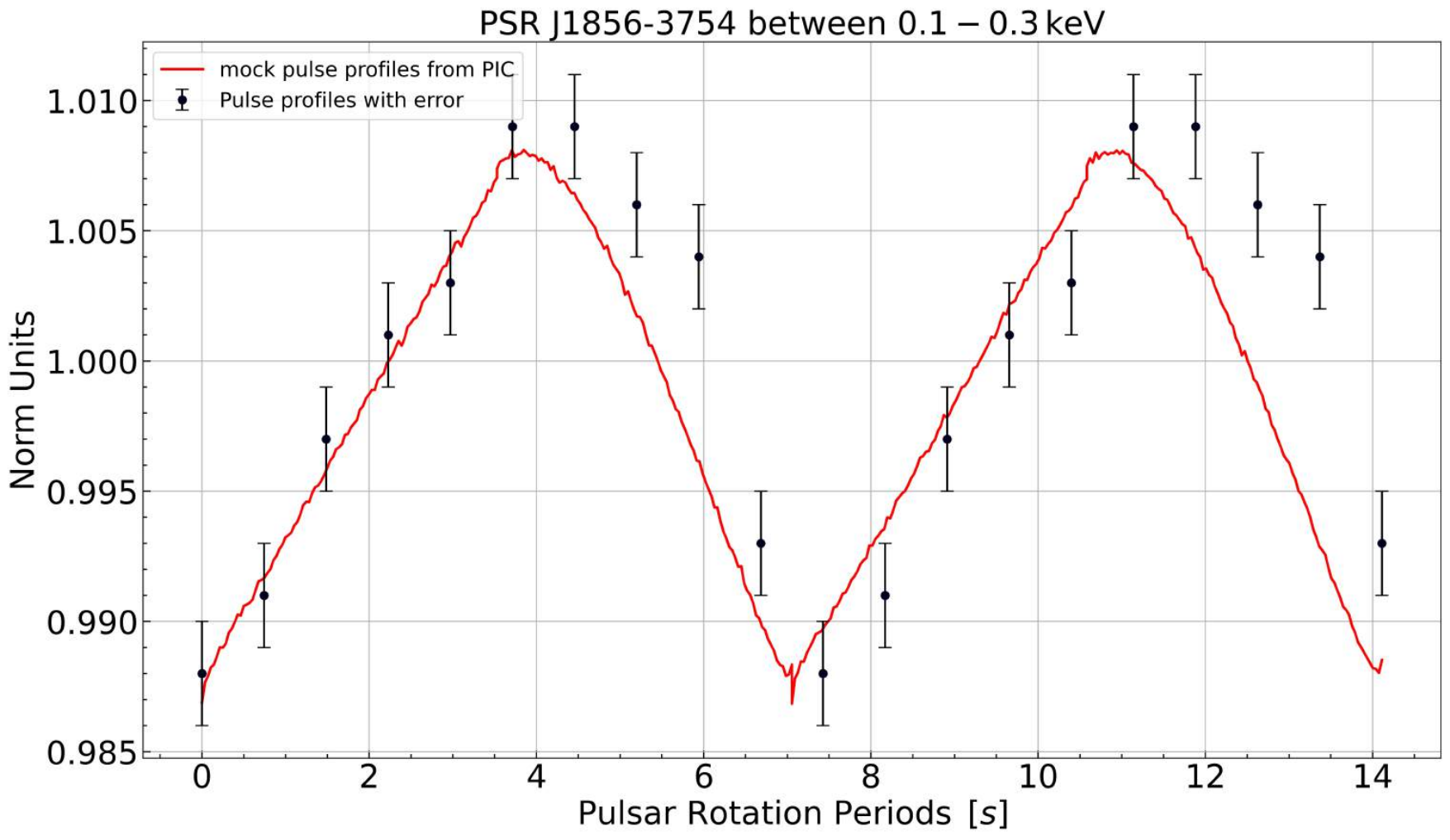}
	\includegraphics[width=0.3\linewidth]{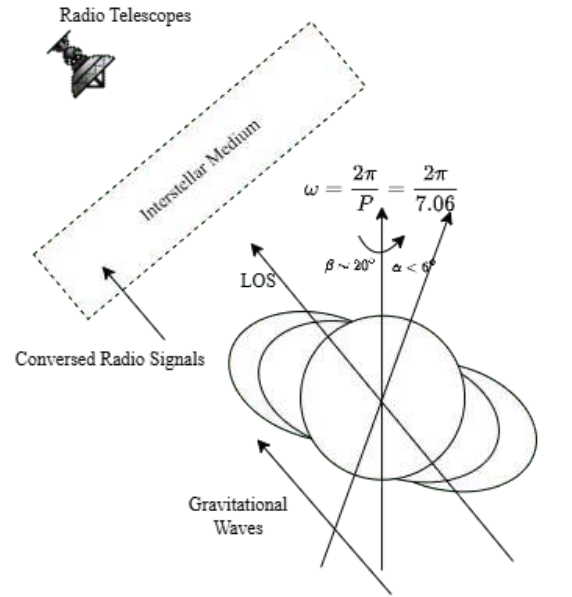}
	
	\caption{PIC simulation results of pulsar magnetospheres of PSR J1856-3754. The picture is divided into two parts. And there are several parts for each pulsar: the top panel shows a slice of the entire simulation in the plane at z = 0 for the three magnetic field directions, and the middle panel shows the electromagnetic field around the radius of the neutron star zoomed in on the top panel. Both panels show the magnetic fields in the order $Bx$, $By$, and $Bz$ in that order. The left side of the bottom panel shows the data used for the examination of the pulsar integral profile and the results of the comparison between the pulsar profile generated by our PIC simulation. In contrast, the right side shows the pulsar parameters and a schematic of the gravitational waves crossing the magnetic field in the direction of the observational LOS.}
	\label{fig:PIC-1}
\end{figure*}

\begin{figure*}
	\centering
	\includegraphics[width=0.3\linewidth]{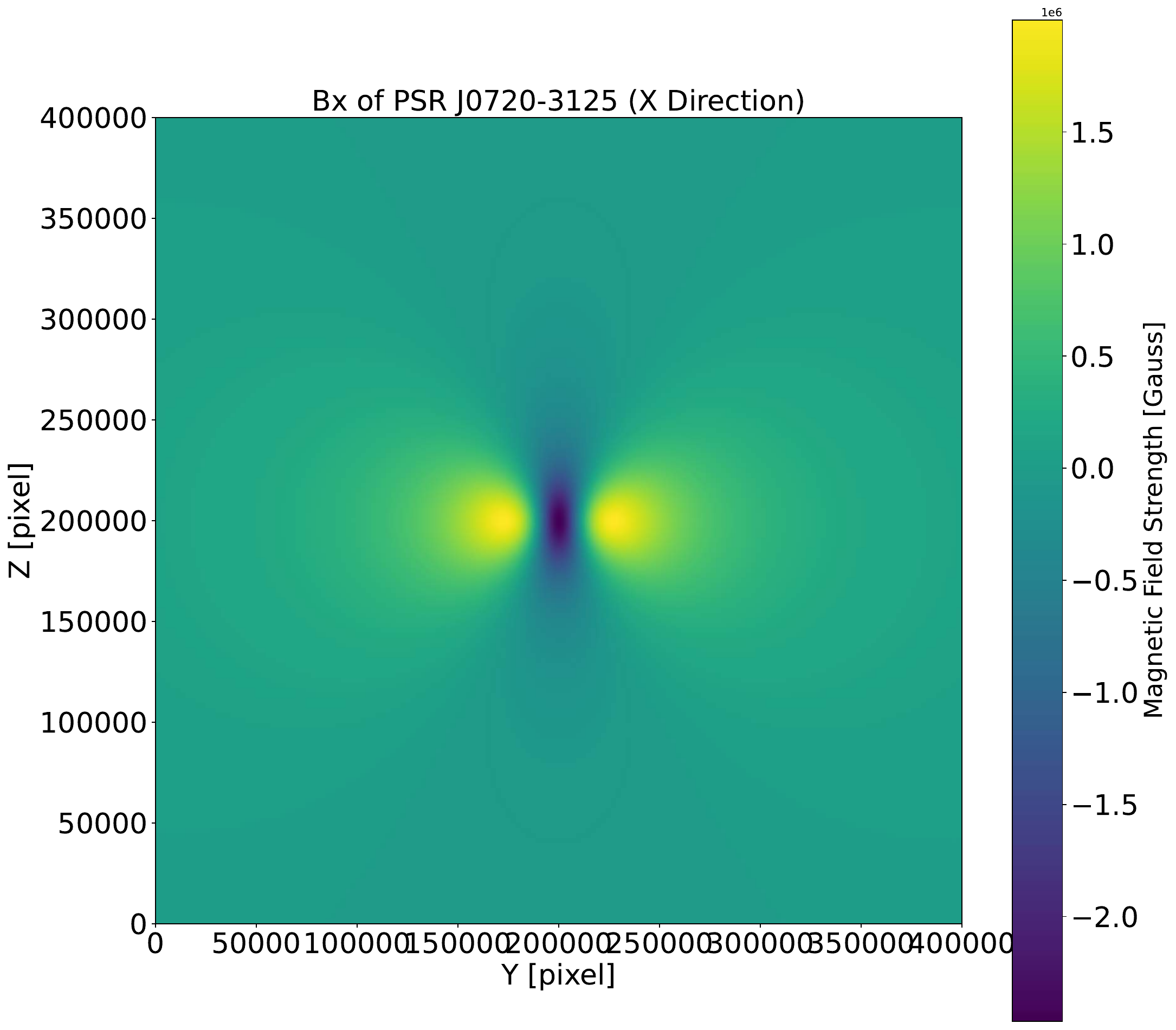}	
	\includegraphics[width=0.3\linewidth]{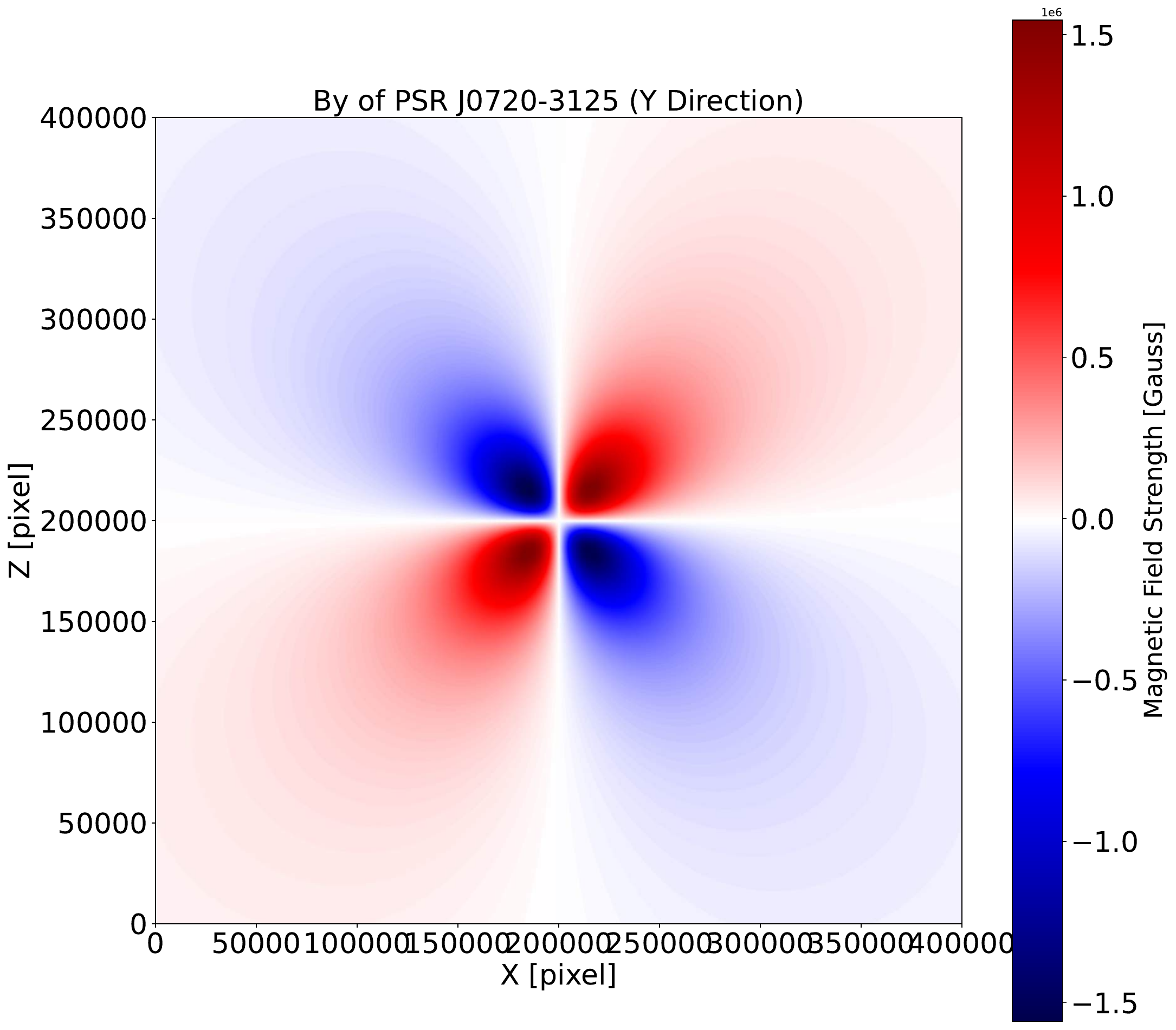}	
	\includegraphics[width=0.3\linewidth]{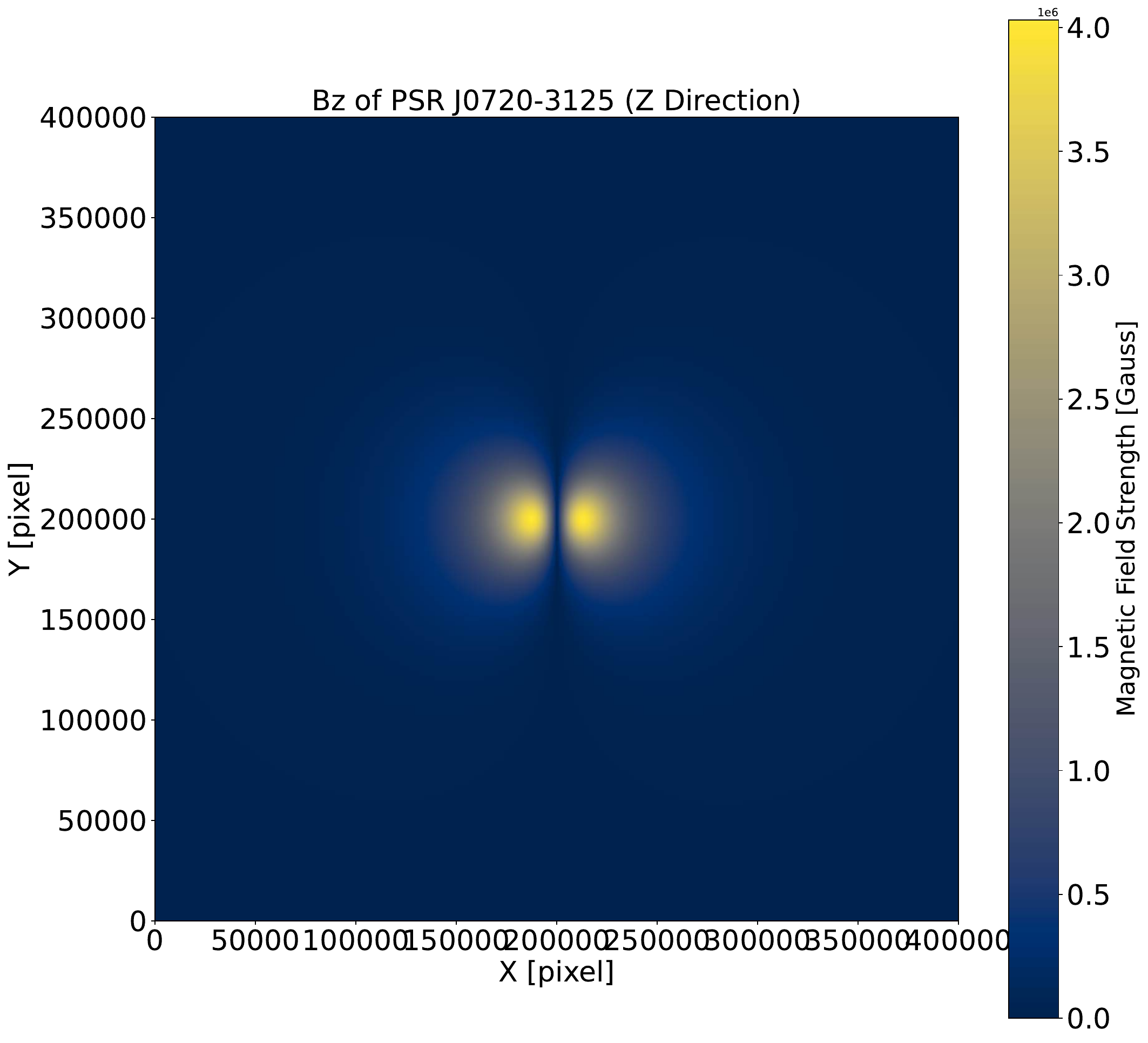}	
	\includegraphics[width=0.3\linewidth]{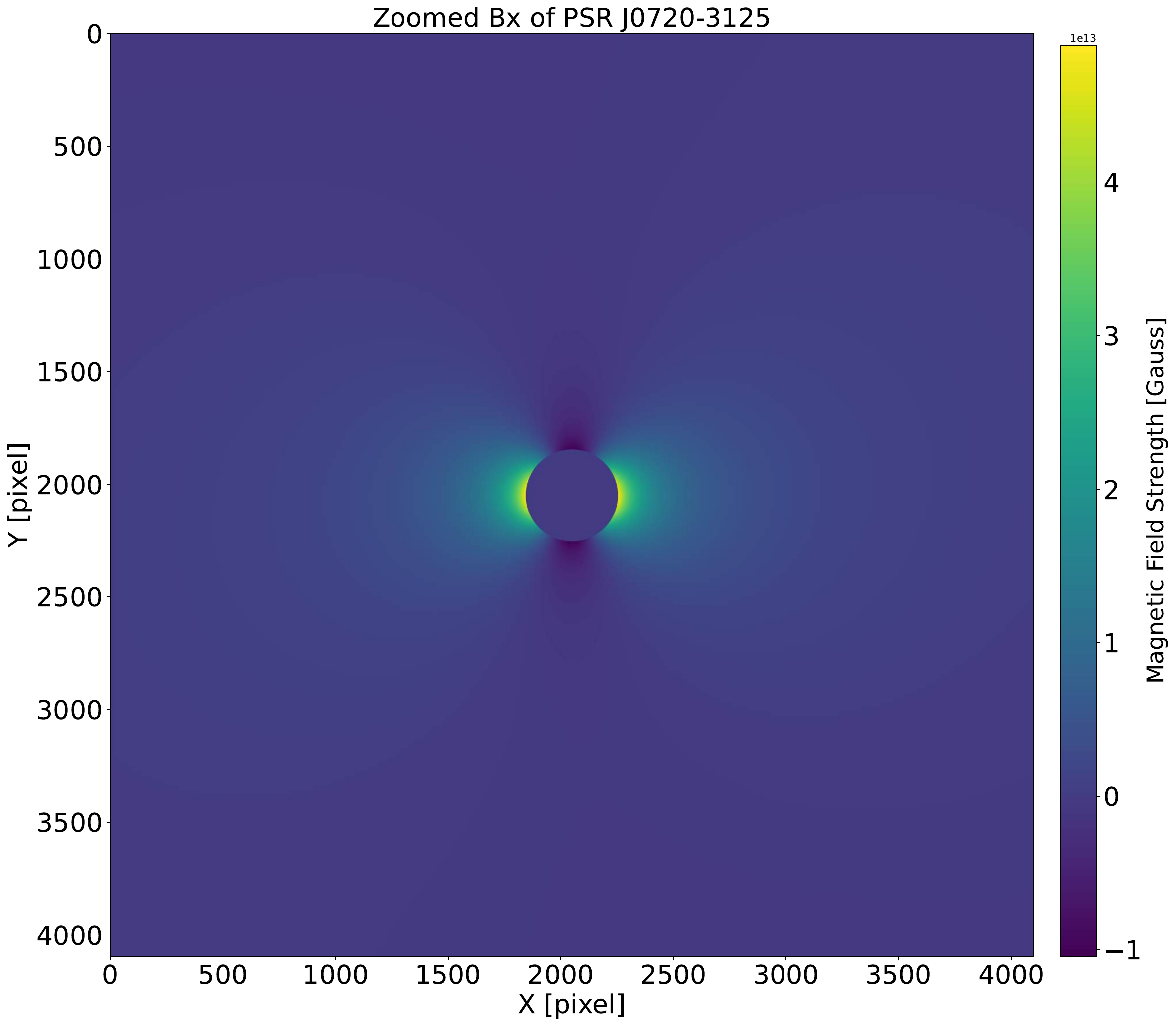}	
	\includegraphics[width=0.3\linewidth]{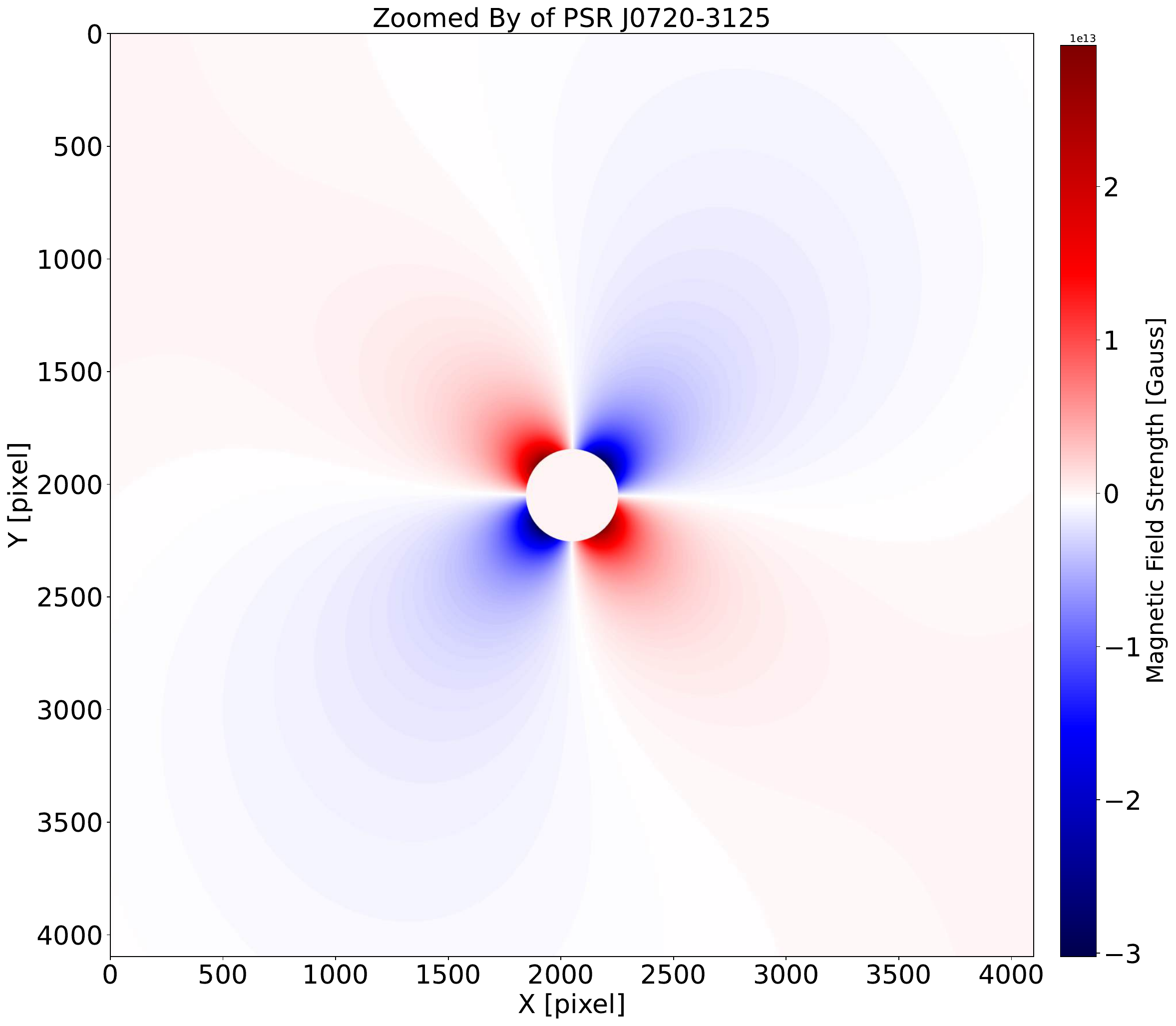}	
	\includegraphics[width=0.3\linewidth]{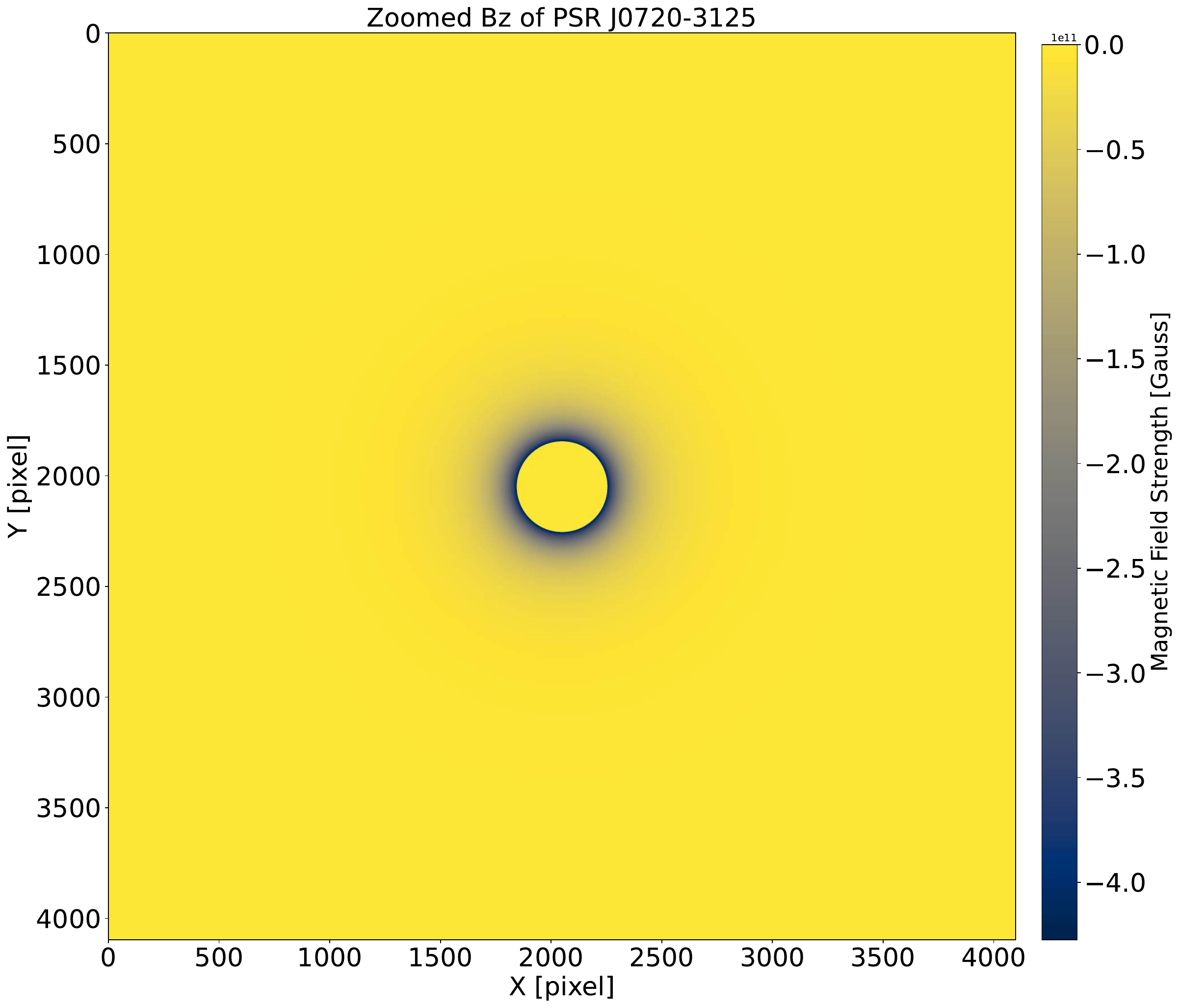}	
	\includegraphics[width=0.4\linewidth]{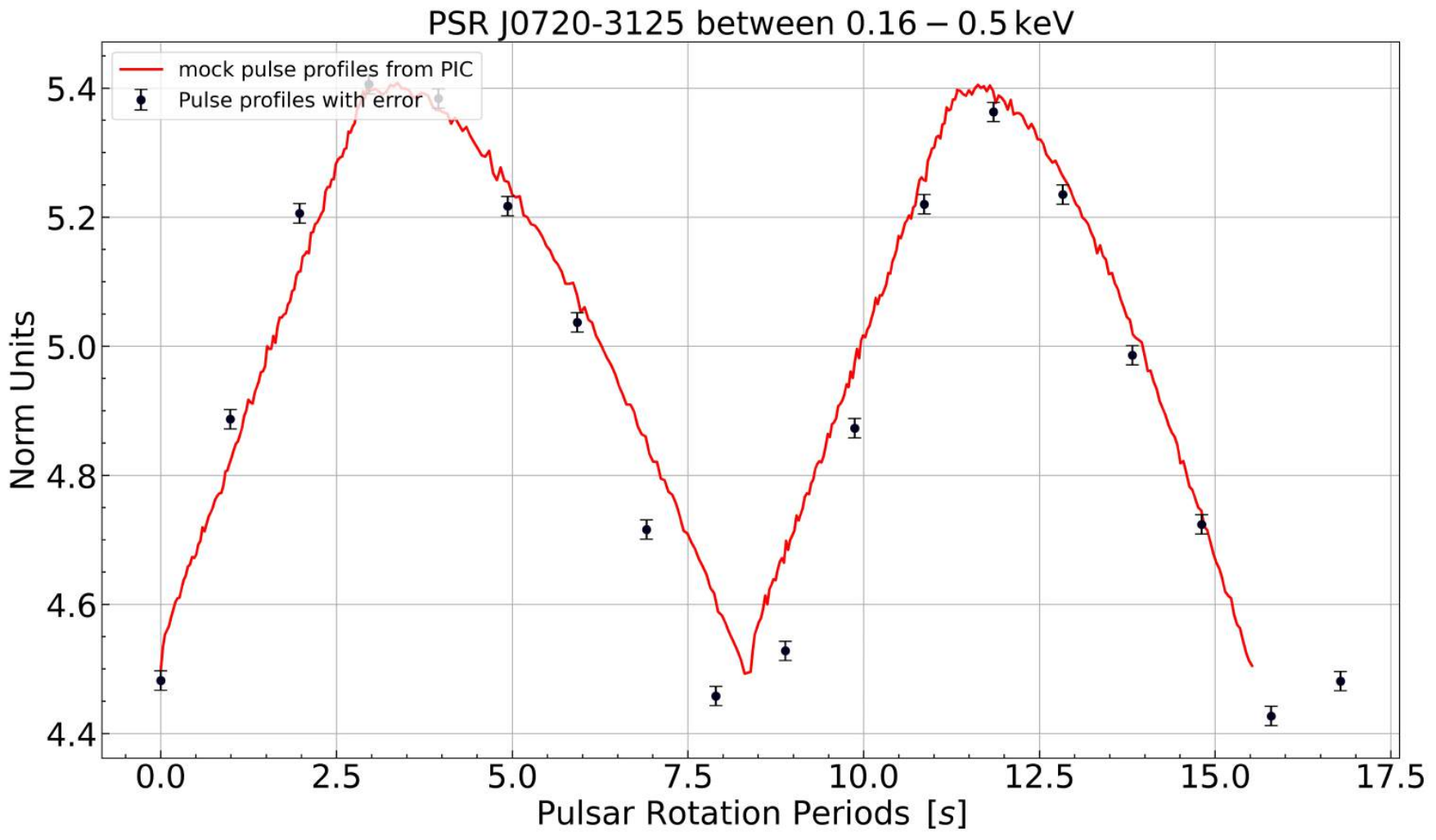}
	\includegraphics[width=0.3\linewidth]{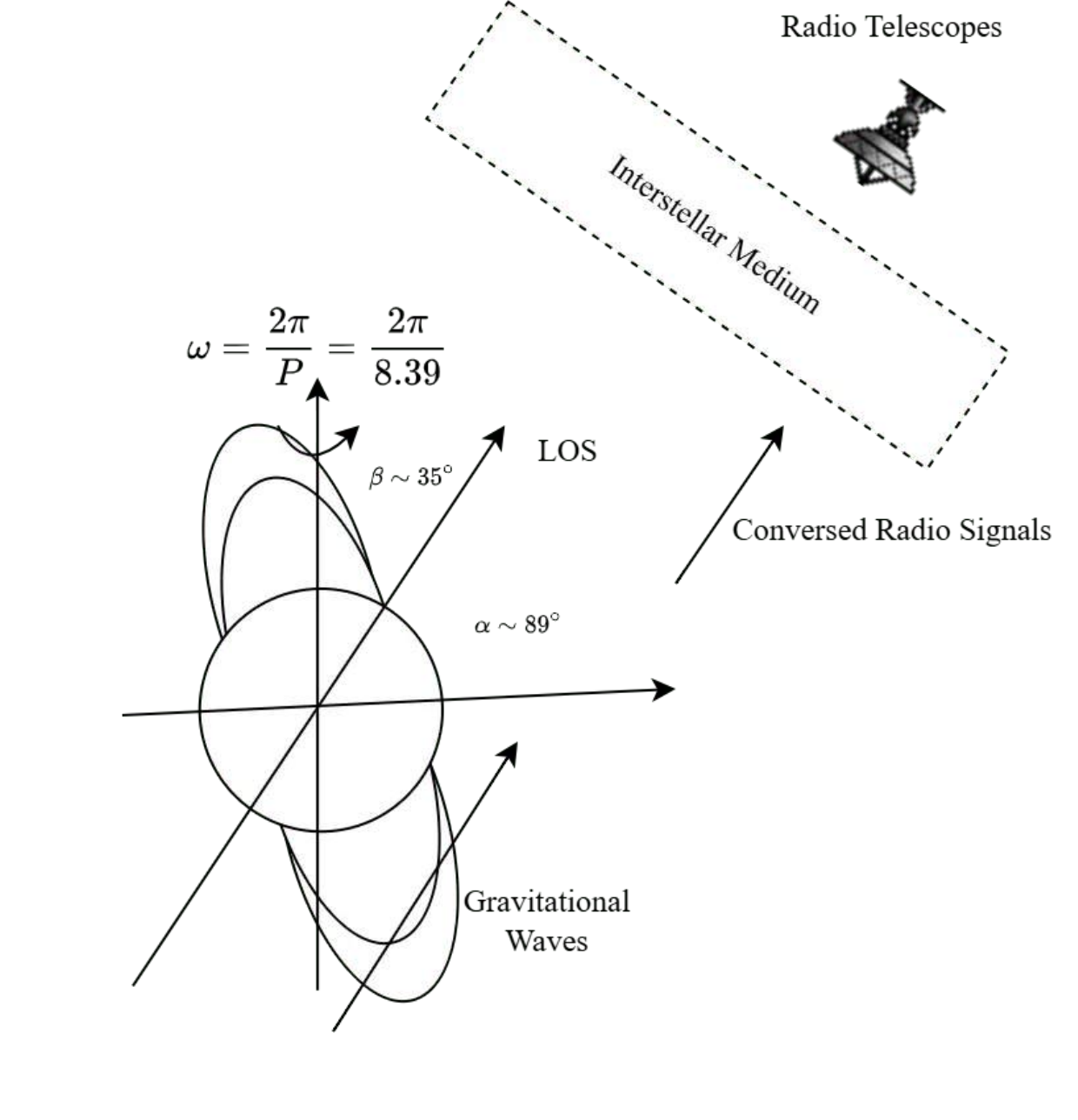}
	\caption{PIC simulation results of pulsar magnetospheres of PSR J0720-3125. This figure shows the results in the same order as Figure 1.}
	\label{fig:PIC-2}
\end{figure*}

We present the simulation and test results in Figures \ref{fig:PIC-1} and \ref{fig:PIC-2}. Since our primary focus lies in the magnetic field variations of pulsars, we present a series of simulation results illustrating both the global structure and localized magnified regions of the pulsar’s magnetic field. Each magnetic field snapshot is sampled at the moment corresponding to the second full rotation period of the pulsar. To enable a meaningful comparison with observational data of normalized pulse profile from X-ray satellites \citep{2022MNRAS.516.4932D,2009A&A...500..861C,1997AA...326..662H,2003ApJ...590.1008K,2017A&A...601A.108H}, the particle energy spectra obtained through particle binning diagnostics module are processed through a phase-resolved normalization procedure. Specifically, the pulsar’s rotational period is first mapped onto the interval $[0,1]$ to define the phase axis, which is then incrementally accumulated over time. The corresponding radiated particle spectra are subsequently normalized within each phase bin. This approach emphasizes the morphology and phase structure of the pulse profile, thereby facilitating a direct comparison between the simulated results and observational data. Finally, we illustrate schematic diagrams of radio observations for the two pulsars under study, incorporating their actual astrophysical parameters and observational parameters along the LOS. The simulation results exhibit a reasonable agreement with observational pulse profiles, particularly in terms of the overall phase structure and the relative positions of emission peaks for PSR J1856-3754 at $0.1-0.3~\mathrm{keV}$ and for PSR J0720-3125 at $0.16-0.5~\mathrm{keV}$. By sampling the magnetic field distribution at the second rotational period and performing phase-resolved normalization of the radiated particle energy spectra, we are able to capture the key morphological features observed in X-ray data. The normalized synthetic profiles successfully reproduce the broad characteristics of the observed pulses, including the relative amplitudes and phase separations of individual components. While discrepancies remain, potentially due to simplified assumptions such as the absence of (a) multipolar field structures, (b) plasma anisotropies, and (c) insufficient spatial resolution, the overall consistency between simulation and observation supports the physical validity of the modeled emission geometry and magnetospheric dynamics. These results demonstrate that the current model captures the essential features of pulsar emission, providing a solid foundation for further refinement of the simulation model and parameters and calculating the conversion probability of gravitational waves in the radio band in the magnetic field.

\subsection{Graviton-Photon Resonance in Simulated Neutron Star Magnetosphere.}
\label{sec:GP-resonance-SNSM}
Having obtained the 3D magnetic field structure from the PIC simulations, we proceed to compute the electromagnetic signals arising from graviton-photon resonance within these fields. Let's briefly summarize the calculation process \citep{Hong:2024ofh}. Starting from the Heisenberg--Euler effective action \citep{Heisenberg:1936nmg,Schwinger:1951nm} and investigate the electromagnetic response induced by the resonance between GWs and an external magnetic field.
	\begin{equation}
		S= \frac{1}{16\pi G}\int d^4 x \sqrt{-g}\left[R+L_{\mathrm{eff}}\right],\label{effect-action}
	\end{equation}
here $R$ denotes the Ricci scalar and $g$ represents the determinant of the metric $g_{\mu \nu}$. The effective Lagrangian can be written as $L_{\mathrm{eff}}=L_{\mathrm{sta}}+L_{\mathrm{ext}}$, where $L_{\mathrm{sta}}=-\frac{1}{4}F_{\mu\nu}F^{\mu\nu}$ corresponds to the standard Maxwell Lagrangian, while $L_{\mathrm{ext}}$ represents the quantum correction to the gauge field $A_\mu$, given by $\frac{im_e^4}{2\hbar^3}\int_0^{\infty}\frac{d\tau}{\tau} \mathrm{e}^{-\epsilon\tau} \mathrm{e}^{-i\tau m_e^2} \operatorname{tr}\left[\left\langle x\left|\mathrm{e}^{-i \hat{H}\tau}\right| x\right\rangle-\left\langle x\left|\mathrm{e}^{-i \hat{H}_0 \tau}\right| x\right\rangle\right]$, and represents the effective action for the gauge field $A_{\mu}$. In this expression, $\tau$ denotes the proper time, and $\epsilon>0$ is a small regulator introduced for the resummation of the effective action. The operator $\mathrm{tr}$ indicates the trace over the Dirac spinor indices. The Hamiltonian is defined as $\hat{H} = D^2 + \tfrac{1}{2} q_f F^{\mu\nu}\sigma_{\mu\nu}$, with the covariant derivative $D^\mu = \partial^\mu + i q_f A^\mu(x)$, and $\sigma_{\mu\nu} = \tfrac{i}{2}[\gamma_\mu, \gamma_\nu]$. The free Hamiltonian is denoted by $\hat{H}_0 = \partial^2$.

Within the framework of linear perturbation theory, GWs are introduced as perturbations to the background spacetime, and the Maxwell equations are expanded accordingly under a strong-field approximation \citep{Tsai:1974ap,Urrutia:1977xb,Adler:1971wn,Tsai:1974fa,Tsai:1975iz,Melrose:1976dr,Dittrich:1985yb,Dittrich:2000zu}
\begin{equation}
		\begin{aligned}
			&\left(\partial_t^2-\partial_s^2\right) h_{i j}^{\mathrm{TT}}(t, s) =16\pi G\left( F_{i\rho}F^{\rho}_{j}-\frac{\eta_{ij}}{4}F_{\alpha\beta}F^{\alpha\beta}\right), \\
			&\left(\partial_t^2-\partial_s^2+\Delta_\omega^2\right) A_i(t, s) =\delta^{k j}\delta^{l m}\left(\partial_s h_{i j}^{\mathrm{TT}}\right) \epsilon_{ksl} B_m,\label{linearized-eq}
		\end{aligned}
	\end{equation}
where TT denotes the transverse--traceless gauge. In this representation, the GW polarizations are expressed as $h_{\times} = h_{12}^{\mathrm{TT}} = h_{21}^{\mathrm{TT}}$ and $h_{+} = h_{11}^{\mathrm{TT}} = -h_{22}^{\mathrm{TT}}$ which is analogous to the definition of the components of the vector potential $A_i$.

Using the WKB approximation, we can obtain the time-dependent electromagnetic responses at different frequencies $\omega$ traveling a distance $L$ at the polar angle $\theta$ of a pulsar \citep{1926ZPhy...38..518W,1948RvMP...20..399E,1988PhRvD..37.1237R}
\begin{equation}
		\begin{aligned}
			&P_{g \rightarrow \gamma}(\omega,L,\theta) =\left|\left\langle\hat{A}_{\omega,\lambda}(L) \mid \hat{h}_{\omega,\lambda}(0)\right\rangle\right|^2\\
			&=\left|\int_{-L/2}^{L/2} d l' \frac{\sqrt{2}B_\mathrm{eff}(l',\theta)}{2M_{\mathrm{planck}}}\exp \left(-i \int_{-L/2}^{l'} d l'' \frac{-\Delta_\omega^2(l'')}{ 2 \omega}\right)\right|^2.
			\label{convertion-probability}
		\end{aligned}
	\end{equation}
We adopt the conversion formalism derived in our previous theoretical work \citep{Hong:2024ofh}, see especially Eq. (7) and Appendix C. The Euler--Lagrange equations of motion derived from the Heisenberg--Euler effective Lagrangian (\ref{effect-action}) govern the dynamics of the photon and graviton field components, $\hat{A}^\mu$ and $\hat{h}_{ij}$, propagating in the presence of an external magnetic field. And, $B_{\mathrm{eff}}$ denotes the effective magnetic field strength associated with the local magnetic field distribution. The total frequency term is given by $\Delta_\omega^2 = \omega_{\mathrm{plasma}}^2 - \omega_{\mathrm{QED}}^2$, which incorporates both plasma and QED contributions \citep{Adler:1971wn,Heisenberg:1936nmg,Itzykson:1980rh,1960ecm..book.....L}. Finally, $M_{\mathrm{planck}}$ denotes the reduced Planck mass. The resulting expressions allow us to compute the coherent energy flux and total coherent photon conversion probability for various propagation directions and polarization states. The basic theoretical framework of GWs resonating with photons in a pulsar magnetic field is the same as that in our previous theoretical papers \citep{Hong:2024ofh}. 

Because the generated electromagnetic wave propagates with a group velocity different from that of the GW in plasma, a phase mismatch accumulates along the LOS. In our calculation, this effect is already encoded in the phase factor of the WKB LOS integral through the $\Delta_\omega^2$ term, and we therefore evaluate the conversion probability directly from the LOS integral without introducing an additional multiplicative coherence-length factor. Here the PIC cell size enters only as a numerical resolution scale of the resolved LOS background profile, rather than as an independent physical coherence length. The key differences in the present study arise in the conversion probability $P_{g \rightarrow \gamma}^{\mathrm{total}}(\omega,L,\theta)=\frac{1}{\Delta\theta}\int_{\Delta \theta} P_{g \rightarrow \gamma}(\omega, L, \theta) d \theta$, the flux intensity $F_{\gamma,\omega}=\pi I_{\gamma,\omega}\left(\frac{R_{tot}}{d}\right)^2$, the signal timescale, the characteristics of the reference response profile, which are modified by considering GWs propagating through the magnetosphere along the LOS. Therefore, the total probability is defined operationally by sampling a set of rotational phases with PIC snapshots $P_{g \rightarrow \gamma}^{\mathrm{total}}(\omega,L,\theta)\simeq\frac{1}{N_\theta}\sum_{i=1}^{N_\theta}P_{g\rightarrow\gamma}(\omega,L,\theta_i(t_i))$. Here $t_i=t_0+i \Delta t$ labels the discrete simulation outputs used to represent different rotational phases $\phi_i=2 \pi i / N_t$. For each snapshot, we trace $N_\theta$ LOS rays specified by $\theta_i$ through the magnetosphere. Finally, the specific intensity of the photon signal produced in the magnetic field,  $I_{\gamma,\omega}$, can be written as
	\begin{equation}
		\begin{aligned}
			I_{\gamma,\omega} &= \frac{dE_\omega}{dt\, dA\, d\Omega\, d\omega} \\
			&= \frac{3 H_0^2 M_{\mathrm{planck}}^2}{4 \pi \omega}\,
			\Omega_{\mathrm{GW}}(\omega)
			P_{g \rightarrow \gamma}^{\mathrm{total}}(\omega,L,\theta),
			\label{specific-intensity}
		\end{aligned}
	\end{equation}
where $H_0$ denotes the Hubble constant \citep{2020AA...641A...6P}. This expression indicates that the photon intensity is proportional to the GW energy density spectrum $\Omega_{\mathrm{GW}}(\omega)$, weighted by the total conversion probability $P_{g \rightarrow \gamma}^{\mathrm{total}}$. In this subsection, we discuss only the case in which the signal has just penetrated the boundary of the pulsar magnetic field, while the results of its travel through the interstellar medium will be discussed in detail in the next subsection.

Using the magnetic field of the pulsar in the 3D Cartesian coordinate system obtained in the PIC simulation and considering the GWs moving along the observation LOS direction, the magnetic field affecting the conversion probability of the GWs can be regarded as the field of a GW undergoing a $SO(3)$ group rotation at the corresponding observation angle as the GW passes perpendicularly through the equatorial plane of the pulsar. In our previous study, we identified two types of temporal signals resulting from the conversion of GWs into electromagnetic waves \citep{Hong:2024ofh}. By accounting for the confusion effect \citep{1957AZh....34..349V,1971R&QE...14..425I,2004P&SS...52.1357L,2025arXiv250111872W}, we can calculate the most optimistic minimum characteristic scales of the telescopes for these two types of signals. For radio-observed transient events, the sensitivity of the telescope, considering the region of occurrence closest to the pulsar and the superposition of the coherent states of the electromagnetic waves, is
\begin{equation}
	\begin{aligned}
		h_c(\omega)=&\sqrt{\frac{3H_0^2\Omega_{\mathrm{GW}}(\omega)}{2\pi^2\omega^2}}\\=&\sqrt{\frac{2H_0^2d^2 F_i}{\pi^2\omega R_{tot}^2M_{\mathrm{planck}}^2P_{g \rightarrow \gamma}^{\mathrm{total}}}}\\
		= &2.26 \times 10^{-17}\left(\frac{d}{1 \mathrm{~kpc}}\right)\left(\frac{10^2 \mathrm{~km}}{R_{orc}}\right)\left(\frac{10^6 \mathrm{~Hz}}{\omega}\right)^{1/2}\\
		&\times\left(\frac{10^{-5}}{P_{g \rightarrow \gamma}^{\mathrm{total}}(\omega,L,\theta)}\right)^{1/2}\left(\frac{F_i}{1 \mathrm{~Jy}}\right)^{1/2},
	\end{aligned}
\end{equation}
where, we define $R_{\mathrm{orc}}\equiv r_{\mathrm{occur}}$ as the inner radial boundary of the effective conversion region \citep{Hong:2024ofh}
\begin{equation}
	\begin{aligned}
		r_{\mathrm{occur}}= & 2.24 \times 10^4\left|3 \cos \theta \boldsymbol{m} \cdot \boldsymbol{r}-\cos \theta_m\right|^{1 / 3} \\
		& \times\left(\frac{r_0}{10 \mathrm{~km}}\right)\left(\frac{B_0}{10^{14} \mathrm{Gauss}}\right)^{1 / 3} \\
		& \times\left(\frac{1 \mathrm{~s}}{P}\right)^{1 / 3}\left(\frac{10^6 \mathrm{~Hz}}{\omega}\right)^{2 / 3} \mathrm{~km},
	\end{aligned}
\end{equation}
when the estimated $r_{\mathrm{occur}}<r_0$ we set $R_{\mathrm{orc}}=r_0$. And $F_i$ denotes the measured energy flux at an S/R of $i \sigma$. We consider the two standard levels $i=1$ and $i=5$, corresponding to the $1 \sigma$ and $5 \sigma$ confidence limits, and present results for both cases throughout the work. For a single-dish telescope such as FAST, the S/R is defined as the ratio of the flux $F$ of the radio source to the measurement error $\Delta F$: 
	\begin{equation}
		\mathrm{S/R} = \frac{F}{\Delta F} 
		= \frac{F \sqrt{n_\mathrm{pol}\,\Delta\nu_{\rm eff}\,\Delta t_{\rm eff}}}{\mathrm{SEFD}},
		\label{snr-single}
	\end{equation}	
where $n_{\mathrm{pol}}$ is the number of polarization channels, $\Delta\nu_{\rm eff}$ is the effective frequency bandwidth of the observation, and $\Delta t_{\rm eff}$ is the effective integration time. For a persistent signal observed over a dwell time $\Delta t_{\rm int}$, we take $\Delta\nu_{\rm eff}=\Delta\nu$ and $\Delta t_{\rm eff}=\Delta t_{\rm int}$. The latter is defined as $\Delta t_{\mathrm{int}} = t_{\mathrm{sur}} \tau$, where $t_{\mathrm{sur}}$ is the total survey duration and $\tau = \lambda_\gamma / 2\pi D$ gives the fractional transit time of a source on the celestial equator across the telescope’s field of view. SEFD denotes the system-equivalent flux density \citep{Yu:2013bia}. For a transient event, $\Delta t_{\rm eff}=\min\!\left(\Delta t_{\rm int},\,\tau_{\rm in\mbox{-}band},\,\tau_{\rm smear},\,\Delta t_{\rm res}\right)$, where $\tau_{\rm in\mbox{-}band}$ is the intrinsic time the signal remains in the analyzed band, $\tau_{\rm smear}$ captures propagation and instrumental broadening, and $\Delta t_{\rm res}$ is the time resolution of the data product.The effective bandwidth $\Delta\nu_{\rm eff}$ depends on how the signal occupies frequency. For a narrowband signal with intrinsic width $\Delta f_{\rm sig}\ll \Delta\nu$, $\Delta\nu_{\rm eff}=\min\!\left(\Delta\nu,\,\max(\Delta f_{\rm sig},\,\Delta f_{\rm chan})\right)$, where $\Delta f_{\rm chan}$ is the channel width. A “quality factor” for the signal can be defined as $Q_{\rm sig}\equiv f/\Delta f_{\rm sig}$, giving $\Delta\nu_{\rm eff}\sim f/Q_{\rm sig}$ for a narrowband case. For chirping transients, a minimal intrinsic width is set by $\Delta f_{\rm sig}\sim \dot f\,\Delta t_{\rm eff}$, which guides our choice of $\Delta\nu_{\rm eff}$ in simulations and filtering \citep{2004hpa..book.....L,2016era..book.....C,2017isra.book.....T}.

For a radio interferometer such as SKA, the signal is treated as a collection of synthesized beams. We assume that the observed signal follows a Gaussian distribution, that different beams are statistically independent, and that only one type of GW is present. Under these assumptions, the likelihood function for the observed energy fluxes $F_{\mathrm{obs},i}$ is
\begin{equation}
	\begin{aligned}
	\tilde{\mathcal{L}}(\boldsymbol{\theta}_{\mathrm{set}}) 
	&= P\!\left(F_{\mathrm{obs},i} \mid \boldsymbol{\theta}_{\mathrm{set}}\right) \\
	&= \left(\prod_i^{N_{\mathrm{syn}}} \frac{1}{\sqrt{2\pi\sigma_i^2}} \right) 
	\exp\!\left(-\frac{\chi^2}{2}\right),
		\label{likelihood}
	\end{aligned}
\end{equation}
where $\boldsymbol{\theta}_{\mathrm{set}}$ denotes the model parameters, including the GW power spectral density $S_h(\omega)$, the energy density $\Omega_{\mathrm{GW}}$, and the characteristic strain $h_c$. The corresponding $\chi^2$ statistic is defined as
\begin{equation}
	\chi^2 = \sum_i^{N_{\mathrm{syn}}} 
	\frac{\left[F_{\mathrm{model},i}(\boldsymbol{\theta}_{\mathrm{set}}) - F_{\mathrm{sys},i} - F_{\mathrm{obs},i}\right]^2}{\sigma_i^2},
\end{equation}
where $F_{\mathrm{sys},i}$ accounts for the predicted background flux density of the $i$-th synthesized beam, including both SEFD and astrophysical foreground contributions. The single-dish error $\Delta F$ is used as an estimate of the measurement error $\sigma_i$. Finally, the S/R for an interferometer can be written as \citep{Hook:2018iia}
\begin{equation}
	\mathrm{S/R} = 
	\sqrt{ n_{\mathrm{pol}} \, \mathrm{Gain}_{\mathrm{array}}^2 
		\sum_{i=1}^{N_{\mathrm{syn}}} 
		\frac{F_{\mathrm{model},i}^2 \, \Delta\nu_i \, \Delta t_{\mathrm{int},i}}
		{T_{\mathrm{sys},i}^2} },
\end{equation}
where $\mathrm{Gain}_{\mathrm{array}} = \mathrm{Gain} \sqrt{N(N-1)}$ represents the array gain determined by the individual antenna performance and the total number of elements $N$. Each synthesized beam is characterized by its flux $F_{\mathrm{model},i}$, integration time $\Delta t_{\mathrm{int},i}$, bandwidth $\Delta\nu_i$, and system temperature $T_{\mathrm{sys},i}$.

Because the GWs propagates through the magnetosphere along the LOS, the observed time‑series response is shaped by two factors: the intrinsic time the GW signal remains within the analyzed frequency band, and the broadening introduced by the magnetospheric transfer response along with other instrumental or propagation effects encoded in our reference profiles.
For a PBH‑like inspiral whose GW frequency reaches the GHz band, the intrinsic in‑band chirp duration can be extremely short. At leading post‑Newtonian order, the time to coalescence from a GW frequency $\omega$ is $t(\omega)\simeq \frac{5}{256}\left(\frac{G\mathcal{M}_c}{c^3}\right)^{-5/3}(\pi \omega)^{-8/3}$,
and the frequency sweep satisfies $\dot \omega=\frac{96}{5}\pi^{8/3}\left(\frac{G\mathcal{M}_c}{c^3}\right)^{5/3}\omega^{11/3}$, where $\mathcal{M}_c$ is the chirp mass. For an equal-mass binary whose ISCO frequency is about 3 GHz -corresponding to a total mass $M \sim 1.5 \times 10^{-6} M_{\odot}$-the intrinsic time spent in the $1-3 \mathrm{GHz}$ band is of order $10^{-8}\mathrm{s}$. Consequently, in our simulated time series the effective burst width that governs detectability is usually set by the instrumental time resolution and by the adopted broadening or transfer response, such as that from scintillation, scattering, or the magnetospheric response embedded in the reference profiles, rather than by the intrinsic GHz-band chirp time itself \citep{1973grav.book.....M,1980RvMP...52..299T,1994PhRvD..49.2658C,2007gwte.book.....M,2009LRR....12....2S}.

For radio-observed persistent events, the sensitivity of the telescope, considering the conversion probability throughout the pulsar region of occurrence and the superposition of the coherent states of electromagnetic waves as the telescope continues to observe the pulsar magnetosphere region, is
\begin{equation}
	\begin{aligned}
		h_c(\omega)= &2.26 \times 10^{-24}\left(\frac{d}{1 \mathrm{~kpc}}\right)\left(\frac{10^6 \mathrm{~km}}{R_{tot}}\right)\left(\frac{10^6 \mathrm{~Hz}}{\omega}\right)^{1/2}\\
		&\times \left(\frac{10^{-5}}{P_{g \rightarrow \gamma}^{\mathrm{total}}(\omega,L,\theta)}\right)^{1/2}\left(\frac{F_i}{1 \mathrm{~Jy}}\right)^{1/2},
	\end{aligned}
\end{equation}
where we define $R_{\rm tot}$ as the effective outer scale of the conversion region that remains dominated by the target magnetosphere and is encompassed by the telescope FoV. The detection capabilities are strongly influenced by the spatial extent of coherent magnetic fields. In our analysis, we adopt $R_{\mathrm{tot}} = 10^6\mathrm{~km}$ as an optimistic yet practical estimate, since this scale lies within the telescope’s field of view and the magnetic field strength remains above $47\,\mathrm{pG}$. Optimistic scenarios assume magnetic regions constrained by the telescope beam width and by typical pulsar magnetosphere dimensions, where the field strength exceeds that of the interstellar background. In more restrictive situations, the interaction volume may be reduced. For example, when the gravitational wave traverses only a fraction of the magnetosphere, which leads to a lower conversion probability and reduced signal amplitude. Additional limiting factors include cases where the converted signal is observable only along a narrow LOS or when the angular size of the GW wavefront cannot be approximated as infinite, for example, for nearby compact binaries. Thus, the results presented here provide a broad perspective on detectability under both idealized and realistic conditions. It should also be noted that actual observations may deviate significantly due to two main effects: (a) the absence of a sharply defined boundary for the neutron star’s magnetic field, particularly under the influence of its environment, and (b) the possibility that radio-band GWs are confined to localized subregions, implying that the effective electromagnetic emission zone must ultimately be determined observationally.

As before, we can simulate the two radio signals generated by conversion in the pulsar magnetic fields of our two primary targets under FAST \citep{Jiang:2019rnj,2020RAA....20...64J,fastref} and SKA2-MID \citep{2019arXiv191212699B,skaref} observation conditions \citep{Hong:2024ofh}. For a transient event, we choose the GW generated by the primordial black holes merging as the signal source and assume that its characteristic scale changes with frequency, that is, with the mass of the binary black hole \citep{2017LRR....20....2R}. For persistent events, we choose a GW generated by the cosmological background as the signal source and assume that its characteristic scale changes with frequency \citep{2017LRR....20....2R}. 

\begin{figure*}
	\centering
	\includegraphics[width=0.42\linewidth]{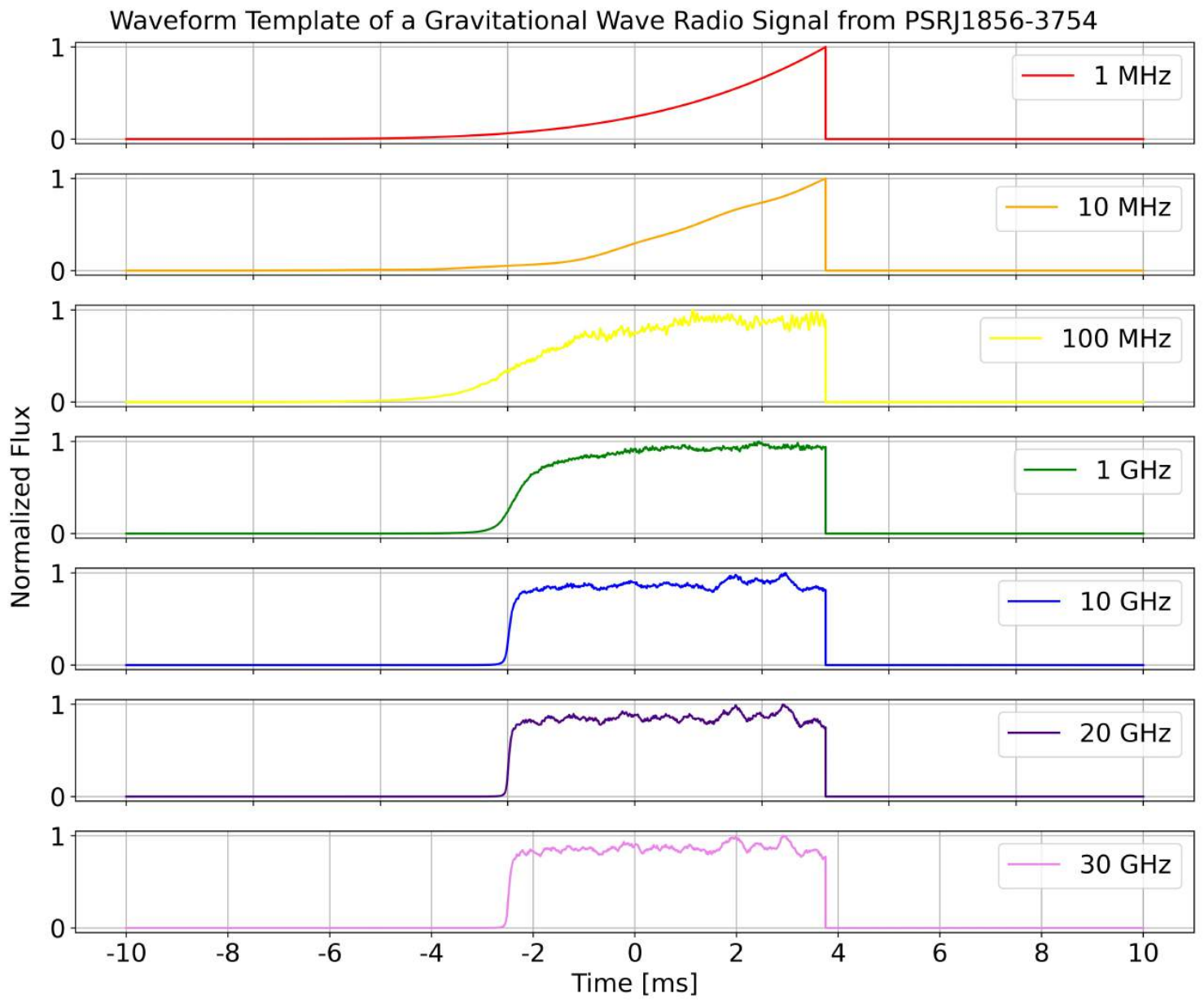}
	\includegraphics[width=0.42\linewidth]{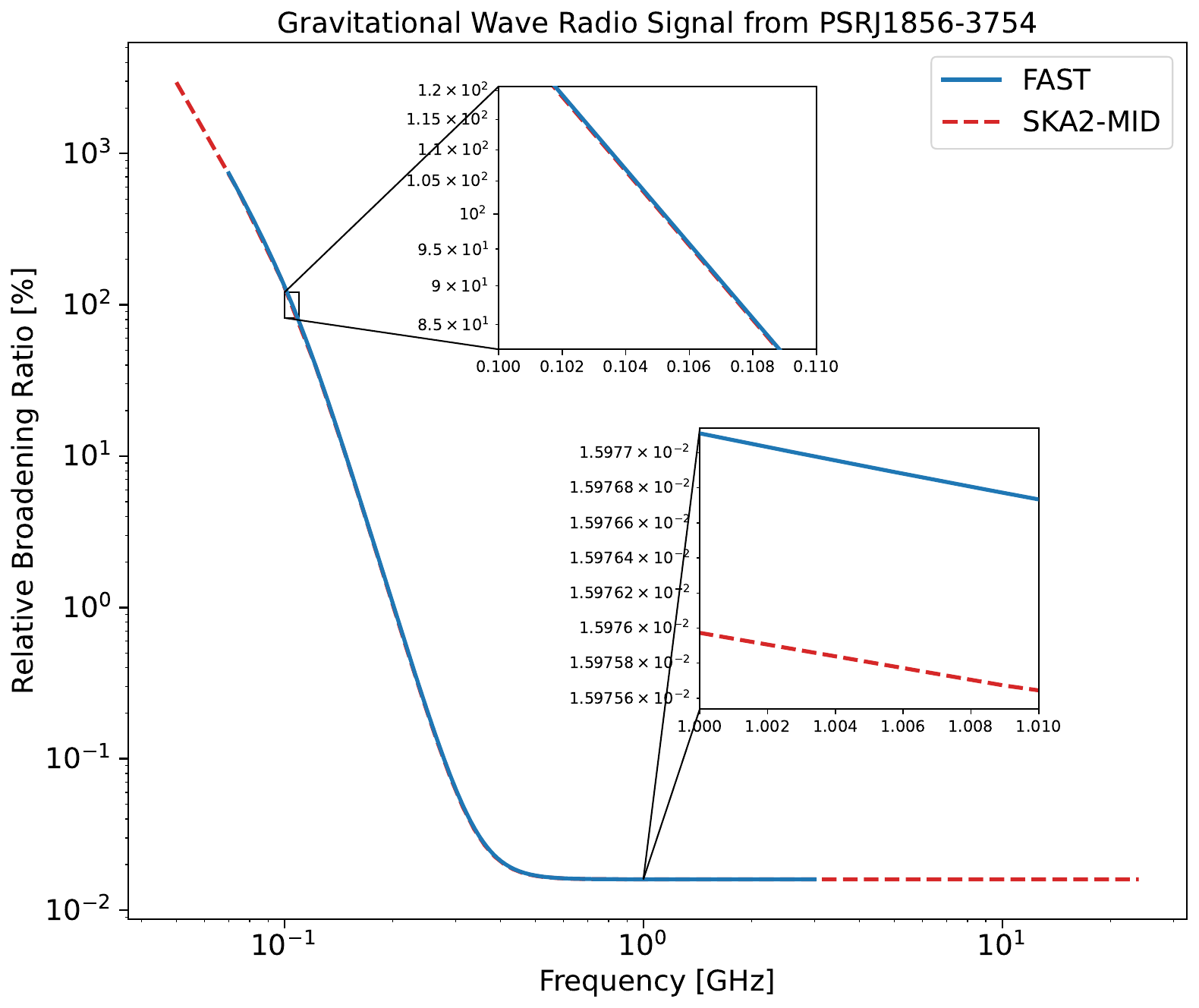}
	\includegraphics[width=0.42\linewidth]{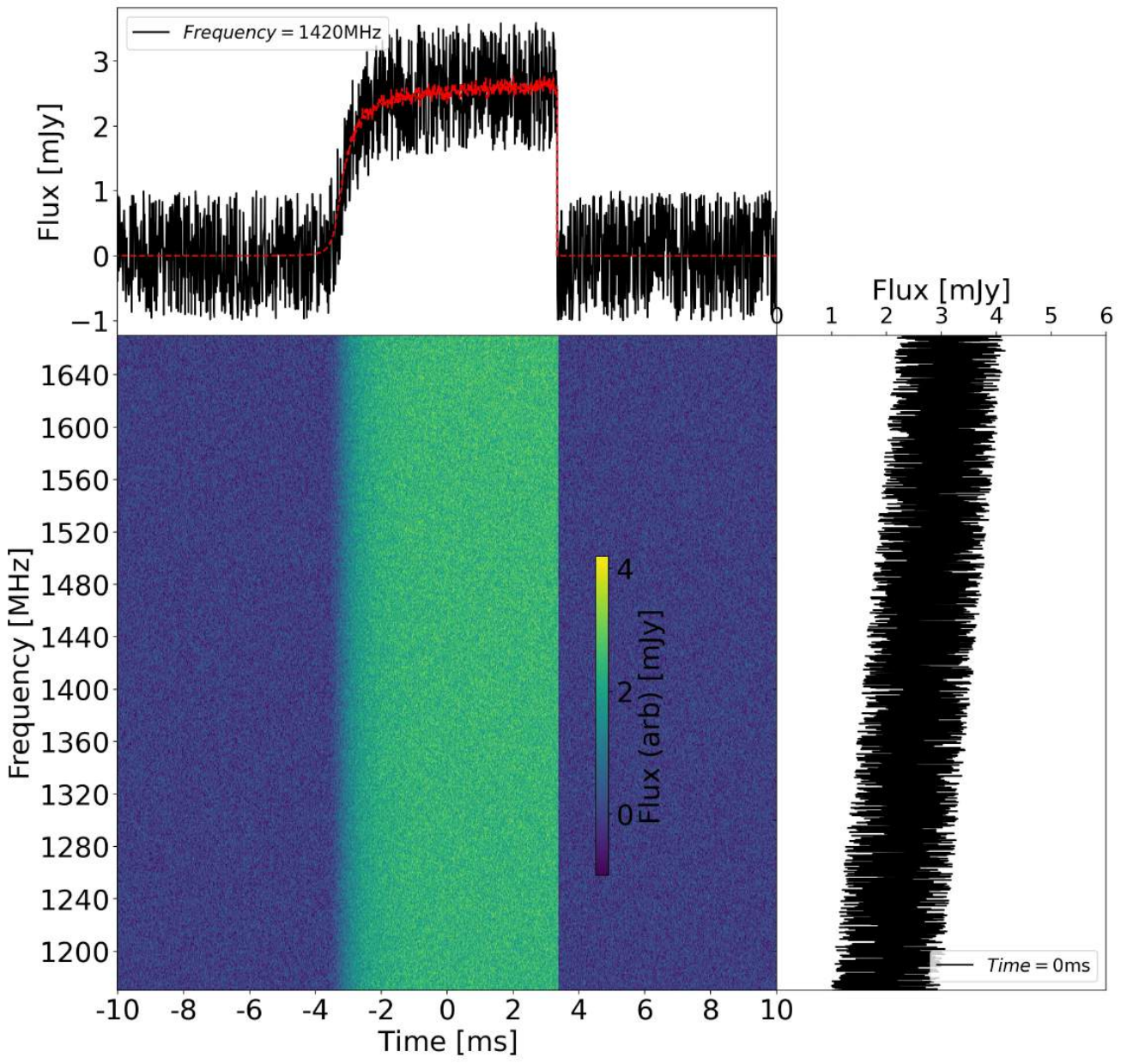}
	\includegraphics[width=0.42\linewidth]{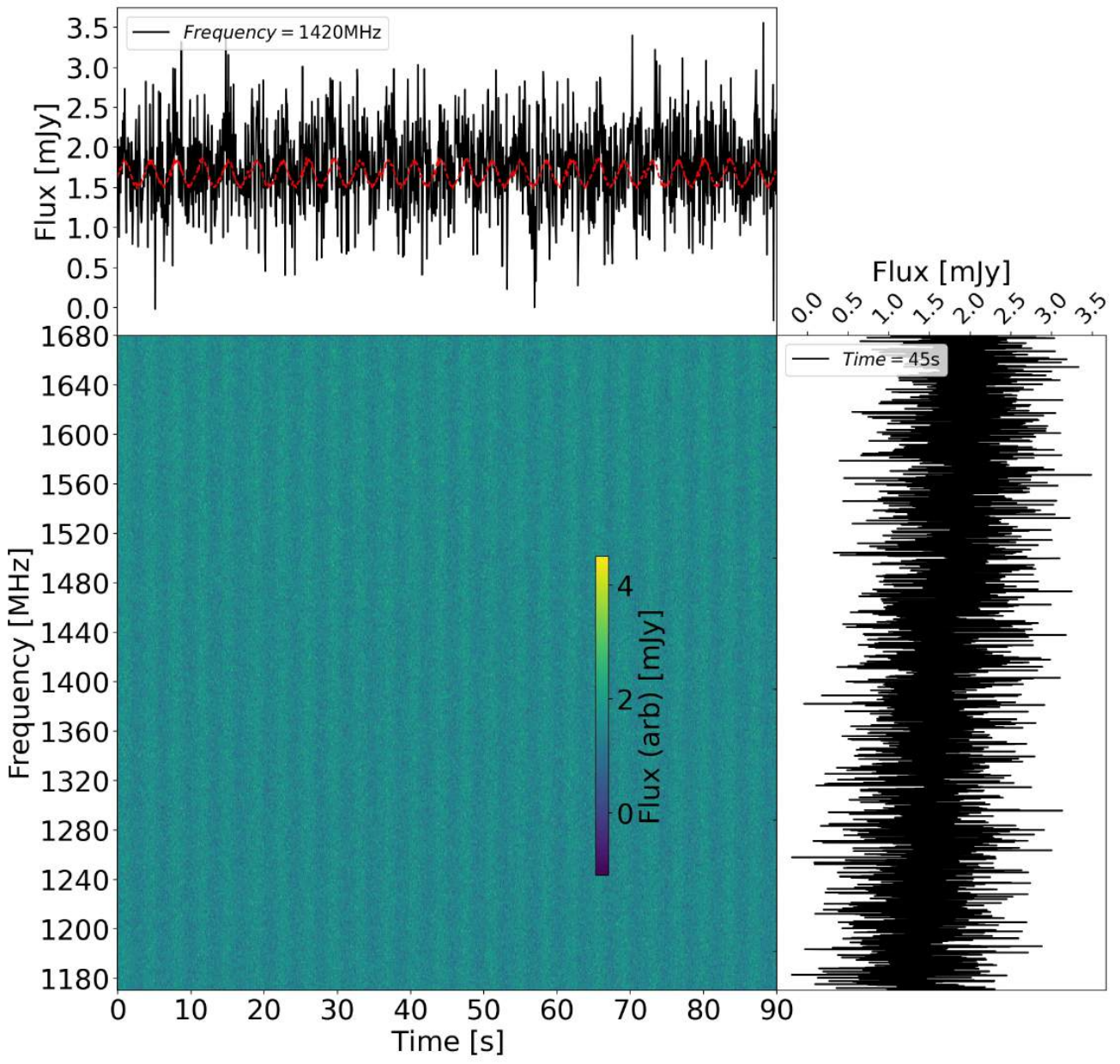}
	\includegraphics[width=0.42\linewidth]{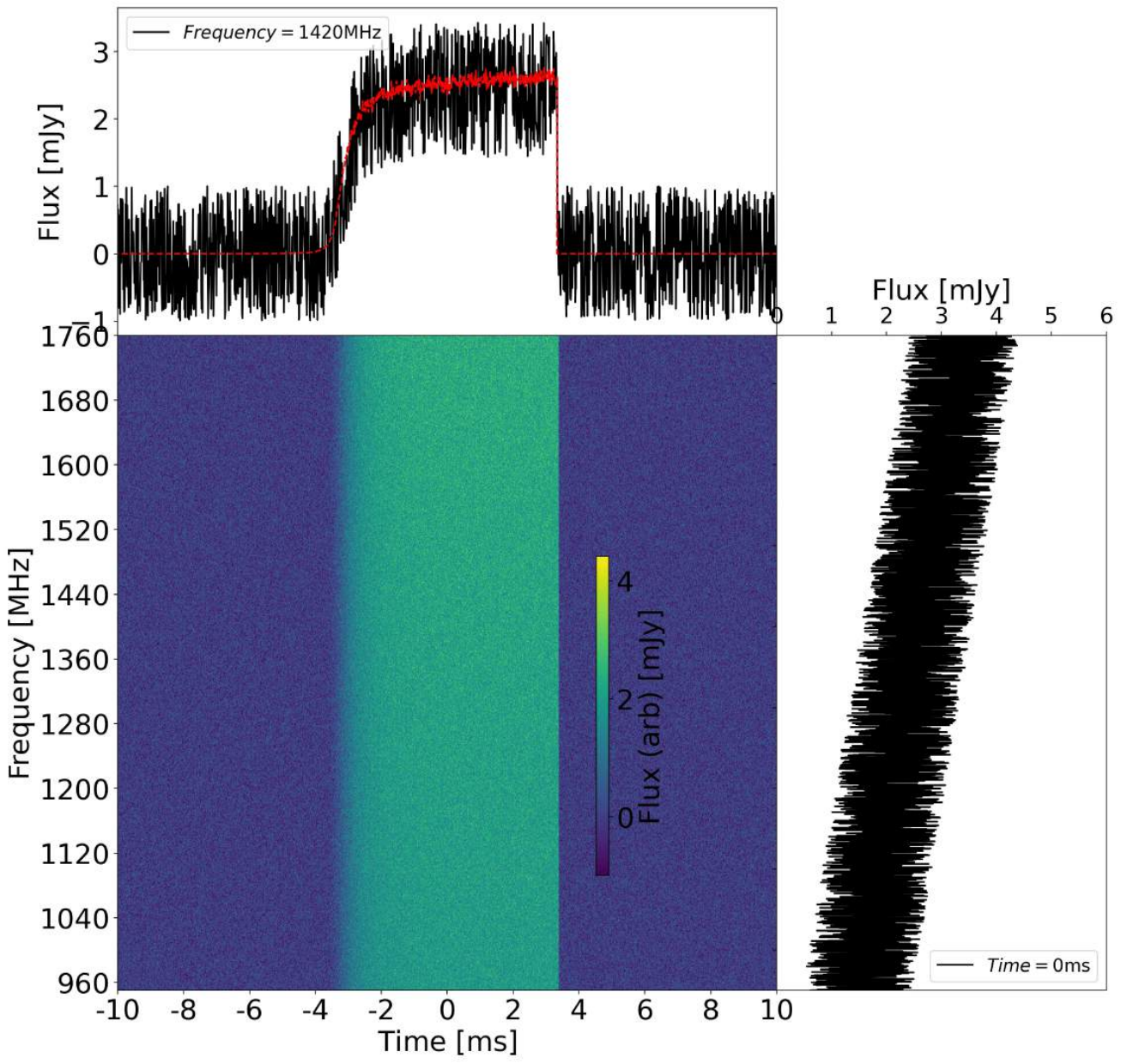}
	\includegraphics[width=0.42\linewidth]{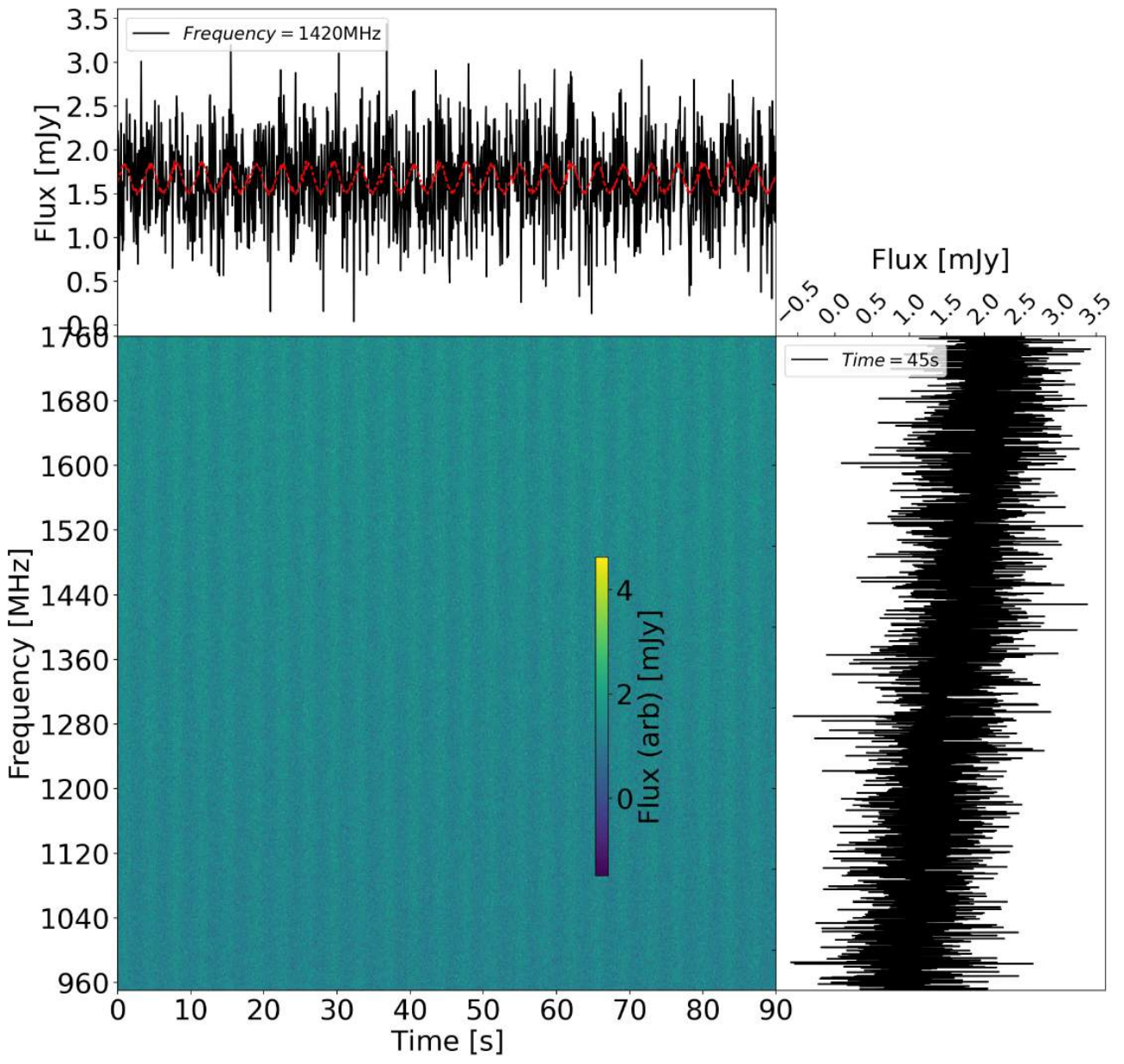}
	\caption{Reference intensity‑response profiles, frequency‑dependent broadening, and simulated time series for GW-EM conversion in pulsar magnetospheres. Top-left: normalised intensity-response profiles for the two pulsars at selected frequencies, shown for the short-transient case at a fixed rotational phase and produced with the filtering pipeline’s bandpass, sampling, and normalisation; these are intensity/flux-density response kernels, not phase-resolved GW-strain templates. Top‑right: frequency‑dependent temporal broadening adopted to model propagation and instrumental effects such as interstellar scintillation and scattering, shown for FAST and SKA2‑MID. Middle and bottom: simulated intensity time series for PSR~J1856-3754 observed with FAST (middle) and SKA2‑MID (bottom). Left column displays transient injections as short, band‑limited excesses; right column shows persistent or stochastic‑background cases in which long integrations reveal the rotation‑modulated intensity envelope from the phase‑dependent magnetospheric response.}
	
	\label{fig:anticipated-signal-1}
\end{figure*}

\begin{figure*}
	\centering
	\includegraphics[width=0.42\linewidth]{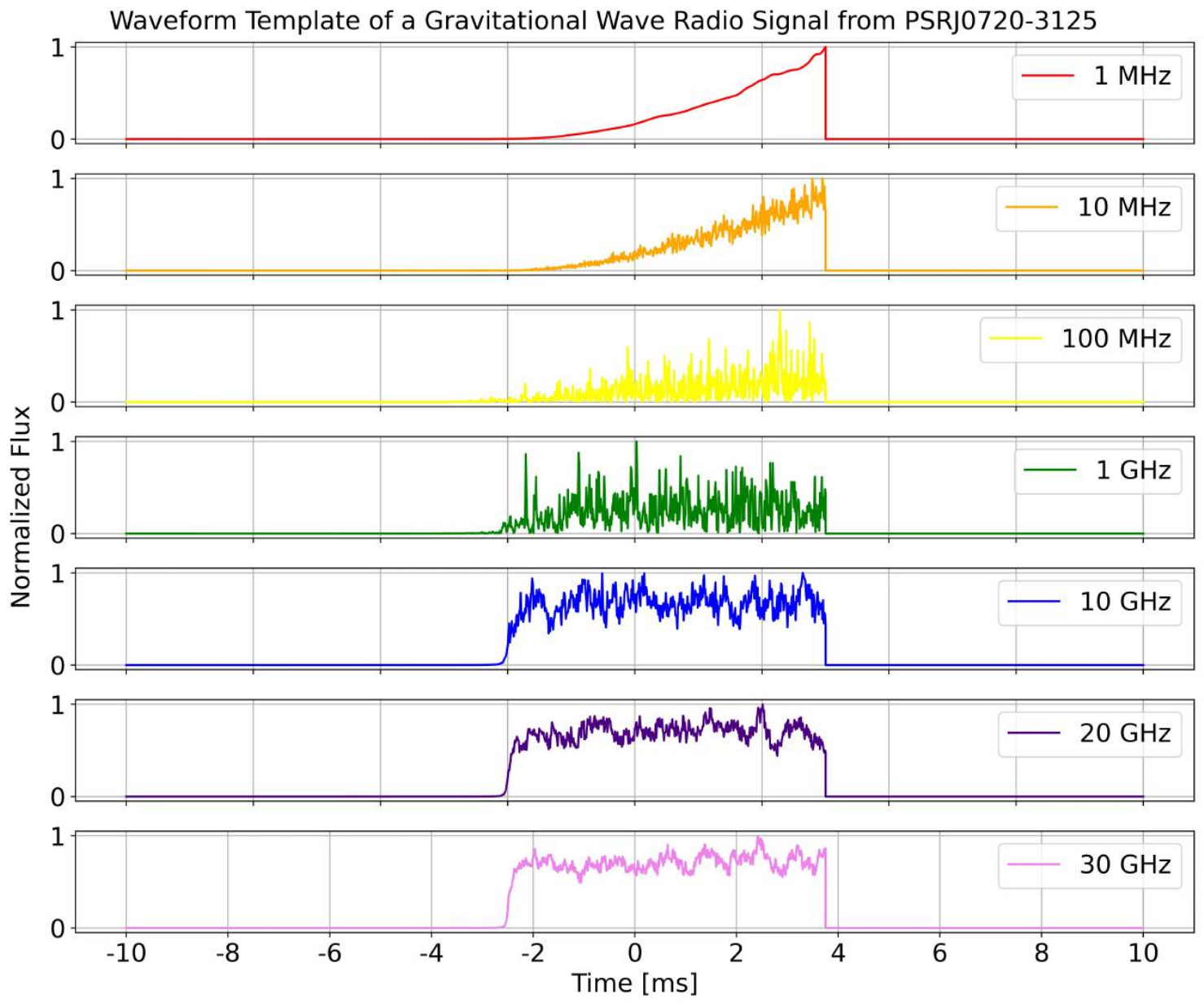}
	\includegraphics[width=0.42\linewidth]{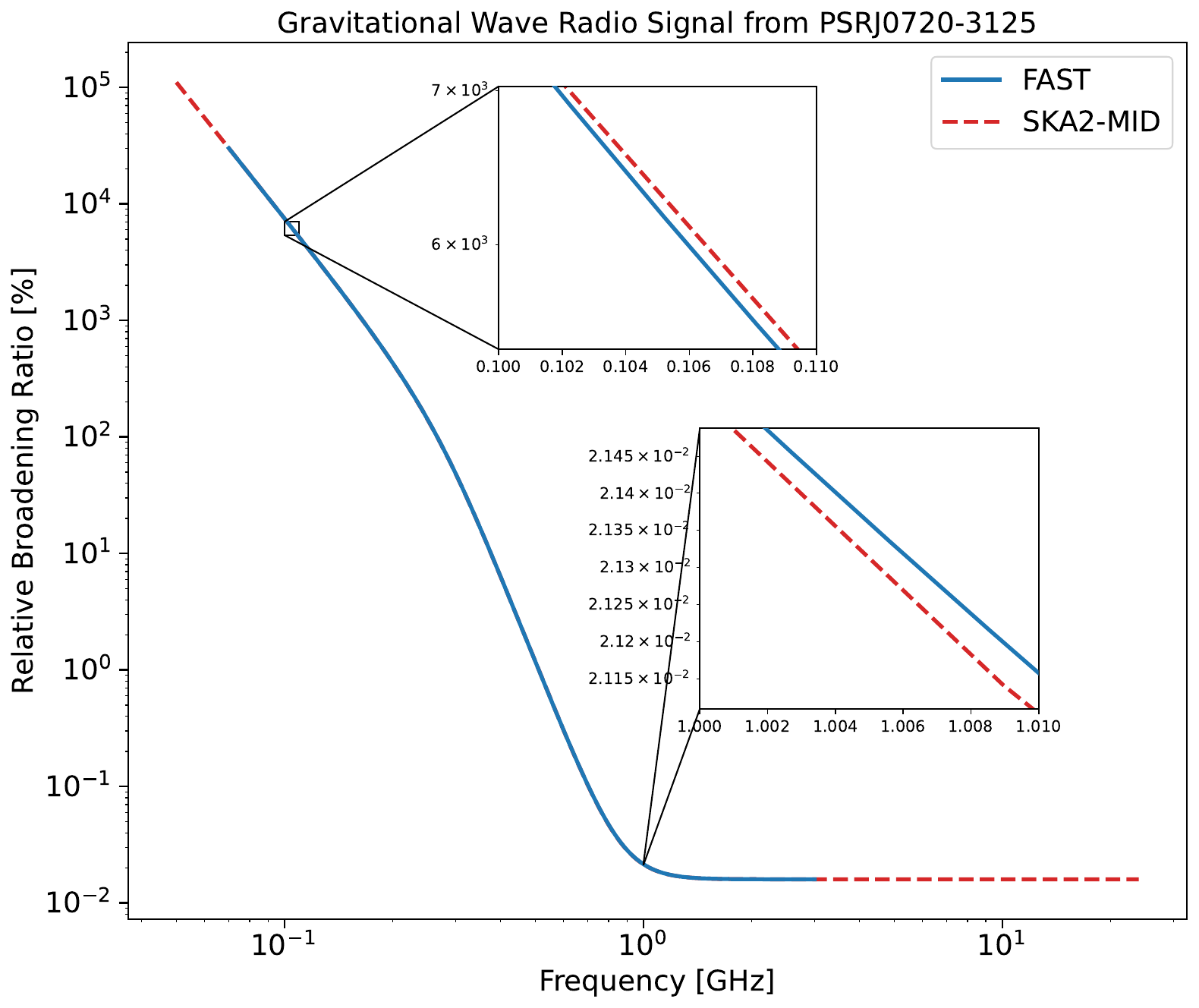}
	\includegraphics[width=0.42\linewidth]{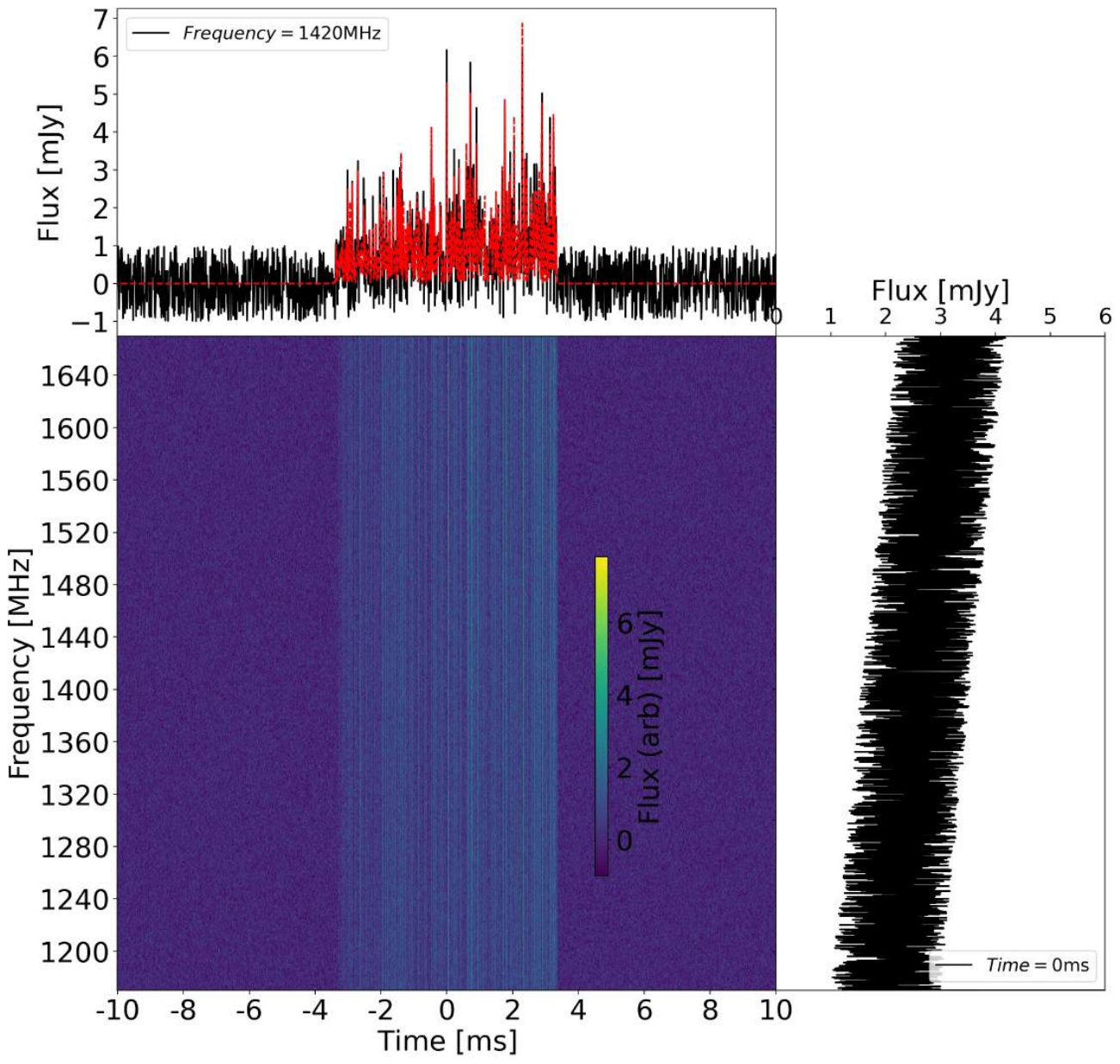}
	\includegraphics[width=0.42\linewidth]{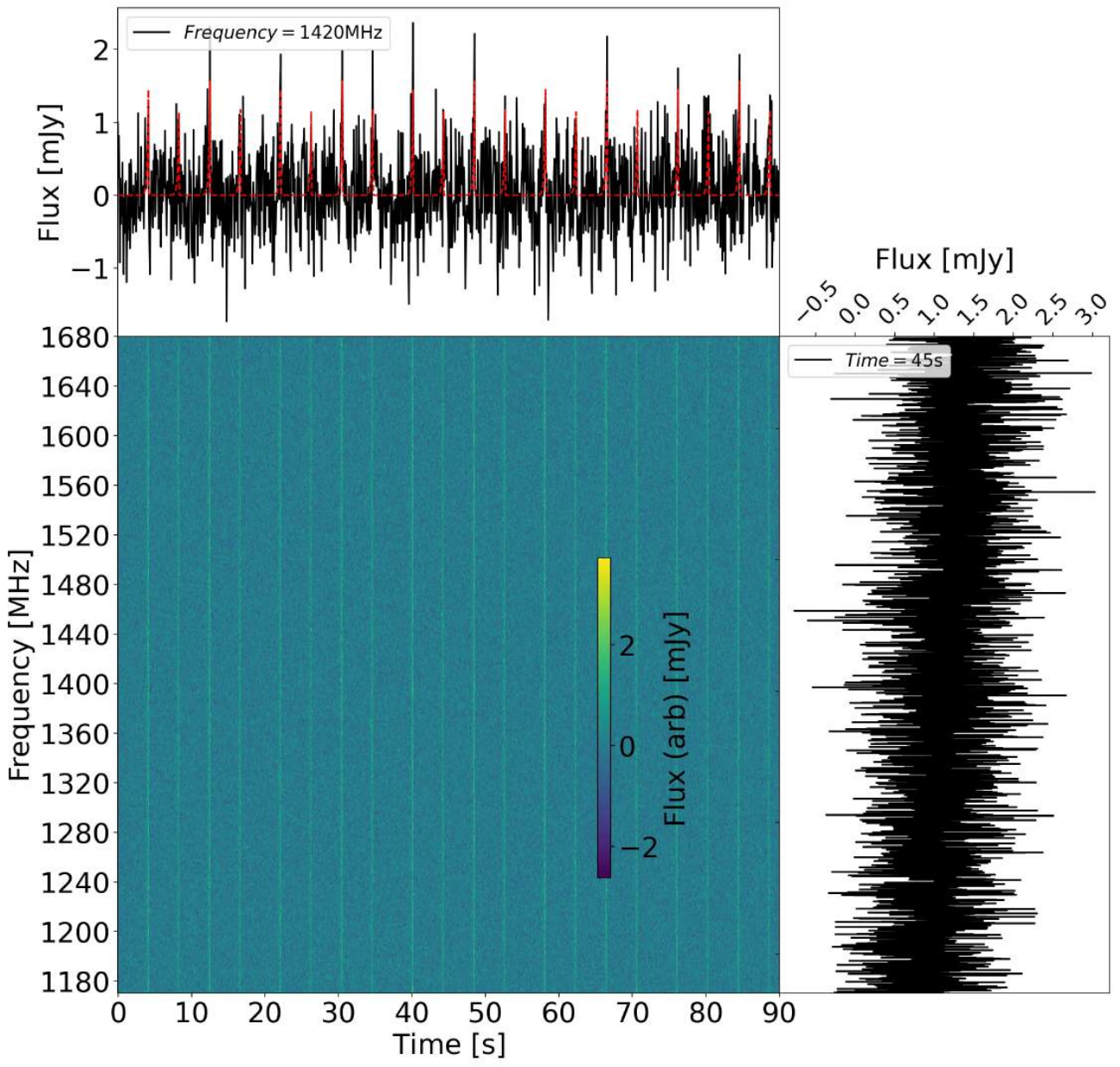}
	\includegraphics[width=0.42\linewidth]{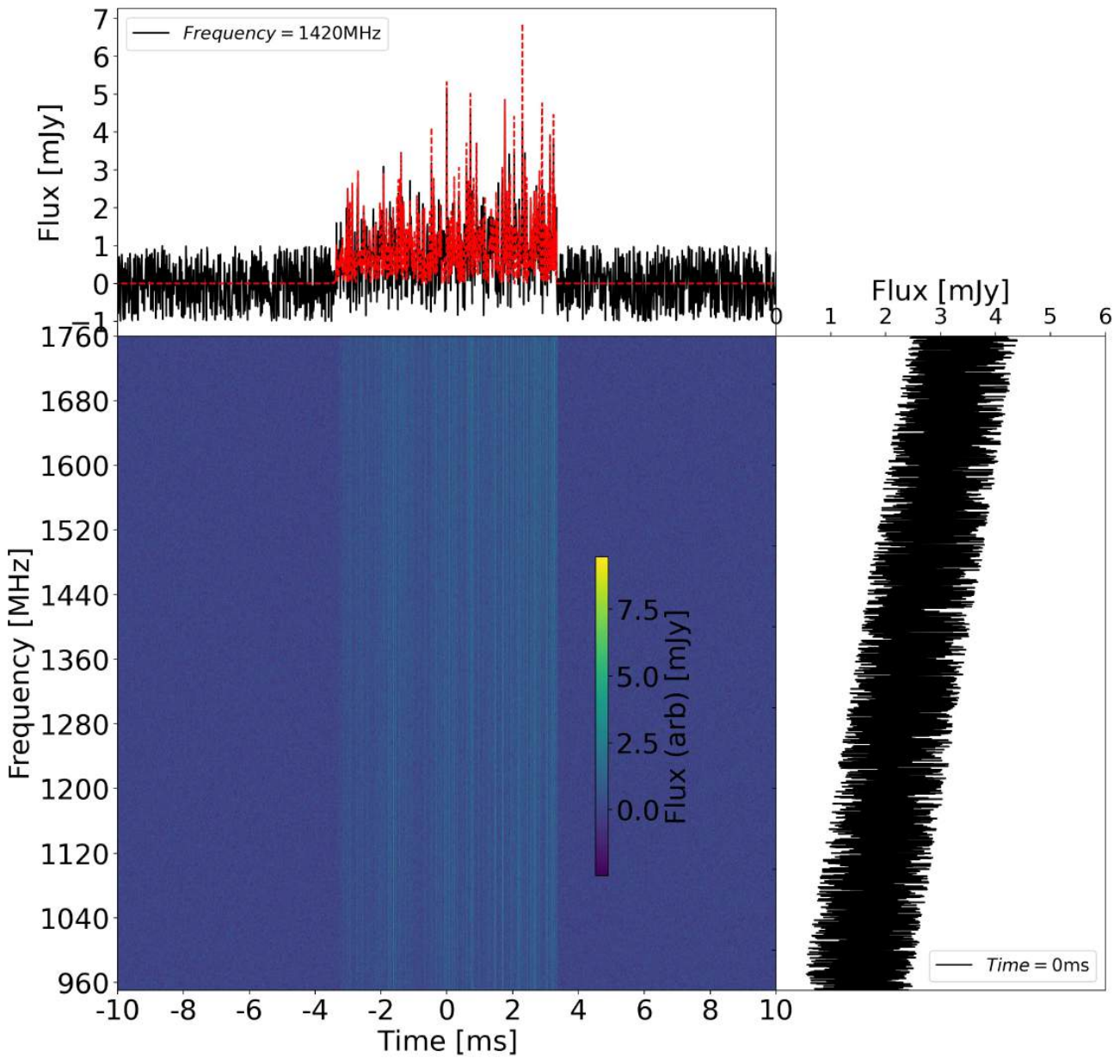}
	\includegraphics[width=0.42\linewidth]{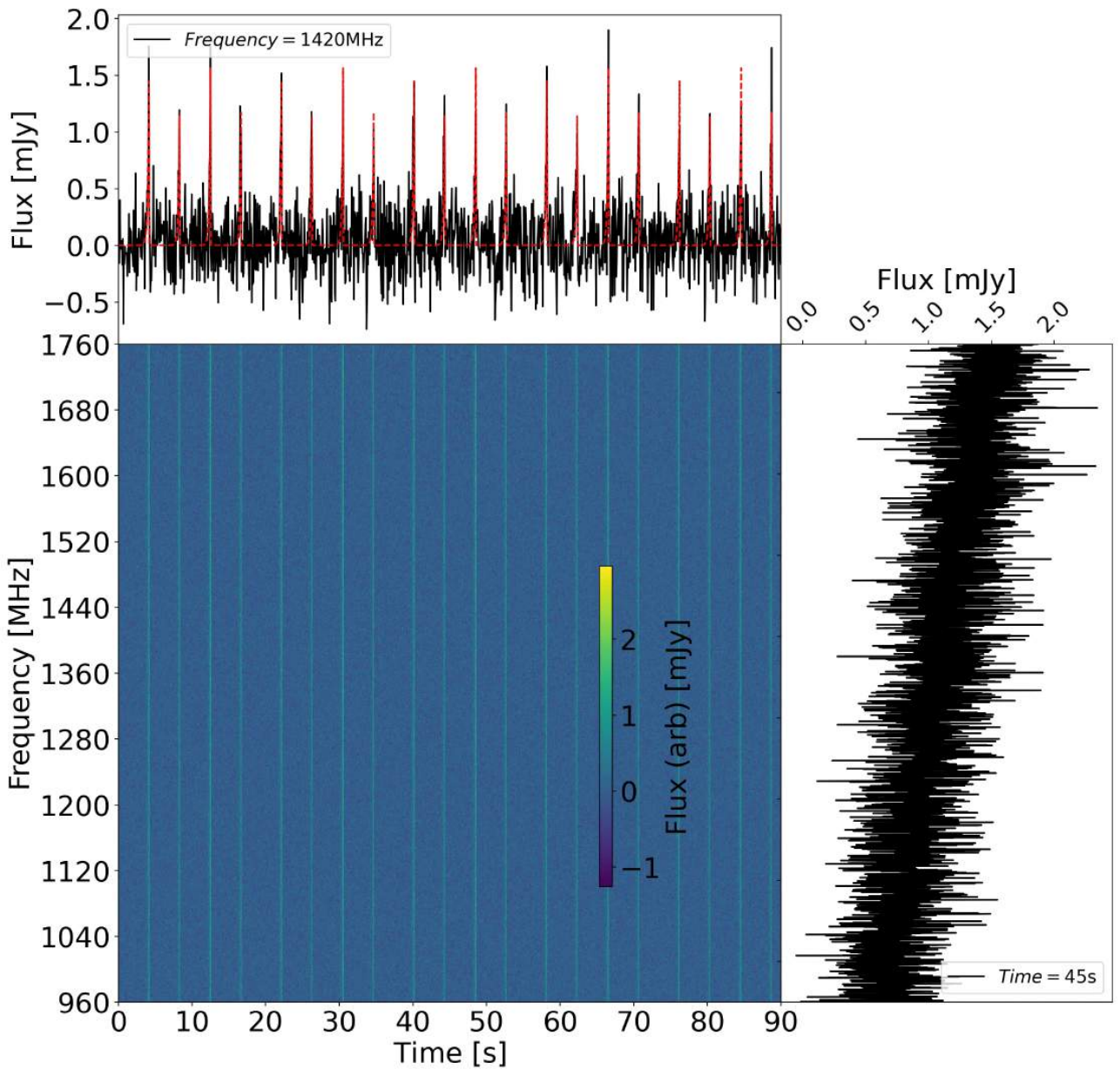}
	\caption{Reference intensity‑response profiles, frequency‑dependent broadening, and simulated time series for GW-EM conversion in pulsar magnetospheres upon reception by FAST and SKA2-MID of PSR J0720-3125. This figure shows the results in the same order as Figure 3.}
	\label{fig:anticipated-signal-2}
\end{figure*}

The simulation results of a single, noiseless, valid observation are shown in Figures \ref{fig:anticipated-signal-1} and \ref{fig:anticipated-signal-2} at the left of top panels. There are our reference response profiles, which represent the band‑averaged flux‑density response of the conversion process along the LOS It should be noted that in order to facilitate the subsequent addition of the longer signal time caused by interstellar scintillation on the observation path and to inject different levels of background noise for different observation methods in the Section \ref{sec:results}, the reference response profile presented has been normalized. The three‑dimensional magnetic‑field configuration $\mathbf{B}$ used in the simulation is taken from PIC calculations, while the plasma contribution to the dispersion relation follows the same analytic parametrisation adopted in \citep{Hong:2024ofh}. For the PBH‑like transients considered here, the intrinsic GHz‑band chirp duration is much shorter than the pulsar rotation period; therefore no rotation‑induced modulation of the chirp itself is modelled, and rotation enters only through the instantaneous line‑of‑sight geometry that selects the magnetospheric response at the event time. For persistent or stochastic signals, by contrast, long integrations naturally imprint a phase‑dependent intensity envelope via the magnetospheric response over a full rotation cycle.

The simulation time of transient events is the signal duration plus twice the redundancy, and the simulation time of persistent events is one-sixth the time of a single observation. These signals are shown in Figures \ref{fig:anticipated-signal-1} and \ref{fig:anticipated-signal-2} from second panel to fourth panel, respectively. Here, the methods for simulating these signals are consistent with those in previous theoretical work \citep{Hong:2024ofh}. The specific steps include: (a) First, determine the frequency range of the simulation and the integral time length of the observation. (b) Then select the corresponding reference response profile and combine the conversion probability with the energy density of the gravitational wave to obtain a noise-free radio signal. (c) Finally, according to the system-equivalent flux density of the telescope, the corresponding level of Gaussian noise is injected. This simulation method is also the data simulation method we used when repeatedly adjusting the signal S/R in Section \ref{sec:results} to explore the limit state of the filter. It should be noted that once the integration time of the observation is determined, the SEFD of the telescope's observation of the same pulsar under different SNRs is almost determined. Subsequent sensitivity limit tests will focus on adjusting the energy density of the GW rather than the SEFD of the telescope.

The simulated signals described here provide the basis for evaluating their detectability under realistic instrumental configurations, as discussed in the subsequent section. We call the signal time of both events the intrinsic duration at the source. In addition, because the pulsars calculated in this work are all located in the Milky Way, the observed signal is not affected by the redshift. As in the previous theoretical work \citep{Hong:2024ofh}, there are no strong gravitational objects such as black holes in our observation path, so the spectral line properties of our signal still hold. Moreover, because the inverse GZ effect can be seen as resonance of GWs and electromagnetic fields, the converted electromagnetic wave can be considered to be $100\%$ linearly polarized when it just passes through the pulsar magnetic field, without considering the influence of the pulsar Faraday rotation on this electromagnetic wave.

\subsection{Path Effect on Resonant Radio Signals.}
\label{sec:PE-RRS}
Because the path from the pulsar to the telescope is full of charged particles and turbulent media and there is relative motion between the pulsar, interstellar medium, and radio telescope, the radio signal converted by GWs will be scattered by the interstellar medium and will show a scintillation effect similar to the interstellar scintillation of the pulsar. The phase modulation caused by the medium results in amplitude modulation, which manifests as a change in signal flux on the telescope receiver at different time scales. In addition, the interference between the scintillation signals creates an interference pattern in the telescope plane and moves above the telescope with the relative motion between the pulsar, interstellar medium, and radio telescope. The size of this region is known as the field coherence scale $s_{\mathrm{fc}}$ and can be compared to the Fresnel scale $l_{\mathrm{F}}$. The interstellar scintillation of the signal can be well described by a thin screen model \citep{1968Natur.218..920S}. The intensity of the observed scintillation depends on the magnitude of the perturbation of the total phase over the entire distance. Considering that the variation in electron number density $n_{e}$ in the interstellar medium usually shows a distribution on a certain scale in actual observations and that our observation method accounts for the proper motion of the pulsar, the total length of the interstellar medium $L$ in the thin screen model can be divided by the length distribution represented by the spatial wavenumber spectrum. Given that the pulsars used in our simulation are all in the Milky Way, the free electron distribution in the Milky Way can be modelled using the ``NE2001" model \citep{Cordes:2002wz,Cordes:2003ik}, which considers the extended spatial wavenumber spectrum power law model and the Kolmogorov spectrum with spectral index $\beta$ \citep{1990ARA&A..28..561R}:
\begin{equation}
	P_{\mathrm{n}_{\mathrm{e}}}(q)=\frac{C_{\mathrm{n}_{\mathrm{e}}}^2(z)}{\left(q^2+\kappa_{\mathrm{o}}^2\right)^{\beta / 2}} \exp \left[-\frac{q^2}{4 \kappa_{\mathrm{i}}^2}\right],\label{extended-spatial wavenumber-spectrum}
\end{equation}
where $q = 1 / a$ is the magnitude of the three-dimensional wavenumber; the `inner' and `outer' scales of the turbulence are $\kappa_{\mathrm{i}}$ and $\kappa_{\mathrm{o}}$, respectively. And $C_{\mathrm{n}_{\mathrm{e}}}^2(z)$ denotes the strength of the fluctuations along a given LOS. The inner and outer scales $\kappa_{\mathrm{i}}$ and $\kappa_{\mathrm{o}}$ correspond to cut-offs in scale sizes. For wavenumber $q$ in the range $\kappa_{\mathrm{o}} \ll q \ll \kappa_{\mathrm{i}}$, Eq. (\ref{extended-spatial wavenumber-spectrum}) simplifies to $P_{\mathrm{n}_{\mathrm{e}}}(q)=C_{\mathrm{n}_{\mathrm{e}}}^2 q^{-\beta}$. Thus, we can reflect the scintillation of radio signals converted by GWs at different observation times by simulating the dynamic spectra $S_{\mathrm{spe}}(\nu,t)$ at each observation time.

Given that cross-correlation analyses rely on epoch-to-epoch decorrelation of background noise, we incorporate the source–observer relative motion (proper motion and annual parallax) when updating the time-dependent LOS geometry. We subsequently compute the corresponding dispersion and rotation measures (DM, RM) along the LOS using the “NE2001” model, and employ these LOS parameters to drive our scintillation simulations. The explicit astrometric expressions and scintillation-strength parametrization are provided in Appendix~\ref{app:path-details}. As illustrated in Figure \ref{fig:interstellar scintillation}, our simulations show strong scintillation in the L-band for both FAST and SKA2-MID. The resulting dynamic spectra and their secondary spectra reveal a distinct scintillation arc feature for the simulated GW-converted signals.

\subsection{Signal Reception by Radio Telescopes.}
\label{sec:SR-RT}
After propagating through the interstellar medium (ISM) and the Earth’s ionosphere, the signals are ultimately received by radio telescopes. To account for the propagation effects, we compute the dispersion and rotation measures (DM, RM) along the full path. For the ISM, the DM and RM are derived using the “NE2001” model as described earlier, while for the Earth’s ionosphere, they are calculated with the program “IonFarRot”; the resulting time-dependent ionospheric RM values are shown in Figure \ref{fig:Earth-RM} in Appendix~\ref{app:depol-details}. The procedure for calculating the SNRs and system temperatures for FAST and SKA-2 under different scenarios is the same as that in previous theoretical work \citep{Hong:2024ofh} with CASA \citep{2022PASP..134k4501C} and ULSA \citep{2021ApJ...914..128C}, except that the flux, shape, and width of the received signals change after the effects of the observational paths are considered. The time width $W$ of the received signal is slightly wider than the typical width of the original signal. The received signal width $W=\left(t_{\mathrm{inh}}^2+t_{\mathrm{sca}}^2+t_{\mathrm{ins}}^2\right)^{1 / 2}$ can typically be divided into three components: first, the inherent width of the signal $t_{\mathrm{inh}}$, which is related to the magnetic field distribution of the pulsar; second, the scattering duration of the signal $t_{\mathrm{sca}}$, which can be regarded as being only related to the interstellar medium since the observational pulsar sources are all within the Milky Way galaxy; and last, the instrumental spreading duration of the telescope $t_{\mathrm{ins}}=\left(t_{\mathrm{samp}}^2+\Delta t_{\mathrm{DM}}^2+\Delta t_{\delta \nu}^2\right)^{1 / 2}$, which is related to the sampling time interval $t_{\mathrm{samp}}$, dispersion $\Delta t_{\mathrm{DM}}=8.3\left(\frac{\mathrm{DM}}{1\mathrm{~cm}^{-3} \mathrm{pc}}\right)\left(\frac{\Delta \nu}{1\mathrm{~MHz}}\right)\left( \frac{\nu}{1\mathrm{~GHz}}\right)^{-3} ~\mu\mathrm{s}$ and frequency bandwidth $\Delta t_{\delta \nu}=1\left(\frac{\Delta \nu}{1\mathrm{~MHz}}\right)~\mu\mathrm{s}$. Figures \ref{fig:anticipated-signal-1} and \ref{fig:anticipated-signal-2} show the relative broadening ratio of signal width of GWs from two pulsars as a function of frequency for a given simulated observation at the right of the top panel. It can be found from the figure that the lower the frequency, the greater the signal width, and it decreases as the observation frequency increases. It should be noted that the radio signals generated by resonance in the magnetic fields of these two pulsars have a relatively weak signal broadening effect within the L-band frequency range of radio observations, almost equivalent to the broadening of the original signals.

\begin{figure*}
	\centering
	\includegraphics[width=0.45\linewidth]{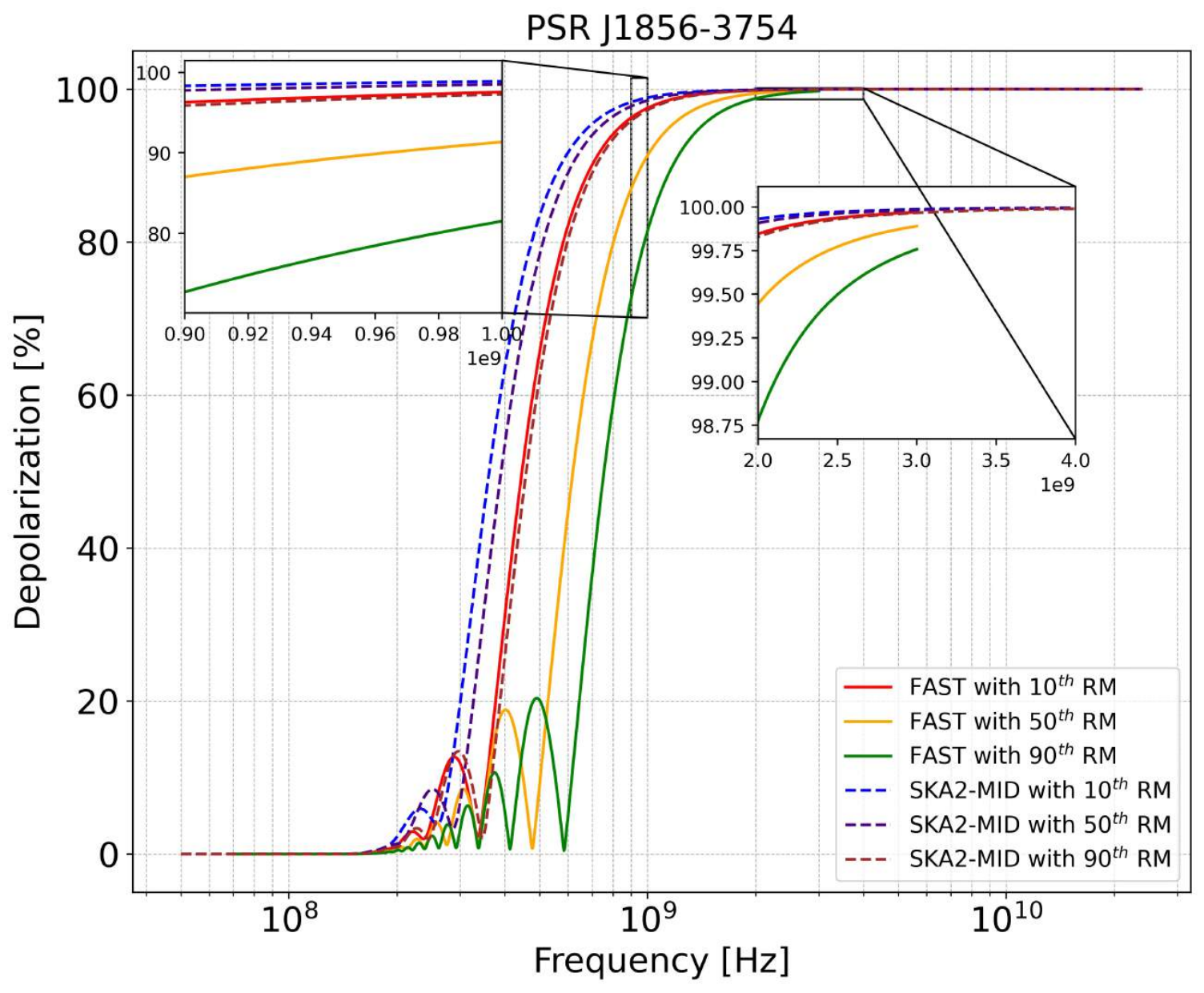}
	\includegraphics[width=0.45\linewidth]{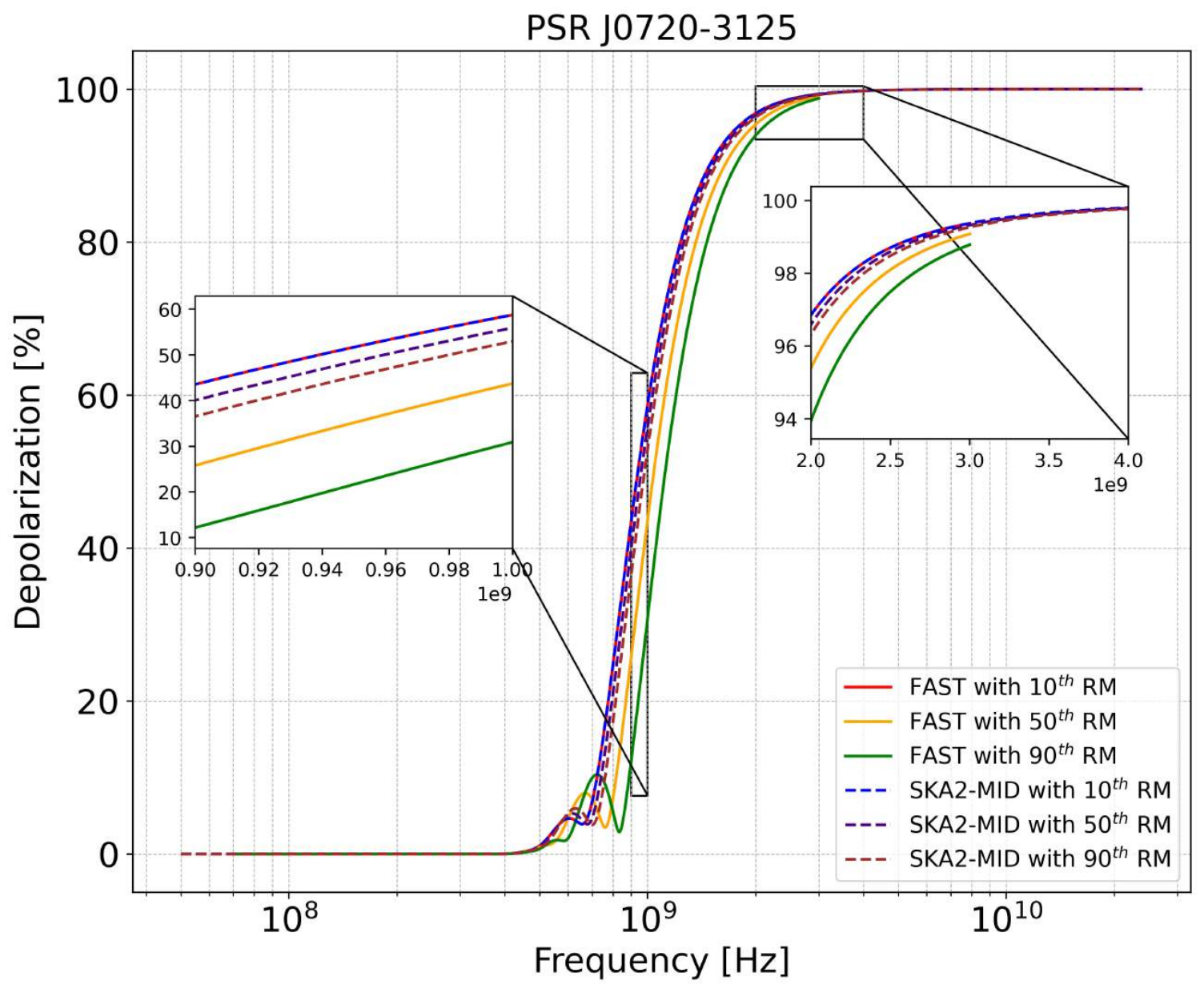}
	\includegraphics[width=0.55\linewidth]{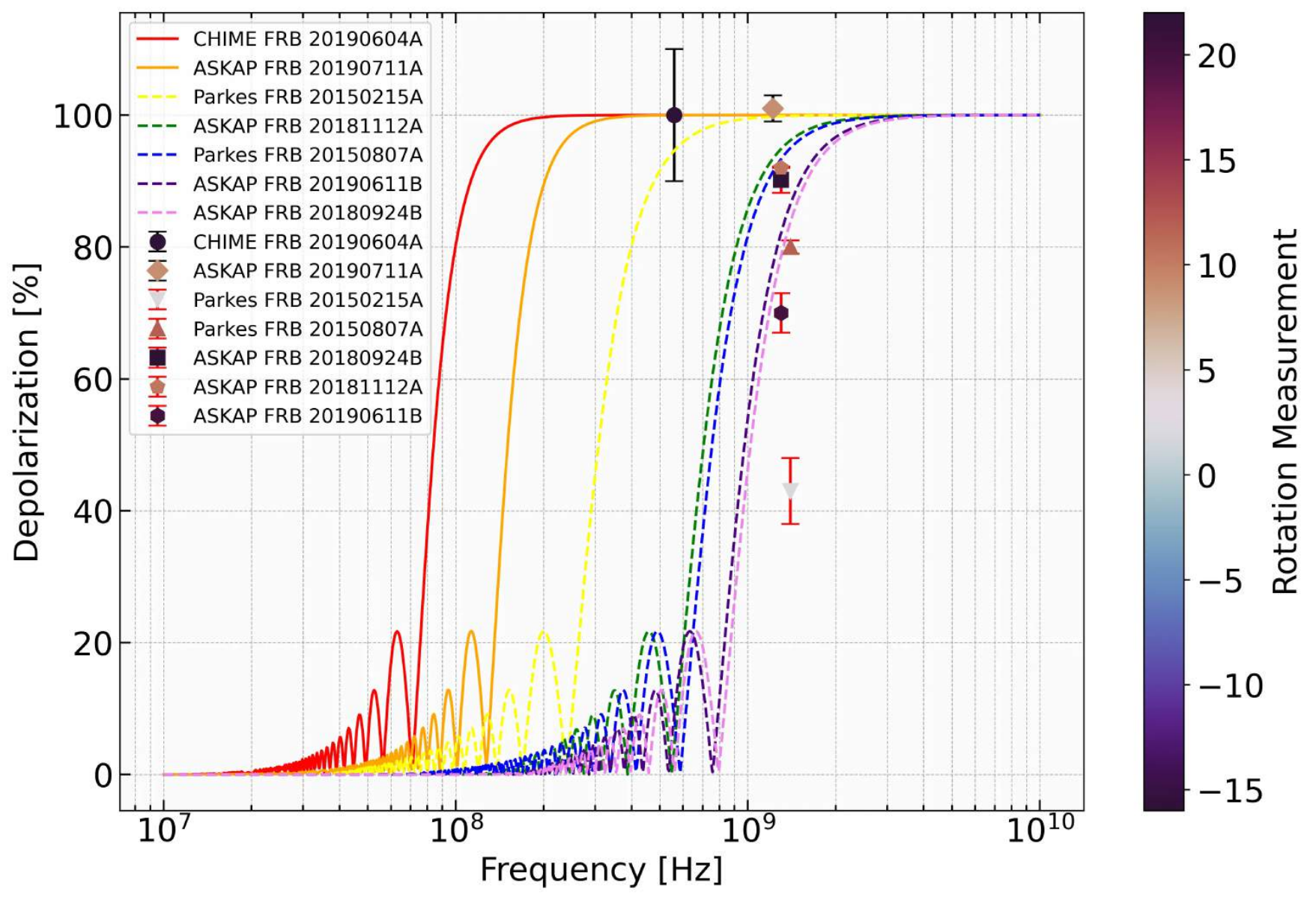}
	\caption{The linear polarization of a GW radio signal varies with frequency. The two diagrams at the top panel show the radio signals from PSR J1856-3754 and PSR J0720-3125, with FAST's signal represented by a coloured solid line and SKA2-MID's signal represented by a coloured dashed line. The bottom panel shows the fit of the depolarisation theory of our signals to several examples of intragalactic FRB signals, where the coloured solid lines indicate the results of our fits to the repeated FRBs, and the coloured dashed lines indicate the results of our fits to the non-repeated FRBs. Different FRBs are represented by different shapes whose colours indicate the magnitude of the RM, with the line-polarised error bars of repeated FRBs shown as black solid lines, and those of non-repeated FRBs shown as red solid lines.}
	\label{fig:depolarization-FRB}
\end{figure*}

We also model polarization-transfer and depolarization effects to assess their impact on signal detectability, which including depth, beam, and bandwidth depolarization mechanisms. Although the radio signals generated via GW conversion are initially linearly polarized, they undergo depolarization due to Faraday rotation along the propagation path. As illustrated in Figure \ref{fig:depolarization-FRB}, our theoretical predictions (upper panels) show how the linear polarization of the magnetosphere-converted signals varies with frequency and telescope parameters. Furthermore, applying the same depolarization model to fit observed data from Galactic FRBs (lower panels) demonstrates good agreement for repeating bursts, effectively revealing the inhomogeneity and turbulence of the intervening magnetic field. The complete treatment of depolarization, including detailed comparisons with FRB observations, is provided in Appendix~\ref{app:depol-details}. In the main text, we focus on the resulting bandwidth- and propagation-driven pulse broadening, which is directly relevant for the matched-filter and cross-correlation analyses that follow.

\subsection{Testing the Stationarity and Gaussianity of the Simulated Observational Data.}
\label{sec:Testing-SG-SOD}
Although determining whether the data are stationary and Gaussian can be challenging because the tests rely on comparisons with models and some models may be better at picking up some forms of nonstationarity and non-Gaussianity, analysing the stability and Gaussianity of the signals received by a radio telescope is important. On the one hand, it can help us choose methods for processing data, such as when using FAST for long-term observations; the baseline will be shifted, the noise will be red, and the signal stability will be worse. The reddened part of the noise must be deducted, and the red noise can usually be modelled with a power-law spectrum $S_{\mathrm{rn}}=A_{\mathrm{rn}} f^\alpha$. On the other hand, it can help us identify some potential regions for finding new phenomena, such as the famous binary neutron star merger GW170817 affected by a glitch \citep{LIGOScientific:2017vwq,Pankow:2018qpo}. Furthermore, since the simulation assumes that the data is stationary and Gaussian, the discussion in this section can also verify the stationarity and Gaussianity of the simulation data used in this work and test the correctness of our data simulation method.

The stationarity of time series data essentially means that the data have statistical characteristics that are independent of time. Stationary time series usually have short-term correlations, and the autocorrelation coefficient tends to rapidly degrade to zero. For nonstationary data, degradation occurs more slowly, or there are changes such as decreases and then increases or periodic fluctuations. Generally, the stationarity of a sequence can be judged by the unit root test, that is, whether there is a unit root in the sequence; if there is, it is a nonstationary sequence, and if there is not, it is a stationary sequence. In this work, we use three test methods: the augmented Dickey--Fuller (ADF) test \citep{Dickey1981LIKELIHOODRS,10.1093/biomet/71.3.599}, the Phillips--Perron (PP) test \citep{10.1093/biomet/75.2.335}, and the Kwiatkowski--Phillips--Schmidt--Shin (KPSS) test \citep{KWIATKOWSKI1992159}. Among them, the ADF test is suitable for the stationarity test of a high-order autoregressive process. The PP test is a nonparametric test method used mainly to solve the potential sequence correlation and heteroscedasticity problems in the residual term. The KPSS test mainly assesses whether the time series is stable around the deterministic trend. This test is particularly useful for confirming the results of ADF and PP tests because it provides a complementary approach to understanding the stationarity of time series. The joint outcomes of these tests for all mock datasets are summarised in Appendix~\ref{app:diagnostics}, Figure~\ref{fig:Stationarity-test}. In Figure \ref{fig:Stationarity-test}, we use red, green, and blue to represent the combination of test results from the three methods. The figure shows that all the simulated data are stable, and only in the simulated data of some transient events does the stable trend phenomenon not appear, which is because the duration of a single transient event is short and the data fluctuate greatly.

The Gaussianity of the data can be tested by the Kolmogorov--Smirnov test \citep{an1933sulla,smirnov1939estimate,smirnov1948table} and D'Agostino's K-square test \citep{272b2fa8-f3b3-371d-b34e-a4c6938458a1,25848ac6-50af-379a-93d8-ad15b9244648}. The former evaluates the distribution characteristics of the data by calculating the maximum difference between the sample distribution and the theoretical distribution. The latter quantifies the difference and asymmetry between the data distribution curve and the standard Gaussian distribution curve by calculating the skewness and kurtosis and then calculates the degree to which these values differ from the expected value of the Gaussian distribution. Similarly, we present the results of our tests in Figure \ref{fig:Gaussianity-test}. The corresponding results are shown in Appendix~\ref{app:diagnostics}, Figure~\ref{fig:Gaussianity-test}, indicating that the errors in our simulated datasets are consistent with Gaussian statistics across all considered cases.

\subsection{BCKA Filter.}
\label{sec:BCKA-filter}
\begin{figure*}
	\centering
	\includegraphics[width=0.9\linewidth]{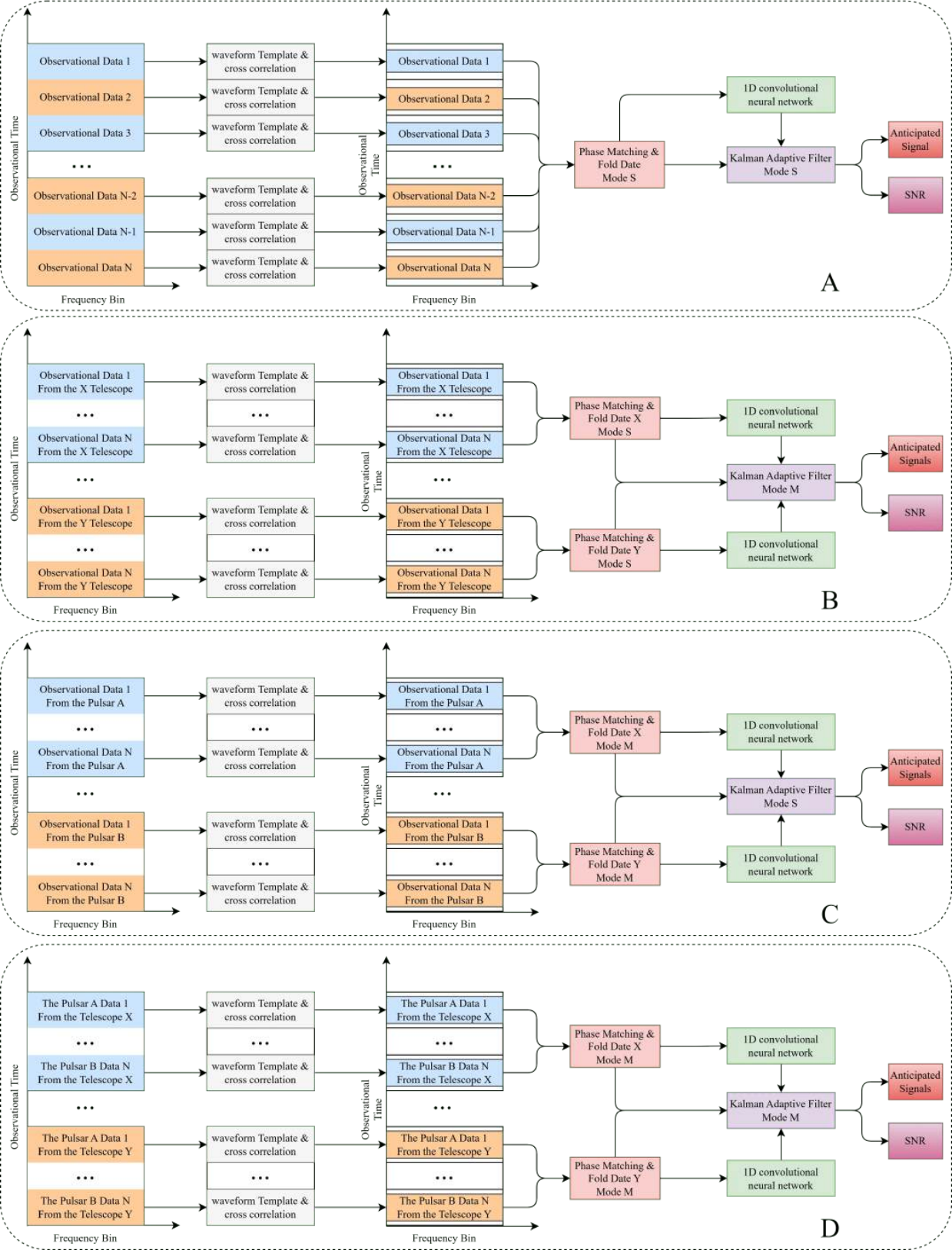}
	\caption{The diagram of the total process of the BCKA filter. The figure shows, from top to bottom, the filters used corresponding to the four observation methods: the (A) SPST method, (B) SPMT method, (C) MPST method and (D) MPMT method. The different coloured squares indicate different modules of the filter, where a change in the width of the square representing the observed data indicates a reduction in data time due to the deletion of non-signal data.}
	\label{fig:BCKA-process}
\end{figure*}

In principle, the processing of radio observation data requires flux calibration first and the deduction of known RFI. However, as mentioned in the subsection \ref{sec:GP-resonance-SNSM}, we assume that the flux calibration is accurate and the simulation of RFI is rather complex. Therefore, here we assume that all known RFI have been deducted. Then, by combining the signal time width varying with frequency obtained from interstellar scintillation in subsection \ref{sec:PE-RRS} and the instrument effect in subsection \ref{sec:SR-RT}, the received signal shape is pushed back to the original signal shape. After the processed simulated observational data are tested for stationarity and Gaussianity, we can obtain the desired VHF GW radio signal through our specific filter. Our filter is essentially a matching filter of response profile, which aims to find the radio signal with correlation according to the reference response profile while maximizing the noise deduction to obtain the optimal S/R. The filter is a combination of Bayesian cross-correlation, a Kalman filter, and adaptive smoothing (we name it the BCKA filter). The hyperparameters of the BCKA filter are optimised by a one-dimensional convolutional neural network (CNN). The flowchart of the whole filter is shown in Figure \ref{fig:BCKA-process}. Our filter has a total of four modes, SS, SM, MS and MM, corresponding to the four cases of the observation method, the difference among which lies essentially in the S/R calculation method.

The observed signal power spectrum $S_{R}^{h}$ and the observed noise power spectrum $S_{R}^{n}$ are subjected to Bayesian cross-correlation analysis through the form of covariance matrix $C$. After carefully deducting the flux during the pulsar pulse window through the fourth section, we find that the observed noise sources primarily consist of the telescope system noise, which includes sky background noise, and the pulsar noise, which comprises radiation from the pulsar outside the pulsar pulse window. The observation mode of target tracking causes the telescope pointing angle and system temperature to fluctuate over time, leading us to assume that there is no correlation between the observed noise. Therefore, we can express the covariance matrix $C$ for a radio signal with only one type of GW as follows:
\begin{equation}
	C=S_{R}^{h} \mathbf{I}_{N\times N}+ \mathrm{Diag}\left(S_{R,1}^{n}, \ldots  ,S_{R,N}^{n} \right),
\end{equation}
where $\mathbf{I}_{N\times N}$ is the identity matrix with the same dimension as the number of observations.

The data processed through cross-correlation template matching still contain significant noise, necessitating additional filtering. The Kalman filter estimates the state of the process by minimizing the mean square error to obtain an optimal estimation of the dynamic system. The Kalman filter, which is based on Bayesian filter theory, can handle nonstationary noise, unlike the Wiener filter, which assumes stationary signals and noise with linear superposition. Moreover, compared with traditional adaptive filtering, Kalman filtering has a faster convergence speed and a lower computational complexity. However, accurate initial prior information needs to be input into the Kalman filter to achieve the optimal filtering result. This initial prior information includes the initial estimation error covariance $P^{-}_{k}$, initial state of a priori estimates $\hat{x}^{-}_{k}$, initial process noise covariance $Q$, and measurement noise covariance $R$. We also use adaptive smoothing to determine the Kalman gain because the state iteration of the Kalman filter is based on the previous step state estimation $\hat{x}^{-}_{k}$, measurement innovation $\left(z_k-H \hat{x}_k^{-}\right)$, and Kalman gain $K$. This allows us to adjust to changes in the covariance of the data. By choosing a certain size of the smoothing window to average the historical gain and combining it with a certain size of adaptive gain weight $K_{weight}$ to update gain $K$, the best possible next-state iteration is obtained. If the predicted Kalman gain increases, this signifies a high estimation uncertainty in the current system, allowing an appropriate reduction in the gain weight. If the predicted Kalman gain decreases, this indicates that the estimation of the system is more accurate, and the gain weight can be appropriately increased. Although our gain weight fluctuates in real time, accurate prior information is still required, which forms the fifth initial prior parameter in the BCKA filter. To accelerate the filtering pipeline, we initialize the BCKA prior parameters with a lightweight one-dimensional CNN trained on simulated folded data. The full network design, hyperparameter optimization, and the distance-function comparison used in training are presented in Appendix~\ref{app:cnn-details}.

We need to select different S/R calculation methods based on the observational data obtained from various observation methods. Notably, only one GW source is considered in this work, so different S/R calculation methods are selected essentially because different magnetic field distributions lead to different reference response profiles; that is, different pulsars are observed. Our S/R calculation is based on Monte Carlo sampling because judging the distribution through the BCKA filter is difficult \citep{Giordano2017AMC,2024arXiv240519747A,optuna_2019}. When we observe only the same pulsar, since the reference response profile is unchanged, the S/R of this signal can be obtained by calculating the mean $\mathbb{E}\left[x_i\right]$ and variance $\mathbb{V}\left[x_i\right]$ of the folding data relative to the data of reference response profiles through sufficient sampling $\mathrm{S/R}\left(x_i \right)= \mathbb{E}\left[x_i\right]/\sqrt{\mathbb{V}\left[x_i\right]}$. When we observe multiple pulsars, due to the different reference response profiles, we need to calculate the mean $\mathbb{E}\left[x_i\right]$ and variance $\mathbb{V}\left[x_i\right]$ of the observations of different pulsars through separate sampling and then consider them as the multiplication of one-dimensional Gaussian distributions equal to the number of pulsars to obtain the mean $\tilde{\mathbb{E}}\left[x_i\right]=\left(\frac{\mathbb{E}\left[x_1\right]}{\mathbb{V}\left[x_1\right]}+ \cdots+\frac{\mathbb{E}\left[x_N\right]}{\mathbb{V}\left[x_N\right]}\right)\tilde{\mathbb{V}}\left[x_i\right]$ and variance $\tilde{\mathbb{V}}\left[x_i\right]=\left( \frac{1}{\mathbb{V}\left[x_1\right]}+ \cdots+\frac{1}{\mathbb{V}\left[x_N\right]}\right)^{-1}$ after multiplication and then obtain the S/R: $\mathrm{S/R}\left(x_i \right)= \tilde{\mathbb{E}}\left[x_i\right]/\sqrt{\tilde{\mathbb{V}}\left[x_i\right]}$. Considering the multiplication of multiple one-dimensional Gaussian distributions, the reason can be analogous to the existence of GW signals in the observed data of pulsar A and the existence of GW signals in the observed data of pulsar B; then, what is the probability of the existence of GW signals in the observed data of pulsars A and B?

Therefore, the complete filtering process involves the following steps. First, we consider the division of the telescope operating frequency performed in our previous work. For a specific frequency bin, we perform cross-correlation matching between the selected reference response profile and the observed data to obtain data slices with high correlation. We then fold these slices together according to the correlation and temporarily discard data slices with a correlation smaller than the normalised correlation threshold. We set the threshold for the normalised correlation at 0.8, indicating a strong correlation. Then, we input the folded data into an optimised 1D convolutional CNN to obtain the initial parameters of the Kalman adaptive filter. The Kalman adaptive filter then receives the initial parameters and folded data, yielding the filtered signal and S/R. Representative examples of the data before and after BCKA filtering, which illustrate the pipeline's effect on folded mock data, are shown in Figures \ref{fig:BCKA-result-examples-1} and \ref{fig:BCKA-result-examples-2} of Appendix~\ref{app:cnn-details}. The corresponding quantitative sensitivity improvements are summarized in the Results section.

Therefore, the complete filtering process involves the following steps. First, we consider the division of the telescope operating frequency performed in our previous work. For a specific frequency bin, we perform cross-correlation matching between the selected reference response profile and the observed data to obtain data slices with high correlation. We then fold these slices together according to the correlation and temporarily discard data slices with a correlation smaller than the normalised correlation threshold. We set the threshold for the normalised correlation at 0.8, indicating a strong correlation. Then, we input the folded data into an optimised 1D convolutional CNN to obtain the initial parameters of the Kalman adaptive filter. The Kalman adaptive filter then receives the initial parameters and folded data, yielding the filtered signal and S/R. To illustrate the effect of the BCKA filter, representative pre- and post-filtering examples are included in Appendix~\ref{app:cnn-details}, specifically in Figures~\ref{fig:BCKA-result-examples-1} and \ref{fig:BCKA-result-examples-2}. The full quantitative outcomes, including the S/R enhancement and sensitivity limits, are reported in the Results section.

\section{Results}
\label{sec:results}

In interpreting the projected sensitivity, several idealized assumptions adopted in our forecasts should be noted, as they likely lead to optimistic detection limits. Specifically: (a) we presume perfect flux calibration and the complete excision of known radio-frequency interference in the simulated data; (b) the magnetospheric plasma and magnetic-field configurations are treated as quasi-static snapshots from particle-in-cell simulations, with the GW–photon conversion probability evaluated via ray propagation through these frozen structures rather than through a self-consistent dynamical evolution; (c) the observing schedule and system-equivalent flux density are simplified, using median values updated semiannually; and (d) the end-to-end pipeline is validated against stationary, Gaussian mock data (see diagnostic tests in Appendix~\ref{app:diagnostics}). Future studies incorporating realistic RFI environments and time-dependent magnetospheric models will be necessary to refine these estimates.

Regarding the gravitational-wave background, a distinction between early- and late-universe sources can be informed by the characteristic strain bound from Big Bang nucleosynthesis (BBN). Cosmological limits from BBN and the cosmic microwave background constrain the total integrated energy density in extra radiation, which sets an upper limit on the logarithmically integrated GW energy density \citep{2005PhRvL..95v1101A,2006PhRvL..97b1301S}. To compare this integrated bound with our sensitivity projections, we convert it into a corresponding characteristic-strain curve. This conversion assumes a reference spectral shape for $\Omega_{\mathrm{GW}}(f)$ and uses the standard relation $\Omega_{\mathrm{GW}}(f) = {2\pi^2}/{3H_0^2} \, f^2 \, h_c^2(f)$ to derive $h_c(f)$.

Finally, these GWs resonate with the magnetic field as they pass through a pulsar, producing potential radio signals that can be detected in the off-pulse window of the pulsar. Therefore, the source selection of pulsars is very important. The criteria for selecting observation candidates are as follows:\\
(a) The inverse Gertsenshtein effect of coherent state superposition is sufficiently strong, and the observed radio signal flux is sufficiently large.\\
(b) Within the observable sky area of the telescope, the astrometric information is sufficient to allow observations over a long period.\\
(c) Cross-correlation is used to distinguish signals from noise, so the power spectra of the radio background noise observed many times need to differ.\\

According to the above three criteria, our observation sources should have the following characteristics: a strong magnetic field, a close distance, and a large proper motion velocity. Based on these characteristics, we select two pulsars, PSR J1856-3754 and PSR J0720-3125, as our primary observation targets, which we believe to be the most sensitive sources for detecting VHF GWs, and then choose pulsars in order of distance, from near to far. However, proper motion information is lacking for most pulsars, and their proper motion at the same time needs to be determined; thus, we set these pulsars as secondary observation targets. Finally, we select some pulsars and magnetars whose magnetic field strength meets the candidate criteria. Although there is insufficient astrometric information and a longer observation time is required, there are more potential observational achievements, so we set these objects as subsidiary observation targets. For readability, we move the full star catalogue and the year-by-year telescope pointing geometry (evaluated on June 1st of each observational year) to Appendix~\ref{app:star_catalog}. The complete lists are given in Tables~\ref{tab:pulsar-list}, \ref{tab:telescope-parameters-1} and \ref{tab:telescope-parameters-2}, and are used throughout the sensitivity forecasts and observing-strategy comparisons presented below.

After selecting the primary observation targets, we use PIC simulations to simulate the magnetosphere of the two pulsars in three dimensions and measured the integrated pulse profiles in their radio bands to compare the simulation results with real observations. The simulation results are presented in Figures \ref{fig:PIC-1} and \ref{fig:PIC-2}, where we focus on showing slices of the magnetic field strength in the coordinate plane for subsequent calculations of the conversion probability from GWs to electromagnetic waves. Moreover, we also show the basic parameters of the selected pulsar, such as the angle between the rotation axis and the magnetic axis, the angle of the radio emission cone, and the angle between the observed LOS and the rotation axis. The basic parameters of the pointing angle observed by the telescope, such as the azimuth and altitude angle of FAST and SKA2-MID on June 1st of each year in the long-term observation plan, are also summarized in Tables \ref{tab:telescope-parameters-1} and \ref{tab:telescope-parameters-2}. As the only standard used to judge the quality of the simulation results, the integrated pulse profiles in the figure show that the simulation results represented by the red dotted line are basically consistent with the observational data represented by the black solid line. Therefore, we use these simulated pulsar magnetospheres as the magnetic field for subsequent calculations.

\begin{figure*}
	\centering
	\includegraphics[width = 0.43\textwidth]{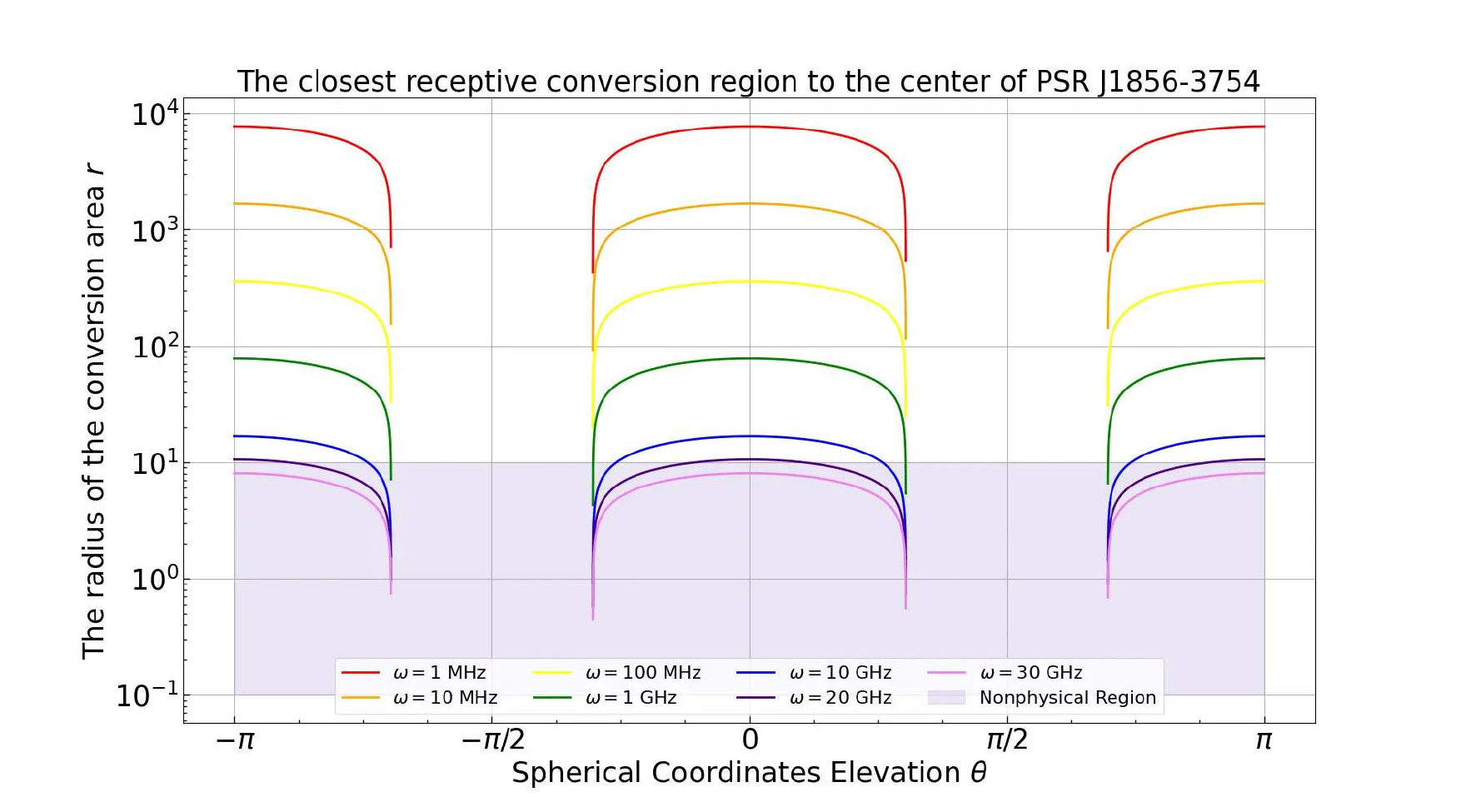}
	\includegraphics[width = 0.43\textwidth]{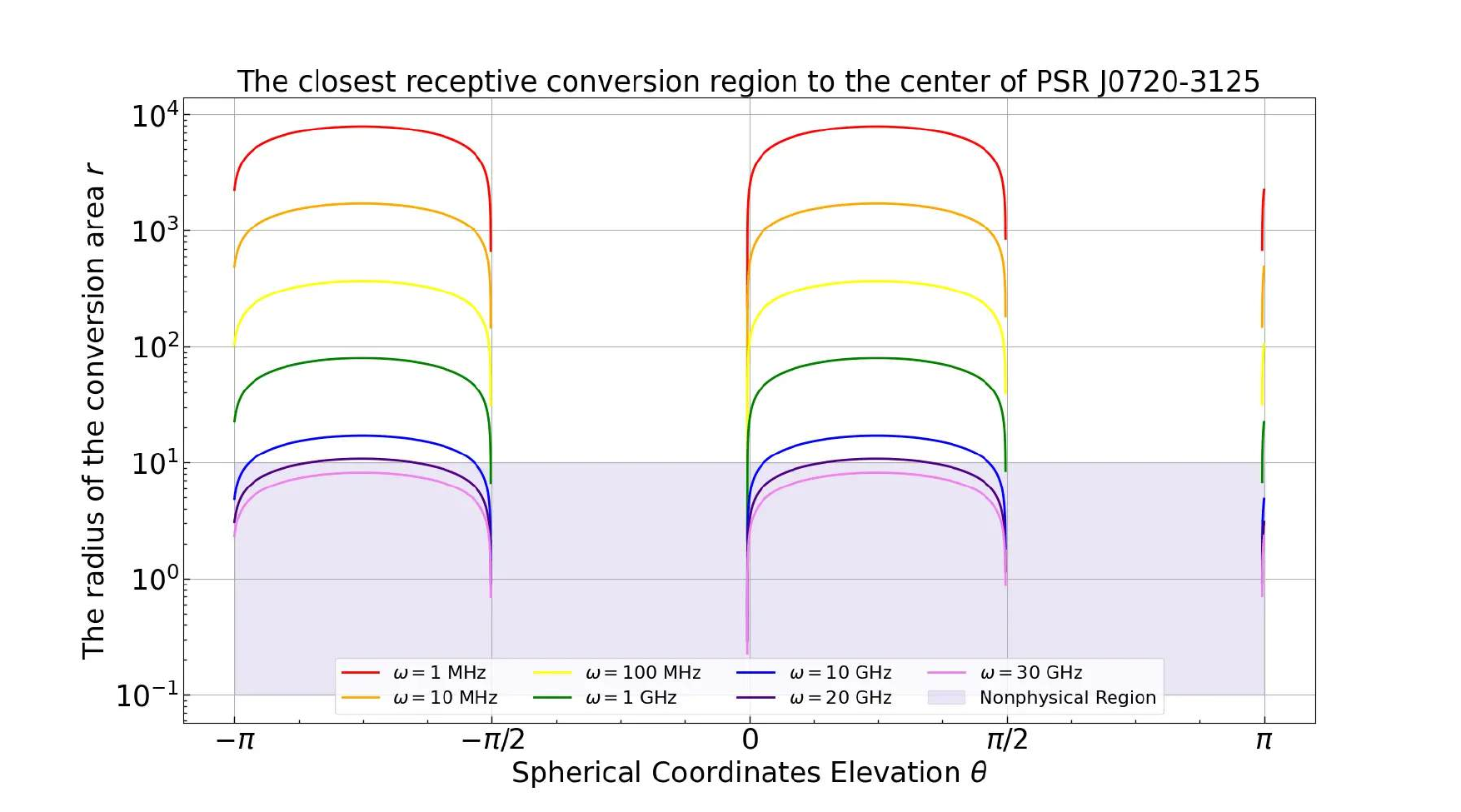}
	\includegraphics[width = 0.41\textwidth]{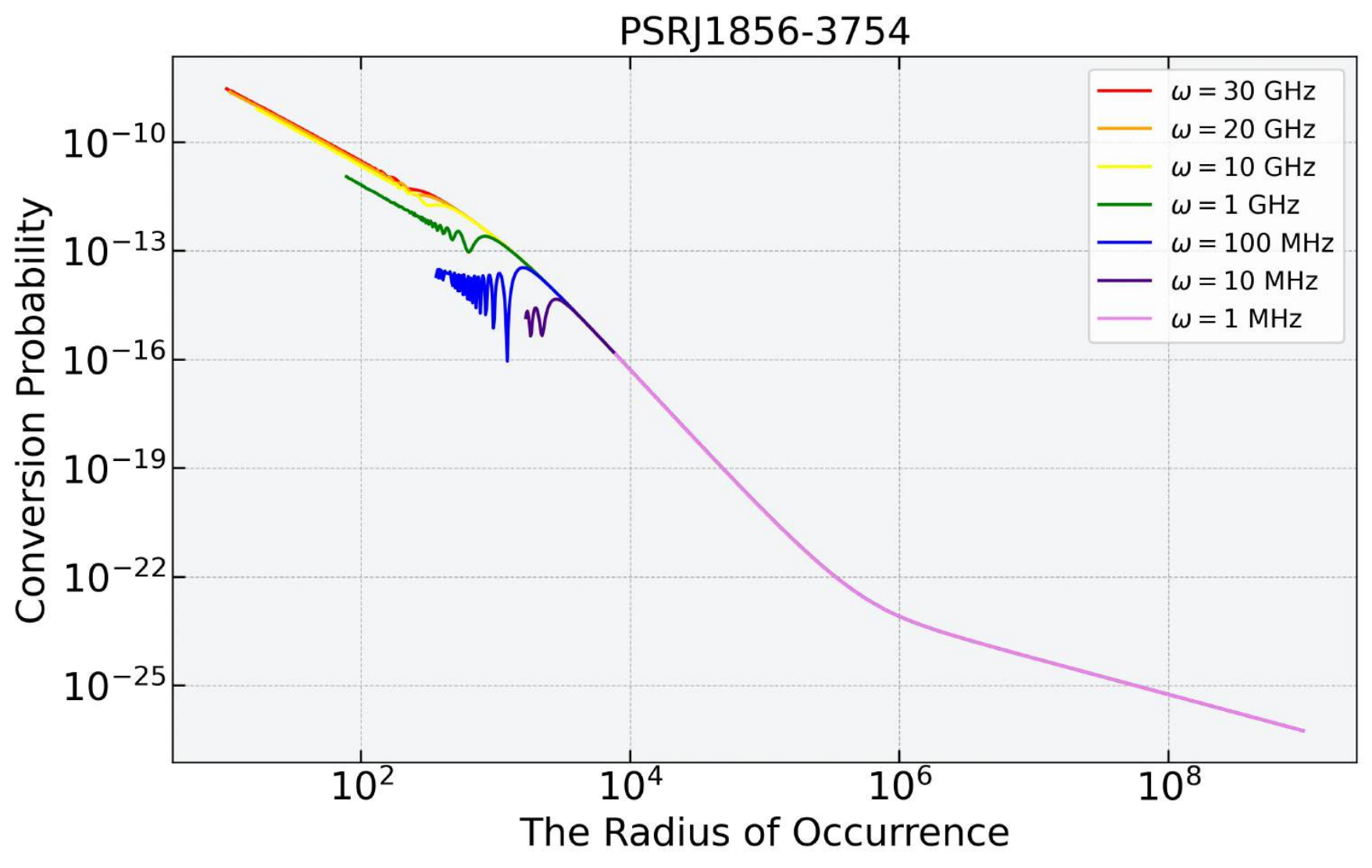}
	\includegraphics[width = 0.41\textwidth]{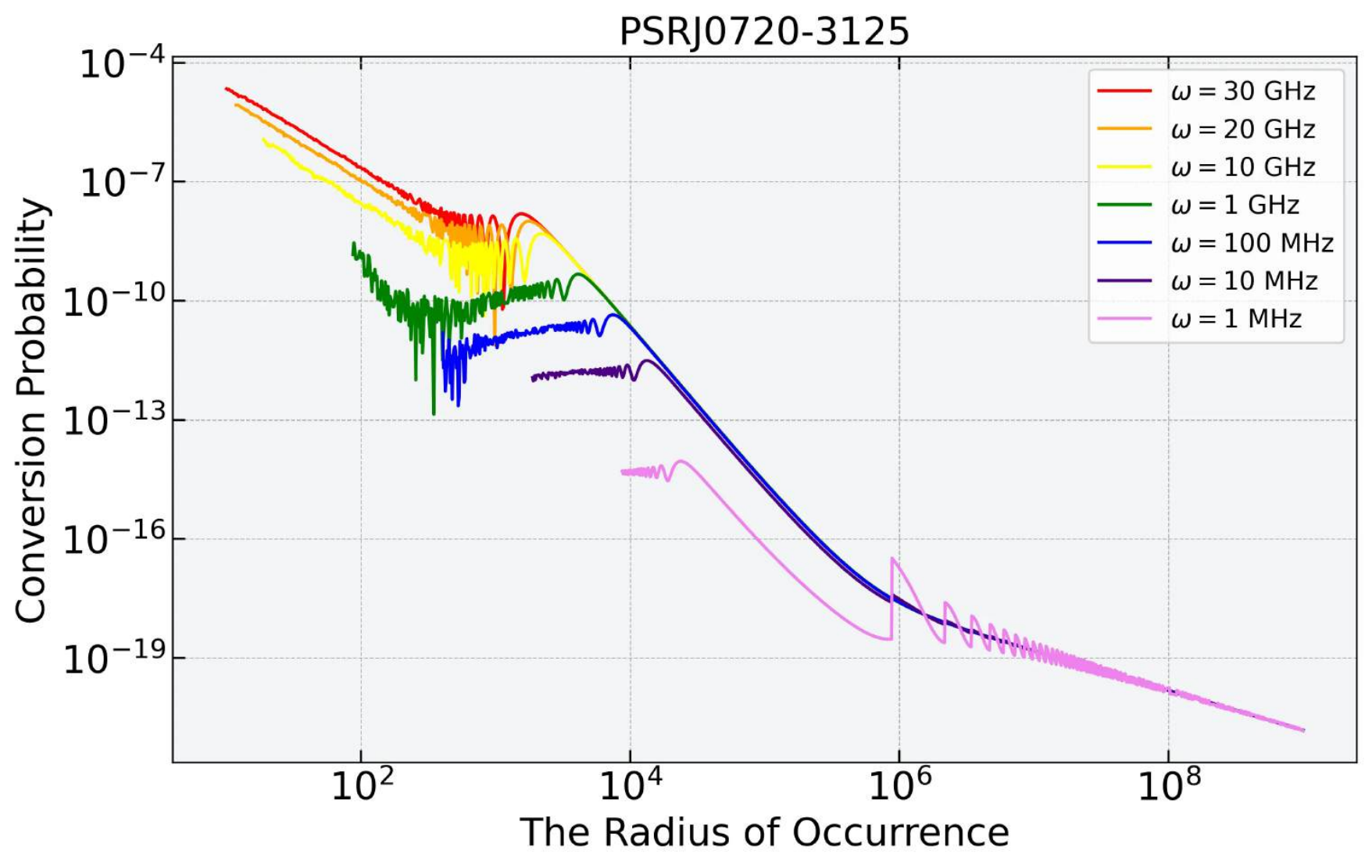}
	\includegraphics[width = 0.41\textwidth]{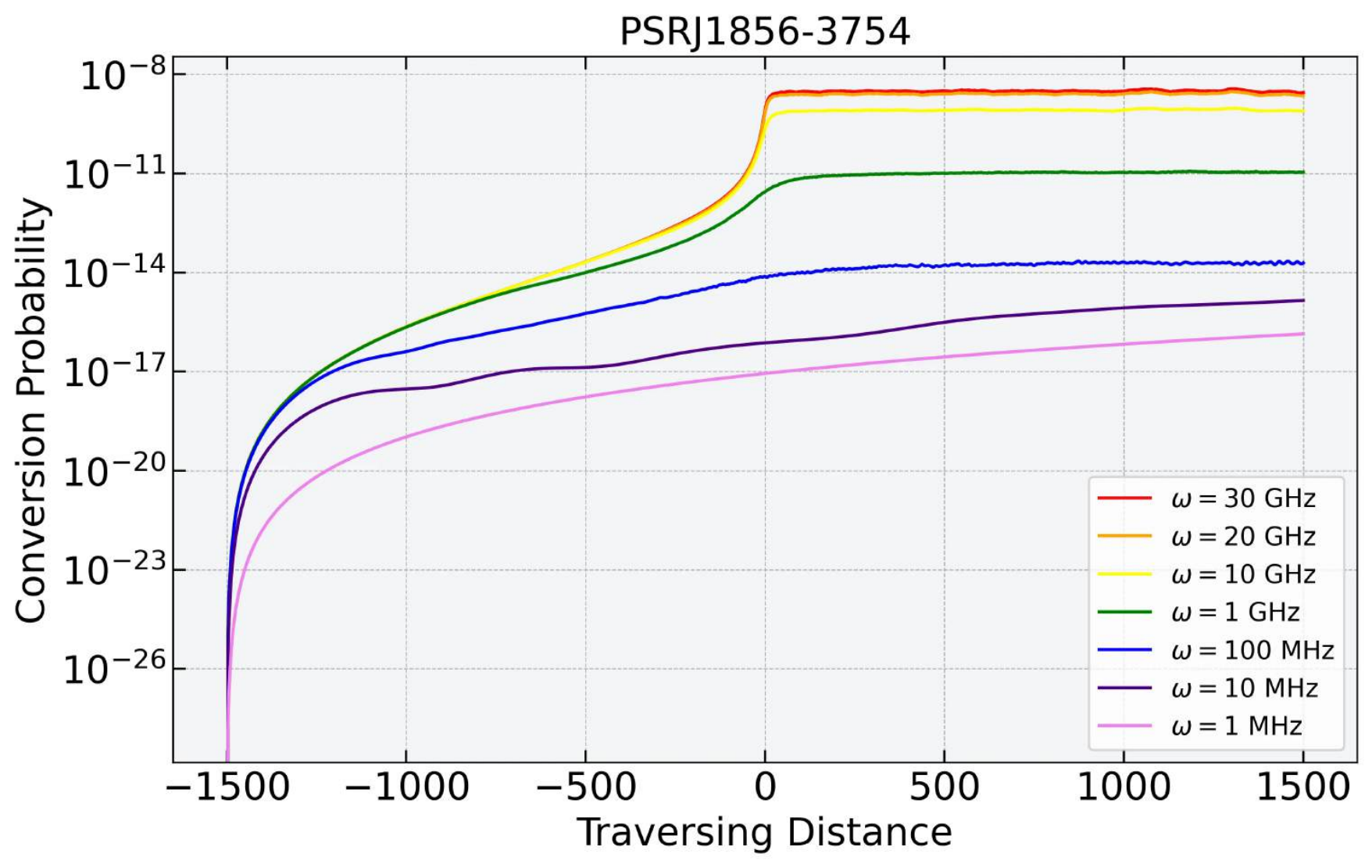}
	\includegraphics[width = 0.41\textwidth]{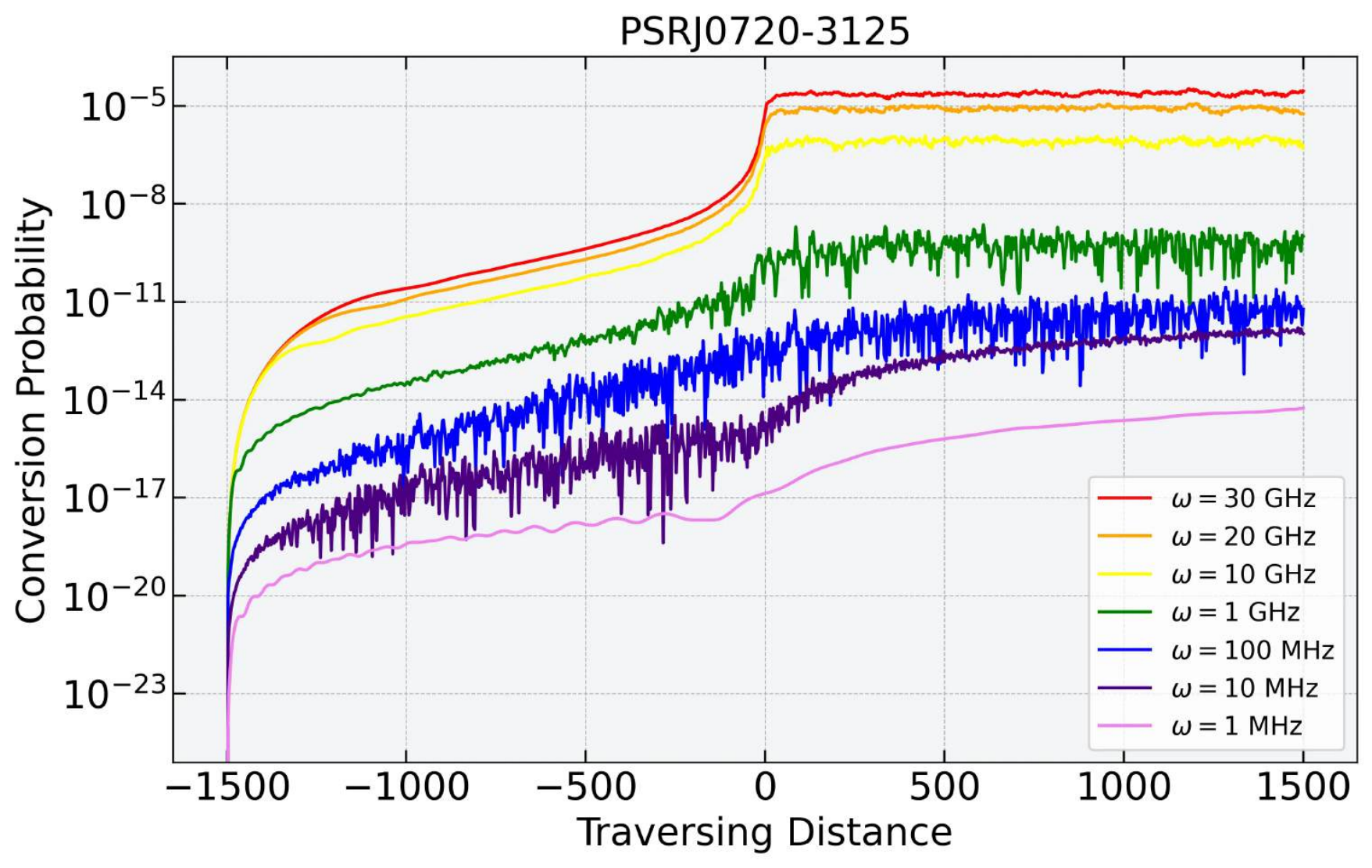}
	\includegraphics[width = 0.41\textwidth]{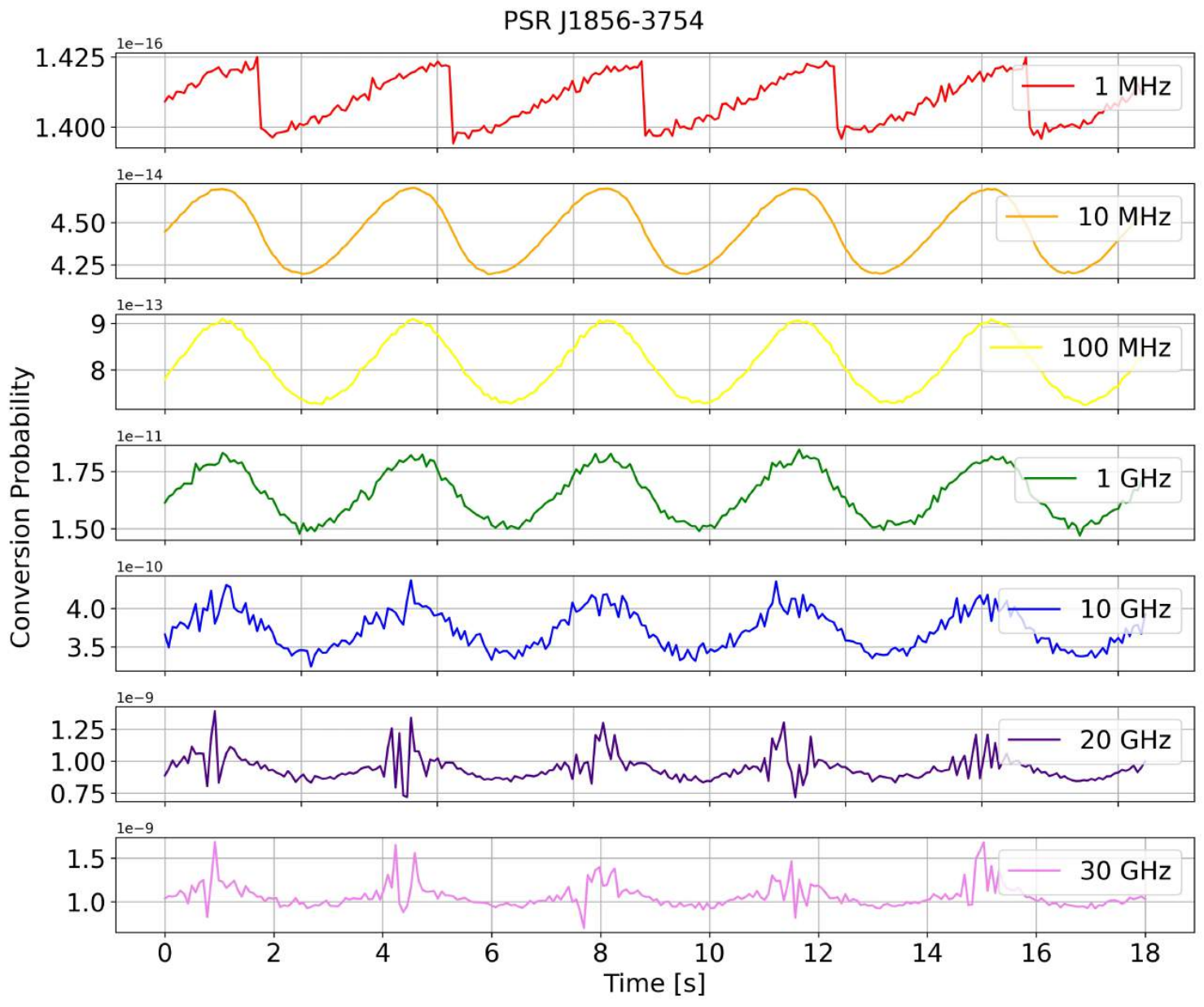}
	\includegraphics[width = 0.41\textwidth]{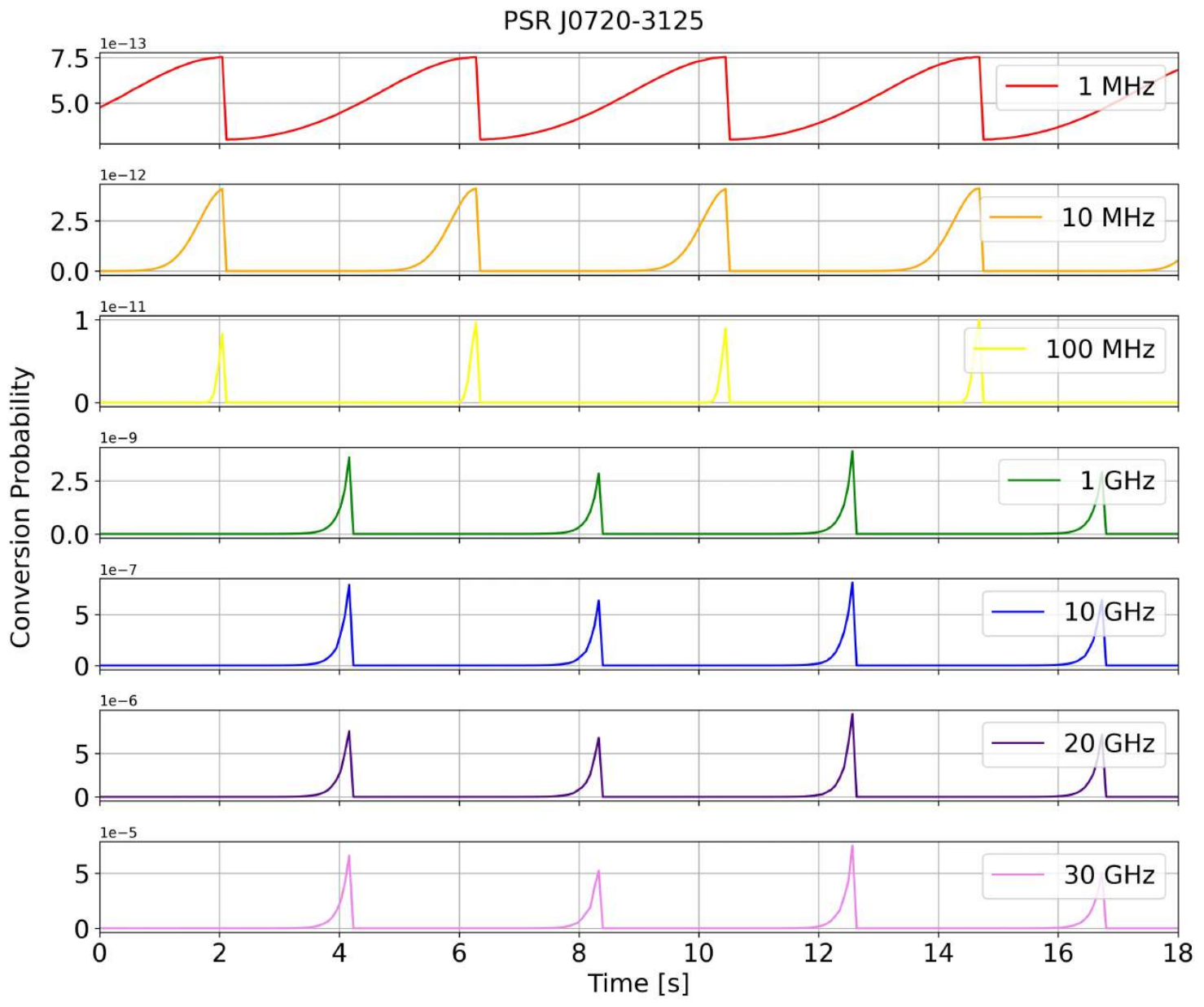}
	\caption{The inverse GZ effect in the radius of the nearest detectable signal-generating region in the pulsar magnetosphere away from the centre of the pulsar and the conversion probability of converting GWs to radio signals. The four left panel plot the results for PSR J1856-3754 and the four right panel depict the results for PSR J0720-3125. The top panel plots the region of permissible signal conversions for the two pulsars; we assume that the neutron star has a radius of $10\mathrm{~km}$, so regions smaller than $10\mathrm{~km}$ we have shaded in purple to indicate unphysical regions. The second panel plots the magnitude of the conversion probability at different positions of the two pulsars, in this plot we assume that the conversion occurs on the x-axis of the PIC simulation. The third panel plots the magnitude of the conversion probabilities at different crossing distances in the magnetic fields of the two pulsars, in this panel we assume that the conversion occurs in the region of occurrence closest to the neutron star. The bottom panel plots the magnitude of the conversion probabilities at different times in the magnetic fields of the two pulsars. In this panel, we assume that the conversion occurs in the region of occurrence closest to the neutron star. The plotted length of time corresponds to at least two rotation periods of the pulsar.}
	\label{fig:convesion-probability}
\end{figure*}

With a simulated magnetic field of the pulsar, we can calculate the conversion probability of a single GW to a single photon along the observed LOS. This calculation is very effective for GW transient events on short timescales since the total intensity superposition within the magnetosphere is usually not considered because the wavelength of the GW in the radio band is much shorter than the thickness of the pulsar magnetosphere, and determining where the GW crosses the pulsar magnetosphere is difficult. Therefore, the single-photon detection sensitivity of the instrument can be used as a lower limit for detecting GW transient events. In conjunction with our previous theoretical work \citep{Hong:2024ofh}, we present in the Methods section a formula for the sensitivity of radio signal detection of a single transient event varying with the distance to the pulsar, radial distance at which the GW passes through the centre of the pulsar, detected frequency, conversion probability, and telescope-equivalent flux density, and we plot the conversion probability of two pulsars as a function of the frequency, position through the pulsar magnetosphere, and distance travelled through the pulsar magnetosphere in Figure \ref{fig:convesion-probability}.

At this point, we can simulate what the radio signal from a GW transient event looks like and how it changes with frequency. To make the path noise in the simulated signal more pronounced, we assume that the characteristic strain of the GW varies from $h_c \approx 1.7 \times 10^{-19}$ to $h_c \approx 2.1 \times 10^{-19}$ and from $h_c \approx 1.4 \times 10^{-19}$ to $h_c \approx 1.7 \times 10^{-19}$ with the frequency of transient events. Considering the operating frequency, bandwidth, resolution, and time sampling interval of FAST and SKA2-MID and assuming that the GW enters the pulsar magnetosphere from the minimum radius of the radio signal, two-dimensional spectral diagrams \ref{fig:anticipated-signal-1} and \ref{fig:anticipated-signal-2} are drawn as an optimistic estimate. The initial signals without noise and not broadened by the interstellar medium and the Earth ionosphere are represented by red dashed lines. The sampled signal of an electromagnetic wave passing through the entire path is represented by a solid black line, and its corresponding time spread is represented by a dashed brown line. The initial signal in this figure, to which noise has not yet been added, is used as one of the reference response profiles of our many signals for subsequent cross-correlation signal extraction. Effects such as interstellar scintillation on the observed path are discussed in detail in the Methods section. The figure promisingly shows that the radio signal converted by GWs into the same frequency is very similar to the signal of fast radio bursts (FRBs) in shape and structure, and the time width of the signal is also very similar to the time width of FRBs. Moreover, in the high-frequency band of the optimistic estimation case, the signal generation region is also within the pulsar light cylinder, which coincides with the potential generation region of FRBs \citep{Zhang:2022uzl}. In addition, because our source is located in the Milky Way galaxy, the GW resonates with the electromagnetic wave and maintains the same degree of linear polarization until the GW penetrates the pulsar magnetosphere; we do not need to consider the dispersion measurement, rotation measurement and redshift effect of the host galaxy, extragalactic space, and observational source but only the dispersion measurement and rotation measurement of the interstellar medium in the Milky Way galaxy and the Earth ionosphere. Therefore, the depolarization model of the radio signal can be expressed as the multiplication of the depth depolarization, beam depolarization, and bandwidth depolarization. The theoretical linear polarization of the GW radio signal obtained from the two pulsars varies with the frequency, instrument conditions, and rotation measurements, as shown in Figure \ref{fig:depolarization-FRB} by different coloured lines. In addition, we select several repeated and single FRBs in the Milky Way with rotation measurements of approximately 20 and used our depolarization model to fit them; the fitting results are shown in Figure \ref{fig:depolarization-FRB}. The figure shows that our physical background and model are in good agreement with the repeated FRBs, but the restriction on the single FRBs is not ideal and is more inclined to the turbulent magnetic field environment. In summary, we believe that the physical process of conversion of VHF GWs into radio signals in pulsar magnetic fields is one of the potential physical origins of the repeated FRBs in the galaxy.

Unlike the calculation of the conversion probability from gravitational to electromagnetic waves for transient events, the detection sensitivity for persistent events due to the SGWB is much higher. Because the GWs produced in the early universe are diffuse, similar to the cosmic microwave background radiation, when they have the same frequency and pass through the pulsar magnetosphere in pairs, the resulting radio signal will be a coherent superposition along the direction of the observed real line in the region of the magnetic field. Under optimistic estimates, this phenomenon is reflected in the conversion probability as a value somewhat higher than the single-photon conversion probability and in the detection sensitivity as the possibility of observing GWs with lower characteristic strain. Similarly, we model what the radio signal produced by a GW long-duration event looks like and how it varies with frequency. With the same telescope setup as that for the observation of transient events, we assume that the characteristic strain of GWs varies with the frequency of the sustained event from $h_c \approx 1.2 \times 10^{-26}$ and $h_c \approx 3.0 \times 10^{-27}$ to $h_c \approx 1.5 \times 10^{-26}$ and $h_c \approx 3.8 \times 10^{-27}$, and present this two-dimensional spectrogram in Figures \ref{fig:anticipated-signal-1} and \ref{fig:anticipated-signal-2}. The subprofile of long-duration events also shows a bell shape with a period consistent with the pulsar rotation period. This behaviour occurs because as the pulsar rotates, the strength of the magnetic field at each location in the magnetosphere changes, the density of the number of charged particles changes, the coherent state of the electromagnetic wave subsequently changes, and the total conversion probability changes with time. The total conversion probability varies with time, and the integration profile periodically fluctuates over a long period, but not as much as that of the pulsar itself. Therefore, extracting it during processing of the observational data is difficult.

With these basic results of a single-simulation observation, we can discuss the use of different observation methods to improve the sensitivity of the detection, improve the S/R of the signal, and reduce the difficulty of extracting the signal from the massive amount of data. Since we did not make real observations, we use simulated data to validate our observation methods. The method for obtaining simulation data in this work is very simple: for a single observation duration of 6 hours, we assume that 1000 observations are planned, with approximately 7 years of total observation time, starting from June 2025 to 2032. Because actual observations also involve flux calibration of the telescope, observing the pulsar magnetosphere 24 hours a day is impossible. Therefore, we inject $1\sigma$ Gaussian observation noise as the initial signal by calculating the equivalent flux density of the telescope at different observation times. Notably, performing equivalent flux density calculations for every hour of every day of these seven years of observations is impractical, and the two pulsars do not move very fast on their own; similar to the approach in the previous theoretical work \citep{Hong:2024ofh}, the 50th quartile equivalent flux density is chosen and injected into the signal with semiannual updates. These noisy signals pass the stability and Gaussianity tests after signal shape pushed back and are used as ``observational data" to verify the four observation methods we propose next. And these estimates represent optimistic upper bounds assuming ideal interference-free environments.

The comparative performance of the four observational strategies, quantified by the S/R enhancement and the derived flux limits, is summarized in Table~\ref{tab:snr-summary}. The complete underlying statistical details are provided in Appendix~\ref{app:snr_scatter}, in the form of scatter distributions from all 500 resampling trials, shown in Figures~\ref{fig:SNR-500-1} through \ref{fig:SNR-500-4}.
\begin{table*}
	\caption{\label{tab:snr-summary}Summary of S/R improvements and minimum detectable flux densities for the four observing strategies. For each strategy we list the mean S/R before and after applying the BCKA filter, the corresponding improvement factor, and the resulting minimum detectable flux densities for the transient (6~s integration) and persistent (6~h integration) cases.}
	\centering
	\begin{threeparttable}
		\begin{tabular}{c c c c c c c}
			\toprule
			Strategy & System & $\langle\mathrm{S/R}\rangle_\mathrm{pre}$ & $\langle\mathrm{S/R}\rangle_\mathrm{post}$ & Improv. Factor & $F_{\min}^{\mathrm{tran}}$ (Jy) & $F_{\min}^{\mathrm{pers}}$ (Jy)\\
			\midrule
			SPST & FAST & 1.0 & 9.7 & 9.7 & $\sim 9.29\times10^{-6}$ & $\sim 3.75\times10^{-9}$ \\
			SPST & SKA2--MID & 1.4 & 13.5 & 9.6 & $\sim 3.72\times10^{-11}$ & $\sim 1.52\times10^{-14}$ \\
			\midrule
			SPMT & PSR~J1856$-$3754 & 1.2 & 14.8 & 12.3 & $\sim 6.84\times10^{-6}$ & $\sim 2.83\times10^{-9}$ \\
			SPMT & PSR~J0720$-$3125 & 1.2 & 16.0 & 13.3 & $\sim 2.64\times10^{-11}$ & $\sim 1.09\times10^{-14}$ \\
			\midrule
			MPST & FAST & 1.0 & 13.3 & 13.3 & $\sim 5.11\times10^{-7}$ & $\sim 4.92\times10^{-10}$ \\
			MPST & SKA2--MID & 1.4 & 18.9 & 13.5 & $\sim 2.64\times10^{-11}$ & $\sim 1.09\times10^{-14}$ \\
			\midrule
			MPMT & FAST$+$SKA2--MID, two pulsars & 1.2 & 21.6 & 18.0 & $\sim 9.82\times10^{-13}$ & $\sim 1.69\times10^{-15}$ \\
			\bottomrule
		\end{tabular}
		\begin{tablenotes}
			\footnotesize
			\item Notes: ``System'' indicates the specific telescope or pulsar to which the reported S/R values correspond (see the full distributions in Appendix~\ref{app:snr_scatter}). The improvement factor is defined as $\langle\mathrm{S/R}\rangle_\mathrm{post}/\langle\mathrm{S/R}\rangle_\mathrm{pre}$. Flux limits are derived by iteratively increasing the signal--noise PSD contrast until the detection threshold is reached across the simulated observation set.
		\end{tablenotes}
	\end{threeparttable}
\end{table*}

\subsection{Single Pulsar with a Single Telescope (SPST) Method.}
From the point of view of the observation operation, this method is the simplest method: we only need to continuously align the telescope to the magnetic field region of a certain pulsar, and we call this the ``Single Pulsar with a Single Telescope" (SPST) Method. Considering that the time scales of different types of signals vary, the typical integration time of each signal is also different. At the $1\mathrm{~GHz}$, the minimum detectable fluxes of FAST are $\sim3.10\times10^{-2}\mathrm{Jy}$ and $\sim2.98\times10^{-5}\mathrm{Jy}$, and the minimum detectable flux of SKA2-MID are $\sim1.20\times10^{-7}\mathrm{Jy}$ and $\sim1.15\times10^{-10}\mathrm{Jy}$, for detecting transient events with 6 seconds of integration time and persistent events with 6 hours of integration time. In other words, the minimum detectable flux density not only depends on the minimum flux density of the telescope, but is also closely related to the integration time of the observation. This means that, considering the $1 \sigma$ S/R, the minimum detectable flux density will be affected by the observation duration. Assuming that the power spectrum of the signal in the ``observed data'' is equal to the power spectrum of the noise after two consecutive ``observations", we randomly select two sets of data from the generated 1000 sets of data with the same power spectrum level, input them into the initialized Bayesian cross-correlation Kalman adaptive (BCKA) filter and select the SS mode of the filter. After 500 rounds of nonrepeated sampling of 1000 sets of observational data, the S/R of the processed signal is significantly improved, and the average S/R is increased from 1 and 1.4 to 9.7 and 13.5. The full distribution of results from the 500 resamplings is provided in Appendix~\ref{app:snr_scatter} (Figure~\ref{fig:SNR-500-1}), while the key summary statistics are compiled in Table~\ref{tab:snr-summary}. Thus, we can further intensify the difference in the power spectrum between the signal and noise, resimulate the observed data, and then apply the BCKA filter again to obtain a new S/R. We repeat this procedure until we have used up the 1000 sets of observational data and can no longer intensify the difference in the signal and noise power spectra. At this moment, the signal flux density at SNRs of 1 and 5 is recorded as the lowest detectable fluxes. Therefore, the minimum detectable fluxes in the SPST method, respectively, are $\sim9.29\times10^{-6}\mathrm{Jy}$, $\sim3.75\times10^{-9}\mathrm{Jy}$, $\sim3.72\times10^{-11}\mathrm{Jy}$ and $\sim1.52\times10^{-14}\mathrm{Jy}$, and the results of the variation in the minimum detectable GW characteristic strain with the pulsar parameters, telescope parameters, and frequency are shown in Figure \ref{fig:sensitivity}.
\subsection{Single Pulsar with Multiple Telescopes (SPMT) Method.}
The basic process is similar to that of the SPST method, except that when observing with multiple telescopes, the overall S/R and detection sensitivity for the same type of GW are determined by the lower sensitivity of the multiple telescopes. We call this method as ``Single Pulsar with Multiple Telescopes" (SPMT) Method, and it requires separate initialization of the optimal filter parameters and processing of the data using the SM mode of the BCKA filter. Similarly, after randomly selecting the observational data twice, the signal S/R is significantly improved, and the mean S/R increases from 1.2 and 1.2 to 14.8 and 16.0. For the analysis below, we refer to the summary results in Table~\ref{tab:snr-summary}. The underlying detailed distribution (Figure~\ref{fig:SNR-500-2}) from the resampling ensemble is included in Appendix~\ref{app:snr_scatter}. In addition, by continuously adjusting the S/R of the observed data, the minimum detectable flux in the SPMT method can be obtained as $\sim6.84\times10^{-6}\mathrm{Jy}$, $\sim2.83\times10^{-9}\mathrm{Jy}$, $\sim2.64\times10^{-11}\mathrm{Jy}$ and $\sim1.09\times10^{-14}\mathrm{Jy}$, respectively. Moreover, the results of the variation in the minimum detectable GW characteristic strain with the pulsar parameters, telescope parameters, and frequency are shown in Figure \ref{fig:sensitivity}.
\subsection{Multiple Pulsars with a Single Telescope (MPST) Method.}
This method can be considered an extension of the SPST method, in which the GW reference response profiles corresponding to different pulsars are used to cross-verify the results and improve the reliability of signal screening. We call this method as ``Multiple Pulsars with a Single Telescope" (MPST) Method. In this case, the overall S/R and detection sensitivity for the same GW are determined by the low conversion probability in multiple pulsars and the distance from Earth. We find that after two random selections of observational data, the S/R of the signal is significantly improved, and this improvement is more obvious than that in the SPST method. The mean S/R increases from 1 and 1.4 to 13.3 and 18.9, We show the complete distribution from the 500 resamplings in Appendix~\ref{app:snr_scatter} (Figure~\ref{fig:SNR-500-3}). The key values used in the following discussion are extracted from this distribution and are summarized in Table~\ref{tab:snr-summary}. The S/R of the data processed by the BCKA filter in MS mode is more concentrated in the high S/R region. Similarly, we can respectively obtain the lowest detectable fluxes in the MPST method as $\sim5.11\times10^{-7}\mathrm{Jy}$, $\sim4.92\times10^{-10}\mathrm{Jy}$, $\sim2.64\times10^{-11}\mathrm{Jy}$ and $\sim1.09\times10^{-14}\mathrm{Jy}$, and the corresponding minimum detectable GW characteristic strain curves are shown in Figure \ref{fig:sensitivity}.
\subsection{Multiple Pulsars with Multiple Telescopes (MPMT) Method.}
\begin{figure*}
	\centering
	\includegraphics[width = 0.45\textwidth]{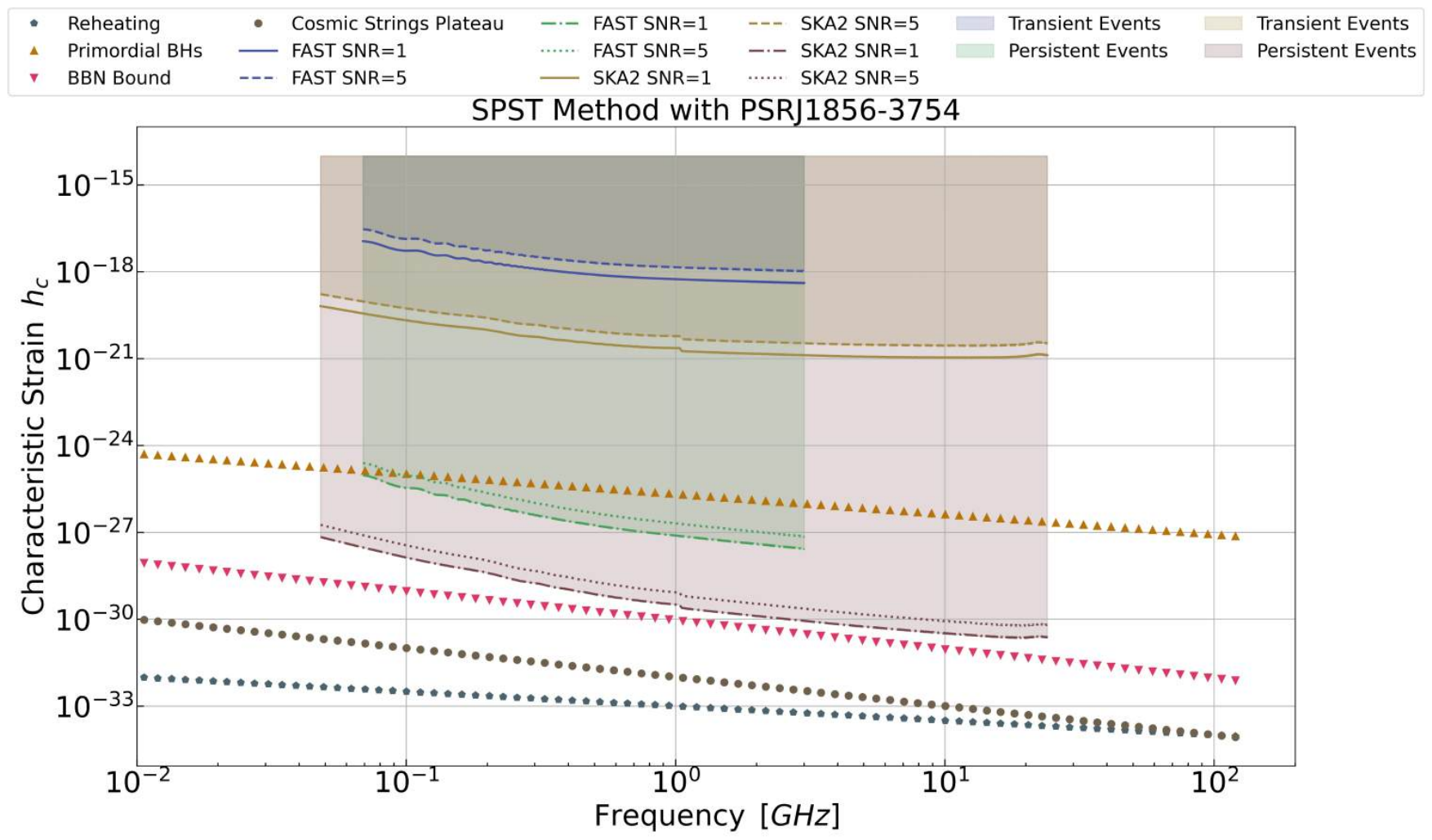}
	\includegraphics[width = 0.45\textwidth]{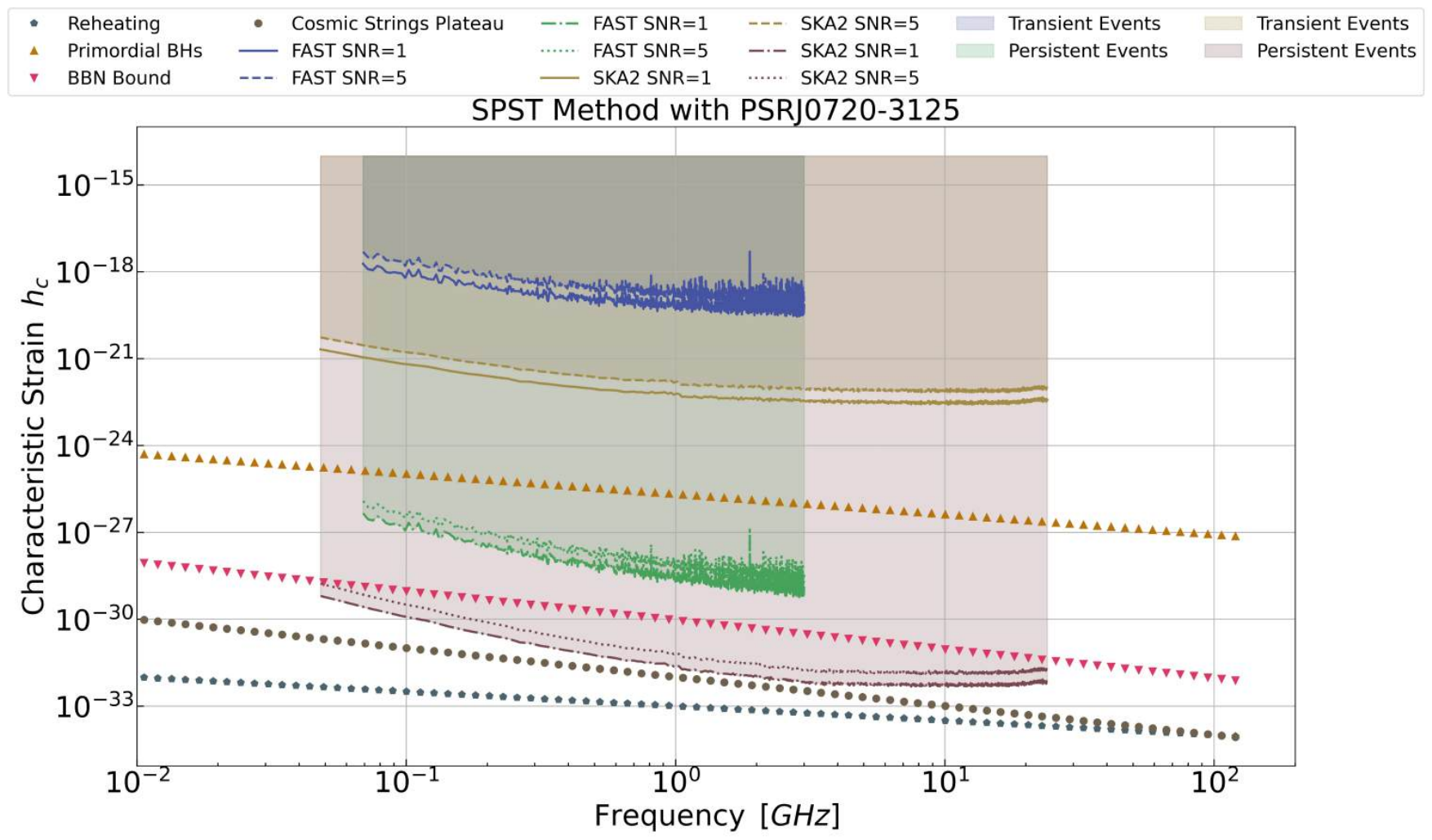}
	\includegraphics[width = 0.45\textwidth]{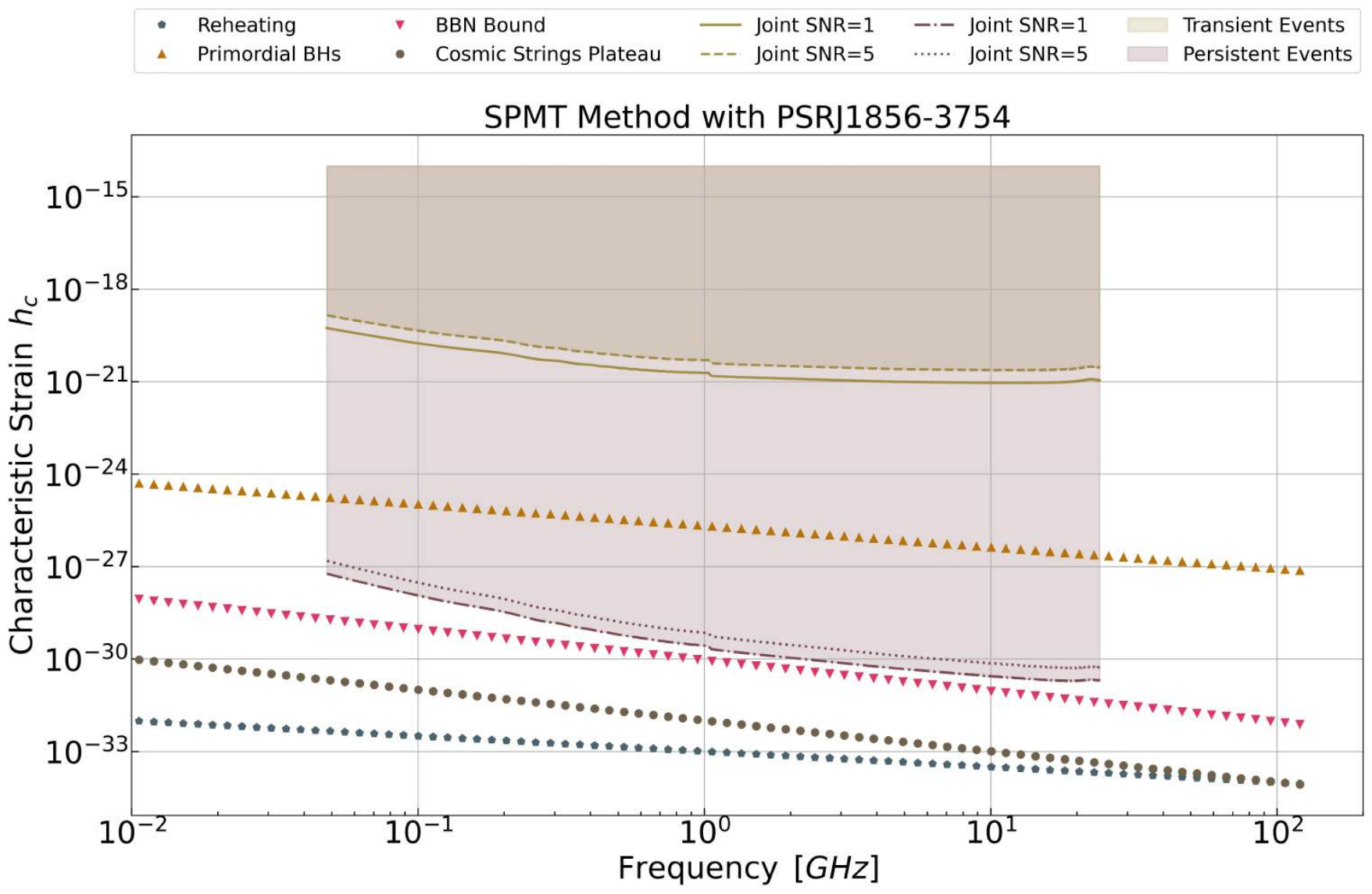}
	\includegraphics[width = 0.45\textwidth]{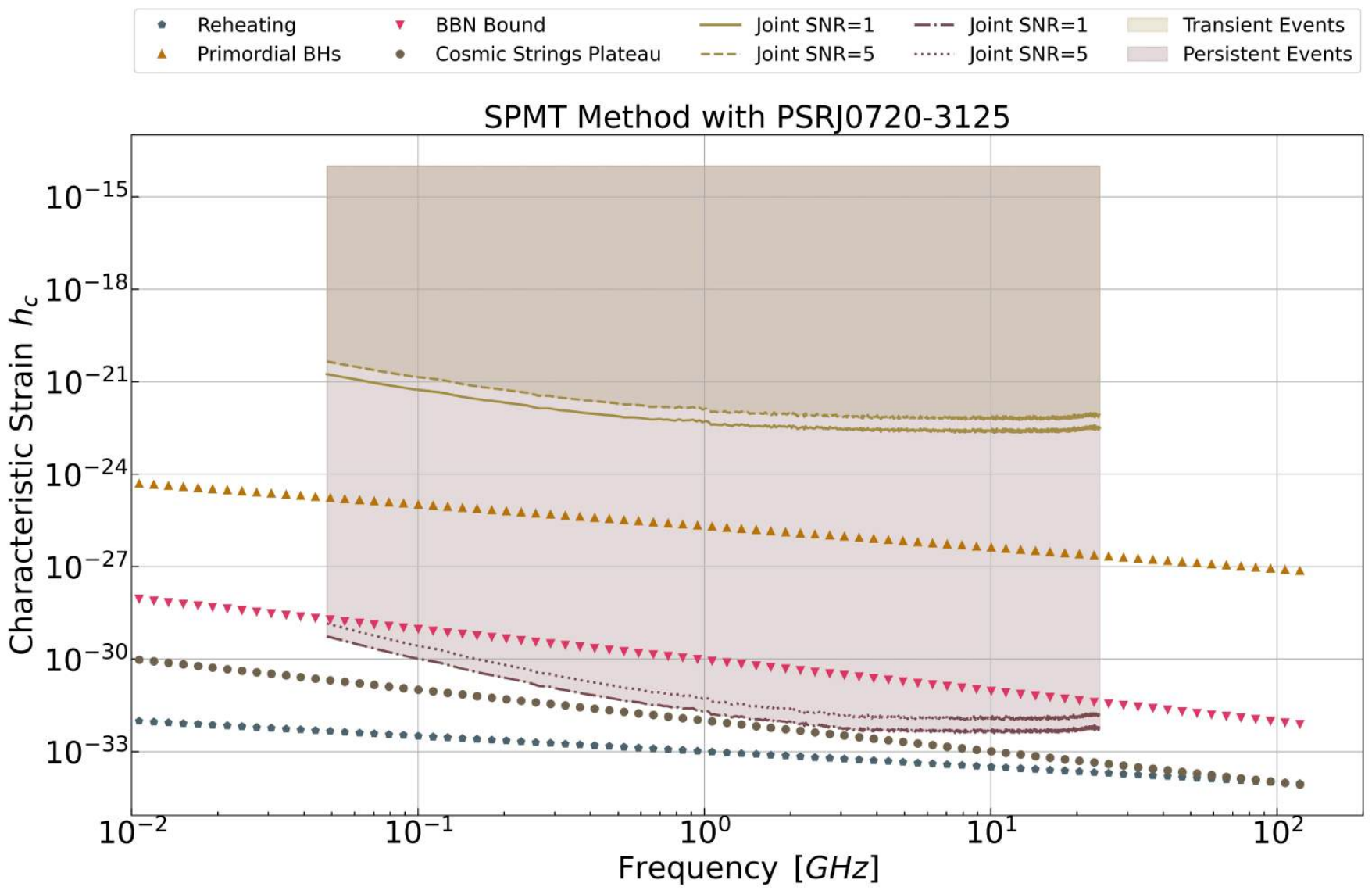}
	\includegraphics[width = 0.45\textwidth]{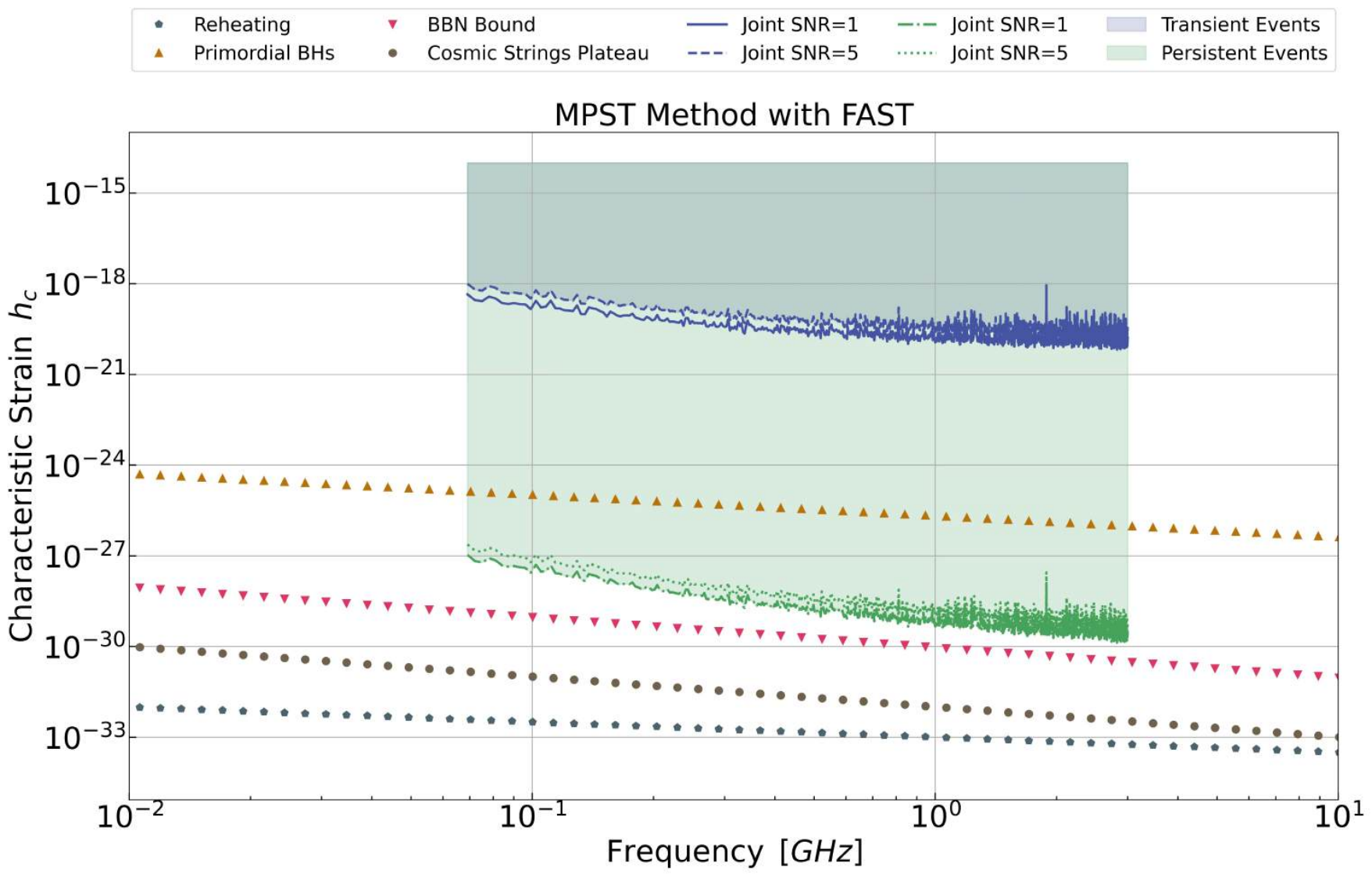}
	\includegraphics[width = 0.45\textwidth]{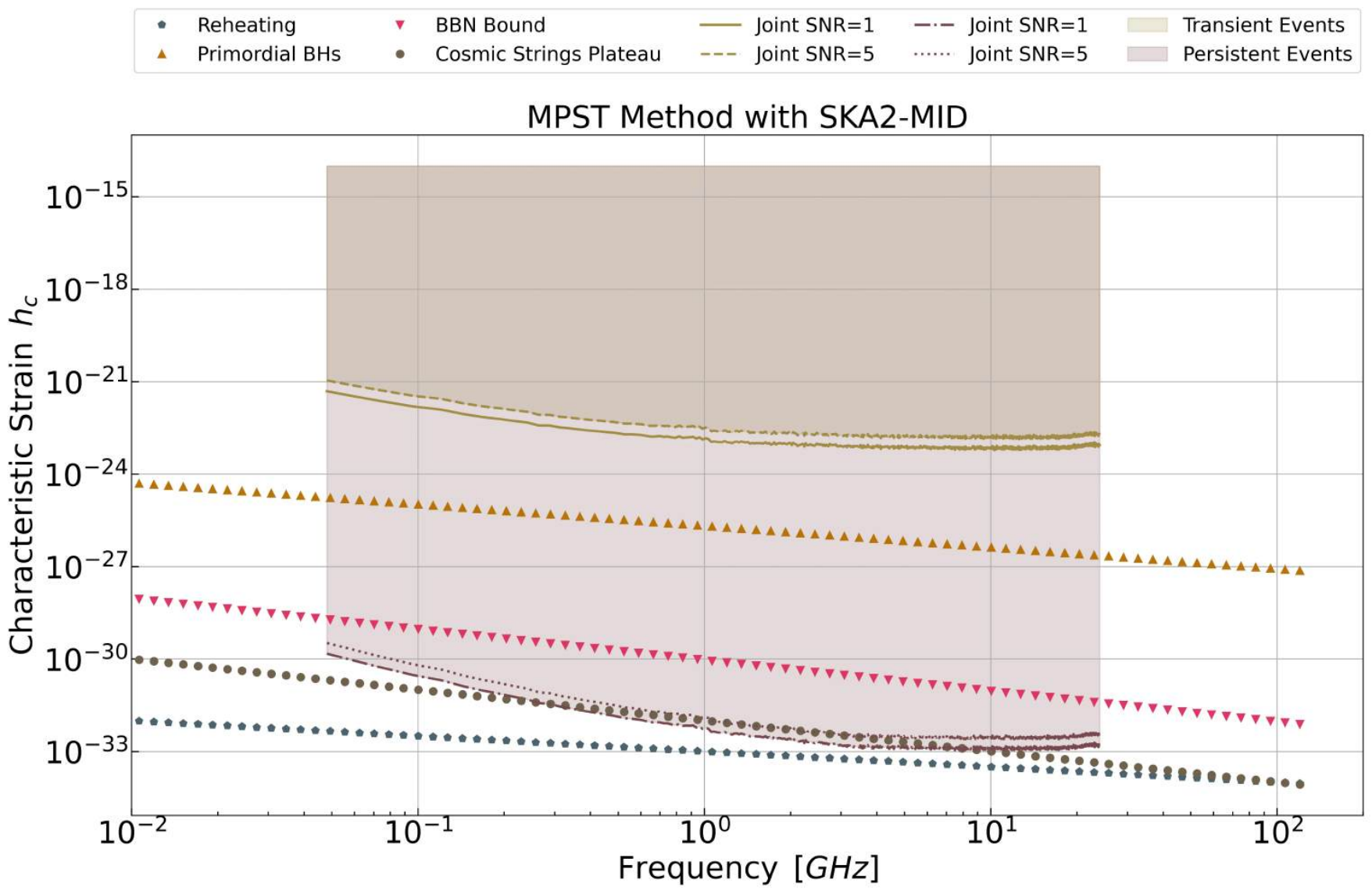}
	\includegraphics[width = 0.5\textwidth]{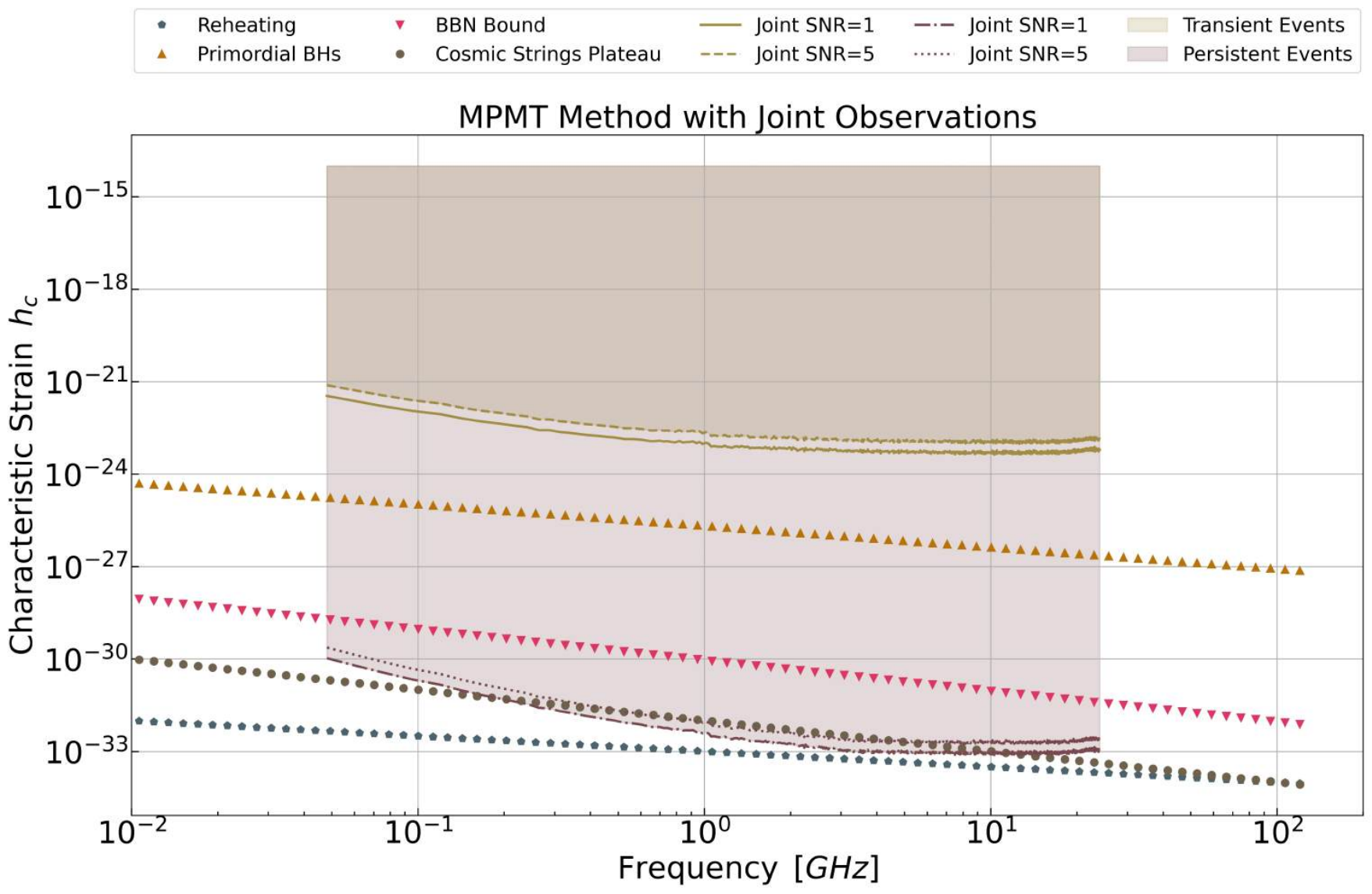}
	\caption{Upper bounds on the VHF GWs derived from four observational method for transient and persistent events. From the top panel to the bottom panel, we show the sensitivity of the four methods SPST, SPMT, MPST, and MPMT in turn. The four typical VHF GW sources are represented by lines of different colours and shapes in the figure; shaded areas of different colours indicate different types of events, while solid, dashed, and dotted lines of different colours indicate the detection sensitivity in different cases.}
	\label{fig:sensitivity}
\end{figure*}

This method is the most time-consuming of the four methods, as every telescope must observe every pulsar, but it most significantly improves the S/R of the final signal, and the additional observations are richer and more statistically relevant. This method is called ``Multiple Pulsars with Multiple Telescopes (MPMT)" Method. In essence, this method is a combination of the SPMT and MPST methods, so the minimum detectable flux is determined by both the pulsar source and the telescope. For example, in our current study, the lowest GW detection sensitivity is determined by PSR J1856-3754 observed by FAST. In this method, the observed data are processed by the BCKA filter in MM mode, and the S/R is calculated. Since both the pulsar and the telescope increase the sensitivity at the same time, the S/R of the signal is significantly improved after two random observations are selected without repeating the data. To provide full context for the summary statistics in Table~\ref{tab:snr-summary}, the corresponding distribution from the resampling procedure is presented in Appendix~\ref{app:snr_scatter} (Figure~\ref{fig:SNR-500-4}), and the mean S/R increases from 1.2 to 21.6. Finally, the lowest detectable fluxes, $\sim9.82\times10^{-13}\mathrm{Jy}$ and $\sim1.69\times10^{-15}\mathrm{Jy}$, in the MPMT method can be obtained, and the corresponding minimum detectable GW characteristic amplitudes are plotted together in Figure \ref{fig:sensitivity}. It is in this conditional sense that our Abstract claims the forecasts “approach sensitivity”. The claim refers to the fact that, under the most favorable integration and efficiency assumptions adopted in this study, our projected sensitivity limits reach into the parameter space occupied by several benchmark theoretical models presented in the comparison figures.

\section{Discussion and Conclusions}
\label{sec:discussion}
In this work, we establish a reproducible theoretical and numerical framework for the detection of radio-band gravitational waves through GW-electromagnetic wave conversion in pulsar magnetospheres. By connecting first-principles plasma simulations with realistic telescope performance, we provide quantitative predictions for FAST and SKA2-MID and evaluate tailored observational strategies that substantially lower the detection threshold for both transient and stochastic VHF GWs. Our analysis demonstrates that statistical validation and custom filtering techniques are essential for extracting weak signals embedded in astrophysical noise. Beyond methodological advances, this study also highlights broader scientific implications: probing primordial GWs from the early universe, constraining compact object mergers such as primordial black holes, and offering a potential physical explanation for repeating fast radio bursts. Together, these results not only enhance the feasibility of radio-band GW detection but also open new avenues for exploring fundamental questions in cosmology and astrophysics.

The methodological choices behind our sensitivity projections become clearer when compared with other approaches in this field. This area of research is growing rapidly, with several different ideas being explored. One common approach uses the average radio signal from the whole sky, together with models of large-scale cosmic magnetic fields, to set limits on gravitational waves \citep{Domcke:2020yzq, Domcke:2023qle}. Another, very conservative method, simply checks that the radio photons produced by GW conversion in any magnetic field would not outshine what our telescopes have already seen, giving broad limits across many frequencies \citep{Ito:2023nkq}. Other researchers focus on neutron stars, either by combining signals from many stars in our Galaxy or by using data from different types of telescopes looking at the same object \citep{Dandoy:2024oqg, 2024PhRvD.110j3003M}. While our work also looks at neutron-star magnetospheres, our core idea is different. We don't just reinterpret old data to set a limit. Instead, we plan a new kind of dedicated search: what if we use powerful telescopes like FAST and SKA2-MID to stare at carefully chosen pulsars for a very long time? This changes the problem completely. The main challenge is no longer about perfectly modeling the entire radio sky or every neutron star in the galaxy. Instead, it becomes an engineering and strategy question: how low can we push the noise floor through extremely long, careful observations? This is why our projected sensitivity improves. It comes from focusing everything—telescope time, signal processing, target selection—on the goal of finding this one specific type of weak signal. It's a shift from setting general bounds to planning a precise hunt. A detailed, side-by-side comparison of these different approaches is provided in Table~\ref{tab:comparison} (Appendix~\ref{app:comparison}).

Our observation method essentially increases the sensitivity of detection by utilizing a long observation time. Thus, the more telescopes we can use and the more pulsars we can observe, the greater the improvement. The inverse GZ effect highly depends on the magnetic field structure, so a more refined model of the pulsar magnetosphere is necessary. In this work, we indistinct the fine structure of the magnetosphere to reduce the use of supercomputing resources, which affects the fine structure of the signal converted by GWs into electromagnetic waves, but the overall shape of the signal does not change much because it is a manifestation of the path integral. In turn, the dependence of the inverse GZ effect on the magnetic field may be one of the potential ways to observe the structure of the magnetic field when we are sure that we have observed a certain kind of GW.

The nearly 6,000 hours of observation time recommended by the four observation methods does not necessarily have to be fully utilized. As long as gravitational wave signals are successfully observed in the radio band, the observation can be ended, but the S/R at this time may be lower than expected. Conversely, we can achieve a higher S/R by increasing the observation time, or detect gravitational waves with a lower energy density under a fixed S/R condition. The essence of signal extraction is the matching of reference response profiles, so the qualities of the reference response profile and filter determine the quality of the signal obtained by the final processing. Since radio frequency interference is closely related to actual observations, the robustness of the filter mentioned in this work needs to be further verified in a real environment, which will be one of the focuses of our subsequent work. Furthermore, given that the GZ effect has a strong dependence on the shape and intensity of the magnetic field, how to observe and reconstruct the pulsar magnetic field based on the model is also a key direction for our future research.

Owing to the scarcity of observed extragalactic pulsars, coupled with the inverse decay of the electromagnetic wave intensity with distance squared, we choose pulsars in the galaxy relatively close to the Earth as observation candidates. Therefore, our signal depolarization model can be applied only within the Milky Way. For extragalactic sources, more rotation measure (RM) components need to be added, and more complex propagation path effects need to be considered. In addition, the existing FRB observation results are almost evenly distributed in the observed sky area, whereas the distribution of pulsars in the Milky Way is closely related to the density of stars. Where the density of stars is high, there will be more pulsars, and their distribution is not as uniform as that of FRBs. Therefore, the combination of GW transient events with the inverse GZ effect should only be a potential origin of a class of repeated FRBs. Moreover, because the GW resonating with the magnetic field is linearly polarized, the circularly polarized part of the FRB cannot be explained by the inverse GZ effect, which is also why our model does not fit the nonrepeated FRBs well.

In the upcoming FAST and SKA2-MID joint observations, the extremely low system temperature and exceptionally high resolution will improve the detection environment of GWs in the radio band, provide effective exploration of primordial GWs in the unexplored era of the early universe, and enable effective support for the mass range exclusion line of primordial black holes in exotic celestial bodies. Together, these advances and our analysis provide a clear, reproducible roadmap from calibrated observations to statistically robust filtering, culminating in a definitive detection or substantially tighter constraints.

Our projected improvement relative to existing graviton--photon conversion forecasts in the literature
(e.g., \citep{Domcke:2020yzq,Ito:2023nkq,Herman:2022fau,Dandoy:2024oqg,2024PhRvD.110j3003M})
is driven primarily by the combination of (i) enhanced conversion probability
$\langle P_{g\rightarrow\gamma}\rangle$ in pulsar magnetospheres with strong, structured magnetic fields along the LOS, and
(ii) a long-integration, multi-target/multi-instrument observing strategy (MPMT) together with the BCKA processing pipeline, which lowers the effective flux threshold $F_{\min}$.
This scaling is already explicit in our sensitivity relations (Eqs.~5--6), where $h_c^{\min}$ decreases with decreasing $F_{\min}$ and increasing $\langle P_{g\rightarrow\gamma}\rangle$.
Consistent with the idealizations listed at the beginning of this section, the most optimistic curves in Fig.~8 should be interpreted as upper-bound projections under simplified RFI-free conditions.

\begin{acknowledgments}
	We are grateful for the referee's insightful and useful comments, which greatly helped us improve our manuscript. We thank Fang-Yu Li, Jing Niu, Kang Jiao, and Jie-Feng Chen for useful discussions. Wei Hong is grateful to the staff of the particle-in-cell Smilei for their kind help. We acknowledge Beijing PARATERA Tech Co., Ltd., for providing HPC resources that have contributed to the research results reported within this work. This work was supported by National SKA Program of China, No.2022SKA0110202, and the China Manned Space Program with grant No. CMS-CSST-2025-A01. Shi-Yu Li was supported by Science Program of Beijing Academy of Science and Technology (24CD014). Wei Hong was supported by China Scholarship Council (File No. 202506040057). This research used pulsar data from the ATNF Pulsar Catalogue and the McGill Online Magnetar Catalog and astrometric information from the Gaia mission.
\end{acknowledgments}

\begin{contribution}
	T.-J.Z. and W. H developed the idea of Very High-Frequency Gravitational Waves; P. H,  S.-Y.L.  and P.W. contributed to the scientific analysis and discussion of Very High-Frequency Gravitational Waves observation on FAST; W. H drafted the work; All authors contributed to the final version of the manuscript.
\end{contribution}

\software{
	The code {\tt Smilei} is a Particle-In-Cell code for plasma simulation \href{https://smileipic.github.io/Smilei/}{https://smileipic.github.io/Smilei/}, the codes for simulating neutron star magnetospheres are available at \href{https://zenodo.org/records/15680536}{https://zenodo.org/records/15680536},\\
	the code {\tt CASA} is the Common Astronomy Software Applications package \\ \href{https://casa.nrao.edu/}{https://casa.nrao.edu/}, the codes for simulating atmospheric parameters of the radio telescope station and the system-equivalent flux density are available at \href{https://zenodo.org/records/15680536}{https://zenodo.org/records/15680536},\\
	the {\tt ULSA} code can simulate the radio sky at frequencies below 10 MHz and is available at \href{https://zenodo.org/records/4663463}{https://zenodo.org/records/4663463}, \\
	the {\tt SCINTOOLS} is a package for the analysis and simulation of pulsar scintillation data \href{https://github.com/danielreardon/scintools}{https://github.com/danielreardon/scintools}, 
	the {\tt IonFarRot} is a code that allows us to predict the ionospheric Faraday rotation for a specific line-of-sight, geographic location, and epoch \href{https://sourceforge.net/projects/ionfarrot/}{https://sourceforge.net/projects/ionfarrot/},\\
	and the {\tt Optuna} is framework agnostic and can use it with machine learning \href{https://optuna.org/}{https://optuna.org/}.
}

\appendix
\section{Extended Methods}\label{app:methods-extended}

\subsection{PIC setup and numerical implementation}\label{app:pic-setup}
This subsection provides the numerical details of the PIC magnetosphere setup and diagnostics. To maintain the narrative flow in the main text, these specifics have been condensed from Section~\ref{sec:PIC-simulation-NSM} and are presented here for completeness.

The mass $M$ of a pulsar, its period $P$, its time derivative $\dot{P}$, and the angle $\alpha$ between the rotation axis and the magnetic axis are important observables. Most of the input parameters for the magnetosphere simulation and the formulas for verifying the simulation results can be obtained from these four observables. Therefore, the initial configuration of our PIC simulation is as follows: First, a pulsar with an angular rotation speed of $\Omega=\frac{2\pi}{P}$ with a barycentric period $P$, a magnetic inclination of $\alpha$, and mass $M$ is placed in the centre of the three-dimensional Cartesian coordinate system. It has a surface magnetic field strength of $B=\sqrt{\frac{3 c^3}{8 \pi^2} \frac{M}{R^4\sin^2\alpha} \dot{P} P}=1.01\times 10^{12}\mathrm{~Gauss}\left(\frac{M}{\mathrm{M}_{\odot}}\right)^{\frac{1}{2}}\left(\frac{10\mathrm{~km}}{R}\right)^{2}\left(\frac{1}{\sin \alpha}\right)\left(\frac{\dot{P}}{10^{-15}}\right)^{\frac{1}{2}}\left(\frac{P}{1 \mathrm{~s}}\right)^{\frac{1}{2}}$ and a spin-down power of $L_{\mathrm{sd}}=4 \pi^2 M R^2 \frac{\dot{P}}{P^3}=3.95 \times 10^{31} \mathrm{~erg}\mathrm{~s}^{-1}\left(\frac{M}{\mathrm{M}_{\odot}}\right)\left(\frac{R}{10\mathrm{~km}}\right)^{2} \left(\frac{\dot{P}}{10^{-15}}\right)\left(\frac{P}{\mathrm{1~s}}\right)^{-3}$, where $c$ is the speed of light, $R$ is the radius of the pulsar and $\dot{P}$ is time derivative of barycentric period. The lower limit of the pulsar radius can be estimated by the Schwarzschild radius $R_{\min} \approx 1.5 R_{\mathrm{S}}=\frac{3 G M}{c^2}=6.2 \mathrm{~km}\left(\frac{M}{1.4 \mathrm{M}_{\odot}}\right)$, whereas its upper limit can be obtained by measuring its rotation period $R_{\max } \approx\left(\frac{G M P^2}{4 \pi^2}\right)^{\frac{1}{3}}=16.8 \mathrm{~km}\left(\frac{M}{1.4 \mathrm{M}_{\odot}}\right)^{\frac{1}{3}}\left(\frac{P}{\mathrm{ms}}\right)^{\frac{2}{3}}$, where $G$ is the gravitational constant and $\mathrm{M}_{\odot}$ is the solar mass. For ease of calculation, we uniformly set the neutron star radius to $10\mathrm{~km}$. Moreover, the radius of the light cylinder $R_{\mathrm{LC}}=\frac{c}{\Omega}=4.77 \times 10^4 \mathrm{~km}\left(\frac{P}{\mathrm{1~s}}\right)$ and the ratio of available magnetic energy to its rest mass energy $\sigma =\frac{\omega_B}{\Omega}\left(\frac{R}{R_{\mathrm{LC}}}\right)^2=1.21\times 10^{11}\left(\frac{10\mathrm{~km}}{R}\right)^{2}\left(\frac{\mathrm{1~s}}{P}\right)$ for each particle on the light cylinder can be obtained, where $\omega_B=\frac{e B}{m c}=1.73\times 10^{19}\mathrm{~Hz}\left(\frac{B}{10^{12}\mathrm{~Gauss}}\right)$ is the cyclotron frequency, $m$ is the particle mass and $e$ is the elementary charge. In this work, our initial particles are only considered as positive and negative electrons. Second, considering the force-free electrodynamics approximation $\rho \boldsymbol{E}+\boldsymbol{J} \times \boldsymbol{B} / c=0$ under the GJ model, the initial magnetic field $\boldsymbol{B}_{cor}$ containing the oblique dipole and quadrupole and the initial corotating electric field $\boldsymbol{E}_{cor}$ are obtained \citep{1999ApJ...511..351C,2006MNRAS.367...19K,2012ApJ...746...60L}. Thus, the initial charge density $\rho_{cor}=-\frac{\boldsymbol{\Omega} \cdot \boldsymbol{B}_{cor}}{2 \pi c}=112.14\mathrm{~C} \mathrm{~cm}^{-3}\left(\frac{P}{1\mathrm{~s}}\right)^{-\frac{1}{2}}\left(\frac{\dot{P}}{10^{-15}}\right)^{\frac{1}{2}}$ and the initial electric drift velocity $\boldsymbol{v}_{cor}=\frac{\boldsymbol{E}_{cor} \times \boldsymbol{B}_{cor}}{\left|\boldsymbol{B}_{cor}\right|^2}$ can be obtained \citep{1975Ap&SS..32L...7C,2005ApJ...631..456L}. Third, consider the local plasma skin depth $d_e=\frac{c}{\sqrt{2 \Omega \omega_B}}=2.03\mathrm{~cm}\left(\frac{10^{12}\mathrm{~Gauss}}{B}\right)^{\frac{1}{2}}\left(\frac{P}{1\mathrm{~s}}\right)^{\frac{1}{2}}$, which can be used as our resolution indicator, that is, how many cells there are in the three-dimensional simulation. However, the existing supercomputers cannot perform high-resolution simulations of a pulsar surface combined with the skin effect, so we reduce the resolution of the region within the light cylinder, for which radio observations are currently lacking, to reduce the supercomputer load. Considering the size of the field of view of the telescope and the tendency of the pulsar magnetic field to decay with distance, for our pulsar simulation, we need to include only $10^8$ times the radius of the pulsar in its equatorial plane and $0.01$ times the radius of the light cylinder in the direction of the pulsar magnetic axis, as we can fit the emission height of the pulsar with the KG model at the polar cap of the pulsar $r_{\mathrm{em}}^{\mathrm{KG}}=400 \mathrm{~km}\left(\frac{f}{1\mathrm{~GHz}}\right)^{-0.26}\left(\frac{\dot{P}}{10^{-15}}\right)^{0.07}\left(\frac{P}{1\mathrm{~s}}\right)^{0.30}$ \citep{1998MNRAS.299..855K,2002A&A...392..189K}. Since the main criterion for testing the quality of our simulation results is the goodness of fit of the results with the real integrated pulse profiles, we need to increase the resolution of the simulation in the pulsar radio-emission direction. At the moment, the rate of high-energy photon emission $\frac{d^2 N_\gamma}{d \tau d \chi_\gamma}$ corresponds to the electron quantum parameter $\chi_{e}=\frac{\gamma \hbar e}{m^2 c^3} \sqrt{(\mathbf{E}+\mathbf{v} \times \mathbf{B})^2-(\mathbf{v} \cdot \mathbf{E})^2 / c^2}$, the photon quantum parameter $	\chi_\gamma=\frac{\gamma_\gamma \hbar e}{m^2 c^3} \sqrt{(\mathbf{E}+\mathbf{c} \times \mathbf{B})^2-(\mathbf{c} \cdot \mathbf{E})^2 / c^2}$ and the quantum emissivity $S(\chi_{e}, \xi)=\frac{\sqrt{3}}{2 \pi} \xi\left[\int_x^{+\infty} \mathrm{K}_{5 / 3}(x') d x'+\frac{\xi^2}{1-\xi} \mathrm{K}_{2 / 3}(x)\right]$
\begin{equation}
	\frac{d^2 N_\gamma}{d \tau d \chi_\gamma}=\frac{2}{3} \frac{\alpha^2}{\tau_e} \frac{S\left(\chi_{e}, \chi_\gamma / \chi_{e}\right)}{\chi_\gamma},
\end{equation}
where $\tau_e=r_e/c$ is the time for light to cross the classical radius of the electron, $\alpha$ is the fine-structure constant, $\hbar$ is the reduced Planck's constant, and $\mathrm{K}_{2/3}$ and $\mathrm{K}_{5/3}$ are the modified Bessel functions of the second kind. Besides,  $\gamma=\varepsilon /\left(m_e c^2\right)$ and $\gamma_\gamma=\varepsilon_\gamma /\left(m_e c^2\right)$ are the normalized energies of the radiating particle and emitted photon, respectively. Moreover, $\xi=\frac{\chi_\gamma}{\chi_{e}}$ is the rate of radiating particles, $x=2 \xi /[3 \chi_{e}(1-\xi)]$ is a constant used for simplification. Based on the above discussion, since the Lorentz factor of the radiation particles in the magnetosphere of pulsars is usually large $\gamma_0=\frac{1}{16\pi} \frac{mR^3c^2}{eR_{\mathrm{LC}}\rho_{cor}}>10^{5}$ \citep{2012Natur.482..507A,2023ApJ...943..105H,2023NatAs...7.1341H}, we can consider the Landau-Lifshitz model to simulate this kind of radiation \citep{1971ctf..book.....L}, which the threshold of the electron quantum parameter is $\chi_{e}\gtrsim 10^{-3}$ and our calculation result from initial conditions is $\chi_{e}>0.008$. At the same time, the threshold for activating the weak Breit-Wheeler generation process $\chi_\gamma \gtrsim 0.1$ from the Ritus formulae has also been reached, as our calculation result from initial conditions is $\chi_\gamma>0.15$. Then, the total production rate of pairs is
\begin{equation}
	\frac{d N_{\mathrm{Breit-Wheeler}}}{d t}=\frac{\alpha m_e^2 c^4}{\hbar \varepsilon_\gamma} \chi_\gamma T\left(\chi_\gamma\right),
\end{equation}
where $T\left(\chi_\gamma\right)=\frac{1}{\pi \sqrt{3} \chi_\gamma^2} \int_0^{+\infty} \int_{y}^{+\infty} \sqrt{y'} K_{1 / 3}\left(\frac{2}{3} y'^{3 / 2}\right) d y'-\left(2-\chi_\gamma y^{3 / 2}\right) K_{2 / 3}\left(\frac{2}{3} y^{3 / 2}\right) d \chi_{e}$ with a dimensionless constant $y=\left[\chi_\gamma /\left(\chi_{e} \chi_{pe}\right)\right]^{2/3}$ from the combination of the electron quantum parameter $\chi_{e}$ and the positron quantum parameter $\chi_{pe}$ after pair creation. Therefore, we can set the number of cells $N_Z$ in the $z$-axis direction to $1024$, and each cell represents a distance of approximately $4~\mathrm{km}$. Currently, we can place detectors at the emission heights corresponding to different frequencies to detect the radiation they emit.

Next, the resolution in the $x-y$ plane must be determined. This depends on the available memory of the computing platform, which is directly influenced by the number of injected particles $n_{cor}=\rho_{cor}/\left| e \right|$ and the additional particles $N_{\mathrm{Breit-Wheeler}}$ generated via the Breit-Wheeler pair production process. In this work, we utilize 32 high-memory nodes (each equipped with 3TB of RAM) on the Tianhe-2 supercomputing system.  Under the assumption of an 80\% memory usage threshold, this configuration allows us to employ approximately $N_X\times N_Y\times N_Z\approx8.5\times 10^{11}$ total grid cells in the three-dimensional simulation. If the grid points are evenly distributed along the $x$ and $y$ directions $N_X=N_Y=28716$, the resulting grid configuration yields a spatial resolution of approximately $3.5\times 10^{4}~\mathrm{km}$ per cell in the $x-y$ plane. As seen from the above lattice number settings,  After the spatial step size is determined, each time step of the PIC simulation must be considered. According to the Courant--Friedrichs--Lewy (CFL) conditions, when charged particles move in a PIC simulation, their speed and information transfer cannot exceed the speed of light, so the time step criterion can be expressed as $\Delta t\ll\frac{m^{1/2}\Delta x^{3/2}}{2\pi^{1/2}e}$, where $\Delta x$ is the one-dimensional spatial resolution. Therefore, the time step in our simulation is approximately $1.097\times 10^{-5}~\mathrm{s}$.

We use the PIC simulation code ``Smilei" to complete our above collisionless plasma pulsar magnetic field simulation \citep{2018CoPhC.222..351D,whzencode}. In the Main variables block, we use the 3Dcartesian geometry, set the grid length as $10^8R\times10^8R\times0.01R_{\mathrm{LC}}$ and the number of cells as $28716\times28716\times1024$, respectively. Then, we set the simulation time is three times of the pulsar rotation period with time step is $10.97\mu s$. Next, the boundary condition of the electromagnetic field for the principal variable is the three-dimensional ``Silver-Muller", the Maxwell equation solution is ``Yee", the relativistic Poisson solution is turned on, and the other conditions are preset by Smilei. Under these settings, both the injected and the newly generated particles are allowed to propagate out of the simulation domain, rather than undergoing elastic collisions at the boundaries. This leads to more realistic physical results and prevents a continuous increase in data storage requirements. After configuring Main variables block, we open the Load Balancing and activate the Vectorization with the adaptive mode. Then, we set the Current Filter with binomial model and the Field Filter with Friedman model. At this stage, we can proceed to define the particle species and their initial populations. The electromagnetic field and the particle number density of the GJ model are set according to the above initialization. However, a slight difference in the number densities of electrons and positrons is maintained to drive the formation of the pulsar wind region \citep{2019A&A...622A.161C,2022ApJ...939...42H,2023ApJ...958L...9B,2020PhRvL.124x5101P,2024A&A...690A.170S,annurev}. In the particle species block, we begin by setting only cold electrons, positrons and photons, while enabling the relativistic field initialization conditions. Notably, X-ray observations of some pulsar atmospheres reveal that there may be hydrogen ions, carbon ions, and other particle components in the pulsar atmosphere \citep{2005Natur.434.1104G,2009Natur.462...71H,2011Sci...334...69V,2018A&A...618A..76D,2023Sci...381..761S}. However, to reduce the complexity of the simulation, the parameters of these ions are not set in our simulation. Furthermore, the influence of other charged particle species in the pulsar atmosphere will be addressed in future studies. With regard to radiation processes, we adopt the Landau--Lifshitz model for all particles defined in the particle species block, enabling the self-consistent generation of high-energy photons. In parallel, we activate multiphoton Breit--Wheeler pair production in the photon species block to account for photon-induced pair creation. As for the initial conditions, the particle velocities are assigned based on the drift velocities of electrons and positrons, combined with their maximum Lorentz factor of $\gamma=10^5$ at the light-cylinder surface. Their spatial distribution is randomized within each cell, and their momentum distribution is initialized as cold. To prevent artificial particle accumulation, we apply a ``remove" boundary condition, allowing particles to be removed from the simulation domain upon reaching its edges. Finally, all remaining parameters in this block follow Smilei's default settings, and ionization processes are neglected in this simulation. Following the initialization of particle species, the next step is to initialize the electromagnetic fields.  By combining the observed parameters of the pulsar with its electromagnetic structure described by the GJ model, we can determine both the initial electromagnetic field configuration at $t=0$ and the time-dependent background field at arbitrary moments, which only push the particles but do not participate in the Maxwell solver. Finally, it is necessary to specify the computational methods for radiation reaction block and multiphoton Breit--Wheeler block. In particular, for radiation reaction block, the parameter of ``Niel computation method" is set to the ``fit10" mode for higher calculation accuracy for $\chi$, while all other settings retain their default values.

Before starting the simulation, the empirical formula of the pulsar emission height $r_{\mathrm{em}}^{\mathrm{KG}}$ is used to place a two-dimensional measurement corner in the direction of the observation LOS to measure the radio luminosity $L$ of the simulated pulsar and compare it with the real situation after the simulation. Next, we proceed to utilize the diagnostic modules within Smilei to analyze the simulation results, with a particular focus on the distribution and intensity of the magnetic field. Considering the rotational periods of the two pulsars, PSR J1856-3754 and PSR J0720-3125, used in our simulations, completing three full rotational cycles requires approximately $1.929\times 10^6$ and $2.295\times 10^6$ simulation steps, respectively. First, for the scalar diagnostic module, which generates relatively lightweight data, we can store the results at every simulation step. This module allows us to evaluate globally integrated quantities such as the total energy density $U_{\mathrm{tot}}$, kinetic energy density $U_{\mathrm{kin}}$, electromagnetic field energy density $U_{\mathrm{elm}}$, expected energy variation $U_{\mathrm{exp}}$, balanced energy density $U_{\mathrm{bal}}$, and total radiated energy density $U_\mathrm{rad}$. Second, the field diagnostic module records the three spatial components of the magnetic field $B_x, B_y, B_z$, electric field $E_x, E_y, E_z$, and current density $J_x, J_y, J_z$ at each grid cell. Given the significant volume of data produced, we limit the output to 1000 global samplings, which is equivalent to an output interval of approximately $21~\mathrm{ms}$ for PSR J1856-3754 and $25~\mathrm{ms}$ for PSR J0720-3125. Finally, we employ the particle binning diagnostics module to track the evolution of particle energy spectra and the properties of radiative particles. As this module generates a moderate amount of data, substantially less than the field module, we perform 2000 global samplings, corresponding to output intervals of approximately $11~\mathrm{ms}$ and $13~\mathrm{ms}$ for the respective pulsars.

\subsection{Quantifying differences between the PIC-fit magnetosphere and a GJ dipole baseline}
\label{app:fieldcmp}

To clarify how the choice of magnetospheric field model impacts the GW-EM conversion calculation, we compare the magnetic field obtained from our PIC-based fitting against an idealized oblique vacuum dipole that serves as a controlled GJ baseline. Our purpose is not to assert that the PIC simulation is uniquely correct. Rather, we regard the oblique vacuum dipole as a controlled GJ reference and the PIC simulation as a structured alternative. This approach provides a transparent model‑bracketing exercise that quantifies how much the magnetospheric geometry $\mathbf{B}$ can alter the LOS transfer response and, consequently, the detectability forecasts.

The time-domain ``template'' used in our filtering pipeline is not a phase-resolved intrinsic GW strain waveform, but rather an intensity-response kernel that encodes the magnetospheric transfer response along the LOS. This kernel describes how an incident GW perturbation, after undergoing GW-EM conversion in the magnetosphere and after applying the observational bandpass and sampling, manifests as an observable radio-band flux-density response. Operationally, for each pulsar and frequency we compute the conversion-related response along the LOS using the chosen magnetic-field model, and then apply the same bandpass, sampling, and normalization that are used in the subsequent matched-filtering stage. Therefore, the top-left panel of Figs.~\ref{fig:anticipated-signal-1} and \ref{fig:anticipated-signal-2} should be interpreted as a set of normalized intensity transfer-response kernels, not as a direct template of the incoming GW strain.

For transient PBH-like inspirals whose GW frequency reaches the GHz band, the intrinsic in-band chirp duration can be short (ns--$\mu$s). On such timescales the pulsar magnetosphere is effectively frozen, and the pulsar rotation phase does not evolve appreciably during the burst. Consequently, for transient signals we evaluate the magnetosphere at a fixed phase and treat the transfer response as quasi-static; no rotation-induced envelope modulation is applied to the transient waveform. Rotation-phase modulation becomes relevant only for persistent or stochastic-background signals, where long integrations coherently sample the phase-dependent transfer response, yielding the rotation-modulated intensity envelope shown in the right-hand panels of Figs.~\ref{fig:anticipated-signal-1} and \ref{fig:anticipated-signal-2}.

Because the conversion efficiency depends on both the field strength and the field geometry relative to the propagation direction and the polarization basis, we adopt two complementary, numerically stable diagnostics to quantify differences between the PIC simulation and dipole models. We avoid a direct fractional contrast such as $(|B_{\mathrm{PIC}}|-|B_{\mathrm{dip}}|)/|B_{\mathrm{dip}}|$ because it can spuriously diverge in weak-field regions where $|B_{\mathrm{dip}}|\rightarrow 0$, which would make the visual impression dominated by numerical artefacts. Instead, we use regularized, bounded or logarithmically stabilized quantities that remain well behaved over the entire slice.

To quantify the overall discrepancy between the two vector fields, we use a stabilised logarithmic relative difference,
\begin{equation}
	\Delta_{\rm vec}(\mathbf{x}) \equiv
	\log_{10}\left(
	\frac{\left|\mathbf{B}_{\rm PIC}(\mathbf{x})-\mathbf{B}_{\rm dip}(\mathbf{x})\right|}
	{\left|\mathbf{B}_{\rm dip}(\mathbf{x})\right|}
	\right).
	\label{eq:app_logvecrel}
\end{equation}
$\Delta_{\rm vec}(\mathbf{x})$ compactly captures changes arising from both amplitude and direction, while remaining well-behaved throughout the weak-field region. This quantity corresponds to the diagnostic and directly quantifies the order-of-magnitude deviation of the PIC simulation from the dipole baseline in a vector sense.

To isolate geometry effects independent of overall amplitude rescaling, we additionally evaluate the cosine similarity
\begin{equation}
	\cos\chi(\mathbf{x}) \equiv
	\frac{\mathbf{B}_{\rm PIC}(\mathbf{x})\cdot \mathbf{B}_{\rm dip}(\mathbf{x})}
	{|\mathbf{B}_{\rm PIC}(\mathbf{x})||\mathbf{B}_{\rm dip}(\mathbf{x})|}.
	\label{eq:app_cos}
\end{equation}
By construction, $\cos\chi\simeq 1$ indicates local alignment, $\cos\chi\simeq 0$ indicates near-orthogonality, and $\cos\chi\simeq -1$ indicates anti-alignment. This statistic, corresponding to the diagnostic, therefore highlights directional/structural differences even when $|\mathbf{B}|$ changes are modest.

In our conversion formalism, the effective coupling is controlled by the magnetic-field configuration along the GW propagation path, with sensitivity to both (a) the local field strength and (b) the orientation of $\mathbf{B}$ relative to the propagation direction and polarisation basis. The pair $(\Delta_{\rm vec},\cos\chi)$ therefore provides a concise summary of how the PIC-fit model departs from the dipole baseline in the conversion-relevant region: $\Delta_{\rm vec}$ captures the overall vector-level deviation, while $\cos\chi$ isolates directional differences that can modify the effective transverse component and hence the transfer response.

\begin{figure*}
	\centering
	\includegraphics[width=1\linewidth]{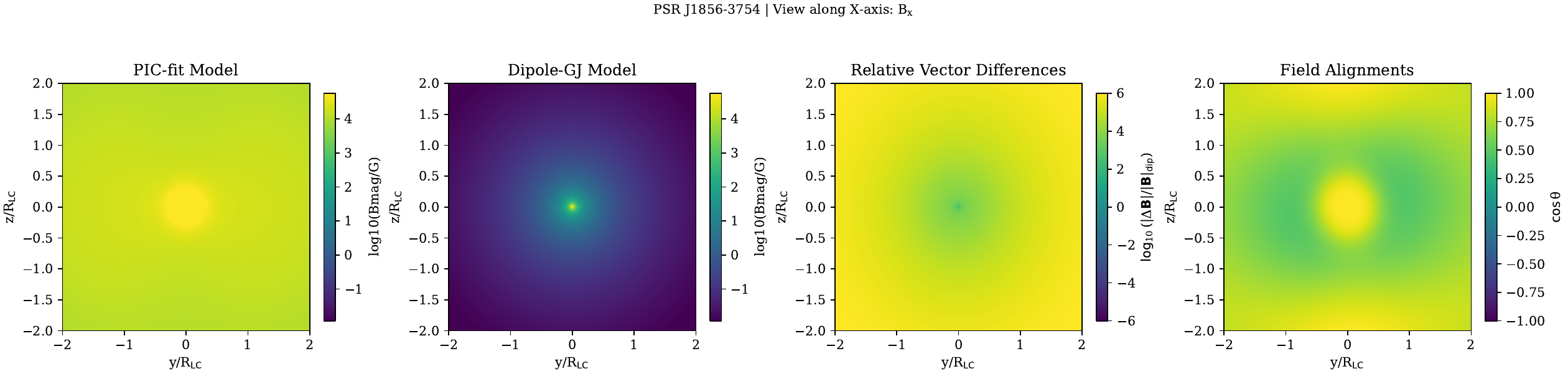}
	\includegraphics[width=1\linewidth]{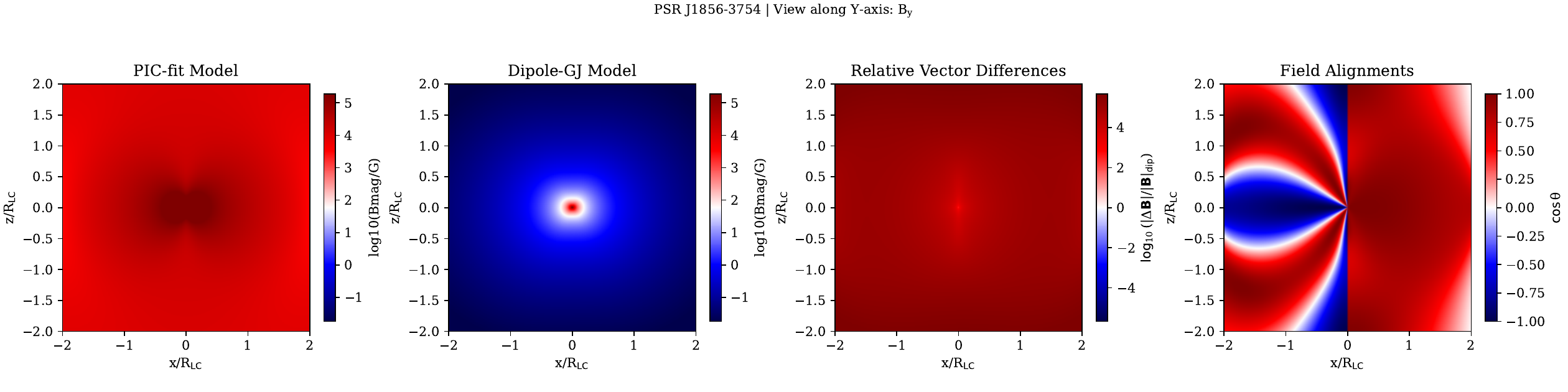}
	\includegraphics[width=1\linewidth]{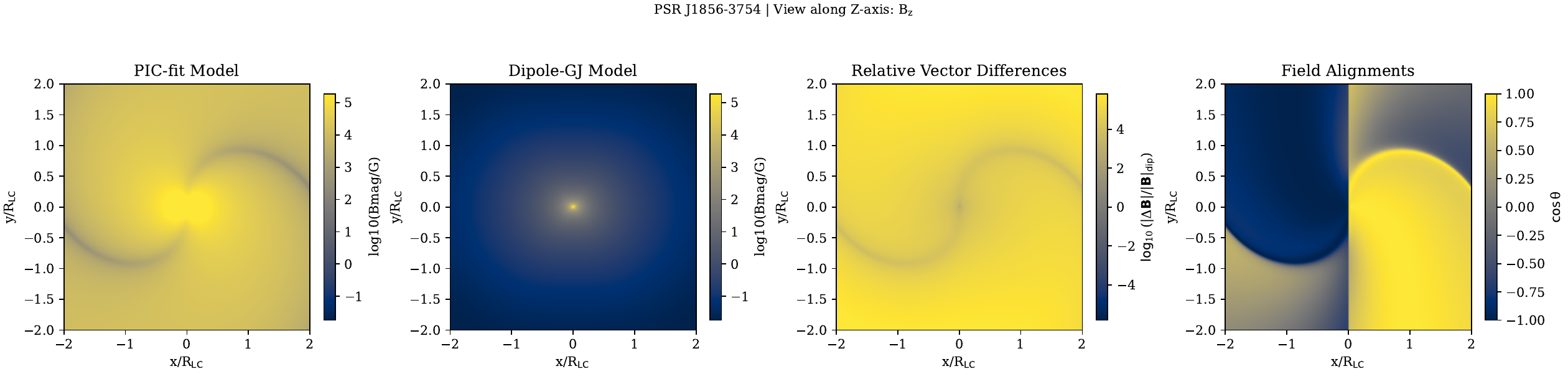}
	\caption{Comparison between the PIC and dipole (GJ-like) baseline magnetospheres on three orthogonal slices aligned with the simulation axes for PSR J1856-3754. From top to bottom: $B_x$, $B_y$, and $B_z$ slices, respectively. For each slice, the four panels show (from left to right): (1) PIC-fit model, (2) dipole baseline, (3) stabilized vector-relative logarithmic contrast $\Delta_{\rm vec}$ defined in Eq.~(\ref{eq:app_logvecrel}), and (4) cosine similarity $\cos\chi$ defined in Eq.~(\ref{eq:app_cos}). All coordinates are in units of $R_{\rm LC}$, and all diagnostics are evaluated at rotation phase 0$^\circ$ on identical grids.}
	\label{fig:app_fieldcmp1}
\end{figure*}

\begin{figure*}
	\centering
	\includegraphics[width=1\linewidth]{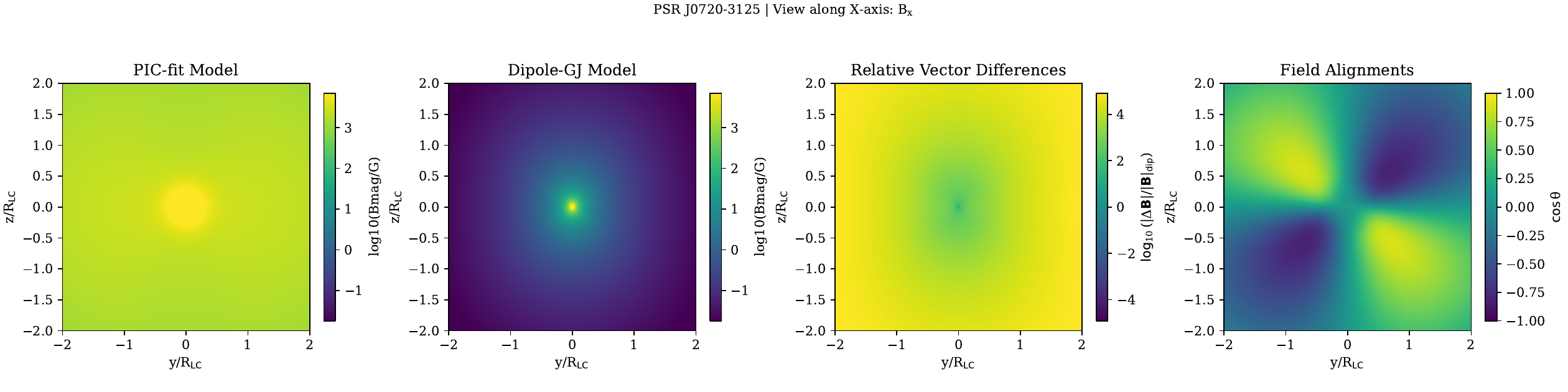}
	\includegraphics[width=1\linewidth]{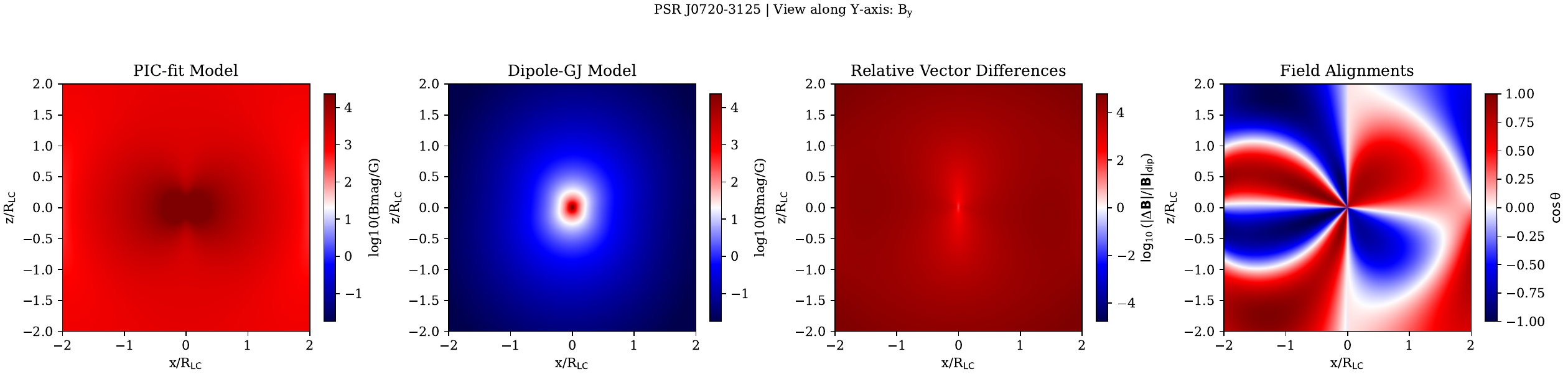}
	\includegraphics[width=1\linewidth]{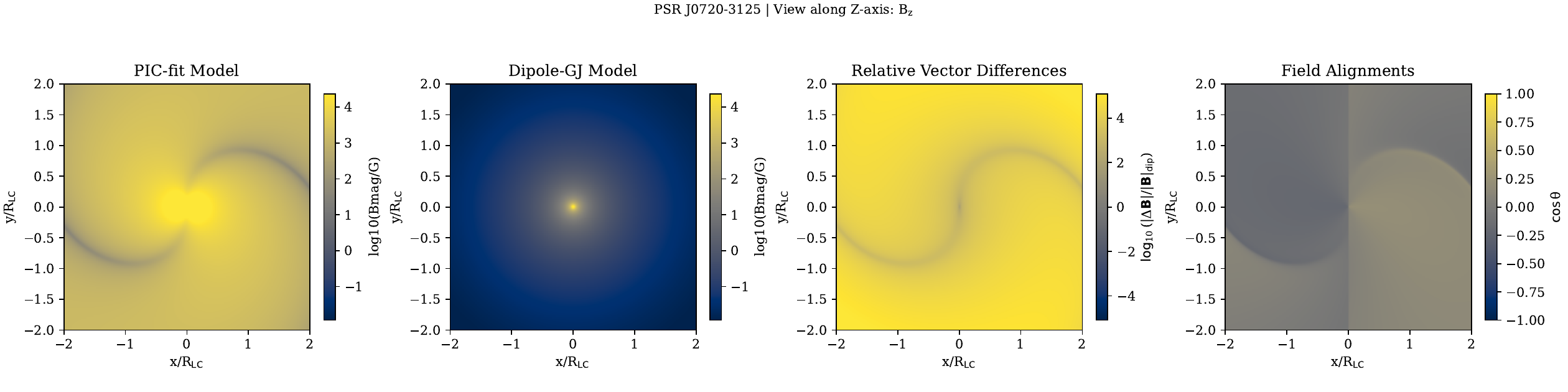}
	\caption{Comparison between the PIC and dipole (GJ-like) baseline magnetospheres on three orthogonal slices aligned with the simulation axes for PSR J0720-3125. This figure shows the results in the same order
		as Figure 9.}
	\label{fig:app_fieldcmp2}
\end{figure*}

For direct comparison with the figures in the main text, we show three orthogonal slices along the simulation axes Figs.~\ref{fig:app_fieldcmp1} and \ref{fig:app_fieldcmp2}: the $xy$‑plane viewed along the $z$‑axis, the $xz$‑plane viewed along the $y$‑axis, and the $yz$‑plane viewed along the $x$‑axis. All coordinates are given in units of $R_{\mathrm{LC}}$, and the PIC simulation and dipole maps are evaluated at the same rotation phase on identical grids.

\subsection{WKB derivation}
\label{app:wkb_derivation}

In this subsection, we derive the WKB LOS expression Eq.~(3) from the wave equation Eq.~(2) and demonstrate its validity for our radio-band LOS calculations. Starting from Eq.~(2)
\begin{equation}
	\begin{aligned}
		&\left(\partial_t^2-\partial_s^2\right) h_{i j}^{\mathrm{TT}}(t, s)
		=16\pi G\left( F_{i\rho}F^{\rho}_{j}-\frac{\eta_{ij}}{4}F_{\alpha\beta}F^{\alpha\beta}\right), \\
		&\left(\partial_t^2-\partial_s^2+\Delta_\omega^2\right) A_i(t, s)
		=\delta^{k j}\delta^{l m}\left(\partial_s h_{i j}^{\mathrm{TT}}\right)\epsilon_{ksl} B_m,
		\label{app:eq:wave_start}
	\end{aligned}
\end{equation}
where $s$ is the LOS coordinate. We adopt the WKB ansatz that write each field as a rapidly oscillating phase factor multiplied by a slowly varying complex amplitude
\begin{equation}
	A_i(t,s)=\tilde{a}_i(s)e^{-i\omega t+i\omega s},\qquad
	h_{ij}^{TT}(t,s)=\tilde{h}_{ij}(s)e^{-i\omega t+i\omega s}.
	\label{app:eq:wkb_ansatz}
\end{equation}
The phase factor $e^{-i\omega t+i\omega s}$ accounts for the rapid oscillations, while $\tilde{a}_i(s)$ and $\tilde{h}_{ij}(s)$ are slowly varying envelopes. Substituting into Eq.~(\ref{app:eq:wave_start}) yields
\begin{equation}
	\begin{aligned}
		\partial_s^2 A_i(t,s)
		&=
		\left[\tilde{a}_i''(s)+2i\omega \tilde{a}_i'(s)-\omega^2 \tilde{a}_i(s)\right]
		e^{-i\omega t+i\omega s},\\
		\partial_t^2 A_i(t,s)
		&=
		-\omega^2 \tilde{a}_i(s)e^{-i\omega t+i\omega s},\\
		\partial_s^2 h_{ij}^{TT}(t,s)
		&=
		\left[\tilde{h}_{ij}''(s)+2i\omega \tilde{h}_{ij}'(s)-\omega^2 \tilde{h}_{ij}(s)\right]
		e^{-i\omega t+i\omega s},\\
		\partial_t^2 h_{ij}^{TT}(t,s)
		&=
		-\omega^2 \tilde{h}_{ij}(s)e^{-i\omega t+i\omega s},\\
		\partial_s h_{ij}^{TT}(t,s)
		&=
		\left[\tilde{h}_{ij}'(s)+i\omega \tilde{h}_{ij}(s)\right]e^{-i\omega t+i\omega s}.
	\end{aligned}
\end{equation}
Dropping the tildes for clarity, the vector-potential equation reduces to
\begin{equation}
	\tilde{a}_i''(s)+2i\omega \tilde{a}_i'(s)+\Delta_\omega^2(s)\,\tilde{a}_i(s)=\mathcal{S}_i(s),
	\label{app:eq:second_order_env}
\end{equation}
with source term
\begin{equation}
	\mathcal{S}_i(s)\equiv
	-\delta^{kj}\delta^{lm}\left[h_{ij}'(s)+i\omega h_{ij}(s)\right]\epsilon_{ksl}B_m.
\end{equation}
The WKB approximation requires that the induced field and background quantities vary on scales much larger than the radio wavelength, i.e.
\begin{equation}
	|\tilde{a}_i''|\ll \omega |a_i'|,\qquad
	\left|\frac{\partial_s \mathcal{S}}{\mathcal{S}}\right|
	=
	\left|\partial_s\ln \mathcal{S}\right|\ll \omega,\qquad
	\left|\partial_s\Delta_\omega^2\right|\ll \omega^2.
	\label{app:eq:wkb_conditions}
\end{equation}
Neglecting $\tilde{a}_i''$ under these conditions gives the first-order transport equation
\begin{equation}
	2i\omega\,\tilde{a}_i'(s)+\Delta_\omega^2(s)\,\tilde{a}_i(s)=\mathcal{S}_i(s).
	\label{app:eq:first_order_env}
\end{equation}
Introducing the integrating factor
\begin{equation}
	\mu(s)=
	\exp\!\left[
	\frac{i}{2\omega}\int_{-L/2}^{s}\Delta_\omega^2(s')\,ds'
	\right]
\end{equation}
and imposing $\tilde{a}_i(-L/2)=0$ yields
\begin{equation}
	\tilde{a}_i(L/2)
	=
	\frac{1}{2i\omega}
	\int_{-L/2}^{L/2}dl'\,
	\mathcal{S}_i(l')\,
	\exp\!\left[
	-\frac{i}{2\omega}
	\int_{-L/2}^{l'}\Delta_\omega^2(l'')\,dl''
	\right].
	\label{app:eq:env_solution}
\end{equation}
Projecting onto polarization $\lambda$ and expressing the source via the effective magnetic field $B_{\rm eff}$, we obtain the conversion probability
\begin{equation}
	P_{g\rightarrow\gamma}(\omega,L,\theta)
	=
	\left|
	\left\langle\hat{A}_{\omega,\lambda}(L)\mid\hat{h}_{\omega,\lambda}(0)\right\rangle
	\right|^2
	=
	\left|
	\int_{-L/2}^{L/2}dl'\,
	\frac{\sqrt{2}B_\mathrm{eff}(l',\theta)}{2M_{\mathrm{planck}}}
	\exp\!\left(
	-i\int_{-L/2}^{l'}dl''\,\frac{-\Delta_\omega^2(l'')}{2\omega}
	\right)
	\right|^2,
	\label{app:eq:amp_projection}
\end{equation}
which is the WKB LOS expression Eq.~(3) used in the main text; it incorporates the dispersion-induced phase mismatch via $\Delta_\omega^2$ in the phase factor.

We now verify that these conditions are satisfied for our pulsar targets, PSR J1856-3754 and PSR J0720-3125. Their periods are $P=7.05520287\,\mathrm{s}$ and $P=8.391115532\,\mathrm{s}$, giving light-cylinder radii $R_{\rm LC}=\frac{c}{\Omega}=\frac{cP}{2\pi}=4.77\times10^4\,{\rm km}\left(\frac{P}{1\,{\rm s}}\right)$. In Appendix~A.1, the simulation box along the magnetic axis spans $0.01R_{\rm LC}$ with $N_Z=1024$ zones, so the resolved scale is $\Delta z=\frac{0.01R_{\rm LC}}{N_Z}$, which is of order $3$--$4\,\mathrm{km}$ for our pulsars. The characteristic emission height follows the KG model $r_{\rm em}^{\rm KG}=400\,{\rm km}\left(\frac{f}{1\,{\rm GHz}}\right)^{-0.26}\left(\frac{\dot P}{10^{-15}}\right)^{0.07}\left(\frac{P}{1\,{\rm s}}\right)^{0.30}$. The estimate for the background-variation scale is $\ell_{\rm background}\gtrsim \Delta z$, as the LOS path may sample longer scales than a single $z$-cell. Using $\Delta z$ gives a lower bound for the WKB validity. Here $\Delta z$ should be understood as the numerical resolution scale of the resolved LOS background, not as an independent physical coherence length. At $f=1\,\mathrm{MHz}$, $\lambda = c/f \approx 300\,\mathrm{m}$, yielding
\begin{equation}
	\frac{\lambda}{\ell_{\rm background}}
	\approx 0.075,
	\qquad
	k_{\rm em}\ell_{\rm background}
	=
	\frac{2\pi \ell_{\rm background}}{\lambda}
	\approx 83.78,
	\label{app:eq:wkb_numeric1}
\end{equation}
where $k_{\rm em}=2\pi/\lambda$. At $f=1\,\mathrm{GHz}$, $\lambda\approx0.3\,\mathrm{m}$, so that
\begin{equation}
	\frac{\lambda}{\Delta z}\approx 7.5\times10^{-5},
	\qquad
	k_{\rm em}\Delta z\approx 8.38\times10^{4}.
	\label{app:eq:wkb_numeric2}
\end{equation}
Comparing $\lambda$ to $r_{\rm em}^{\rm KG}$ yields even stronger scale separation. Hence the LOS backgrounds used in our post-processing vary on scales much larger than $\lambda$, fully satisfying Eq.~(\ref{app:eq:wkb_conditions}).

\begin{figure*}[t]
	\centering
	\includegraphics[width=1\textwidth]{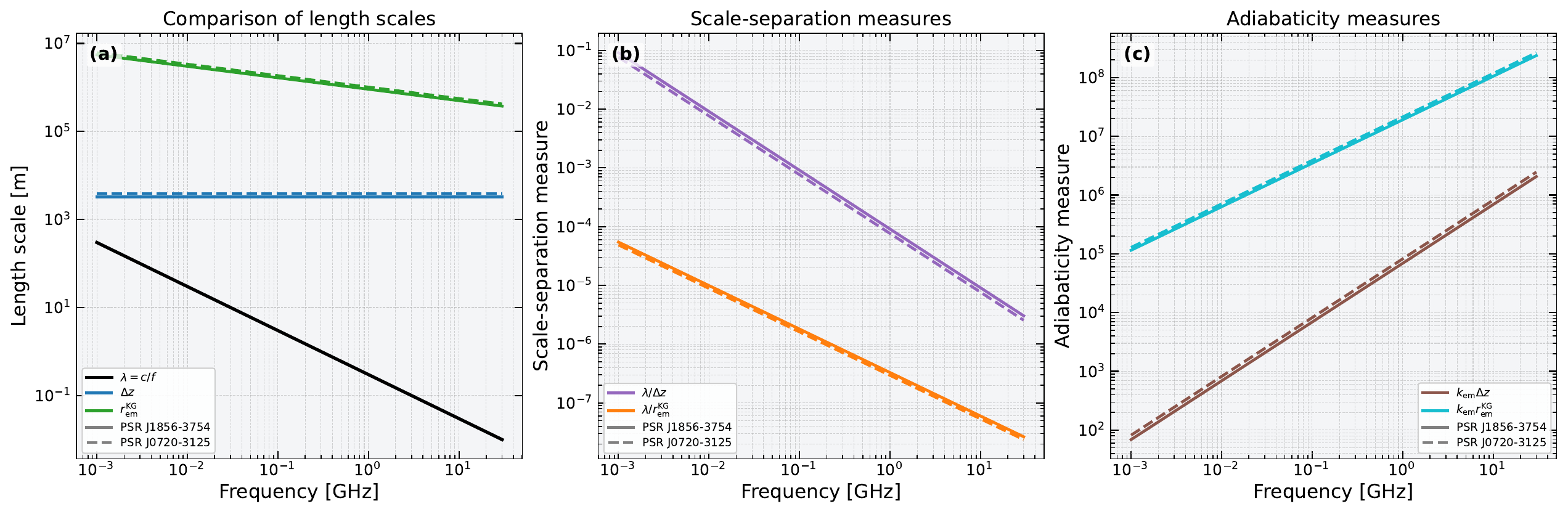}
	\caption{The test of the WKB scale-separation conditions for PSR J1856$-$3754 (solid) and PSR J0720$-$3125 (dashed). (a) Comparison of $\lambda=c/f$ with $\Delta z$ and $r_{\rm em}^{\rm KG}$. (b) Scale-separation measures $\lambda/\Delta z$ and $\lambda/r_{\rm em}^{\rm KG}$. (c) Adiabaticity measures $k_{\rm em}\Delta z$ and $k_{\rm em}r_{\rm em}^{\rm KG}$. Across the MHz--GHz band, $\lambda/\Delta z,\lambda/r_{\rm em}^{\rm KG}\ll1$ and $k_{\rm em}\Delta z,k_{\rm em}r_{\rm em}^{\rm KG}\gg1$, satisfying the WKB ordering in Eq.~(\ref{app:eq:wkb_conditions})}
	\label{fig:wkb_scale_hierarchy}
\end{figure*}

\begin{figure*}[t]
	\centering
	\includegraphics[width=1\textwidth]{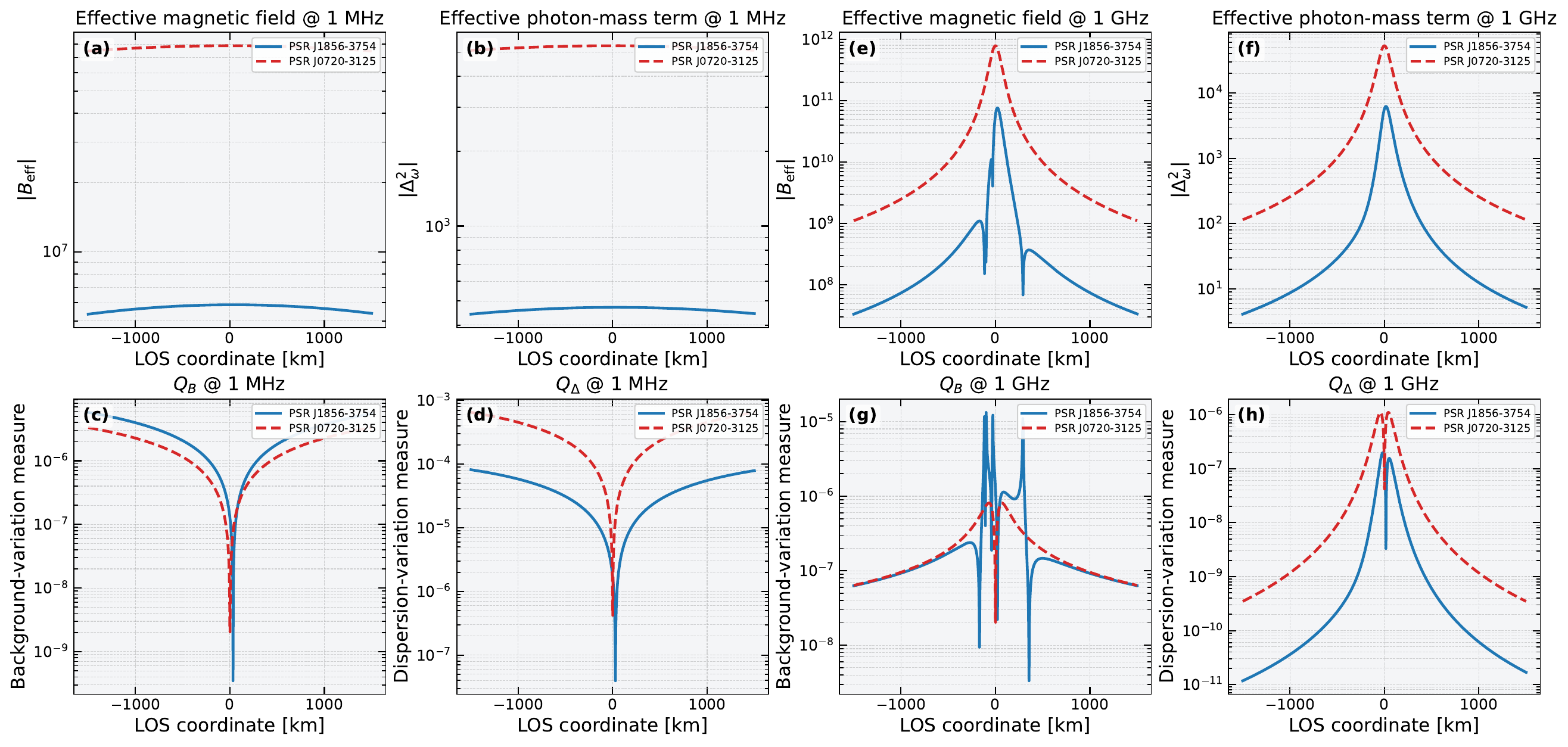}
	\caption{The validity test of WKB evaluated along LOSs for PSR J1856$-$3754 (blue solid) and PSR J0720$-$3125 (red dashed). The left four panels (a–d) correspond to $1\,\mathrm{MHz}$, the right four (e–h) to $1\,\mathrm{GHz}$. For each frequency, the top panels show the effective transverse magnetic field $|B_{\rm eff}|$ and the effective photon-mass term $|\Delta_\omega^2|$ along the LOS, while the bottom panels show the associated WKB diagnostics $Q_B\equiv |\partial_s B_{\rm eff}|/(k_{\rm em}|B_{\rm eff}|)$ and $Q_\Delta\equiv |\partial_s\Delta_\omega^2|/k_{\rm em}^2$. Even at the lower end of the radio band at $1\,\mathrm{MHz}$, both $Q_B$ and $Q_\Delta$ remain below 1, and they drop by several orders of magnitude at $1\,\mathrm{GHz}$.}
	\label{fig:wkb_direct_los}
\end{figure*}

\begin{figure*}[t]
	\centering
	\includegraphics[width=1\textwidth]{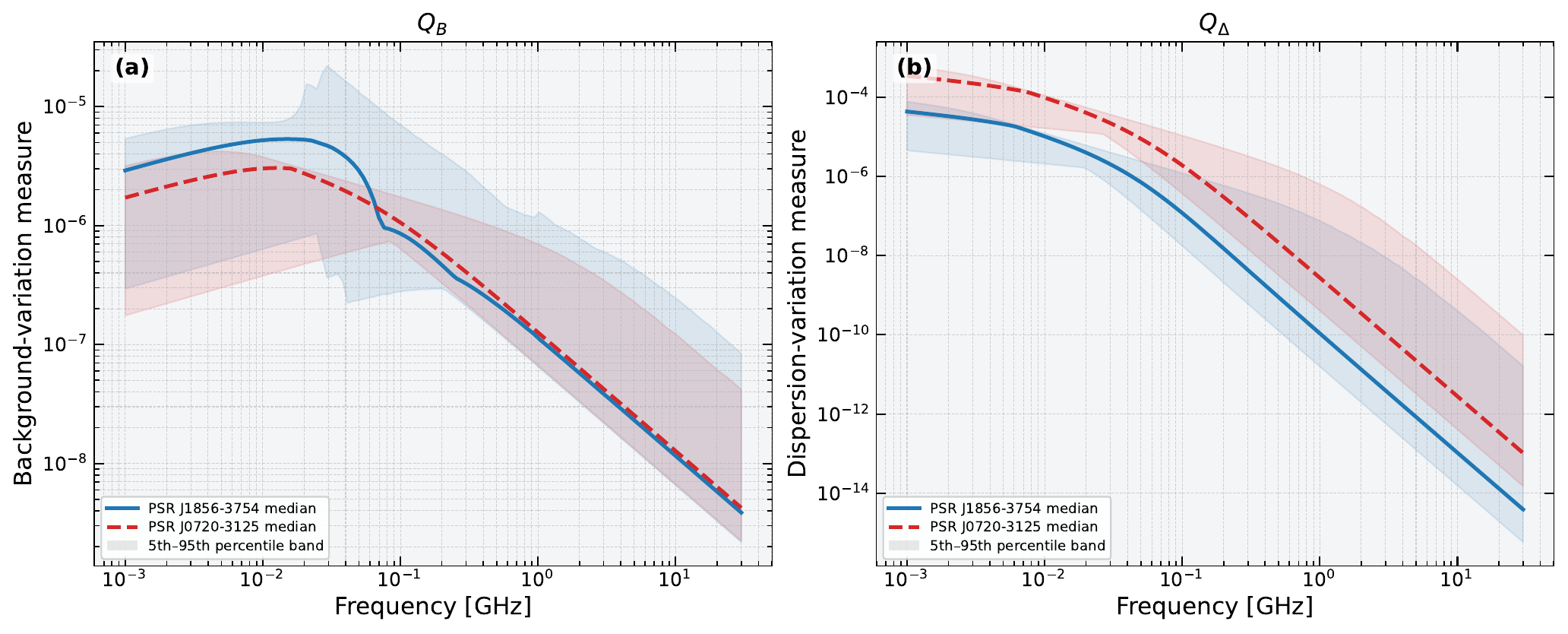}
	\caption{The frequency dependence of the validity test of WKB for PSR J1856$-$3754 (blue) and PSR J0720$-$3125 (red). The two panels show $Q_B$ (a) and $Q_\Delta$ (b) as functions of frequency, with solid/dashed curves denoting median values along the LOS and shaded regions indicating the 5th–95th percentile ranges.}
	\label{fig:wkb_direct_statistics}
\end{figure*}

Figure~\ref{fig:wkb_scale_hierarchy} visually verifies that the scale separation condition relied upon by the WKB approximation holds within the parameter range of this work. It can be observed that throughout the MHz–GHz radio frequency band, the wavelength $\lambda = c/f$ is always much smaller than the two background variation scales: the numerical grid scale $\Delta z$ and the radio emission characteristic height $r_{\rm em}^{\rm KG}$. Here, $\Delta z$ remains basically stable at the kilometer scale, while $r_{\rm em}^{\rm KG}$ is larger and changes slowly with frequency. Therefore, $\lambda/\Delta z$ is even less than $10^{-1}$ at the lowest frequency end and further decreases with increasing frequency, while $\lambda/r_{\rm em}^{\rm KG}$ is smaller throughout the frequency range. Correspondingly, the adiabatic parameters $k_{\rm em}\Delta z$ and $k_{\rm em}r_{\rm em}^{\rm KG}$ are always much greater than 1. This indicates that the WKB conditions in Equation~(\ref{app:eq:wkb_conditions}) are not only satisfied but also have sufficient robustness in the radio frequency band. Figure~\ref{fig:wkb_direct_los} further confirms this from the changes in physical quantities along the LOS. At $1,\mathrm{MHz}$, the profiles of $|B_{\rm eff}|$ and $|\Delta_\omega^2|$ along the LOS are overall smooth, and the corresponding $Q_B$ and $Q_\Delta$ are much smaller than 1: $Q_B$ is approximately in the range of $10^{-6}$ to $10^{-5}$, and $Q_\Delta$ is between $10^{-5}$ and $10^{-4}$. When the frequency rises to $1,\mathrm{GHz}$, although some sharper structures appear near the center of the LOS profile, $Q_B$ and $Q_\Delta$ are still much smaller than 1, with their maximum values differing by several orders of magnitude from the threshold of WKB failure $Q \sim 1$, indicating that even in regions with rapid background variations, the WKB approximation remains valid. Figure~\ref{fig:wkb_direct_statistics} further presents the statistical results along the entire LOS. For two pulsars, $Q_\Delta$ continuously decreases with increasing frequency, from the MHz band down to the minimum value in the GHz band; $Q_B$ changes gradually in the low-frequency band and then significantly decreases at high frequencies. Regardless of the median curve or the 5th–95th percentile interval, $Q_B$ and $Q_\Delta$ are all much lower than 1 throughout the MHz–GHz frequency band. From the above three figures, it can be seen that the LOS background profile adopted in this work numerically fully meets the requirements of the WKB approximation.

\subsection{Astrometry and scintillation-strength parametrization}\label{app:path-details}
This subsection elaborates on the specific astrometric model and the scintillation-strength definitions applied in the simulations of Section~\ref{sec:PE-RRS}, with details provided here for reference and reproducibility.

Because of the relative motion between the observed pulsar and the radio telescope, coupled with the fact that we mainly rely on different background noise for cross-correlation data processing, the pulsar motion needs to be considered. The motion of the pulsar in the celestial coordinate system can be expressed as
\begin{equation}
	\begin{aligned}
		& \alpha_{\mathrm{source}}(t)=\alpha_{\mathrm{source}}(T)+(t-T) * \mu_\alpha^{\mathrm{source}}+P_\alpha * \varpi^{\mathrm{source}}, \\
		& \delta_{\mathrm{source}}(t)=\delta_{\mathrm{source}}(T)+(t-T) * \mu_\delta^{\mathrm{source}}+P_\delta * \varpi^{\mathrm{source}},
	\end{aligned}
\end{equation}
where $\left(\alpha_{\mathrm{source}}(t), \delta_{\mathrm{source}}(t)\right)$ is the location of the pulsar at one time point $t$ and $\left(\alpha_{\mathrm{source}}(T), \delta_{\mathrm{source}}(T)\right)$ is its location at reference epoch $T$, with its astrometric parameters denoted as $\left(\mu_\alpha^{\mathrm{source}}, \mu_\delta^{\mathrm{source}}, \varpi^{\mathrm{source}}\right)$. Here, $\mu_\alpha^{\mathrm{source}}$ is the proper motion in the right ascension (RA) direction, $\mu_\delta^{\mathrm{source}}$ is the proper motion in the direction of declination (Dec), and $\varpi^{\mathrm{source}}$ is the annual parallax. Moreover, $P_\alpha$ and $P_\delta$ are the known parallax factors in the RA and Dec directions, respectively. On this basis, combined with the ``NE2001" model and the conversion of the station coordinate system of the radio telescope to the celestial coordinate system, we can obtain simulation parameters such as the dispersion measure (DM) $\mathrm{DM}= \int\left(\frac{n_e}{1\mathrm{~cm}^{-3}} \right)\frac{\mathrm{~d} l}{1\mathrm{~pc}} \mathrm{~cm}^{-3}\mathrm{~pc} $ and rotation measure (RM) $\mathrm{RM}=0.812 \int\left(\frac{n_e}{1\mathrm{~cm}^{-3}} \right) \left(\frac{B_{\|}}{10^{-6}\mathrm{~G}}\right) \frac{\mathrm{~d} l}{1\mathrm{~pc}} \mathrm{~rad} \mathrm{~m}^{-2}$. For the convenience of calculation, we assume that the average strength of the magnetic field $B_{\|}$ in the galaxy is $3~\mu \mathrm{Gauss}$ \citep{Lai:2000at,Neronov:2010gir,Tavecchio:2010mk,Takahashi:2013lba,Jedamzik:2018itu,Pshirkov:2015tua,Durrer:2013pga}.

\begin{figure*}
	\centering
	\includegraphics[width=0.45\linewidth]{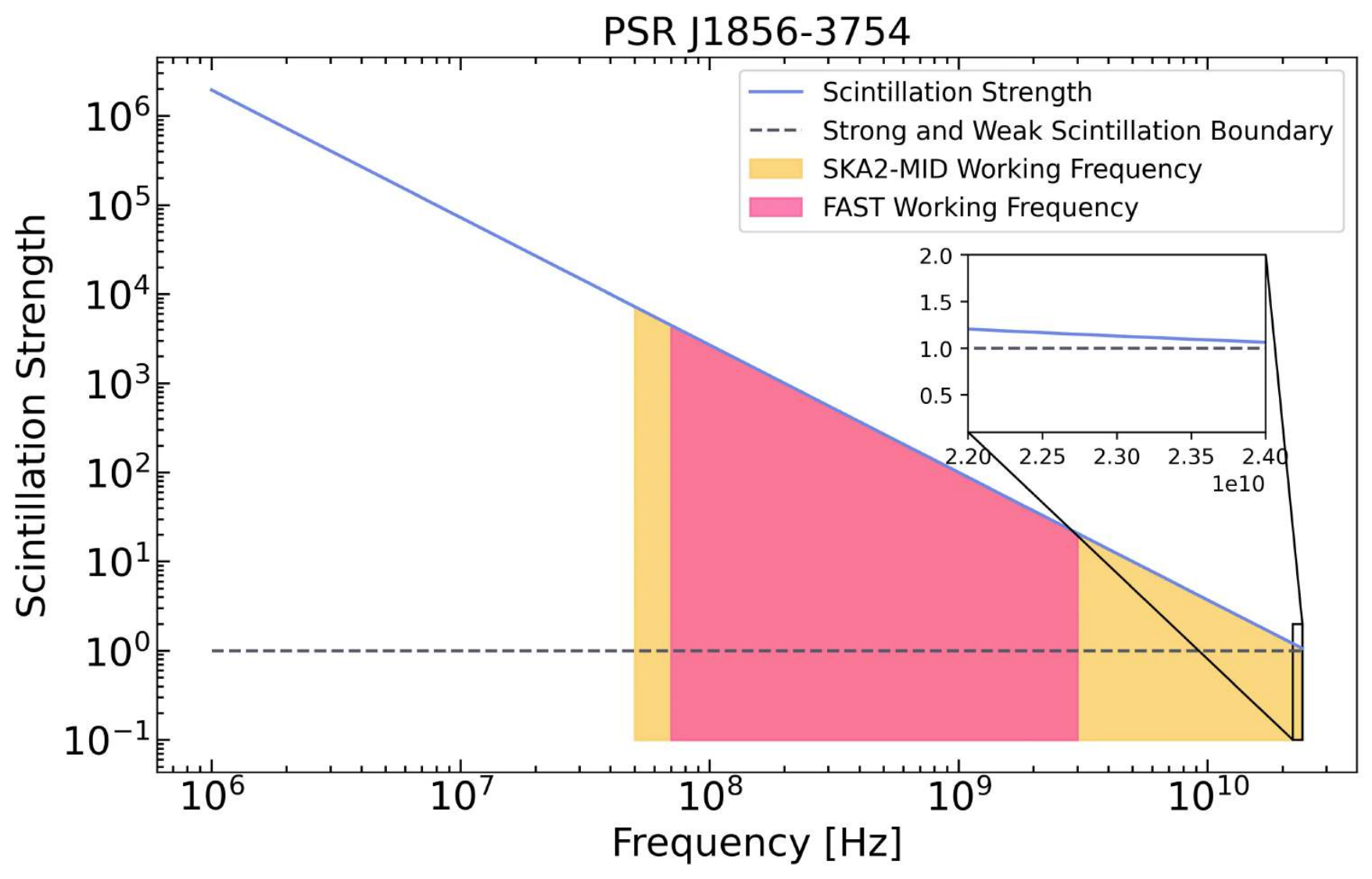}
	\includegraphics[width=0.45\linewidth]{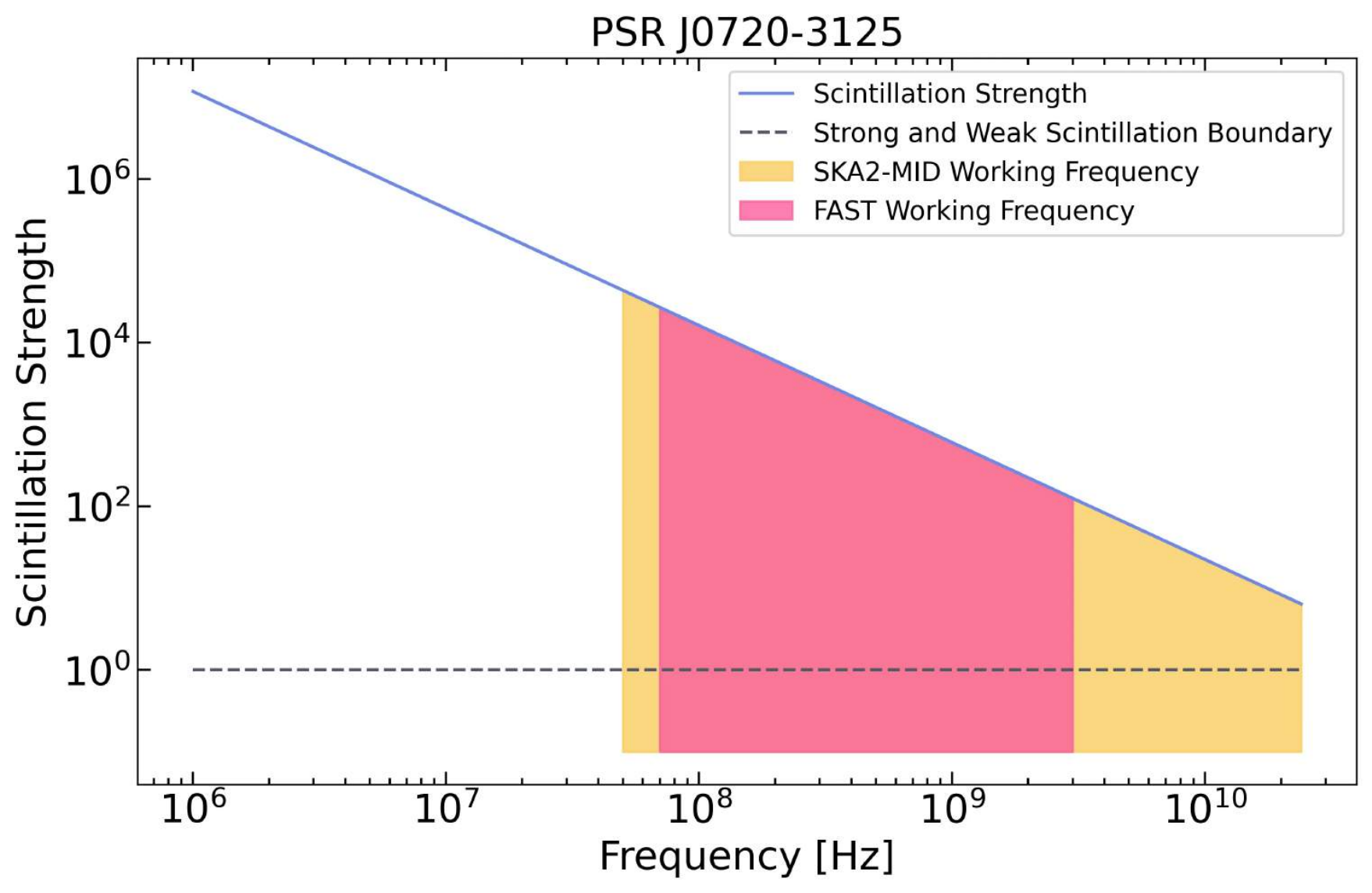}
	\includegraphics[width=0.23\linewidth]{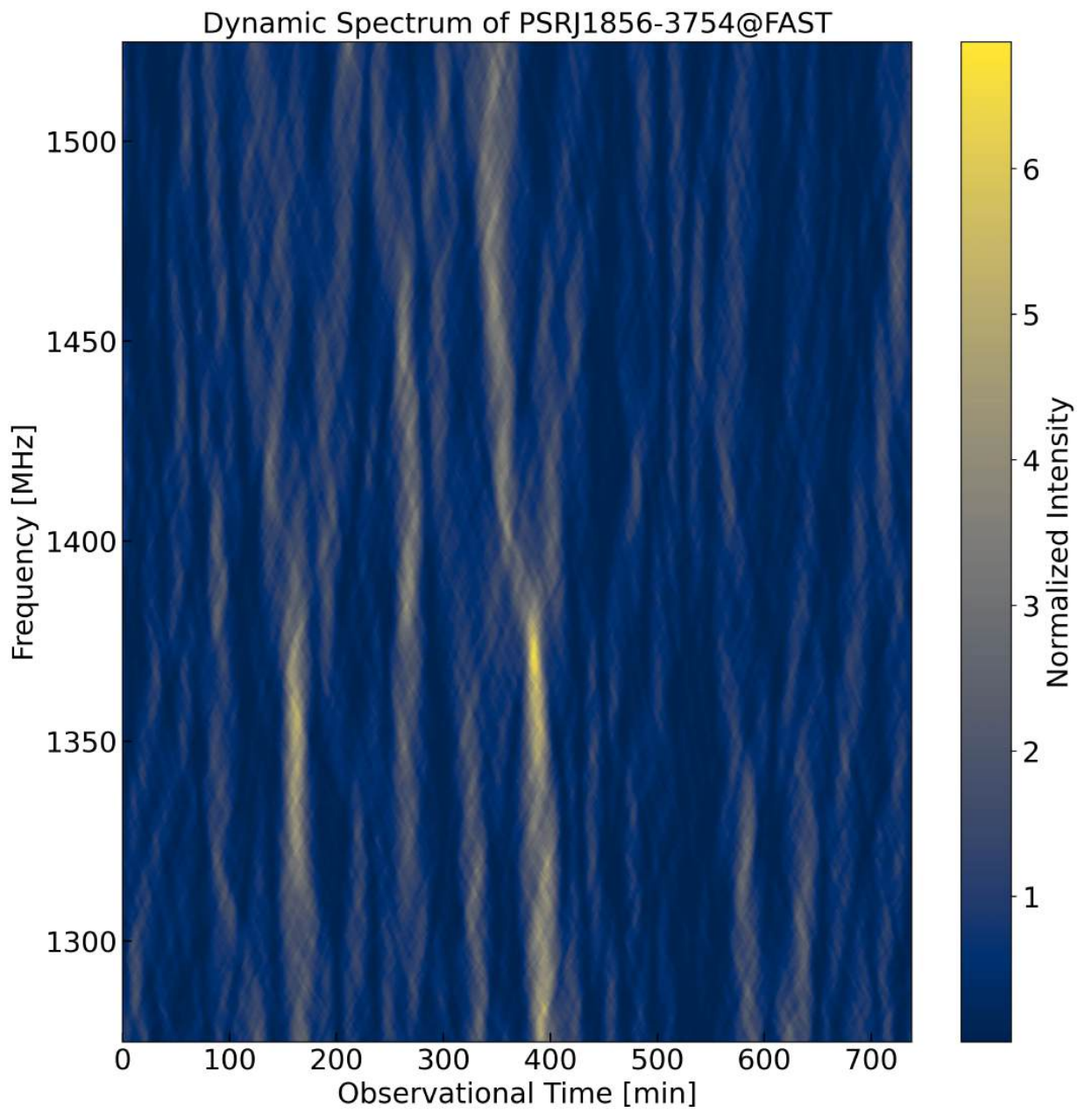}
	\includegraphics[width=0.23\linewidth]{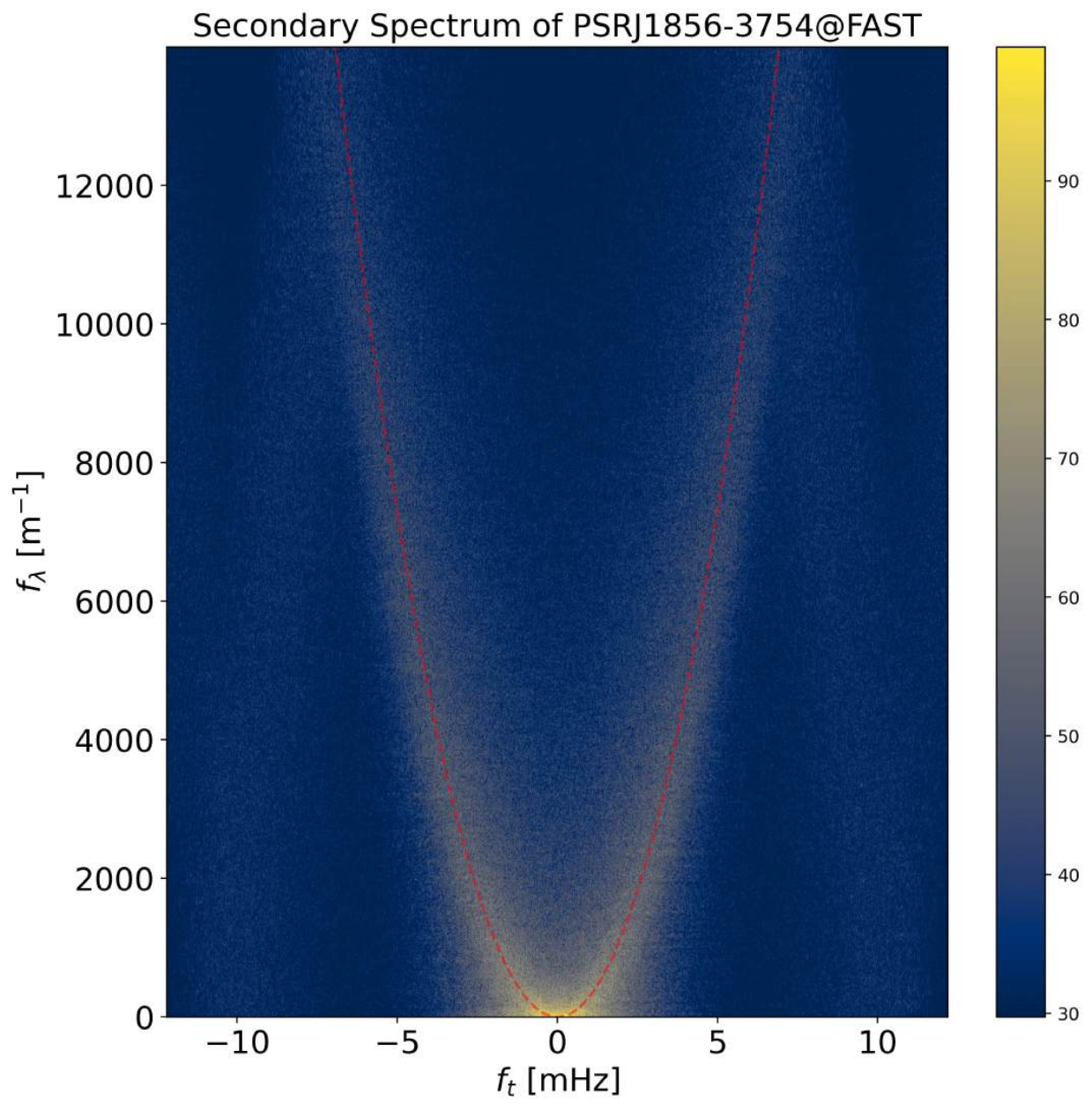}
	\includegraphics[width=0.23\linewidth]{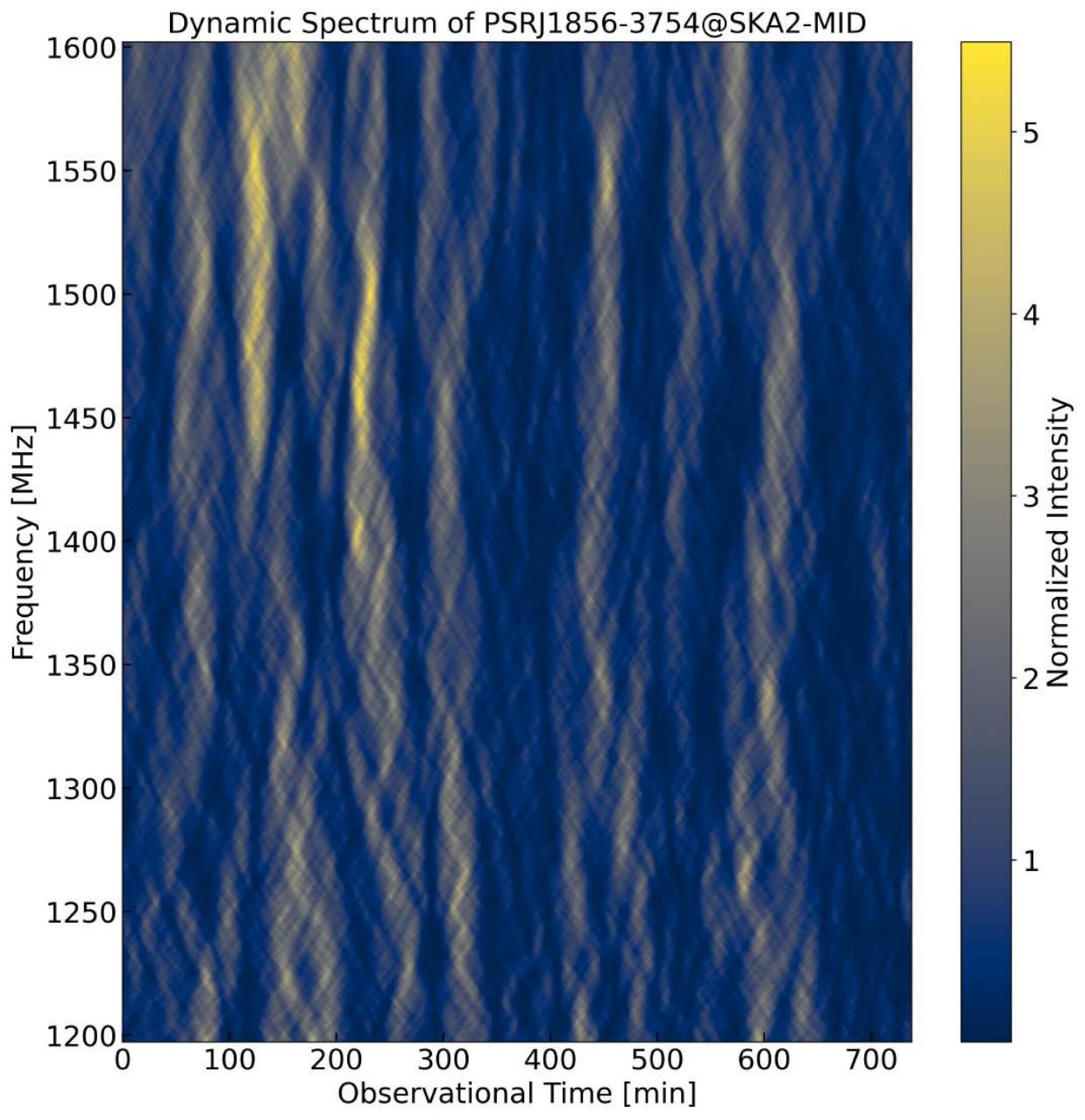}
	\includegraphics[width=0.23\linewidth]{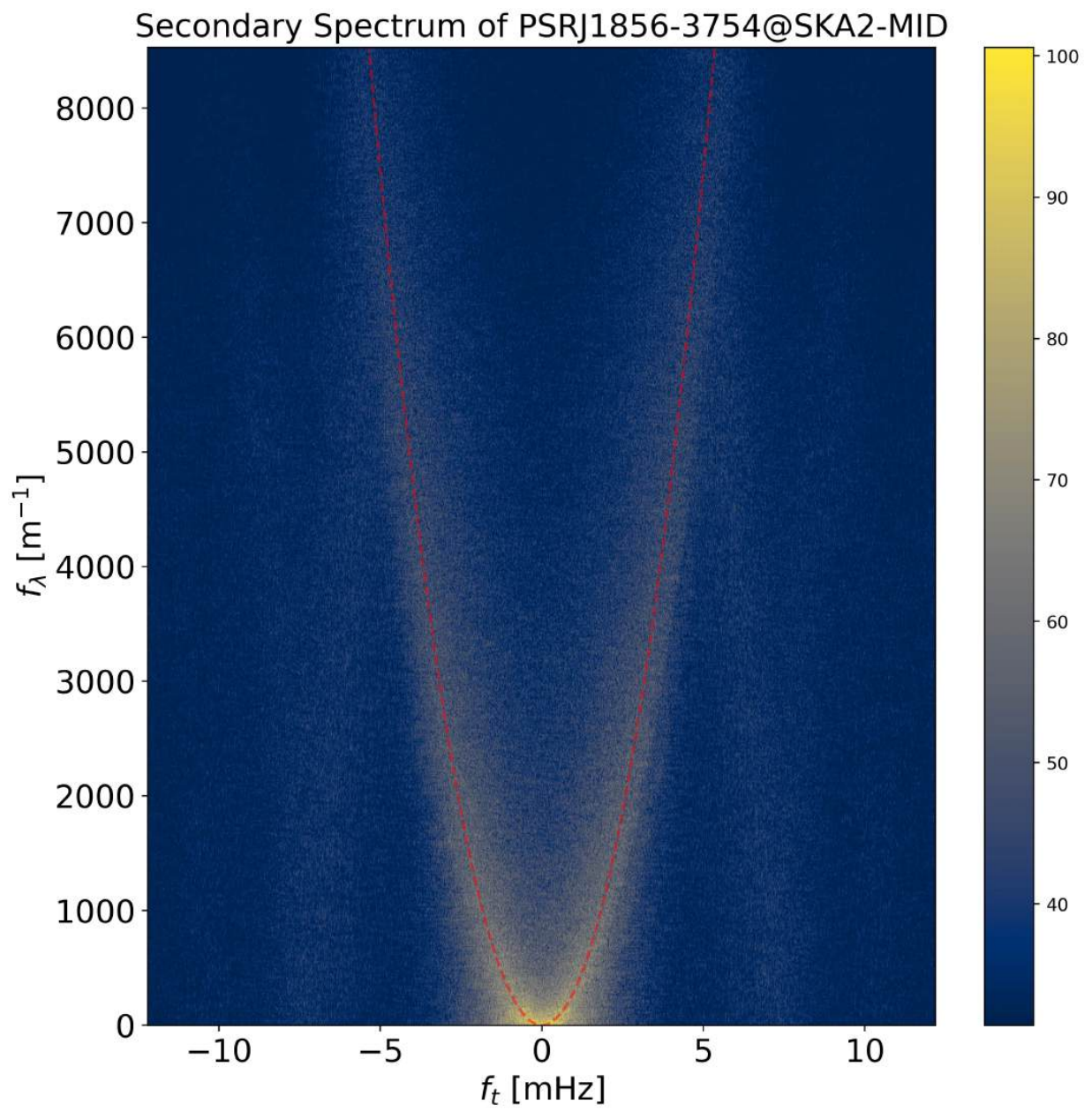}
	\includegraphics[width=0.23\linewidth]{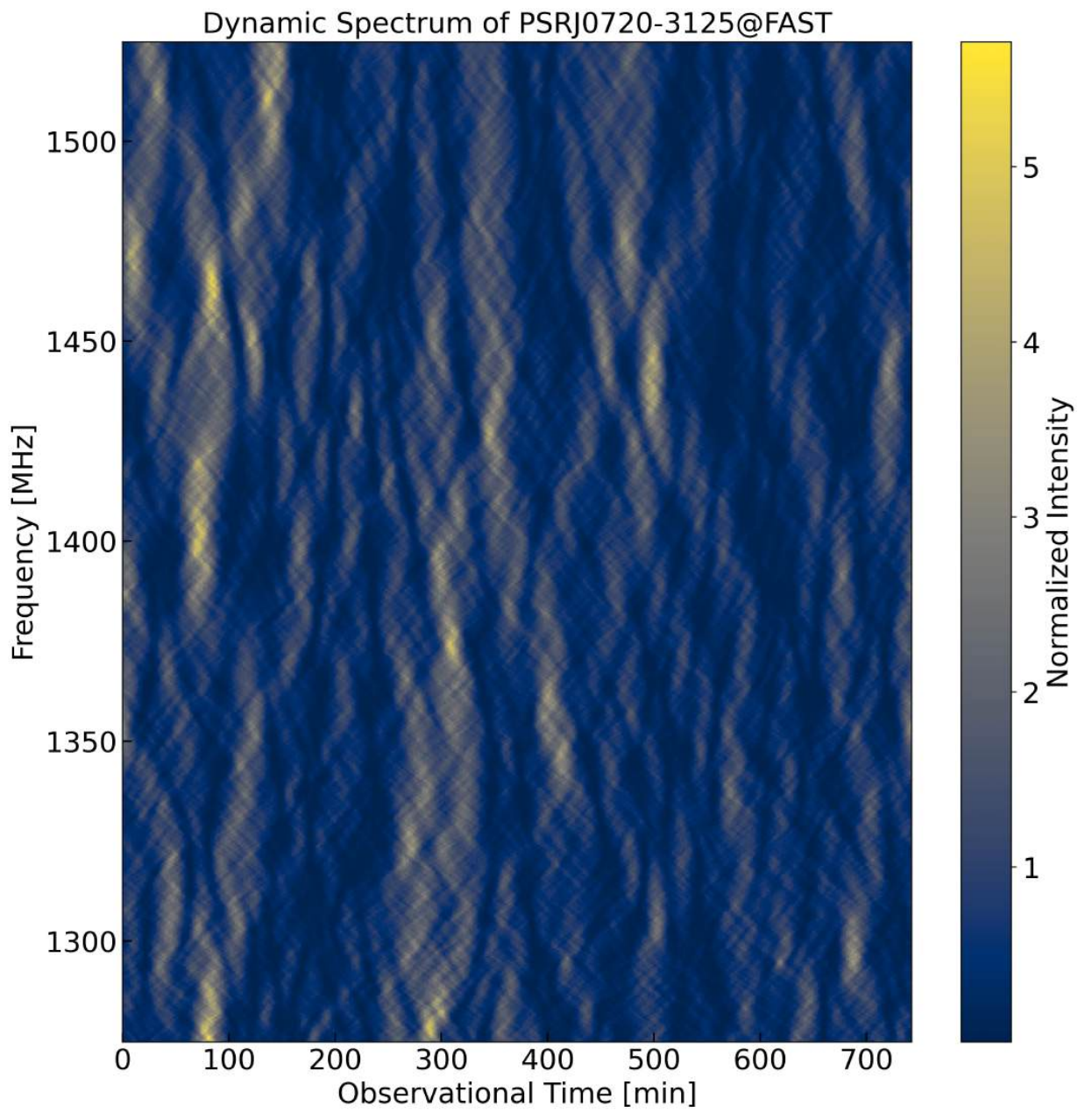}
	\includegraphics[width=0.23\linewidth]{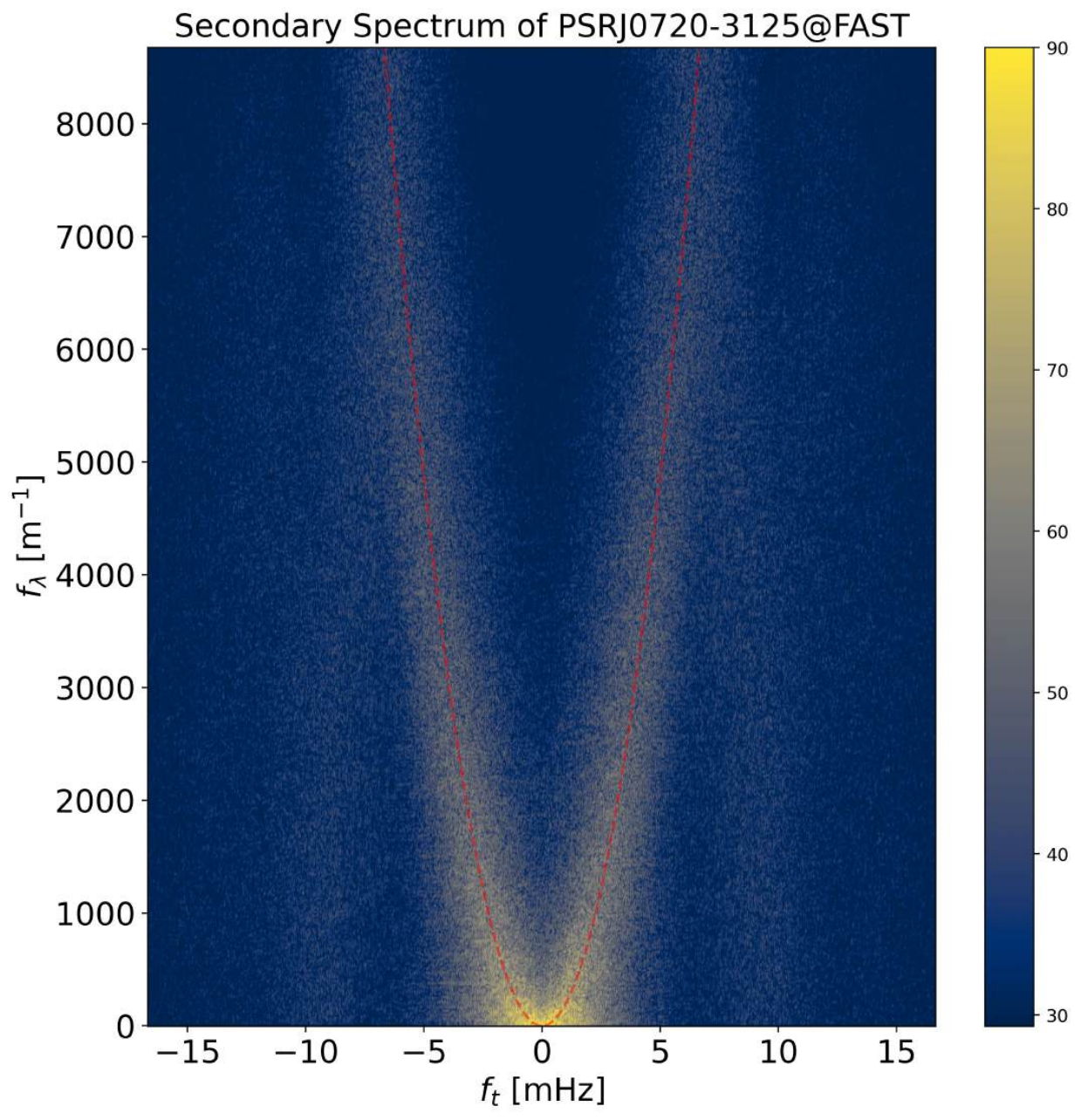}
	\includegraphics[width=0.23\linewidth]{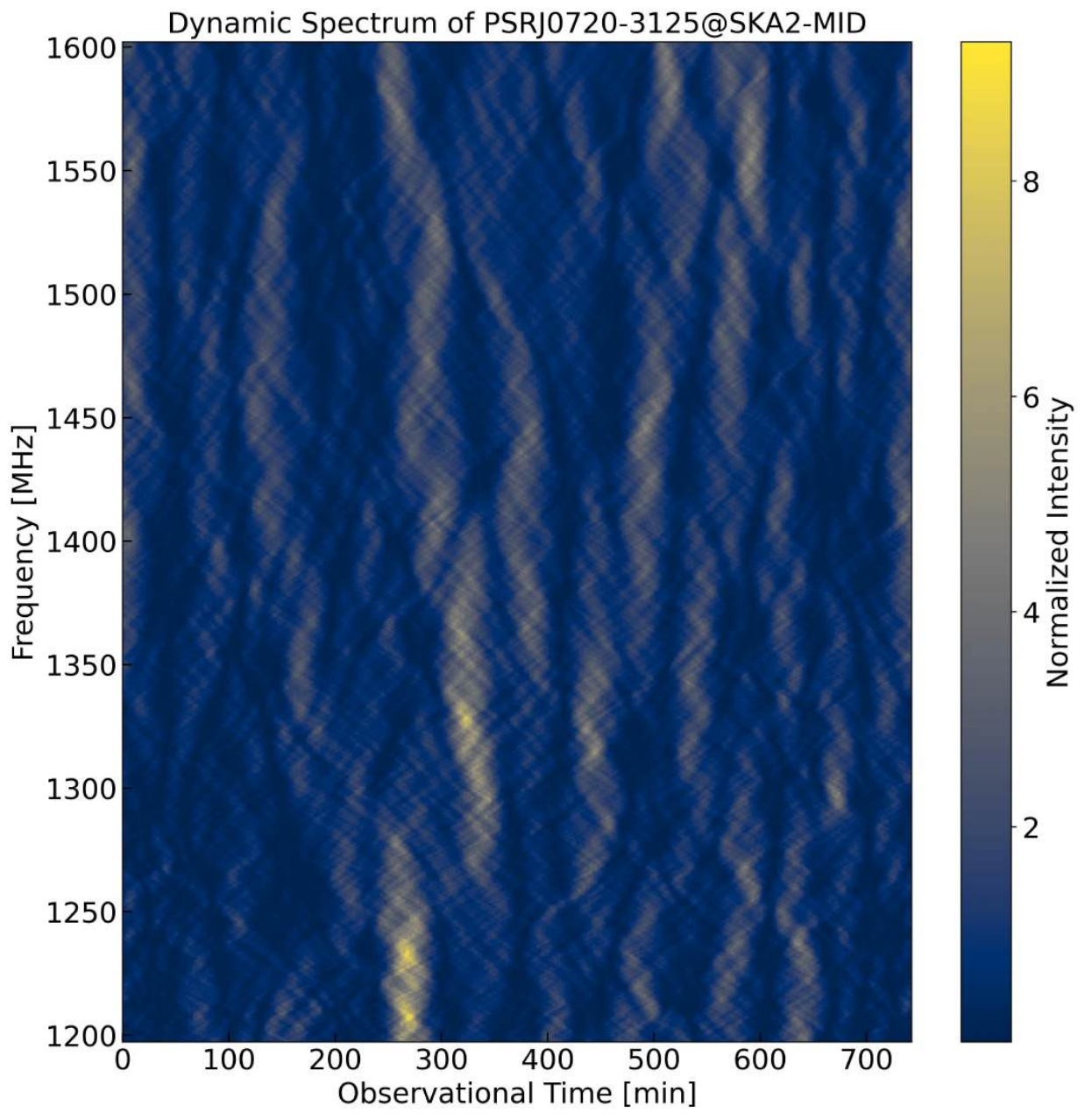}
	\includegraphics[width=0.23\linewidth]{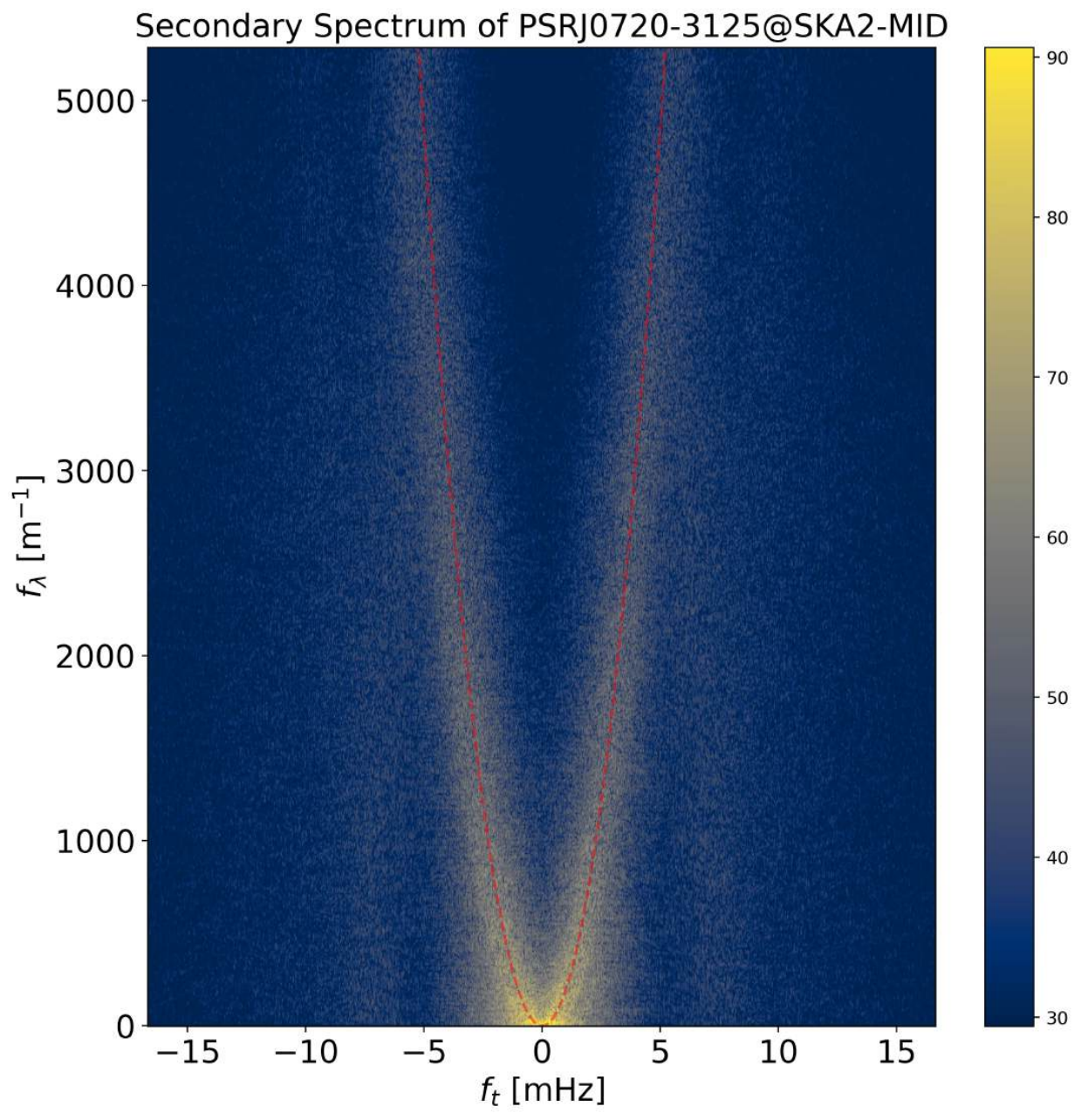}
	\caption{Variation of interstellar scintillation intensity with frequency and simulated observations of several pulsars using different telescopes, incorporating dynamic and second-order dynamic spectra. The upper panel shows the interstellar scintillation intensity versus observation frequency.  The scintillation intensity is shown by the solid blue line, and the operational frequencies of FAST and SKA2-MID are shown by the filled areas of different colours.  The central and lower panels sequentially display the 12-hour dynamic and second-order dynamic spectra of pulsars J1856-3754 and J0720-3125 under FAST and SKA2-MID observational circumstances.}
	\label{fig:interstellar scintillation}
\end{figure*}

The strength of the interstellar scintillation $u$ is related to the ratio of the refraction scale $l_{\mathrm{R}}=\frac{l_{\mathrm{F}}^2}{s_0}$ to the radius of the first Fresnel region $l_{\mathrm{F}}=\sqrt{\frac{d}{k}}$, where $d$ is the distance from the pulsar to the telescope, $k=\frac{2 \pi}{c} \mu f$ is the wavenumber, and $s_0=1 /\left(k \theta_{\mathrm{d}}\right)$ is the field coherence scale. The distance $d$ from the pulsar to the telescope can be obtained by consulting the catalogues; the refractive index of the magnetic fluid in the Milky Way $\mu=\sqrt{1-\frac{f_{\mathrm{p}}^2}{f^2} \mp \frac{f_{\mathrm{p}}^2 f_{\mathrm{B}}}{f^3}}$ is related to the observed frequency $f $, plasma frequency $f_{\mathrm{p}}=\sqrt{\frac{e^2 n_{\mathrm{e}}}{\pi m_{\mathrm{e}}}}$, and cyclotron frequency $f_{\mathrm{B}}=\frac{e B_{\|}}{2 \pi m_{\mathrm{e}} c}$ where $m_{\mathrm{e}}$ is electron mass. And the angular radius $\theta_{\mathrm{d}}=\sqrt{\frac{c\tau_s}{d}}$ of the diffuse disk centred on the pulsar can be calculated by combining the scattering time $\tau_s$ and pulsar distance $d$, where the scattering time can be estimated by the empirical formula $\ln\tau_{\mathrm{s}}=-6.46+0.154 \ln(\mathrm{DM})+1.07(\ln\mathrm{DM})^2-3.86 \ln f$ in relation to the logarithm of the DM \citep{2004ApJ...605..759B}. In Figure \ref{fig:interstellar scintillation}, we use coloured solid lines to show the results of the frequency variation of the interstellar scintillation intensity that corresponds to the two pulsars at the top panel. The regions coloured differently represent the operating frequencies of different telescopes. The scintillation is strong in the L-band of FAST and SKA2-MID. Accordingly, by using the proper motion of the pulsar and annual parallax, combined with the interstellar scintillation simulation and processing program ``Scintools" \citep{2020ApJ...904..104R,2021MNRAS.500.1114S,2022MNRAS.510.4573B,2010ApJ...717.1206C,2023MNRAS.521.6392R}, we can plot in Figure \ref{fig:interstellar scintillation} the scintillation dynamic spectrum and the second-order dynamic spectrum $S_{\mathrm{spe}}^{\mathrm{sec}}(f_t,f_{\lambda})$ of our signal observed for two continuous full-time observations of FAST and SKA2-MID, that is, for 12 hours from the second panel to third panel. The secondary spectrum $S_{\mathrm{spe}}^{\mathrm{sec}}(f_t,f_{\lambda})$ is obtained by performing a two-dimensional Fourier transform of the dynamic spectrum $S_{\mathrm{spe}}(\nu,t)$, resulting in a representation in conjugate fringe-rate $f_t$ and time-delay $f_\lambda$ space. Here, the time-delay encodes differential propagation delays arising from different scattering paths through the interstellar medium, whereas the fringe-rate corresponds to Doppler shifts resulting from relative motions between the pulsar, the scattering screen, and the observer. In this case, the observed dynamic spectra $S_{\mathrm{spe}}(\nu,t)$ can be expressed as the superposition of the original scintillation intensity of the pulsar $S_{\mathrm{spe,pulsar}}(\nu,t)$ and the scintillation intensity of the GW signal $S_{\mathrm{spe,GW}}(\nu,t)$. It can be seen from Figure \ref{fig:convesion-probability}, since the intensity variation period of the radio signal generated by the resonance of GWs with the magnetic field is almost the same as the rotation period of the pulsar, the secondary dynamic spectrum $S_{\mathrm{spe}}^{\mathrm{sec}}(f_t,f_{\lambda})$ obtained through Fourier transform presents a clear scintillation arc. And its intensity can be approximately regarded as the sum of the signal intensities of the pulsar $S_{\mathrm{spe,pulsar}}^{\mathrm{sec}}(f_t,f_{\lambda})$ and the GW $S_{\mathrm{spe,GW}}^{\mathrm{sec}}(f_t,f_{\lambda})$.

\subsection{Polarization transfer and depolarization model}\label{app:depol-details}
This subsection details the complete depolarization modeling and includes an illustrative FRB figure adapted from Section~\ref{sec:SR-RT}, presented here to provide a more comprehensive view of the methodology.

\begin{figure*}
	\centering
	\includegraphics[width=0.45\linewidth]{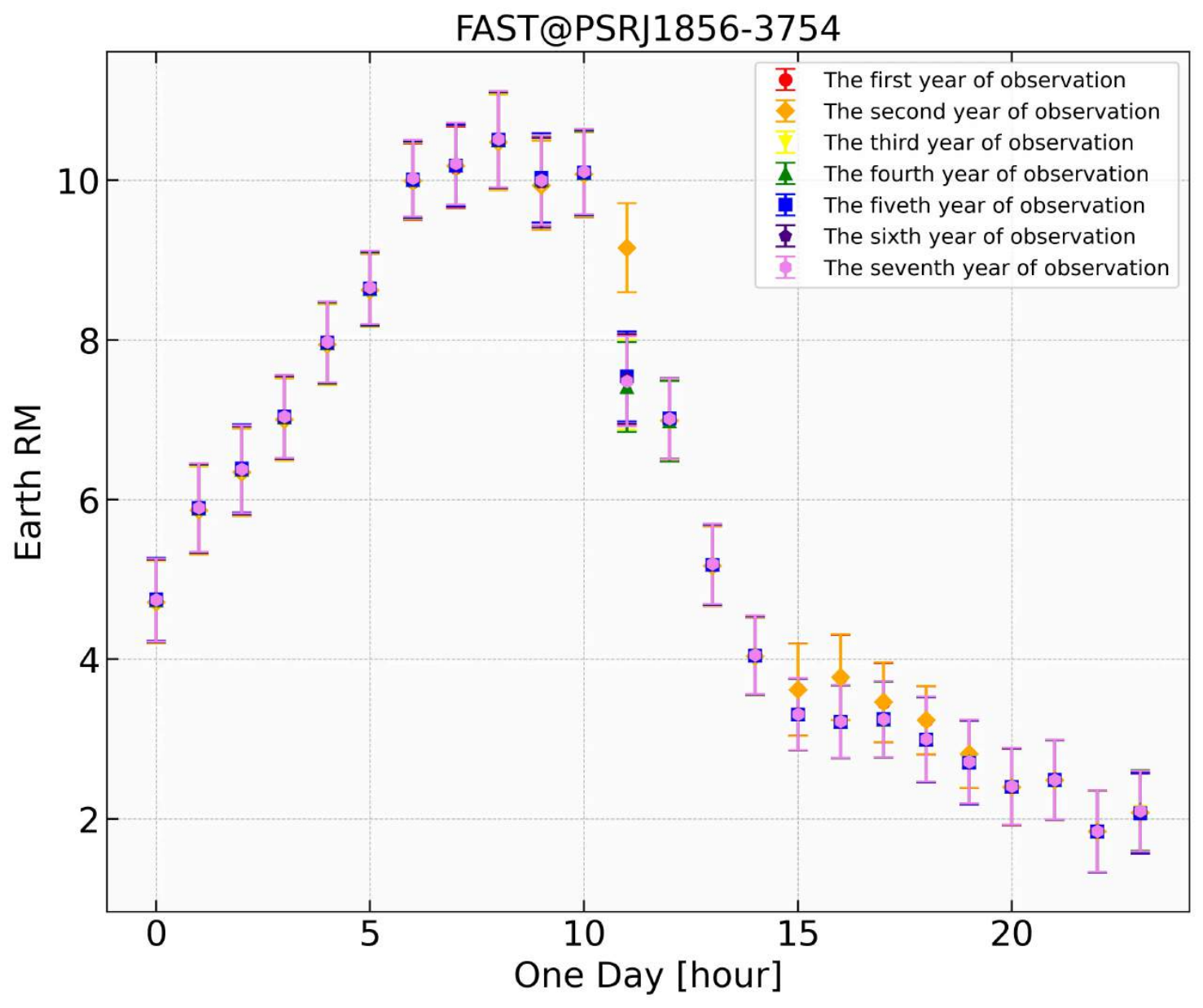}
	\includegraphics[width=0.45\linewidth]{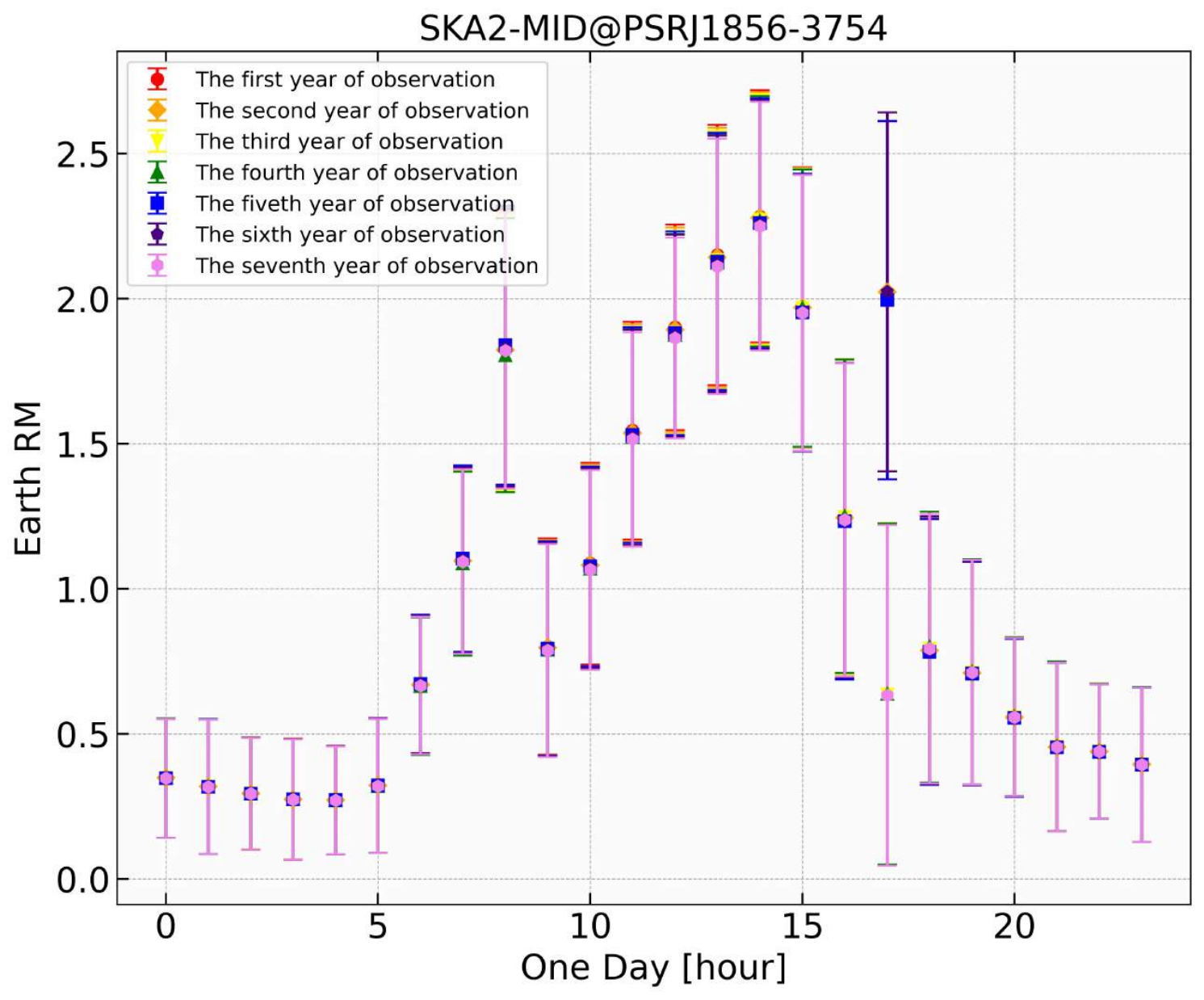}
	\includegraphics[width=0.45\linewidth]{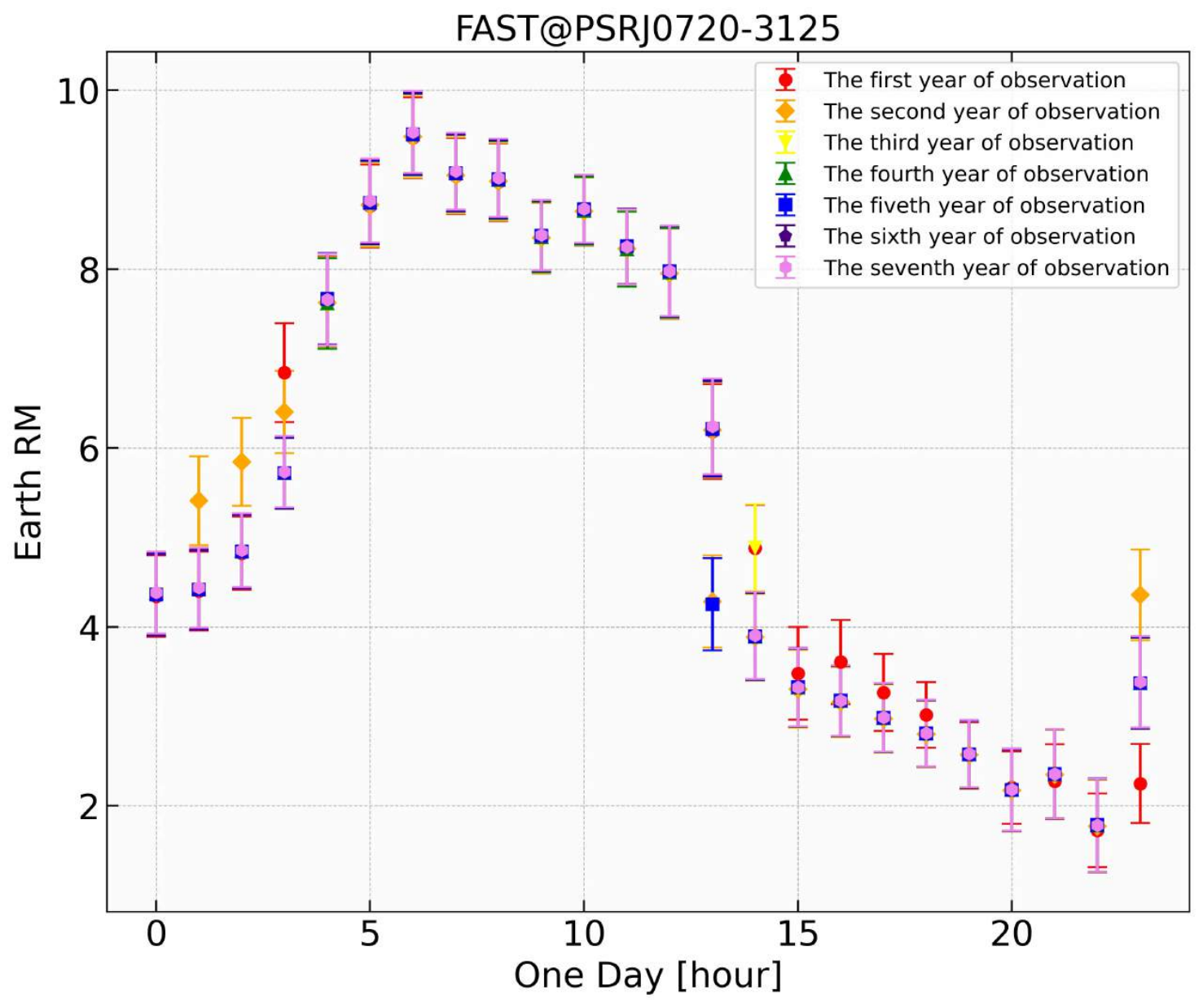}
	\includegraphics[width=0.45\linewidth]{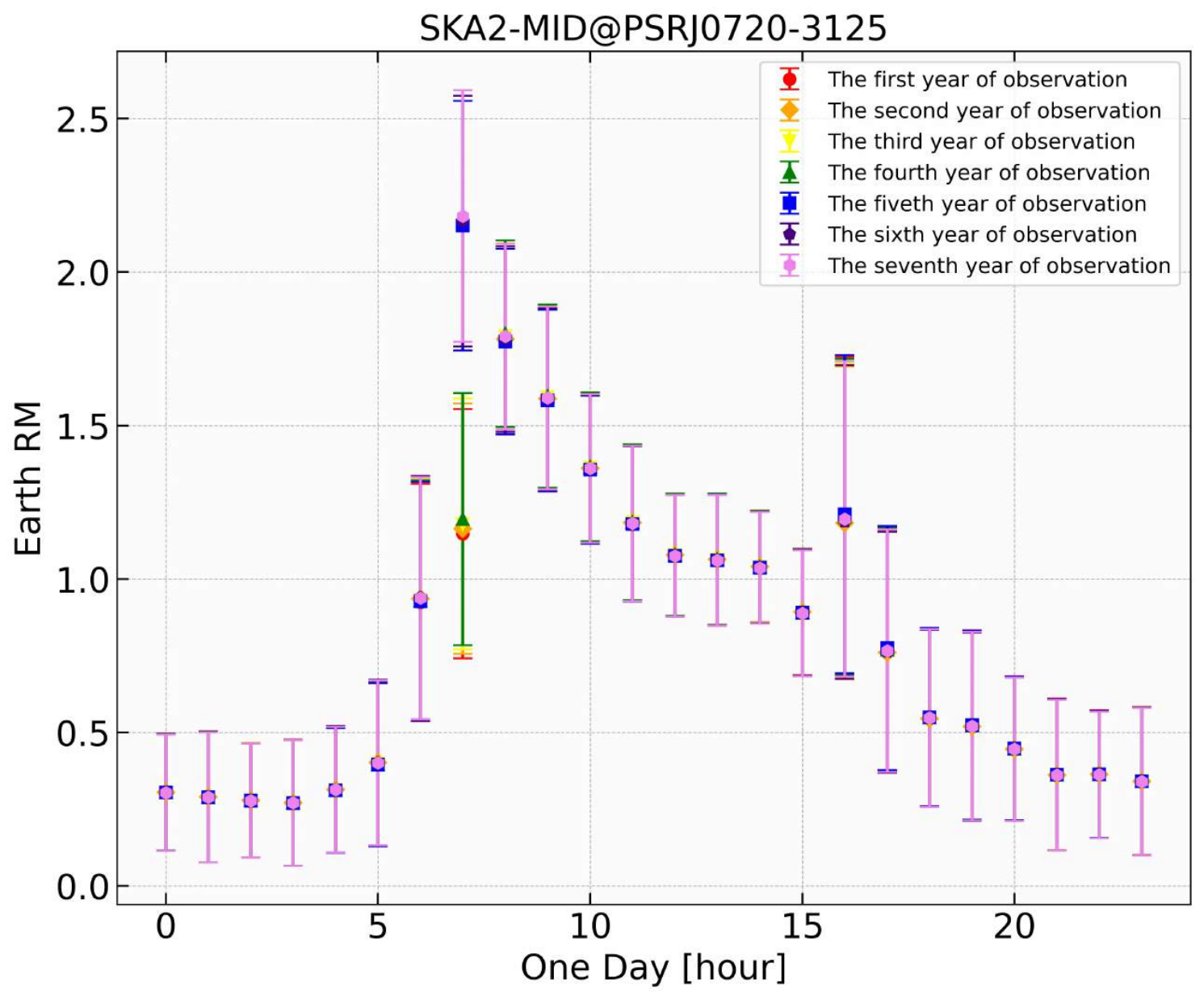}
	\caption{Simulation of the RM of the Earth ionosphere on June 1st of each year of the seven-year observation period. We use different colours and shapes to represent different years of RM, and different years of RM errors are represented by the corresponding colours. The top panel shows the RM in the LOS direction during the observation of pulsar J1856-3754, and the bottom panel demonstrates the RM in the LOS direction during the observation of pulsar J0720-3125. The left panel displays the simulated observation conducted with FAST, whereas the right panel presents the simulated observation performed with SKA2-MID.}
	\label{fig:Earth-RM}
\end{figure*}

Although the linear polarization characteristics of the radio signals of GW enable them to effectively penetrate the magnetosphere of pulsars, the Faraday rotation effect can cause the polarization plane to rotate, resulting in the depolarization of the signal. It is worth noting that during this propagation process, the RM of the signal is only affected by the interstellar medium and the ionosphere of the Earth, and has nothing to do with the RM contribution of the pulsar magnetic field. To quantify the ionospheric contribution, we simulate the time-dependent RM induced by the Earth's ionosphere along the LOS to our target pulsars over the proposed observational period. The results, showcasing the annual variation for both FAST and SKA2-MID telescopes, are presented in Figure \ref{fig:Earth-RM}.

Theoretically, we can calculate the received polarization through the radiation transfer equation containing Stokes parameters:
\begin{equation}
	\frac{d}{d s} S_{\mathrm{t}, i}^{\mathrm{para}}=\epsilon_i-\sum_{j=1}^{j=4}\left[\left(\eta_{\alpha \beta}-\rho_{\alpha \beta}\right)_{i j} S_{\mathrm{t}, j}^{\mathrm{para}}\right],
\end{equation}
where $S_{\mathrm{t}}^{\text {para }}=(I, Q, U, V)^T$ represents the vector of the four Stokes parameters, the labels $i=j=1,2,3,4$ represent the four components $I, Q, U, V$, and $\epsilon_i$ represents the spontaneous emission coefficients. The Faraday rotation and conversion coefficients are represented as the $\rho_{\alpha \beta}$ tensor, and the Stokes parameter absorption coefficients are represented as the $\eta_{\alpha \beta}$ tensor. Then, we can obtain the change in linear polarization. To determine the depolarization, using the complex linear polarization $\mathcal{P}=Q+\mathrm{i} U$ along with the Faraday dispersion function $\mathcal{F}(\phi)$ and the RM propagation function $\mathcal{R}(\phi)$ is easier \citep{1966MNRAS.133...67B}. The complex linear polarization can be expressed by the polarization intensity $P_{0}$ and polarization angle $\chi_0=\frac{1}{2} \arctan\left(\frac{U}{Q}\right)$, and the Faraday rotation acts on the polarization angle $\chi=\chi_0+\mathrm{RM}\times\lambda^2$. In this case, the depolarization can be described by three generation mechanisms: One is depth depolarization, which is a function of the square of the observed wavelength $\lambda^2$, as we assume that the Faraday depth $\phi$ in the LOS direction is fixed \citep{2002ApJ...575..225U}:
\begin{equation}
	\mathcal{P}_{\mathrm{depth}}=P_{0} e^{2 i \chi_0}\left(\frac{1-e^{-S}}{S}\right).
\end{equation}
It is a combination of internal Faraday dispersion, differential Faraday rotation, and depolarization caused by changes in the intrinsic polarization angle $\chi_0$ along the LOS. When $S=2 \sigma_{RM}^2 \lambda^4-2 i \phi \lambda^2$, it contains turbulent and regular magnetic fields together. $\sigma_{RM}=0.812^2 \left( \frac{\Lambda_c}{1\mathrm{~pc}}\right) \int\left(\frac{n_e}{1\mathrm{~cm}^{-3}} \right)^2\left(\frac{B_{\|}}{10^{-6}\mathrm{~G}}\right)^2 \frac{\mathrm{~d} l}{1\mathrm{~pc}} \mathrm{~rad}^2 \mathrm{~m}^{-4}$ is the standard deviation of the RM, and $\Lambda_c$ is the correlation length of the magnetic field \citep{Haverkorn:2004kx,2021MNRAS.502.2220S,2017NatAs...1..621M,Unger:2023lob}. Another mechanism is beam depolarization, which can be represented by Burn's law \citep{1966MNRAS.133...67B}:
\begin{equation}
	\mathcal{P}_{\mathrm{beam}}=P_{0} e^{-2\sigma_{RM}^2 \lambda^4}.
\end{equation}
It is the result of averaging vectors in adjacent directions perpendicular to the LOS within the beam of the same telescope. The last mechanism is bandwidth depolarization, which is caused by the averaging effect of the limited channel bandwidth and the frequency-dependent polarization caused by Faraday rotation, and it usually occurs within a frequency channel \citep{2023MNRAS.520.4822F}:
\begin{equation}
	\mathcal{P}_{\mathrm{bandwidth}}=\frac{P_{0}}{\Delta\nu}\int e^{-2i RM \left( \frac{c}{\nu}\right)^2} d\nu,
\end{equation}
where $\Delta\nu$ is the channel frequency width. We show the combined effect of these three mechanisms $\mathcal{P}_{\mathrm{total}}=\mathcal{P}_{\mathrm{depth}}\mathcal{P}_{\mathrm{beam}}\mathcal{P}_{\mathrm{bandwidth}}$ on our signal and some repeated and nonrepeated FRB signals observed within the Milky Way in Figure \ref{fig:depolarization-FRB}. In the upper part of Figure \ref{fig:depolarization-FRB}, we show how the theoretical linear polarization of the electromagnetic signal generated by conversion in the magnetosphere of two selected pulsars varies with the frequency (RM) and telescope. The linear polarization calculated using the parameters of FAST is represented by coloured solid lines, and the linear polarization calculated using the parameters of SKA2-MID is represented by coloured dashed lines. Different colours correspond to the $10^{th}$, $50^{th}$ and $90^{th}$ percentiles of the total RM. In the lower part of Figure \ref{fig:depolarization-FRB}, we demonstrate the use of our depolarization model to parametrically fit the linear polarization of two repeated FRBs and five single FRBs \citep{Feng:2022ill}. Different graphs in the figure represent different FRB events, corresponding to their polarization degree and observed frequency; the filled colour represents their RM size, and the error bar on the graph indicates the polarization measurement error of the FRB event. We use coloured solid lines to show the fitting results of repeated bursts and coloured dashed lines to show the fitting results of single bursts. The figure demonstrates the good fit of our model for repeated bursts, but it is not ideal for single bursts. However, it effectively captures the inhomogeneity and turbulence of the magnetic field.

\subsection{Diagnostic Tests for Stationarity and Gaussianity}
\label{app:diagnostics}

In this appendix we provide additional diagnostic figures that support the validity of the statistical assumptions adopted in our simulation and analysis pipeline. Throughout the main text we treat the simulated observational time series as approximately stationary and Gaussian, which underpins the use of cross-correlation statistics and subsequent filtering steps. To substantiate these assumptions, we apply multiple, complementary tests to the mock data generated for both pulsars (PSR~J1856$-$3754 and PSR~J0720$-$3125) and for both representative event classes (transient events and stochastic backgrounds), each evaluated on two characteristic time scales as defined in the main text.

\begin{figure*}
	\centering
	\includegraphics[width = 0.4\textwidth]{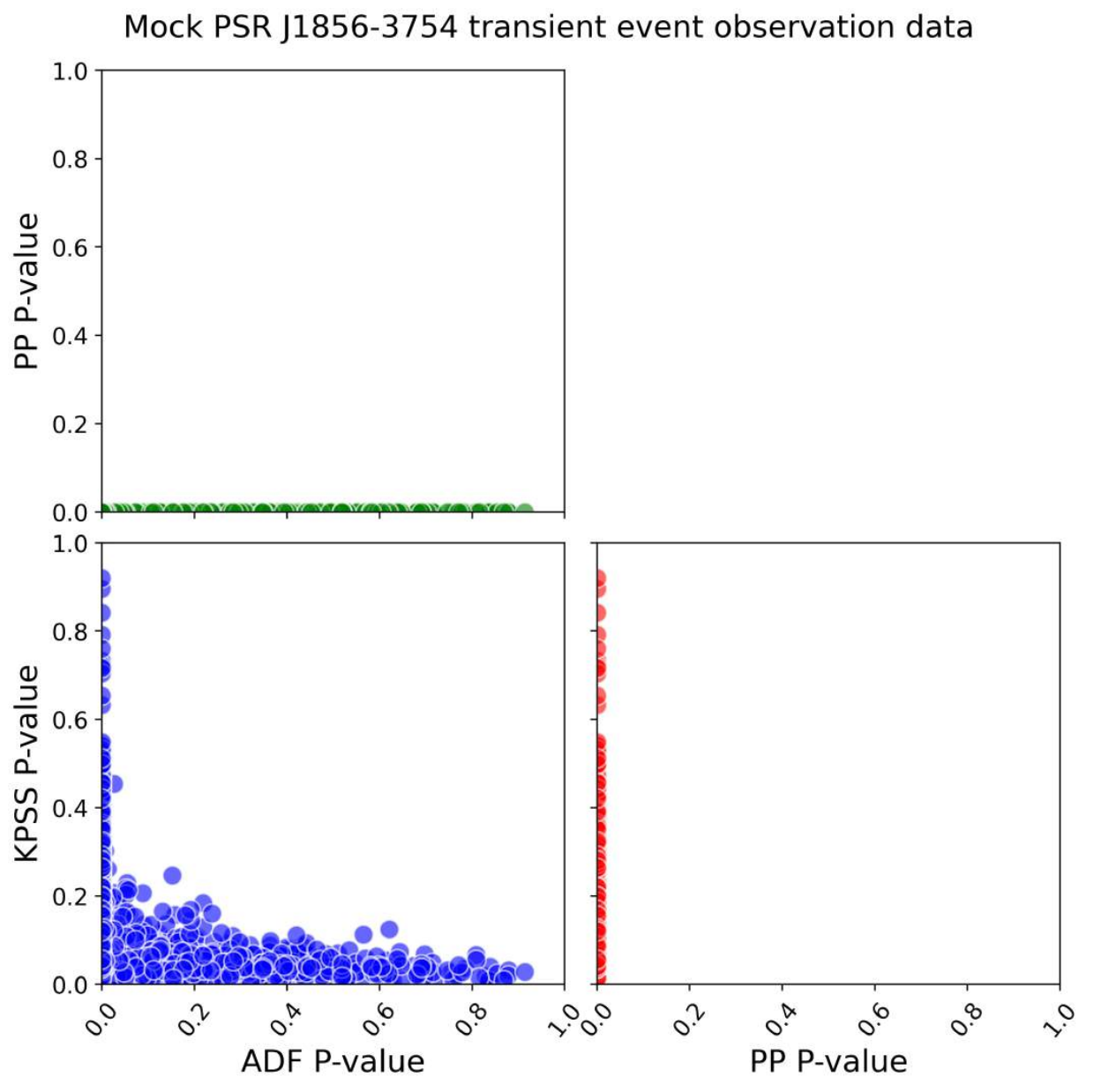}
	\includegraphics[width = 0.4\textwidth]{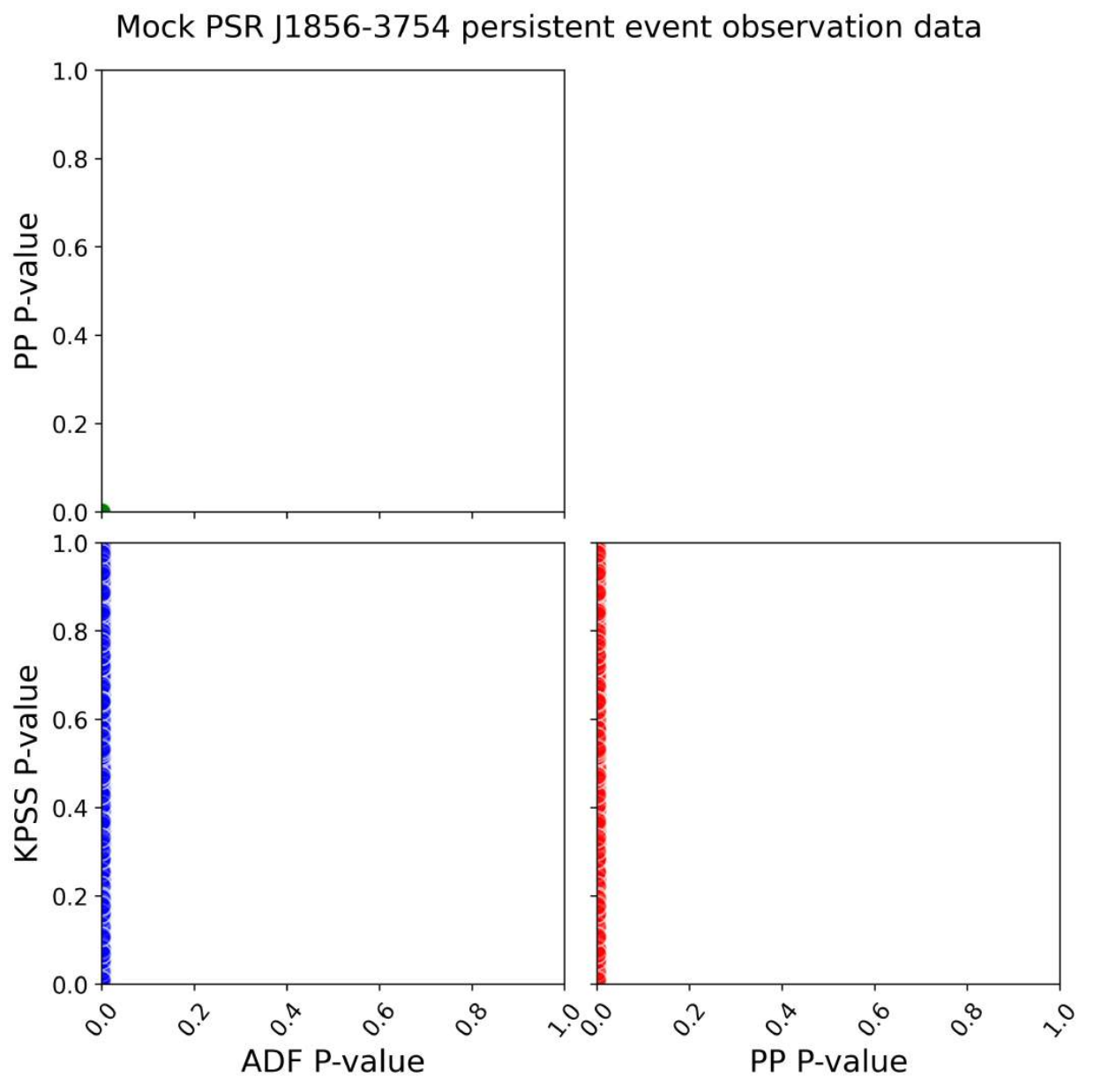}
	\includegraphics[width = 0.4\textwidth]{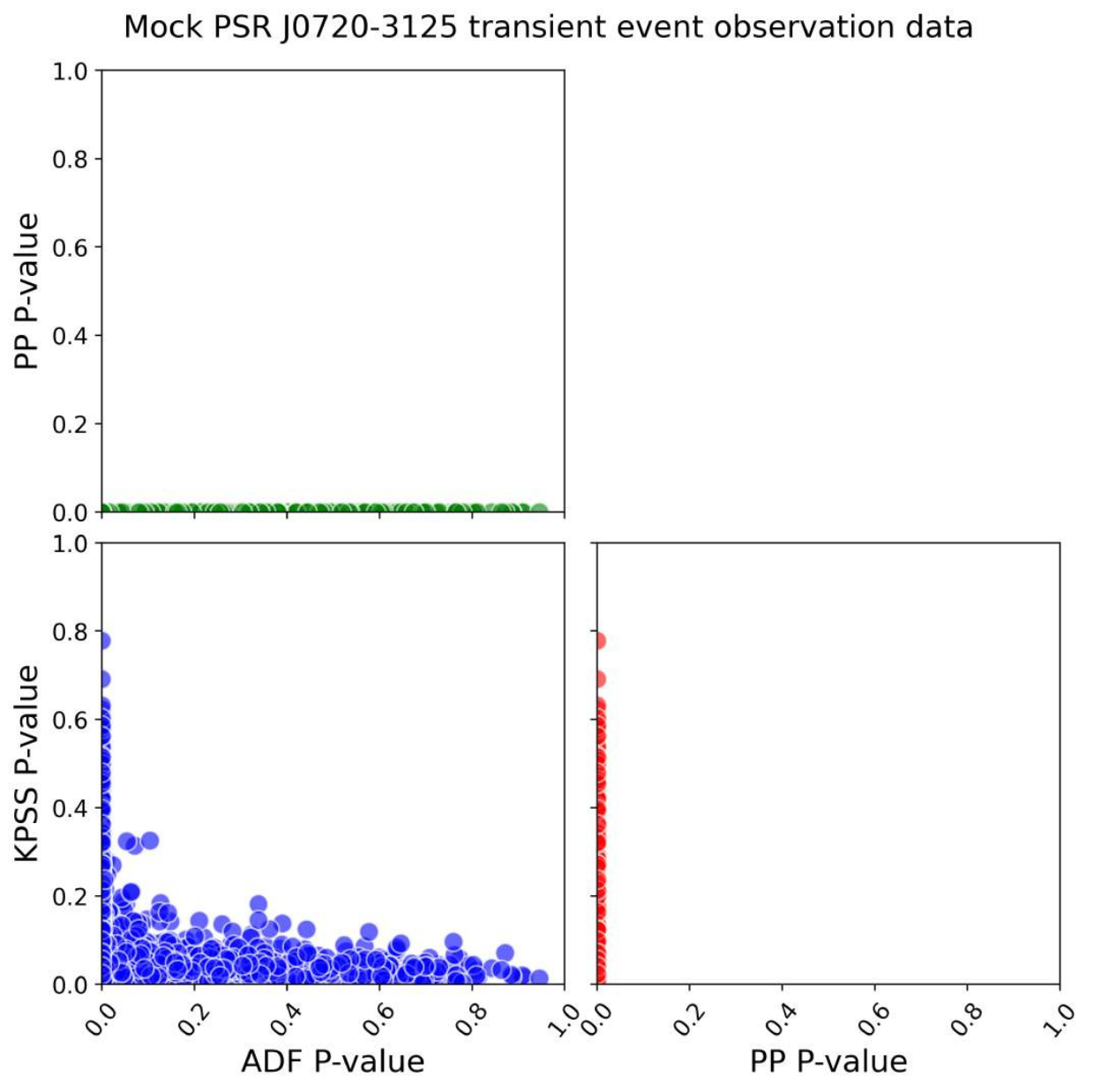}
	\includegraphics[width = 0.4\textwidth]{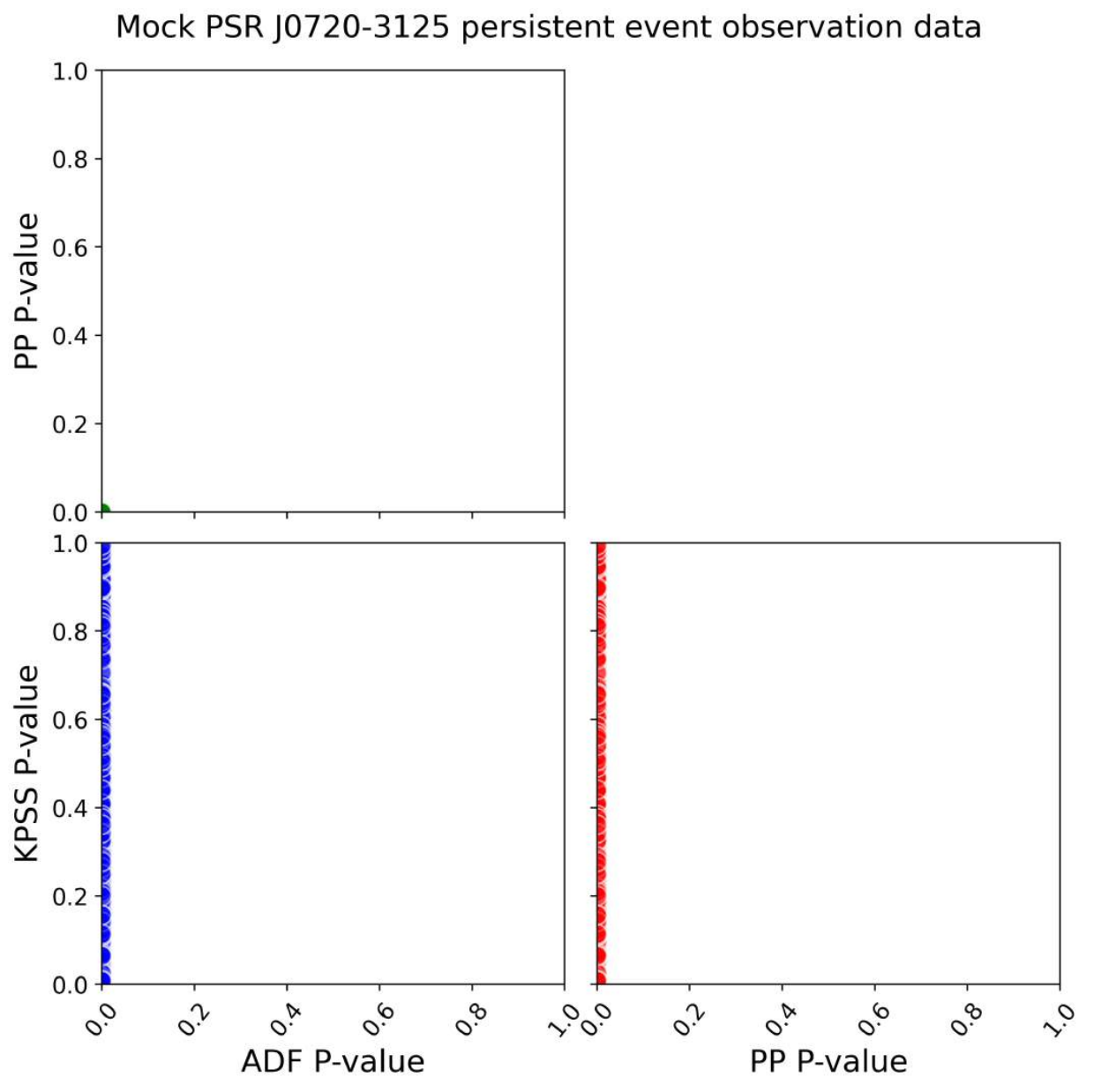}
	\caption{Joint distribution of stationarity test results for mock observations. We show the results of mocked data tests for four sets of events containing two pulsars on two different time scales, where the different joint distributions are represented by scatter plots in different colours.}
	\label{fig:Stationarity-test}
\end{figure*}

Figure~\ref{fig:Stationarity-test} summarises the joint outcomes of the augmented Dickey--Fuller (ADF), Phillips--Perron (PP), and KPSS tests. We visualise the combined decision regions by colour-coding the three-test outcomes to highlight potential tension between different stationarity criteria. Overall, the mock datasets are consistent with stationarity. A small subset of transient realisations exhibits weaker trend-stationarity signatures, which is expected because a short-duration transient can introduce large localised fluctuations relative to the finite observing window.

\begin{figure*}
	\centering
	\includegraphics[width = 0.4\textwidth]{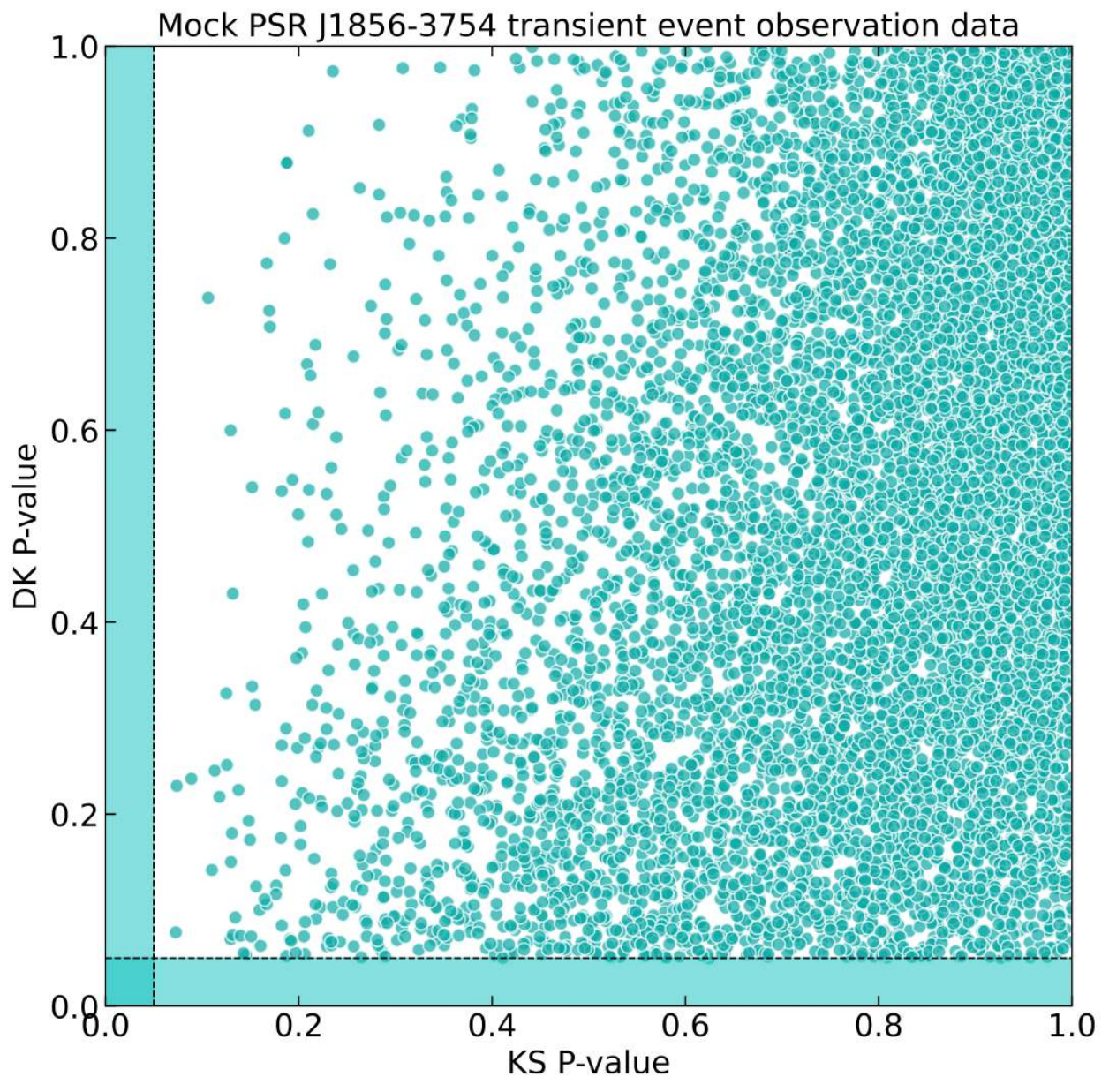}
	\includegraphics[width = 0.4\textwidth]{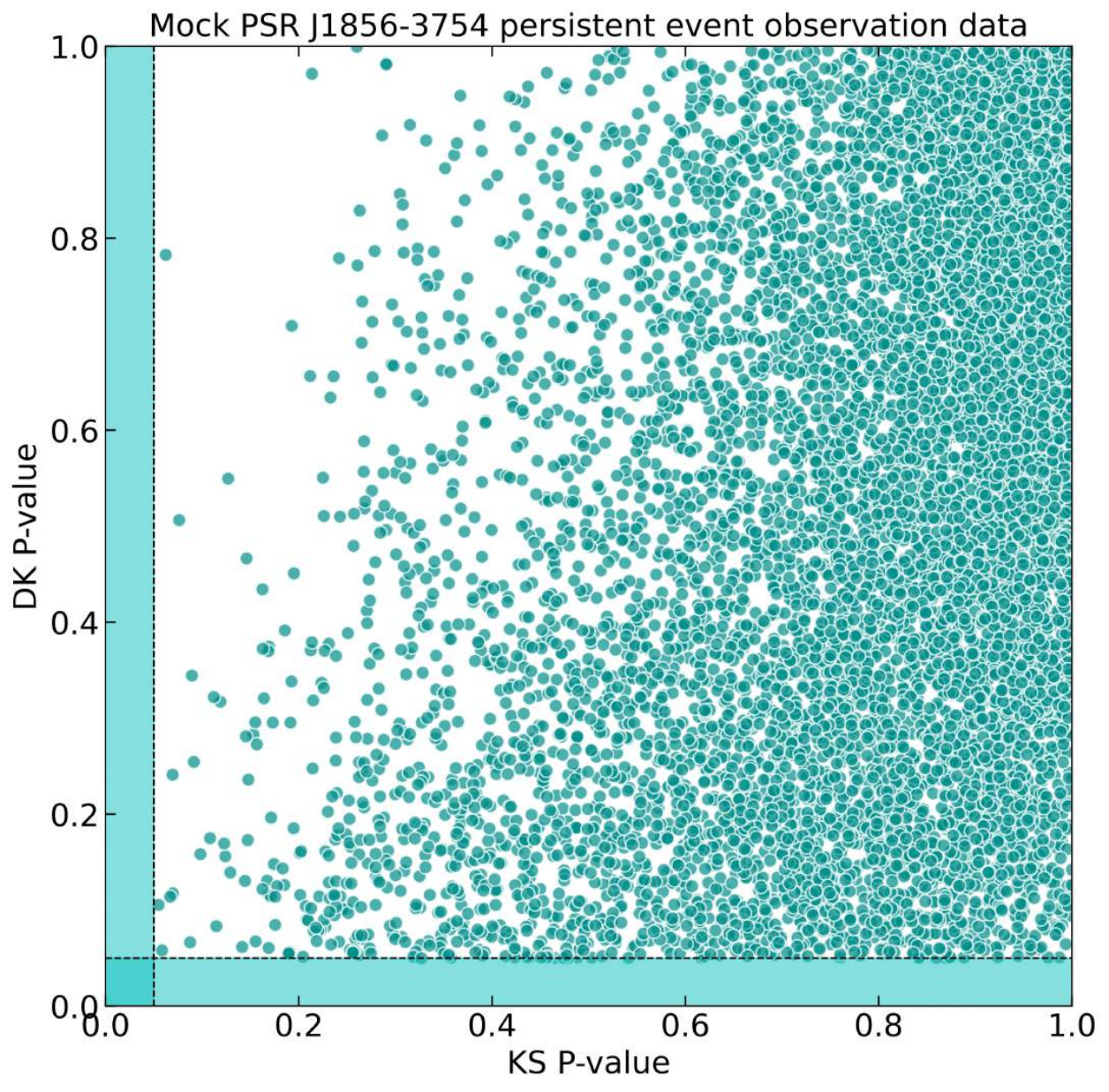}
	\includegraphics[width = 0.4\textwidth]{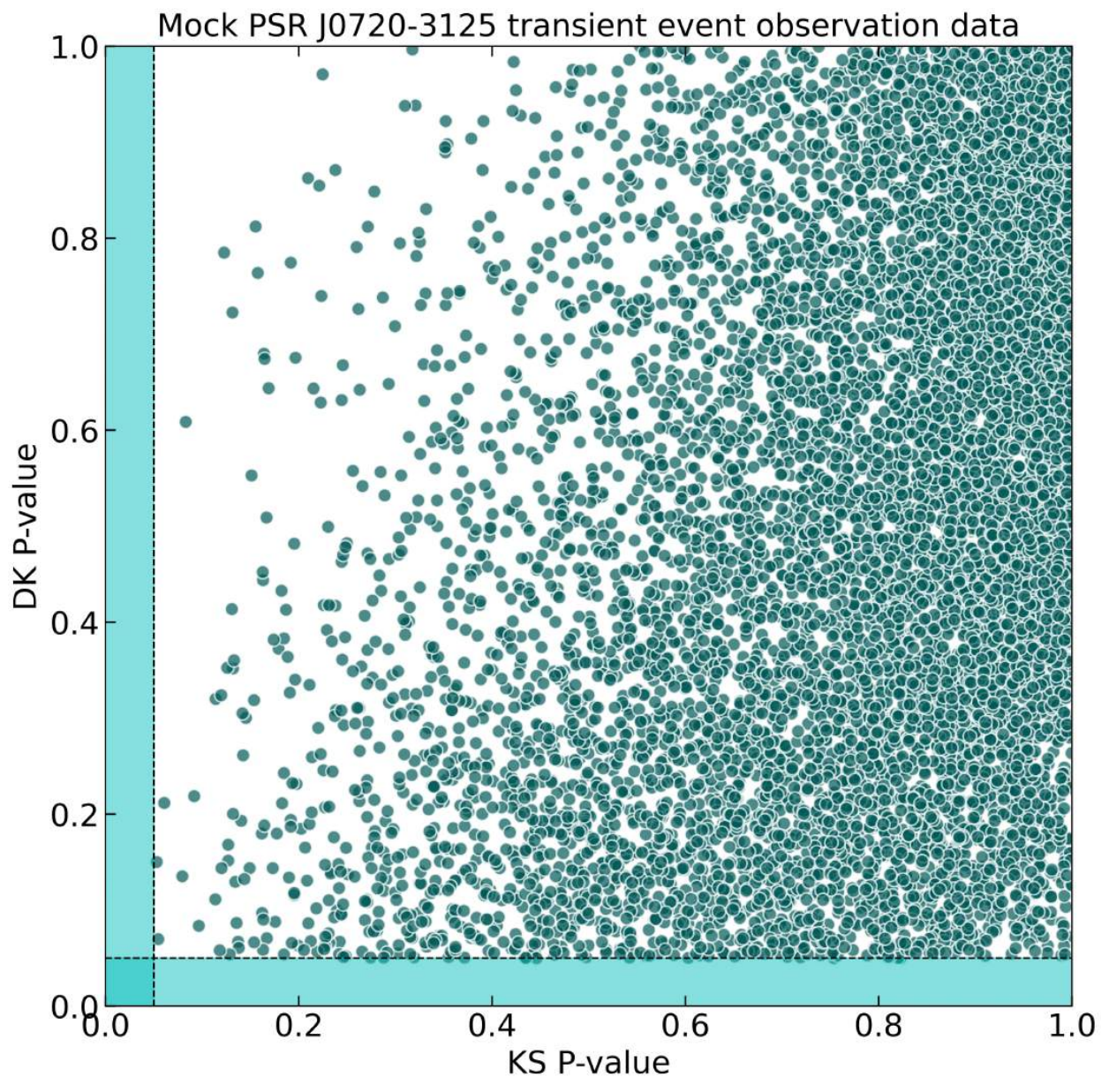}
	\includegraphics[width = 0.4\textwidth]{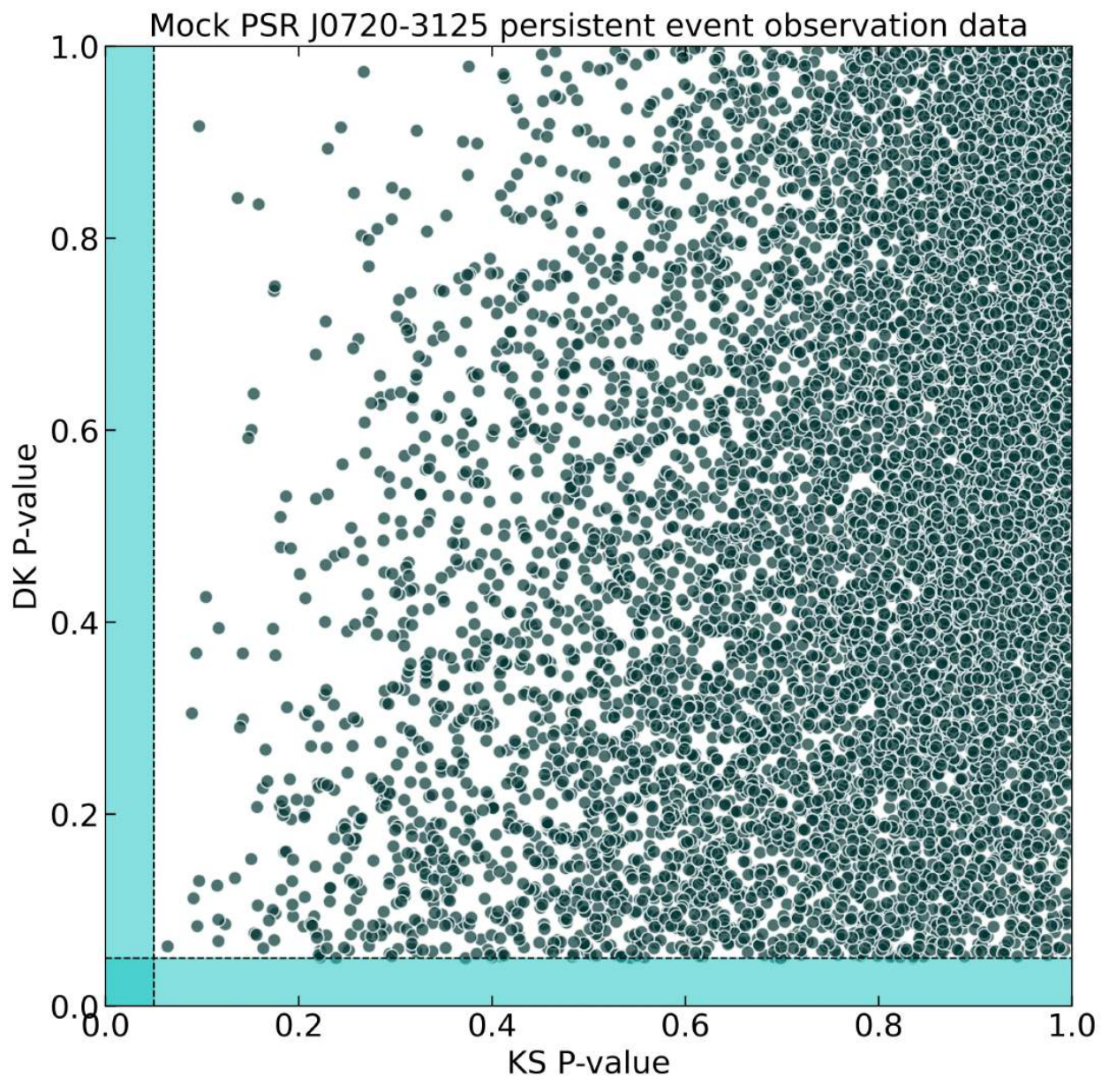}
	\caption{Joint distribution of Gaussian test outcomes for simulated data. We provide the outcomes of simulated data testing for four event sets with two pulsars across two distinct time scales, highlighting locations that deviate from the Gaussian distribution with light blue shading.}
	\label{fig:Gaussianity-test}
\end{figure*}

Figure~\ref{fig:Gaussianity-test} reports the Gaussianity checks using the Kolmogorov--Smirnov (KS) test and D'Agostino's $K^2$ test. The shaded regions mark parameter combinations where the null hypothesis of Gaussian residuals would be rejected at the chosen significance level. We find no evidence for non-Gaussian residuals in the simulated datasets considered here, supporting the Gaussian-noise assumption used in the sensitivity forecasts and strategy comparisons.

\subsection{CNN-based initialization for the BCKA filter}\label{app:cnn-details}
This subsection outlines the CNN hyperparameter optimization and distance-function comparison employed in initializing the BCKA filter in Section~\ref{sec:BCKA-filter}, with full implementation details provided here for clarity and reproducibility.

\begin{figure*}
	\centering
	\includegraphics[width = 0.4\textwidth]{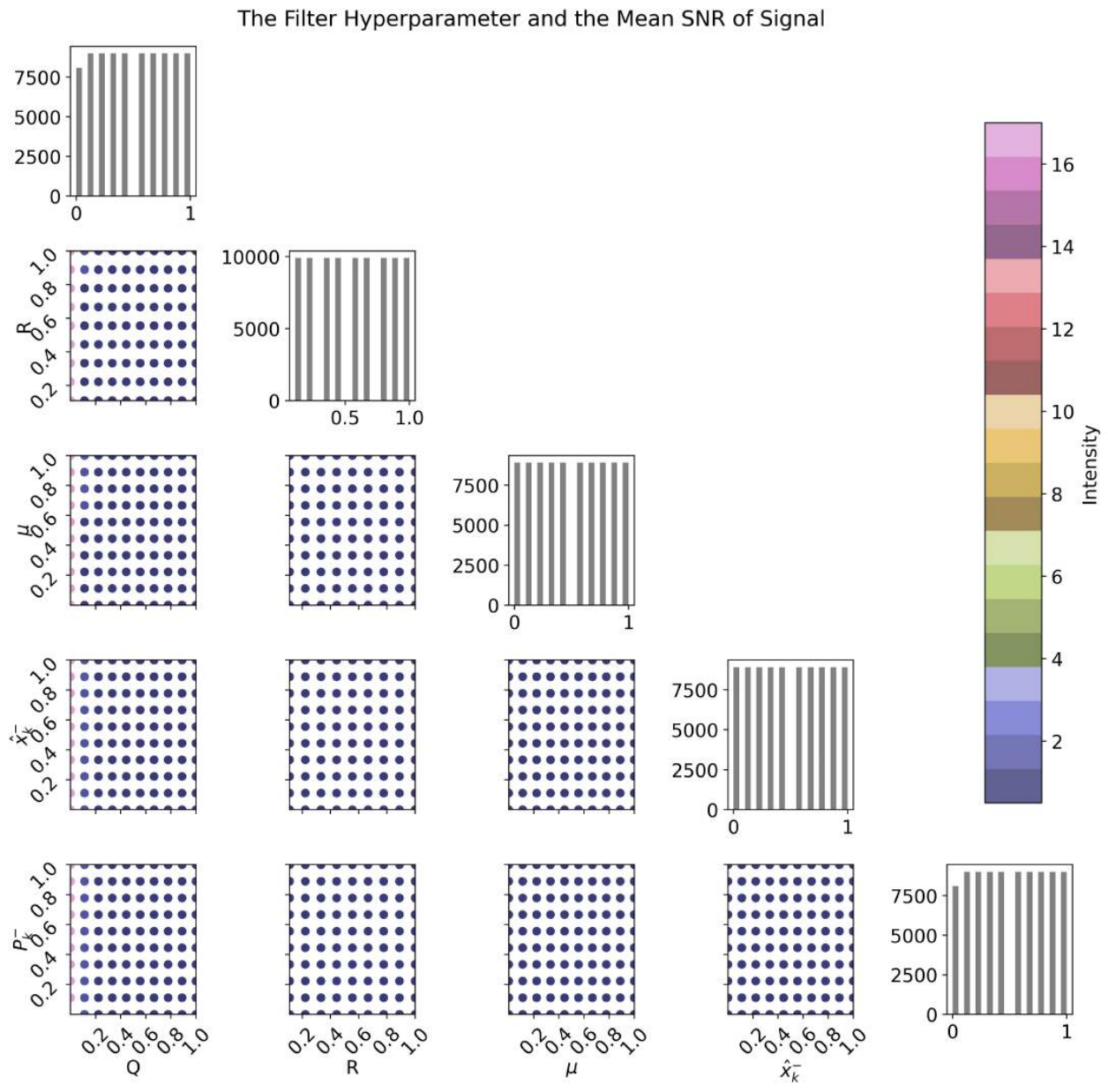}
	\includegraphics[width = 0.4\textwidth]{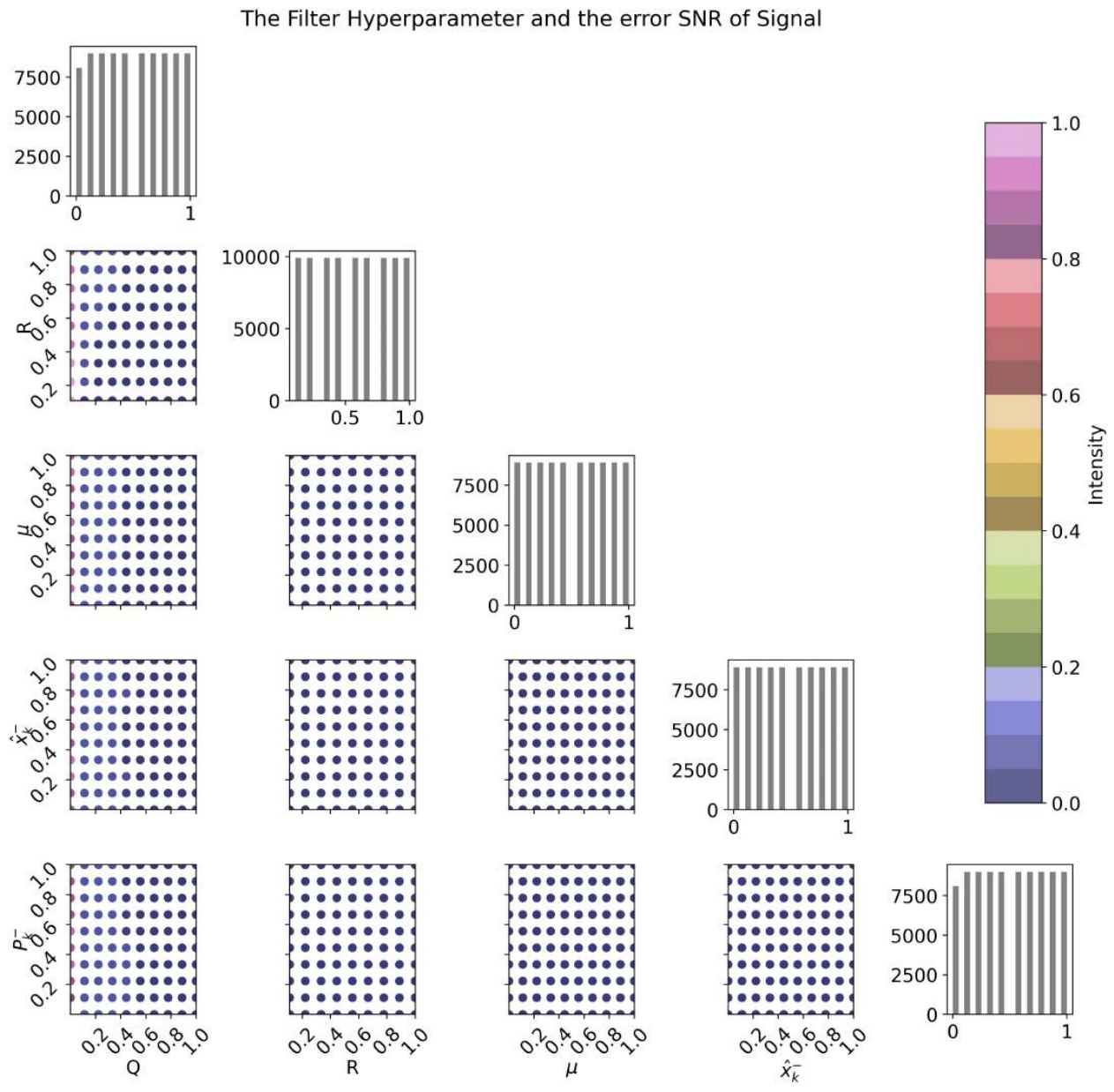}
	\includegraphics[width = 0.32\textwidth]{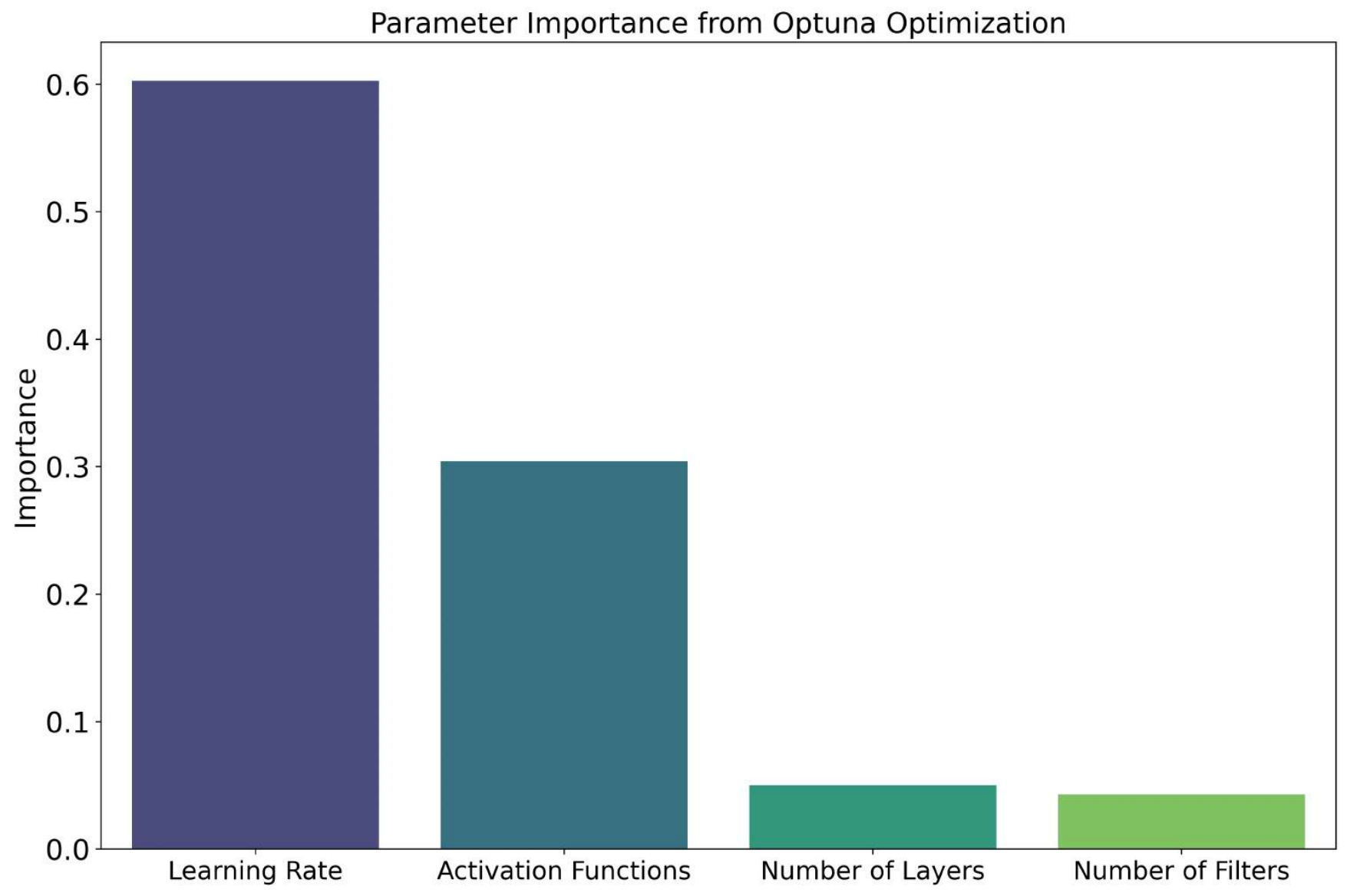}
	\includegraphics[width = 0.32\textwidth]{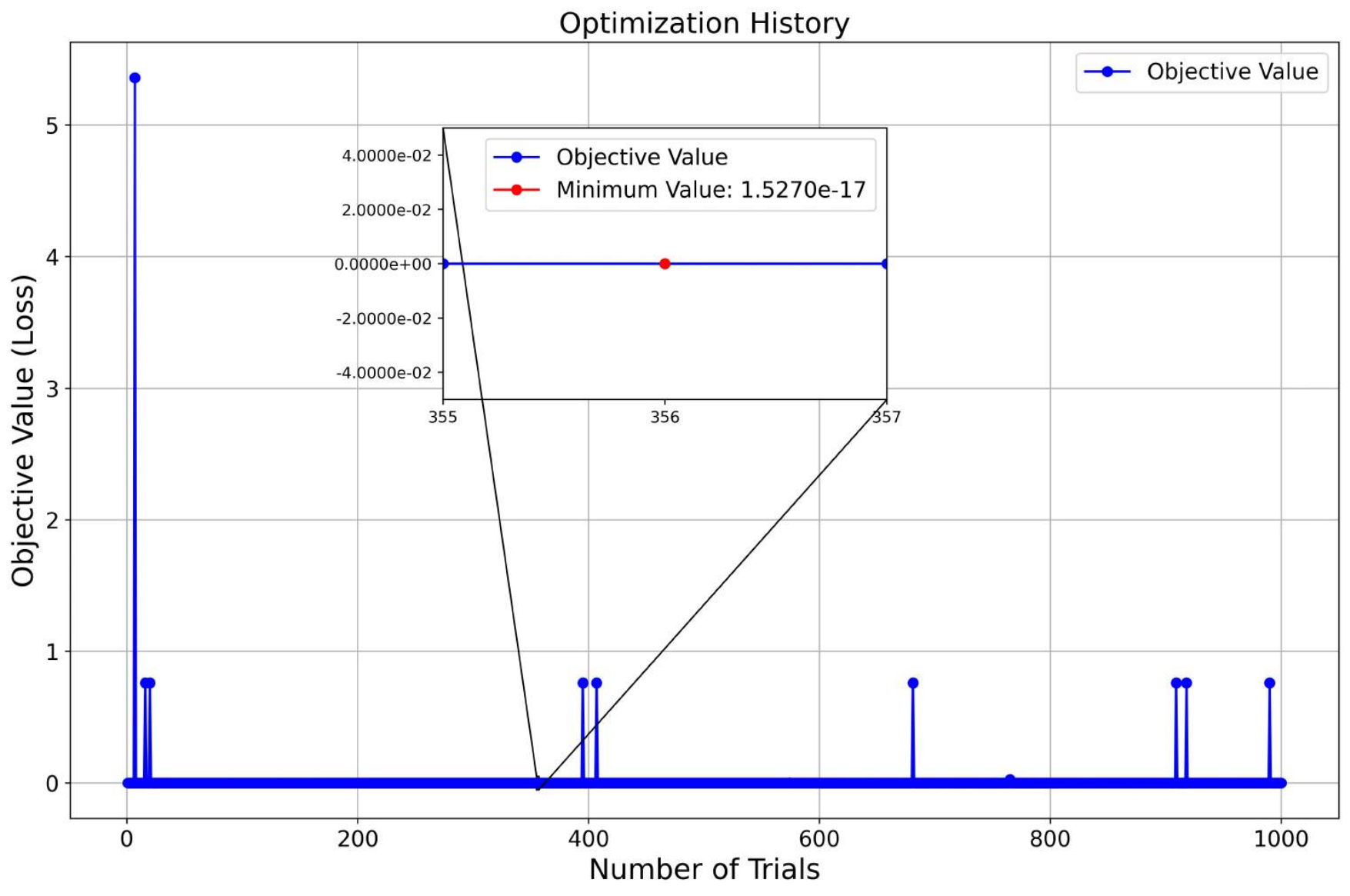}
	\includegraphics[width = 0.32\textwidth]{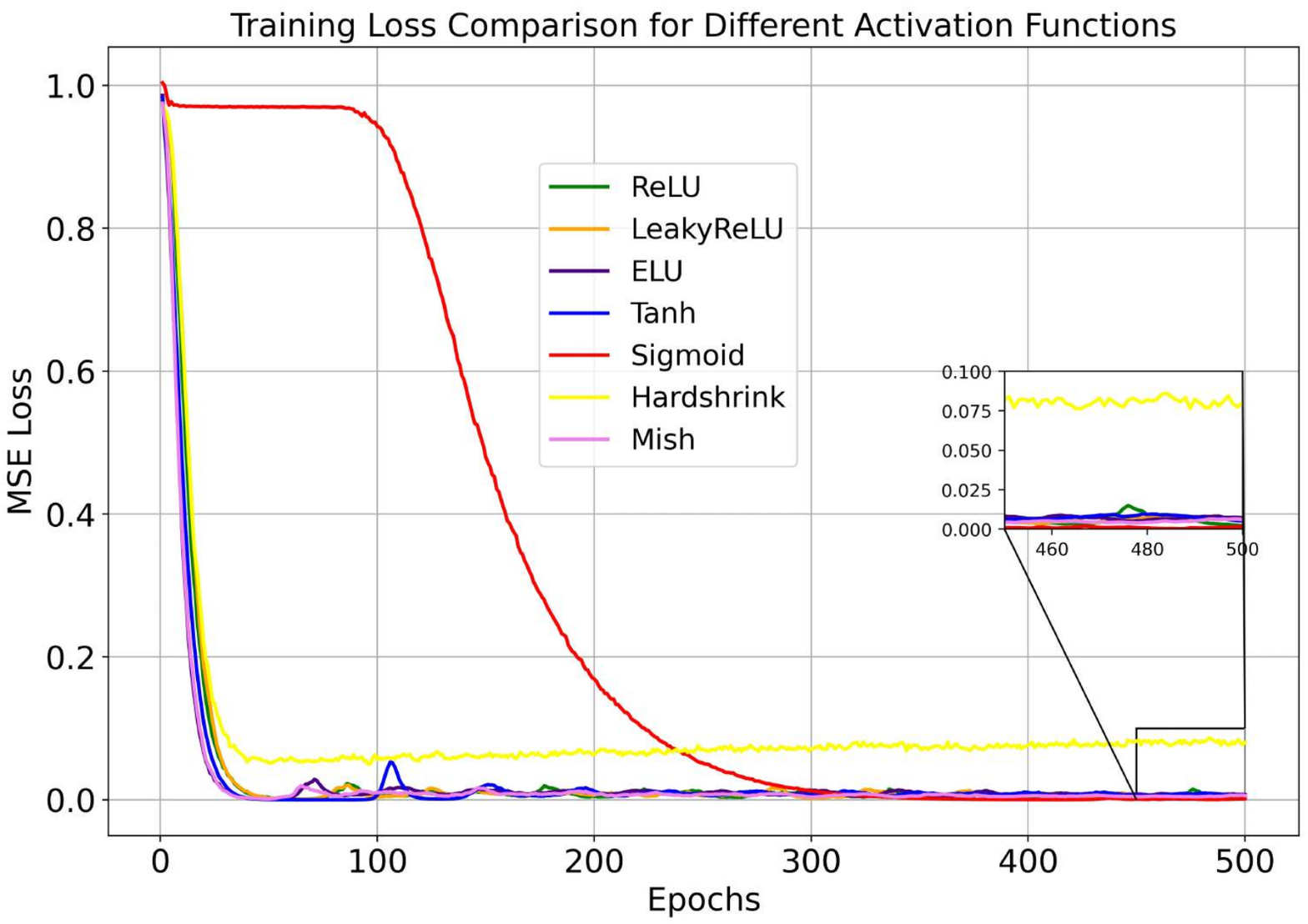}
	\includegraphics[width = 0.4\textwidth]{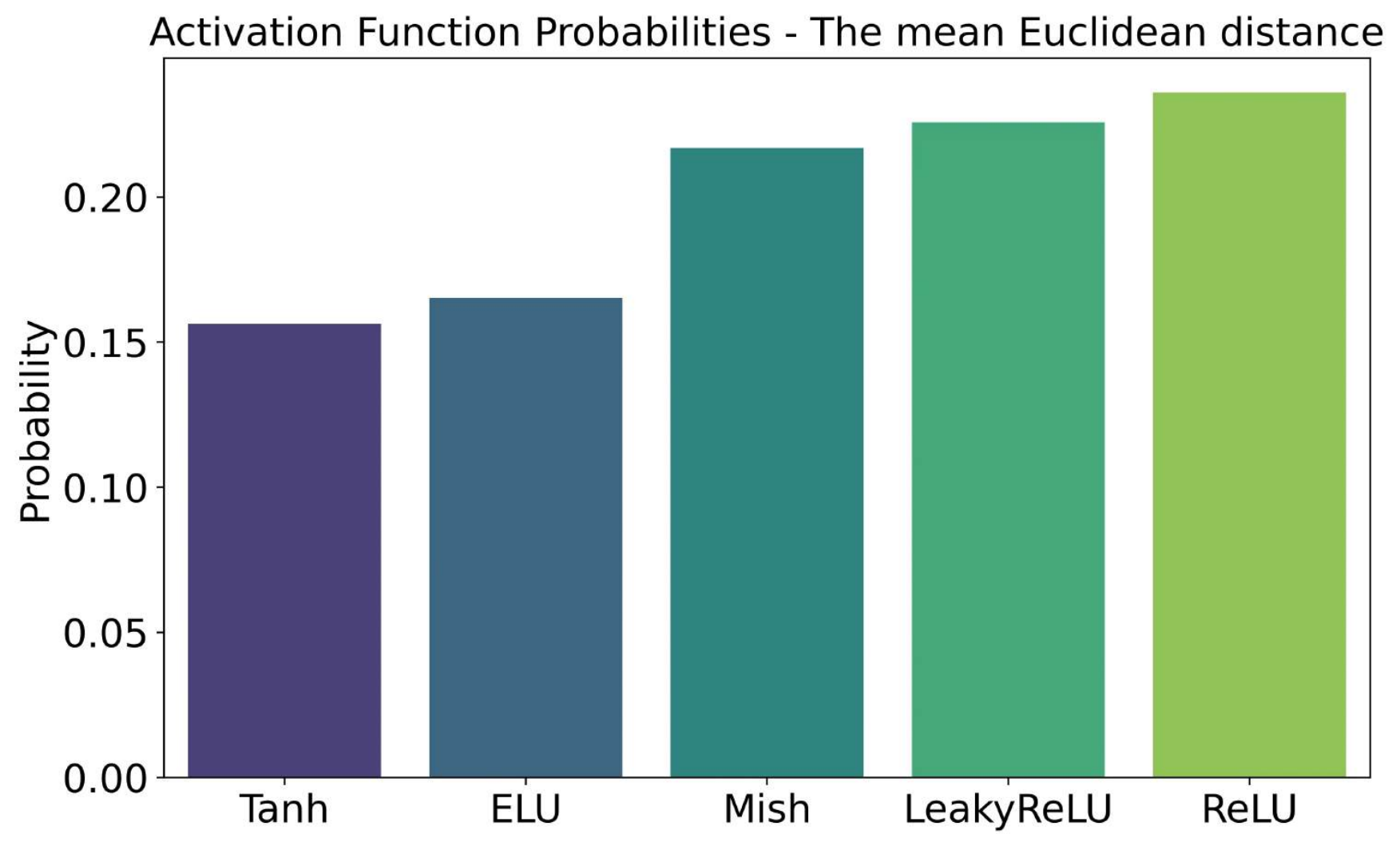}
	\includegraphics[width = 0.4\textwidth]{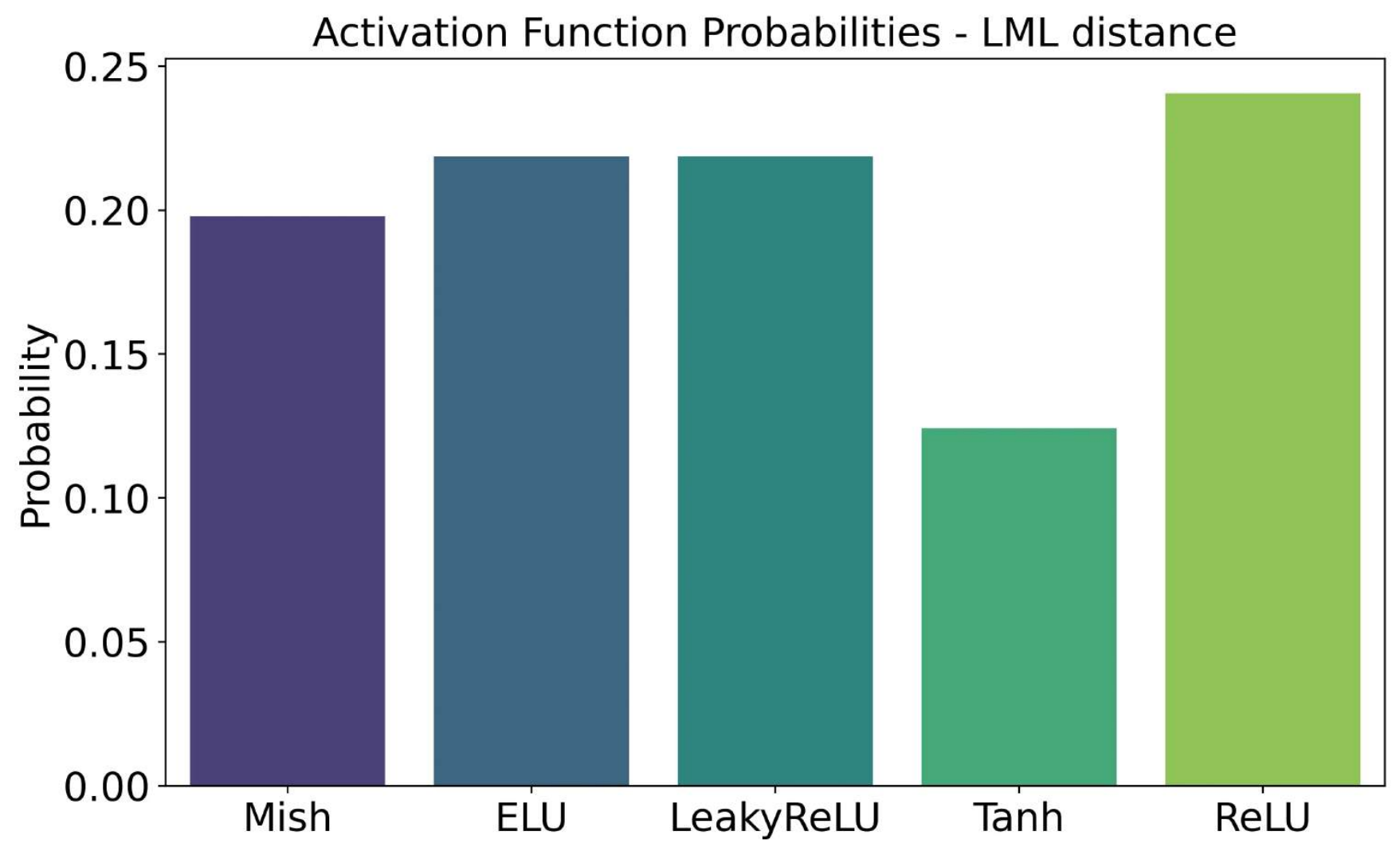}
	\includegraphics[width = 0.4\textwidth]{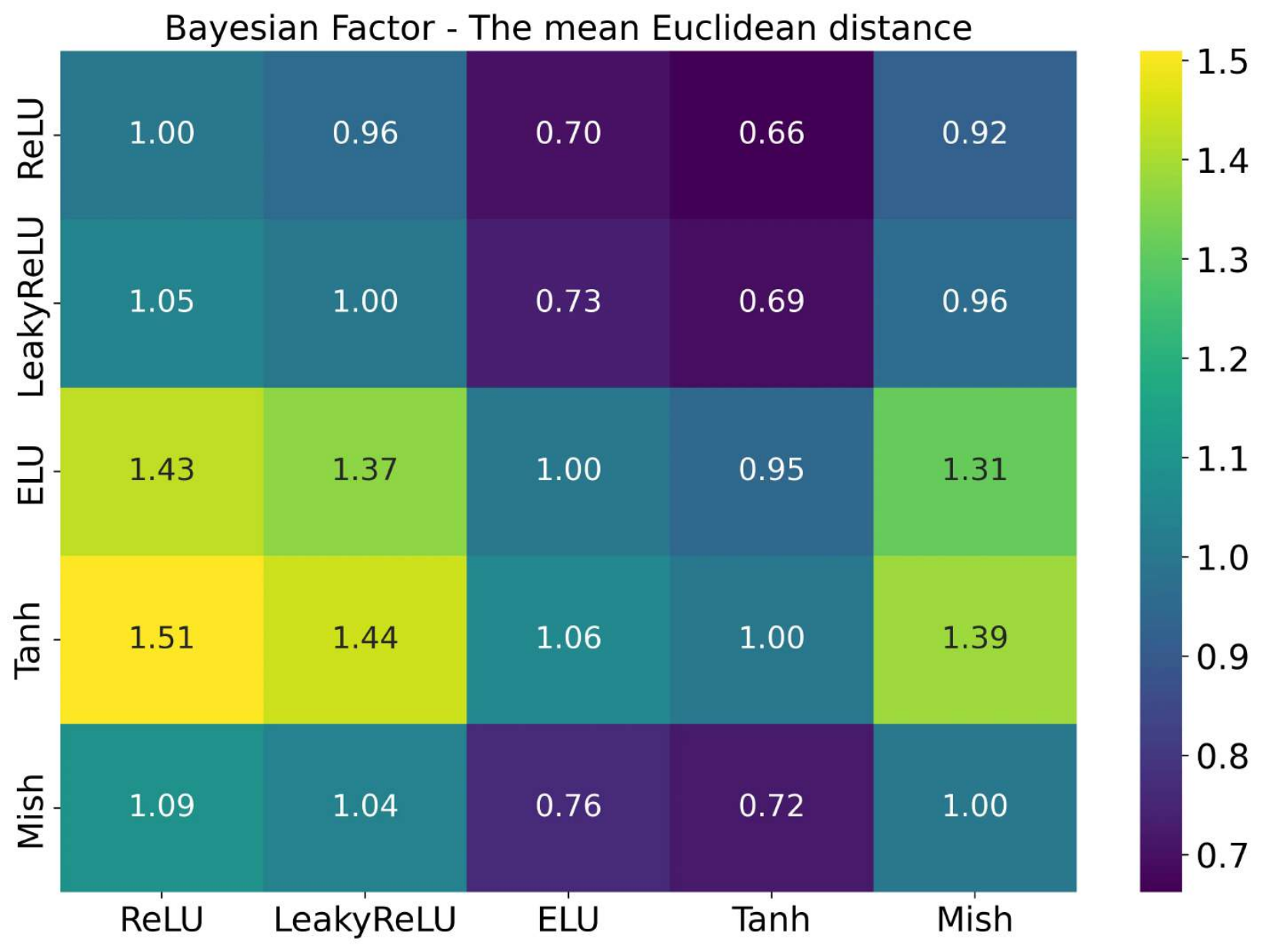}
	\includegraphics[width = 0.4\textwidth]{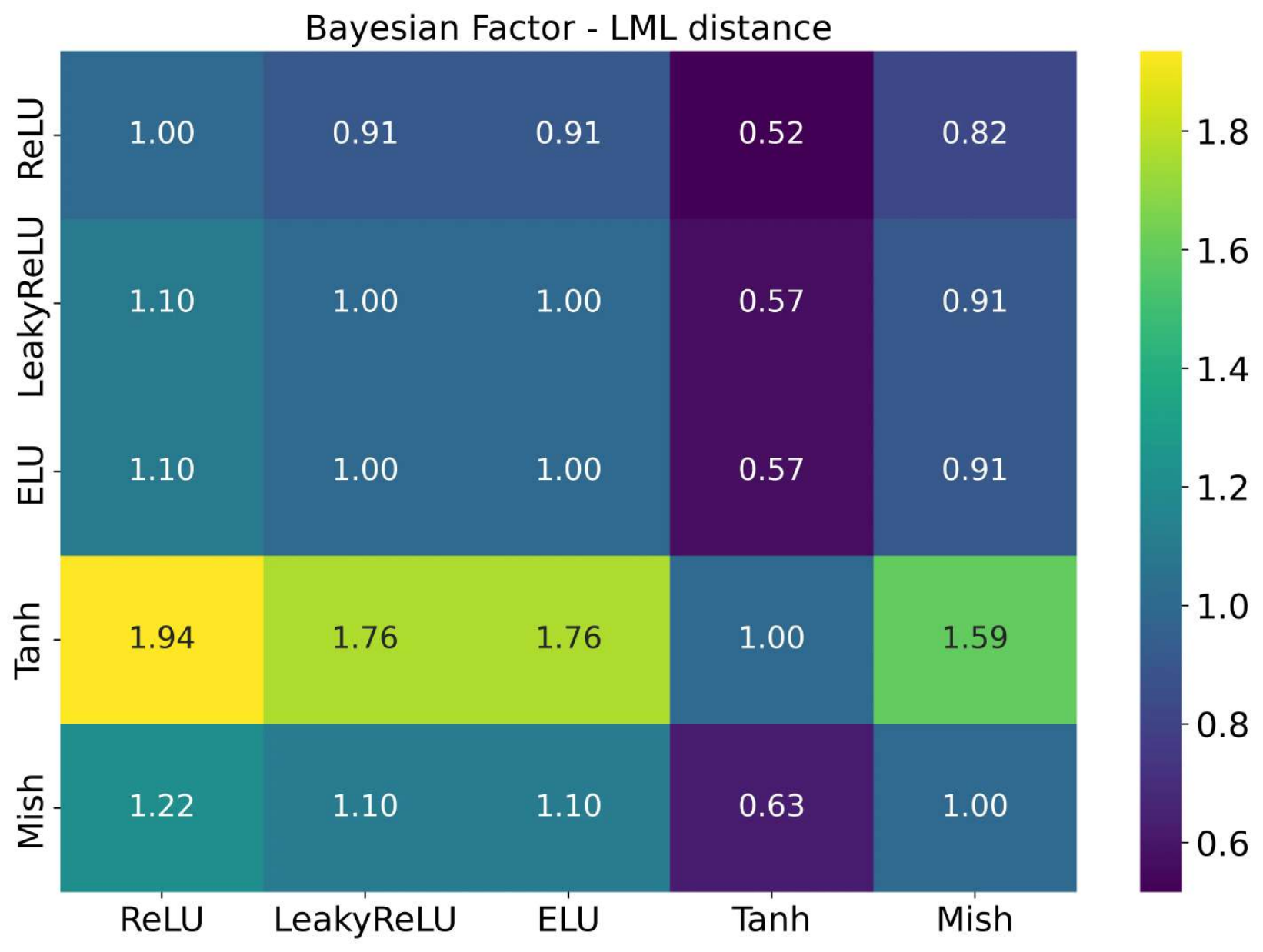}
	\caption{Hyperparameter optimisation of one-dimensional CNN in BCKA filters. The two diagrams in the upper panel illustrate the impact of various hyperparameter configurations of the filter on the S/R and S/R error of the data processing outcomes, which are represented by differently coloured scatter plots. The subgraph indicates that we randomly selected two observations from a total of 1000 for filtering. The left subfigure in the second panel illustrates the significance of optimising the parameters of our 1D CNN using various coloured histograms. The middle subfigure presents the optimisation history of the neural network, indicating that the lowest loss value, which signifies superior optimisation, is denoted by a red dot on the graph. The right subfigure depicts the fluctuation of the loss value of the selected activation function as a function of epoch, represented by different coloured solid lines. The change from the third panel to the fourth panel shows the Bayes factor and the priority probability between two instances, using the activation function at two different distances. The graphic indicates that ReLU is the superior activation function.}
	\label{fig:ML-optimize}
\end{figure*}

The job of optimizing these five initial parameters $\boldsymbol{f}=\left\lbrace P^{-}_{k},\hat{x}^{-}_{k}, Q,R,\eta \right\rbrace $ is tedious and $\eta$ is the learning rate of CNN. In theory, every ``folding" data collection has a set of optimal initial parameters. Despite our use of simulated observations in this work, we can quickly obtain the optimal parameters by exploring all the parameter combinations, whereas real observations typically yield a larger dataset. For example, if we use the 19 beams of FAST and the target tracking mode to continuously observe a pulsar for 20 minutes, we obtain approximately 2.2 TB of data. Our approach involves extending the observation time and conducting multiple observations to identify potential signals and enhance the S/R, resulting in a significantly large data volume. Therefore, to improve the data processing efficiency, we use a simple, one-dimensional CNN to optimize the five initial parameters. The structure of this supervised learning regression CNN consists of the following components: (1) the input layer, which receives a set of ``folding" data; (2) the convolutional layer, in which the number of layers and filters for each layer are determined through Bayesian optimization; (3) the activation function, in which the appropriate activation function is selected by comparing the approximate Bayesian computation rejections of two different distance functions; (4) a maximum pooling layer; (5) a fully connected layer; and (6) an output layer used to output the initial parameters of the optimal five BCKA filters. We randomly generate a thousand sets of ``folding'' data with Gaussian noise and input different initial parameter combinations into the BCKA filter through iterative sampling to obtain the parameter combinations with the highest SNRs, forming a thousand relationship pairs $\left\lbrace \boldsymbol{X},\boldsymbol{f} \right\rbrace $, where $\boldsymbol{X}$ represents the folding data consisting of $n$ subdata and $\boldsymbol{f}$ represents the initial parameters. Next, we randomly select 800 relationship pairs as the CNN training data and 200 relationship pairs as the CNN test data. We then optimize our CNN using these training data. Regarding the two kinds of distances used by the activation function in the CNN, one is the log marginal likelihood (LML) \citep{bishop2006pattern,2015arXiv150205700S}: 
\begin{equation}
	\begin{aligned}
		\ln \mathcal{L}&=\frac{1}{2} \ln \operatorname{det}\left(\boldsymbol{\Sigma}_{\boldsymbol{W}}\right)+\frac{1}{2} \ln \operatorname{det}\left(\boldsymbol{\Sigma}_{\boldsymbol{E}}\right) +\frac{1}{2} \ln \operatorname{det}\left(\boldsymbol{\Lambda}_p\right)\\
		&+\frac{1}{2}\|\boldsymbol{f}-\boldsymbol{f}^{*}\|_{\boldsymbol{\Sigma}_{\boldsymbol{E}}^{-1}}^2+\frac{1}{2}\|\overline{\boldsymbol{w}}\|_{\boldsymbol{\Sigma}_{\boldsymbol{W}}^{-1}}^2+\frac{n}{2} \ln (2 \pi),
	\end{aligned}
\end{equation}
where $\boldsymbol{\Sigma}_{\boldsymbol{W}}$ is the prior variance of the activation function, $\boldsymbol{\Sigma}_{\boldsymbol{E}}$ is the variance of additive noise, $\boldsymbol{\Lambda}_p$ is the feature matrix for the training data, $\boldsymbol{f}^{*}$ is the predicted output initial parameters, and $\overline{\boldsymbol{w}}=\boldsymbol{\Lambda}_p^{-1} \boldsymbol{\Phi}^{\top} \boldsymbol{\Sigma}_{\boldsymbol{E}}^{-1} \boldsymbol{t}$ is the weights of the output layer with the feature matrix for training data $\boldsymbol{\Phi}$. The other variable represents the mean Euclidean distance between the output results:
\begin{equation}
	d\left(\boldsymbol{f}, \boldsymbol{f}^{*}\right)=\frac{1}{n} \sum_{i=1}^n \sqrt{\sum_{j=1}^5\left(\boldsymbol{f}_{i j}-\boldsymbol{f}^{*}_{i j}\right)^2}.
\end{equation}
Figure \ref{fig:ML-optimize} displays all the hyperparameter optimization results of the CNN, along with the learning rate and optimization algorithm. In this way, each time we input a set of ``folding" data, we obtain a set of data-dependent BCKA filter initial parameters.

\begin{figure*}
	\centering
	\includegraphics[width = 0.33\textwidth]{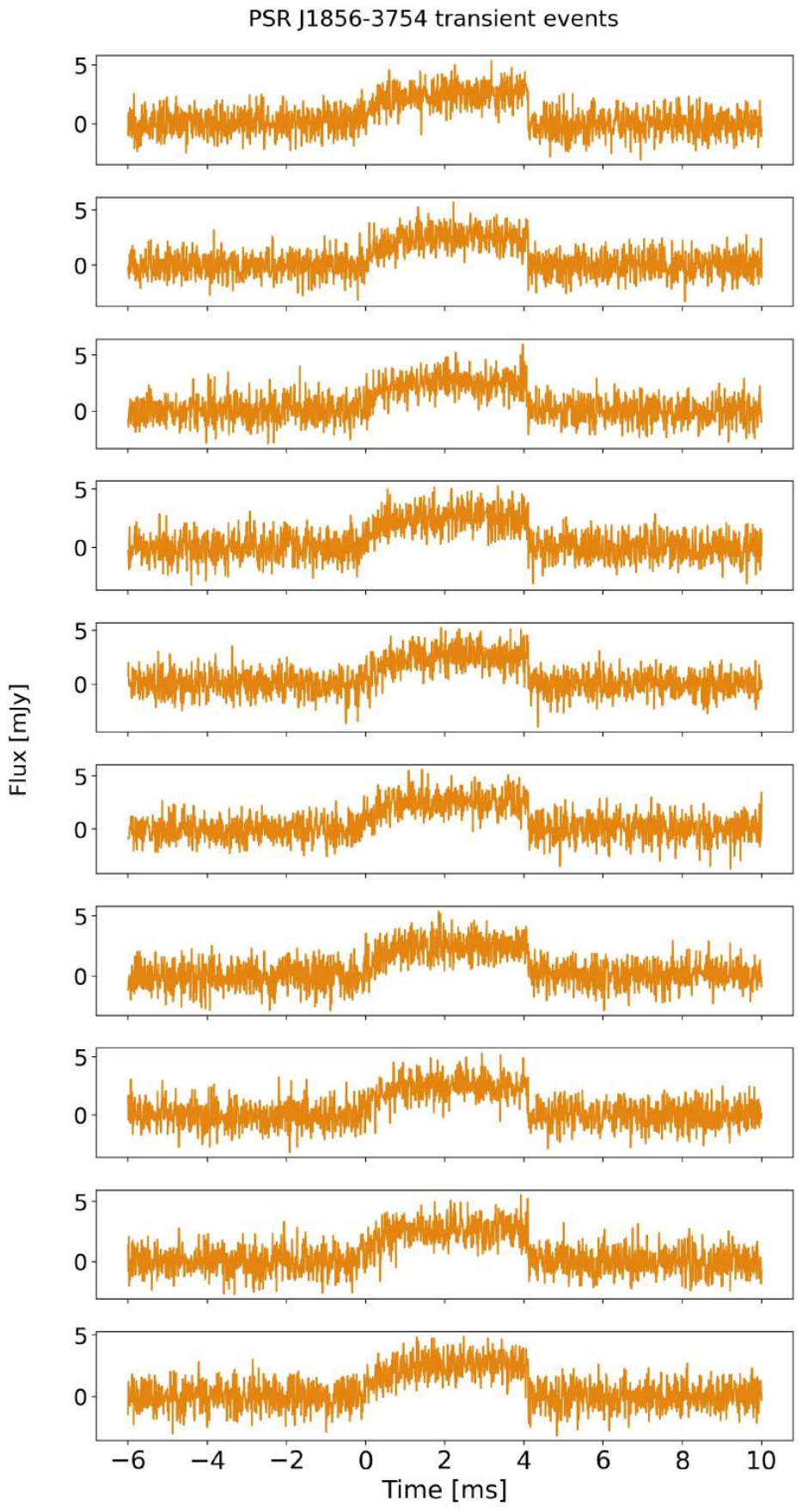}
	\includegraphics[width = 0.33\textwidth]{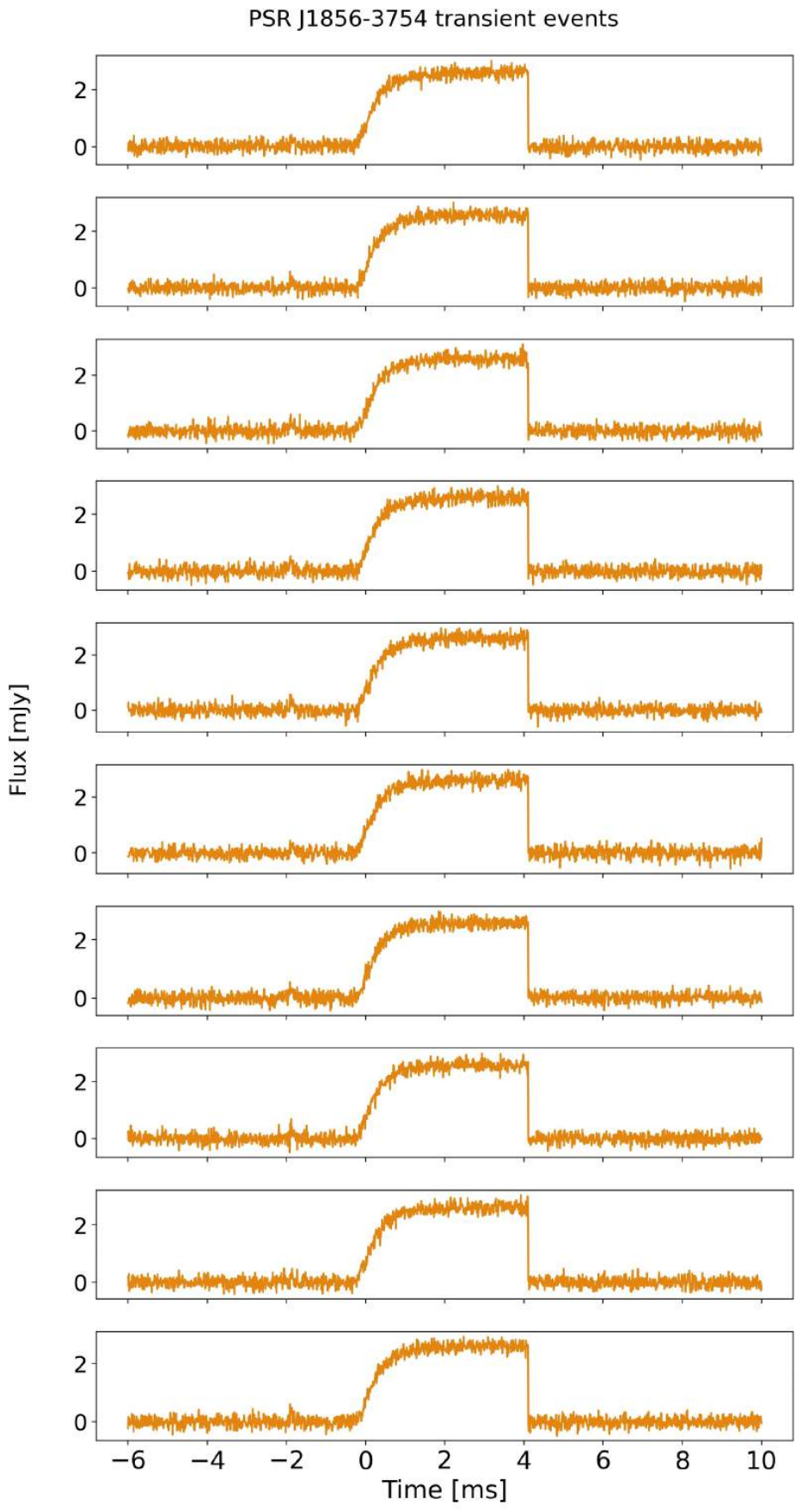}
	\includegraphics[width = 0.33\textwidth]{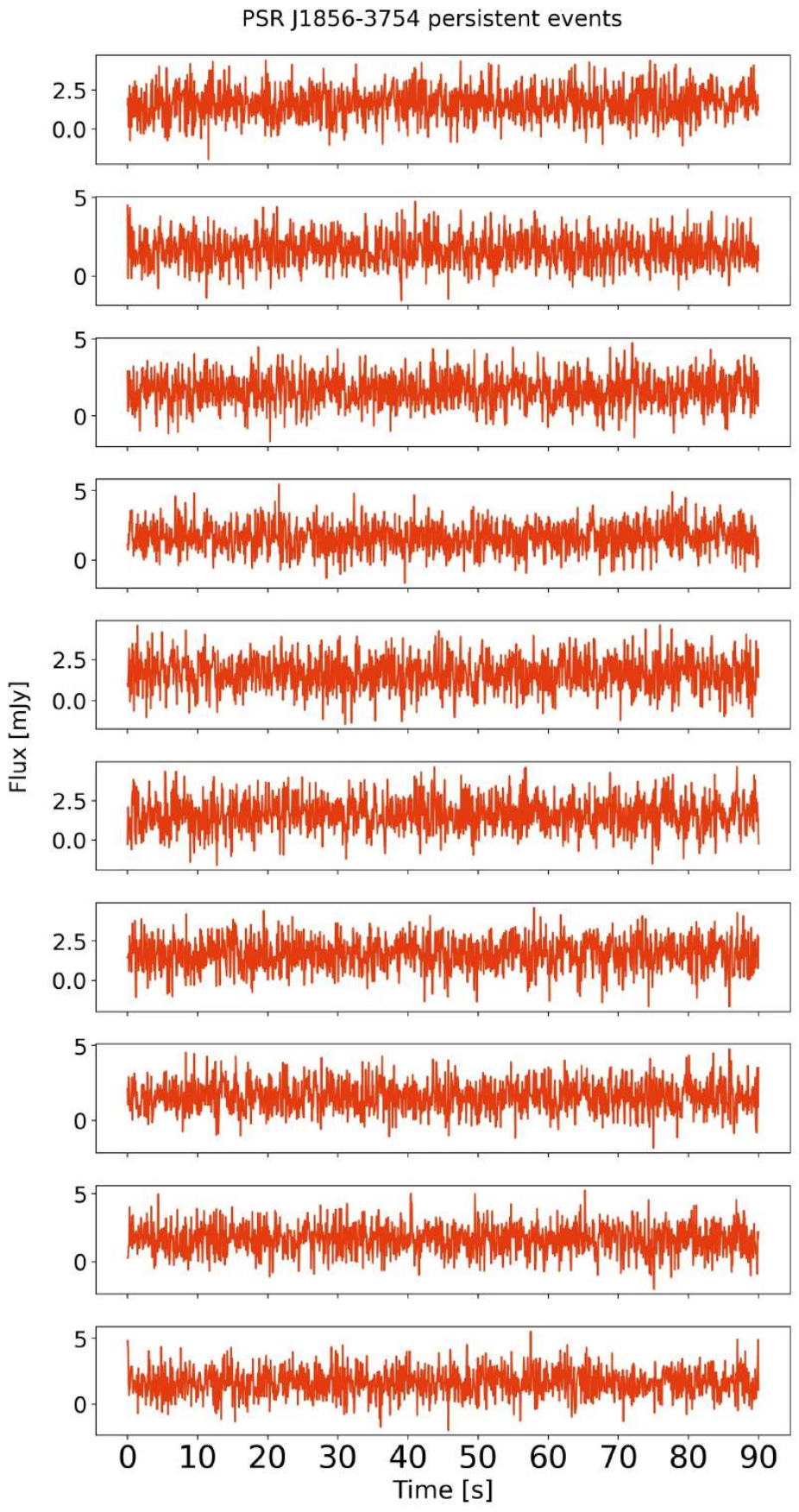}
	\includegraphics[width = 0.33\textwidth]{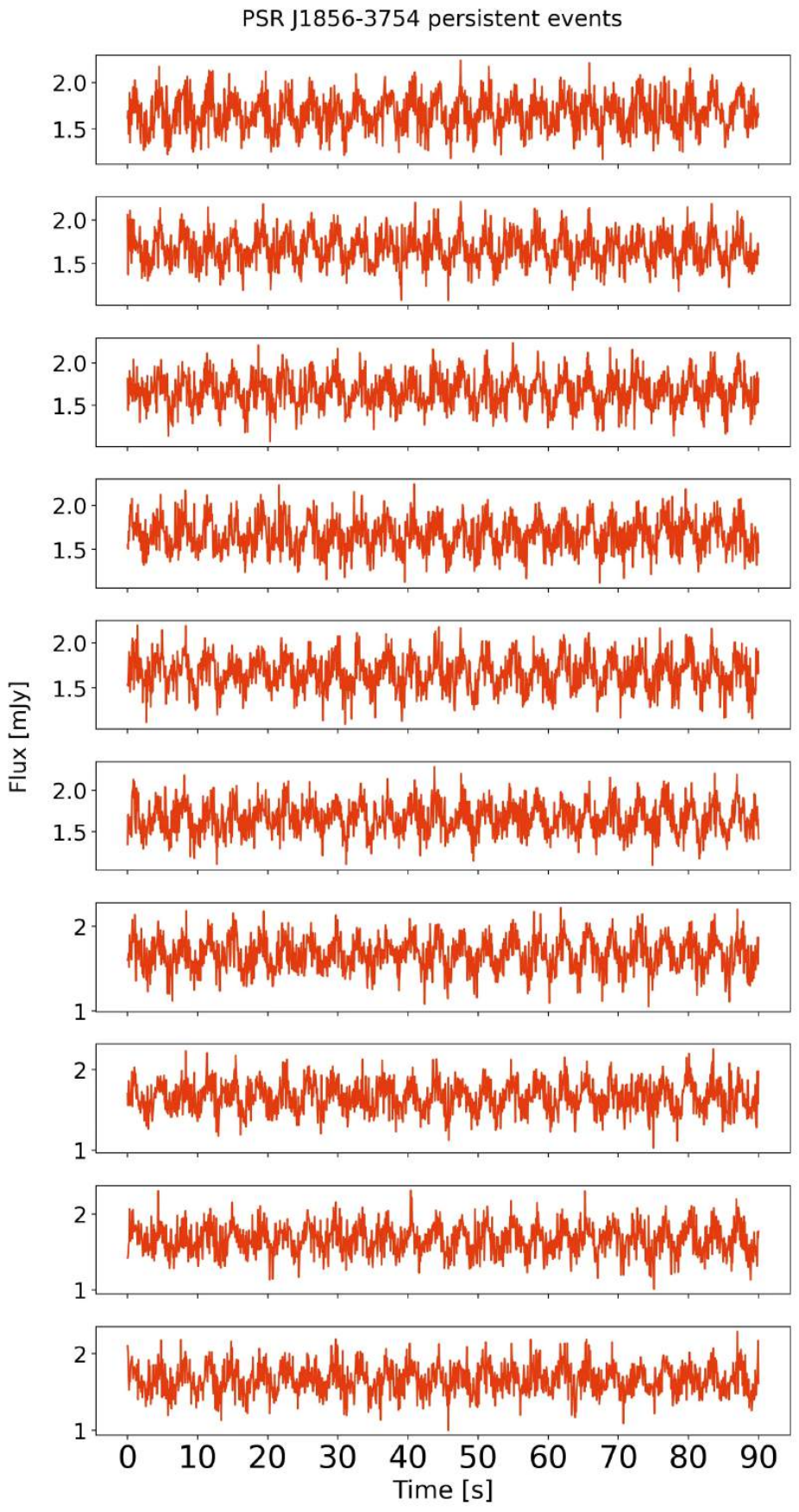}
	\caption{Comparison of the BCKA filter pre- and post-filtering of PSR J1856-3754. The left panel displays the forty-folded sets of simulated observations prior to filtering, while the right panel illustrates the outcomes of the corresponding simulated observations subsequent to filtering. To demonstrate and quantify statistical significance, we randomly chose transient and persistent events from two pulsars without replacement, aggregating 100 observations into a single folded dataset each time.}
	\label{fig:BCKA-result-examples-1}
\end{figure*}

\begin{figure*}
	\centering
	\includegraphics[width = 0.33\textwidth]{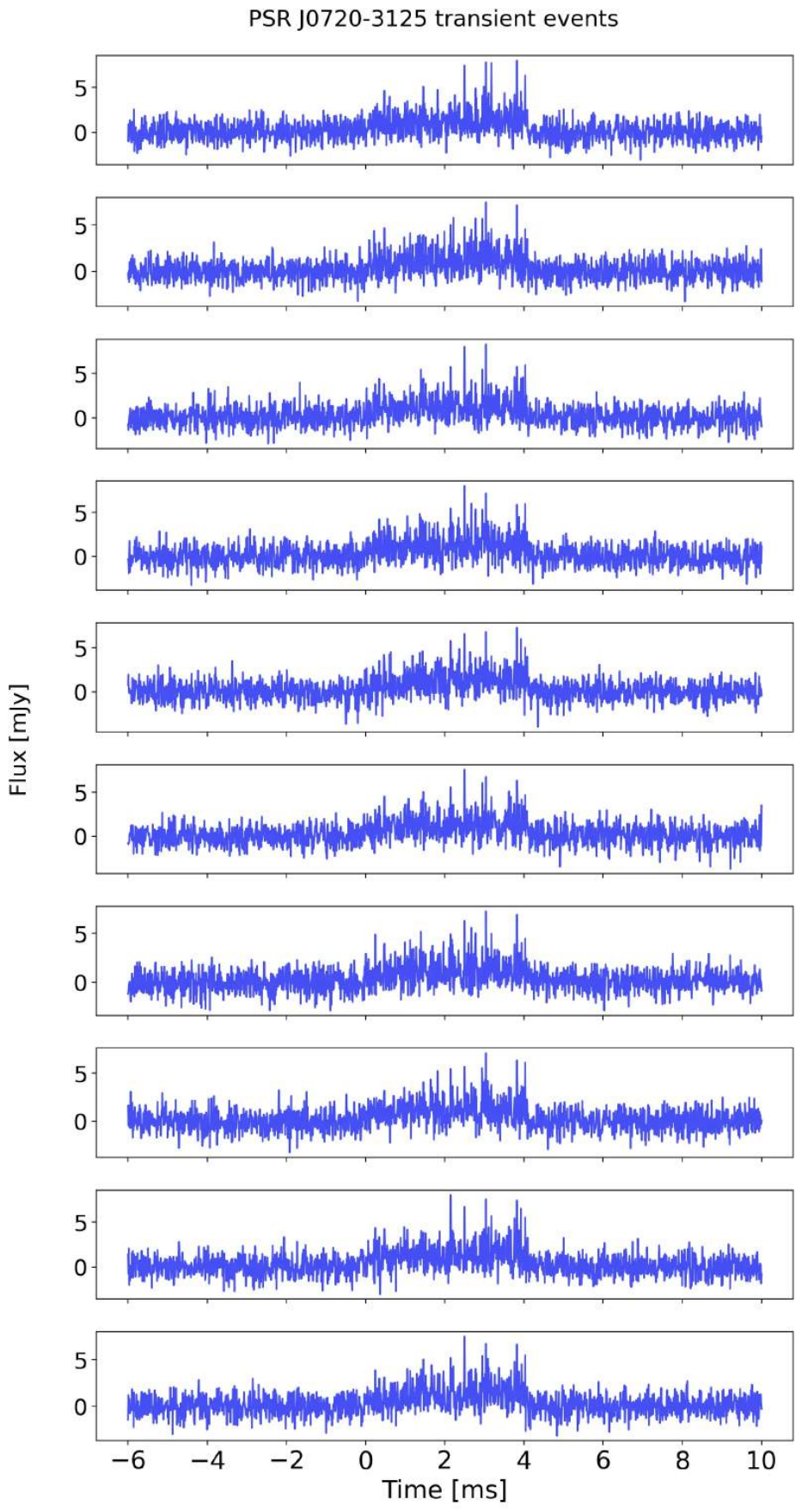}
	\includegraphics[width = 0.33\textwidth]{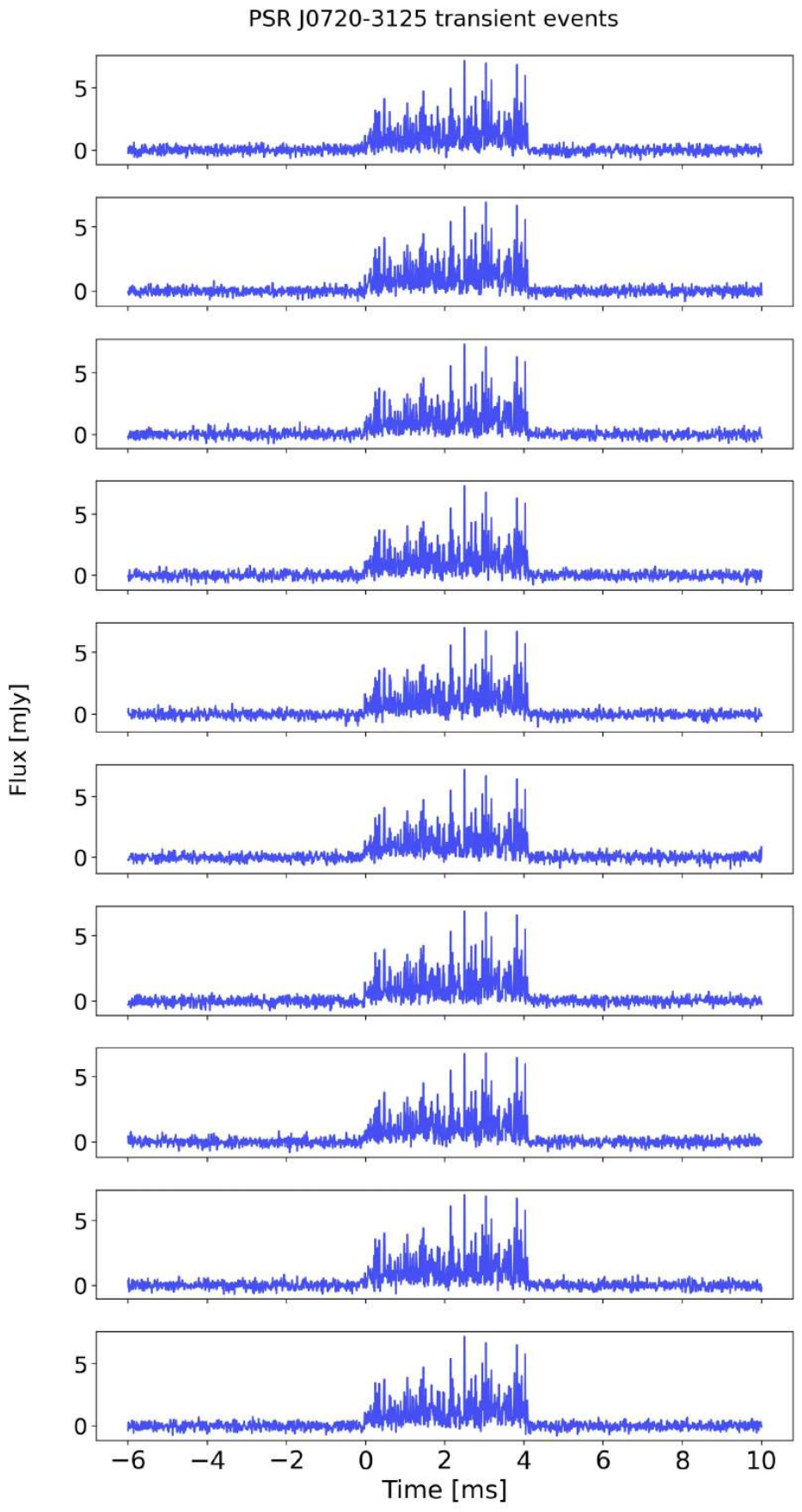}
	\includegraphics[width = 0.33\textwidth]{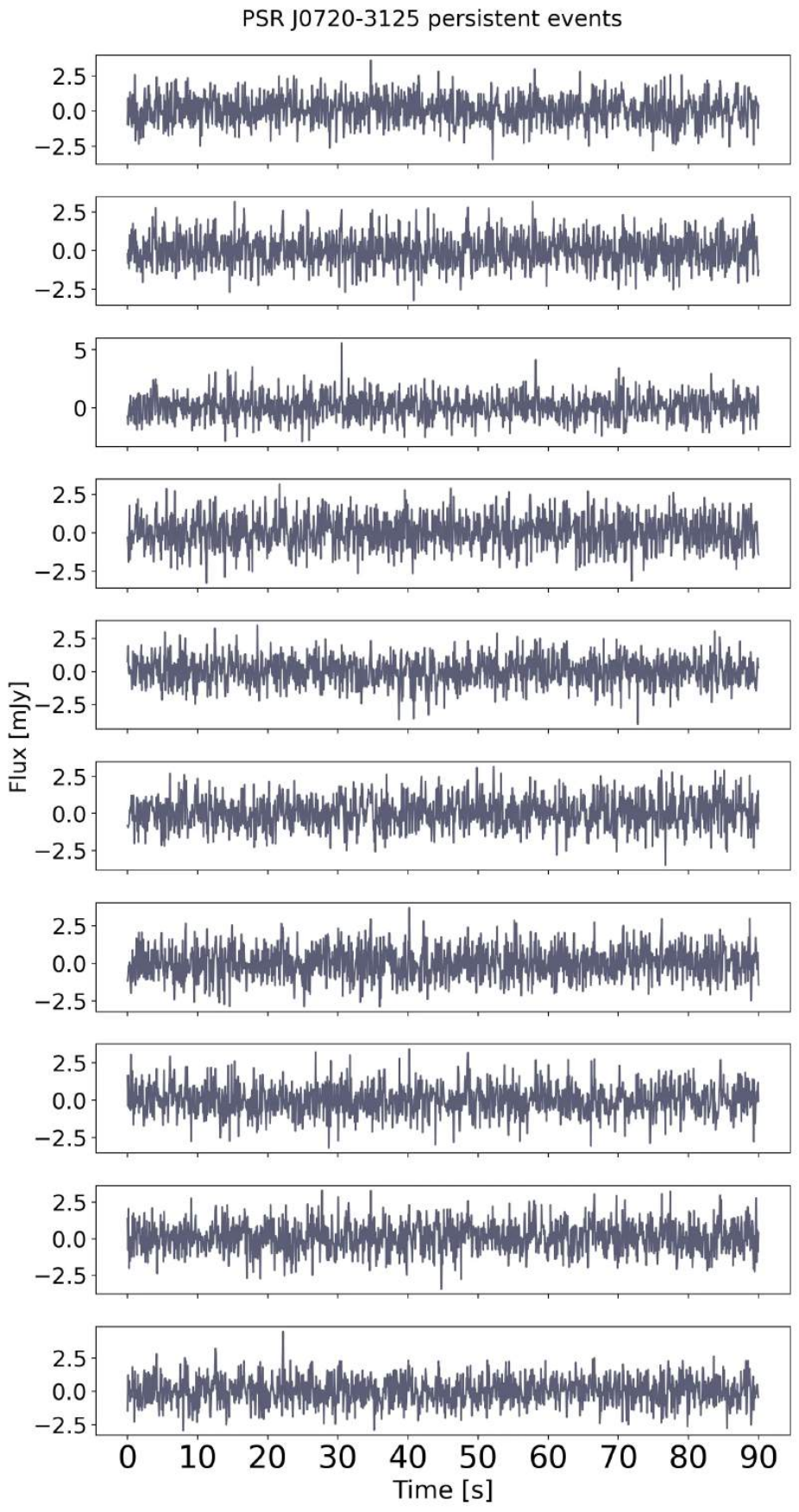}
	\includegraphics[width = 0.33\textwidth]{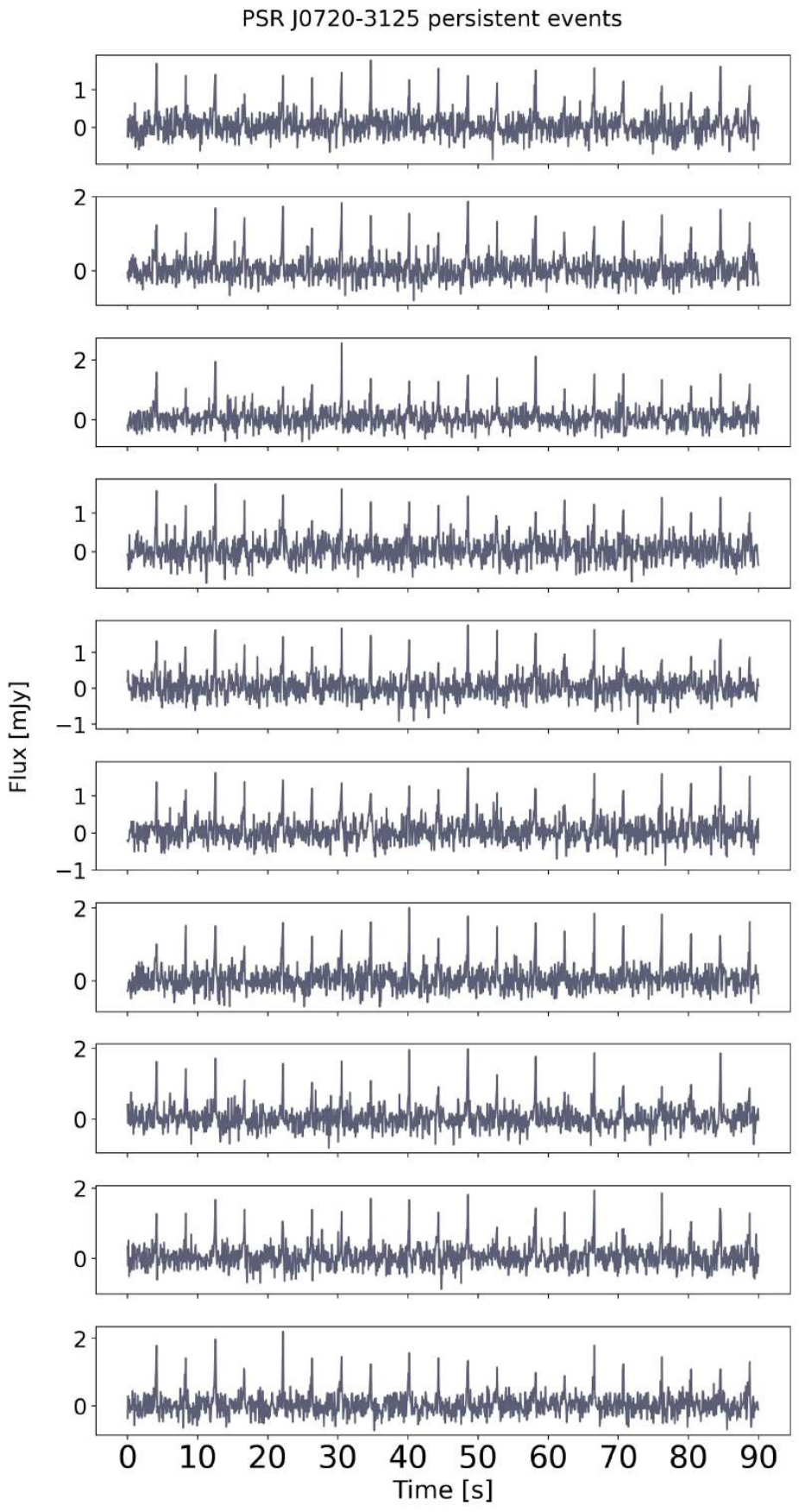}
	\caption{Comparison of the BCKA filter pre- and post-filtering of PSR J0720-3125. This figure shows the results in the same order as Figure \ref{fig:BCKA-result-examples-1}.}
	\label{fig:BCKA-result-examples-2}
\end{figure*}

To complement the implementation details of the CNN component above, we also provide representative visual examples of the BCKA pipeline's output. These examples, shown in the appendix, serve as illustrative diagnostics to demonstrate how the folded data are cleaned by the Kalman-adaptive stage following cross-correlation template matching shown in Figures~\ref{fig:BCKA-result-examples-1} and \ref{fig:BCKA-result-examples-2}. The main text retains its focus on the quantitative performance metrics, namely the S/R enhancement and sensitivity forecasts.

\section{Star Catalogue and Annual Pointing Geometry}
\label{app:star_catalog}
This section presents the complete catalogue of neutron-star targets utilized in our analysis. For each target, we list the essential astrometric information (see Table~\ref{tab:pulsar-list}) alongside the corresponding year-by-year telescope pointing geometry derived for our simulation epoch. The catalogue is organized into primary, secondary, and subsidiary targets, reflecting the observational priority and selection criteria detailed in Section~\ref{sec:results}. The key parameters of the observing facilities used in these simulations are summarized in Tables~\ref{tab:telescope-parameters-1} and~\ref{tab:telescope-parameters-2}.

\begin{table*}
	\caption{\label{tab:pulsar-list}A compilation of neutron stars for estimation in this work.}
	\resizebox{1.1\linewidth}{!}{
		\begin{tabular}{cccccccccc}
			\toprule
			Observation &Pulsar Name&Right Ascension&Declination&PM&PM&Annual&Barycentric&Pulsar&Surface Magnetic\\
			Importance&J2000&J2000&J2000&RA&Dec&Parallax&Period&Distance&Flux Density\\
			(level)&&(hh:mm:ss.s)&(+dd:mm:ss)&(mas/yr)&(mas/yr)&(mas)&(s)&(kpc)&(Gauss)\\
			\midrule
			Primary&PSR J1856-3754 \citep{2007ApJ...657L.101T} &18:56:35.41&-37:54:35.8&325.9&-59.22&8.2&7.05520287&0.160&1.47e+13\\
			Targets&PSR J0720-3125 \citep{1997AA...326..662H} &07:20:24.9620&-31:25:50.083&-93.9&52.8&2.8&8.391115532&0.400&2.45e+13\\
			\midrule
			&PSR J1731-4744 \citep{Large:1968mi} &17:31:42.160&-47:44:36.26&73&-132&-&0.82982878524&0.700&1.18e+13\\
			&PSR J0157+6212 \citep{1972Natur.240..229D} &01:57:49.9434&+62:12:26.648&1.57&44.80&0.56&2.35174493646&1.820&2.13e+13\\
			&PSR J0528+2200 \citep{1968Sci...162.1481S} &05:28:52.264&+22:00:04&-20&7&-&3.74553925030&1.215&1.24e+13\\
			Secondary&PSR J0210+5845 \citep{Sanidas:2019stw} &02:10:56.409999&+58:45:17.718237&-1.116&-0.495&-&1.766218932442&1.952&1.66e+13\\
			Targets&PSR J2301+5852 \citep{1981Natur.293..202F} &23:01:08.29&+58:52:44.45&-6.4&-2.3&-&6.9790709703&3.300&5.8e+13\\				
			&PSR J0146+6145 \citep{1994ApJ...433L..25I} &01:46:22.42&+61:45:02.8&-4.1&1.9&-&8.6889941&3.600&1.33e+14\\
			&PSR J1809-1943 \citep{2004ApJ...609L..21I} &18:09:51.08696&-19:43:51.9315&-6.60&-11.7&-&5.54074283&3.600&1.27e+14\\
			&PSR J1745-2900 \citep{2013ApJ...770L..23M} &17:45:40.1662&-29:00:29.896&2.5&5.89&-&3.763733080&8.300&2.6e+14\\				
			\midrule
			&PSR J1808-2024 \citep{Kouveliotou:1998ze} &18:08:39.337&-20:24:39.85&-4.5&-7&-&7.55592&13.000&2.06e+15\\
			&PSR J0736-6304 \citep{2010MNRAS.402..855B} & 07:36:20.01&-63:04:16&-&-&-&4.8628739612&0.104&2.75e+13\\
			&PSR J1740-3015 \citep{1986Natur.320...43C} &17:40:33.82&-30:15:43.5&-&-&-&0.60688662425&0.400&1.7e+13\\
			&PSR J1848-1952 \citep{1978MNRAS.185..409M} &18:48:18.03&-19:52:31&-&-&-&4.30818959857&0.751&1.01e+13\\
			Subsidiary&SGR J0501+4516 \citep{2014MNRAS.438.3291C} &05:01:06.76&+45:16:33.92&-&-&-&5.7620695&2.000&1.9e+14\\
			Targets&PSR J0501+4516 \citep{2008GCN..8118....1G} &05:01:06.76&+45:16:33.92&-&-&-& 5.76209653&2.000&1.85e+14\\
			&SGR J0418+5729 \citep{2013ApJ...770...65R} &04:18:33.867&+57:32:22.91&-&-&-&9.07838822&2.000&6.1e+12\\
			&PSR J1809-1943 \citep{2004ApJ...609L..21I} &18:09:51.08696&-19:43:51.9315&-&-&-&5.540742829&3.600&1.27e+14\\
			&PSR J1550-5418 \citep{2007ApJ...666L..93C} &15:50:54.12386&-54:18:24.1141&-&-&-&2.06983302&4.000&2.22e+14\\
			&SGR J1935+2154 \citep{2016MNRAS.457.3448I} &19:34:55.598&+21:53:47.79&-&-&-&3.2450650&-&2.2e+14\\
			\bottomrule
		\end{tabular}
	}
\end{table*}

\begin{table*}
	\caption{\label{tab:telescope-parameters-1}Summary of astrometric information and telescope inclination of PSR J1856-3754 on June 1st of each observational year.}
		\begin{threeparttable}
			\begin{tabular}{ccccccc}
				\toprule
				Observational & Astrometric  & Astrometric & FAST  & FAST & SKA2-MID  & SKA2-MID \\
				Chronology$^{a}$ & RA & Dec & Azimuth & Altitude & Azimuth & Altitude \\
				(year) & (hh:mm:ss.ss) & (+dd:mm:ss.ss) & (degrees) & (degrees) & (degrees) & (degrees) \\
				\midrule
				1st & 18:56:35.98 & -37:54:37.36 & 228.6180981 & -2.458248045 & 126.0323905 & 76.63241336 \\ 
				2nd & 18:56:36.01 & -37:54:37.42 & 228.5152131 & -2.284169534 & 125.6173211 & 76.45481421 \\ 
				3rd & 18:56:36.03 & -37:54:37.47 & 228.8137176 & -2.77688693 & 126.8049255 & 76.96194246 \\ 
				4th & 18:56:36.05 & -37:54:37.53 & 228.7120911 & -2.602353632 & 126.3648459 & 76.7865003 \\ 
				5th & 18:56:36.07 & -37:54:37.59 & 228.6100523 & -2.428255795 & 125.9387386 & 76.61017439 \\ 
				6th & 18:56:36.09 & -37:54:37.65 & 228.5074768 & -2.254582406 & 125.5263472 & 76.43289539 \\ 
				7th & 18:56:36.11 & -37:54:37.71 & 228.8062576 & -2.747765067 & 126.7096881 & 76.94093649 \\ 
				\bottomrule
			\end{tabular}
			\begin{tablenotes}
				\footnotesize
				\item[a] For simplicity of the simulation, we assume that the pulse profile of this pulsar, whose pulse profile data come from observations in the X-ray band, remains periodic during the observation period \citep{2022MNRAS.516.4932D,2009A&A...500..861C}.
			\end{tablenotes}
		\end{threeparttable}
\end{table*}

\begin{table*}
	\caption{\label{tab:telescope-parameters-2}Summary of astrometric information and telescope inclination of PSR J0720-3125 on June 1st of each observational year.}
		\begin{threeparttable}
			\begin{tabular}{ccccccc}
				\toprule
				Observational & Astrometric  & Astrometric & FAST  & FAST & SKA2-MID  & SKA2-MID \\ 
				Chronology$^{a}$ & RA & Dec & Azimuth & Altitude & Azimuth & Altitude \\ 
				(year) & (hh:mm:ss.ss) & (+dd:mm:ss.ss) & (degrees) & (degrees) & (degrees) & (degrees) \\ 
				\midrule
				1st & 7:20:24.80 & -31:25:48.69 & 112.3667856 & -32.59965808 & 198.2749241 & -25.12027244 \\ 
				2nd &7:20:24.79 & -31:25:48.63 & 112.3168845 & -32.80704674 & 198.4955365 & -25.05061609 \\ 
				3rd & 7:20:24.78 & -31:25:48.58 & 112.4778661 & -32.19296649 & 197.8350564 & -25.24574791 \\ 
				4th & 7:20:24.78 & -31:25:48.53 & 112.4276399 & -32.40051188 & 198.0566353 & -25.17727286 \\ 
				5th & 7:20:24.77 & -31:25:48.47 & 112.3776151 & -32.60820737 & 198.2779216 & -25.1080678 \\ 
				6th & 7:20:24.77 & -31:25:48.42 & 112.3277026 & -32.81604099 & 198.4989435 & -25.03820493 \\ 
				7th & 7:20:24.76 & -31:25:48.37 & 112.4881197 & -32.20232808 & 197.8391391 & -25.23352957 \\ 
				\bottomrule
			\end{tabular}
			\begin{tablenotes}
				\footnotesize
				\item[a] For simplicity of the simulation, we assume that the pulse profile of this pulsar, whose pulse profile data come from observations in the X-ray band, remains periodic during the observation period \citep{1997AA...326..662H,2003ApJ...590.1008K,2017A&A...601A.108H}.
			\end{tablenotes}
		\end{threeparttable}
\end{table*}

\section{S/R Resampling Diagnostics for the Four Observing Strategies}
\label{app:snr_scatter}
In the main text we summarize the S/R improves and flux limits for the SPST/SPMT/MPST/MPMT strategies in Table~\ref{tab:snr-summary}. Here we provide the full distributions of S/R improvements obtained from 500 nonrepeated resamplings of the simulated observation set, which serve as a diagnostic of robustness against accidental noise realisations and template mismatches.

\begin{figure*}
	\centering
	\includegraphics[width = 0.4\textwidth]{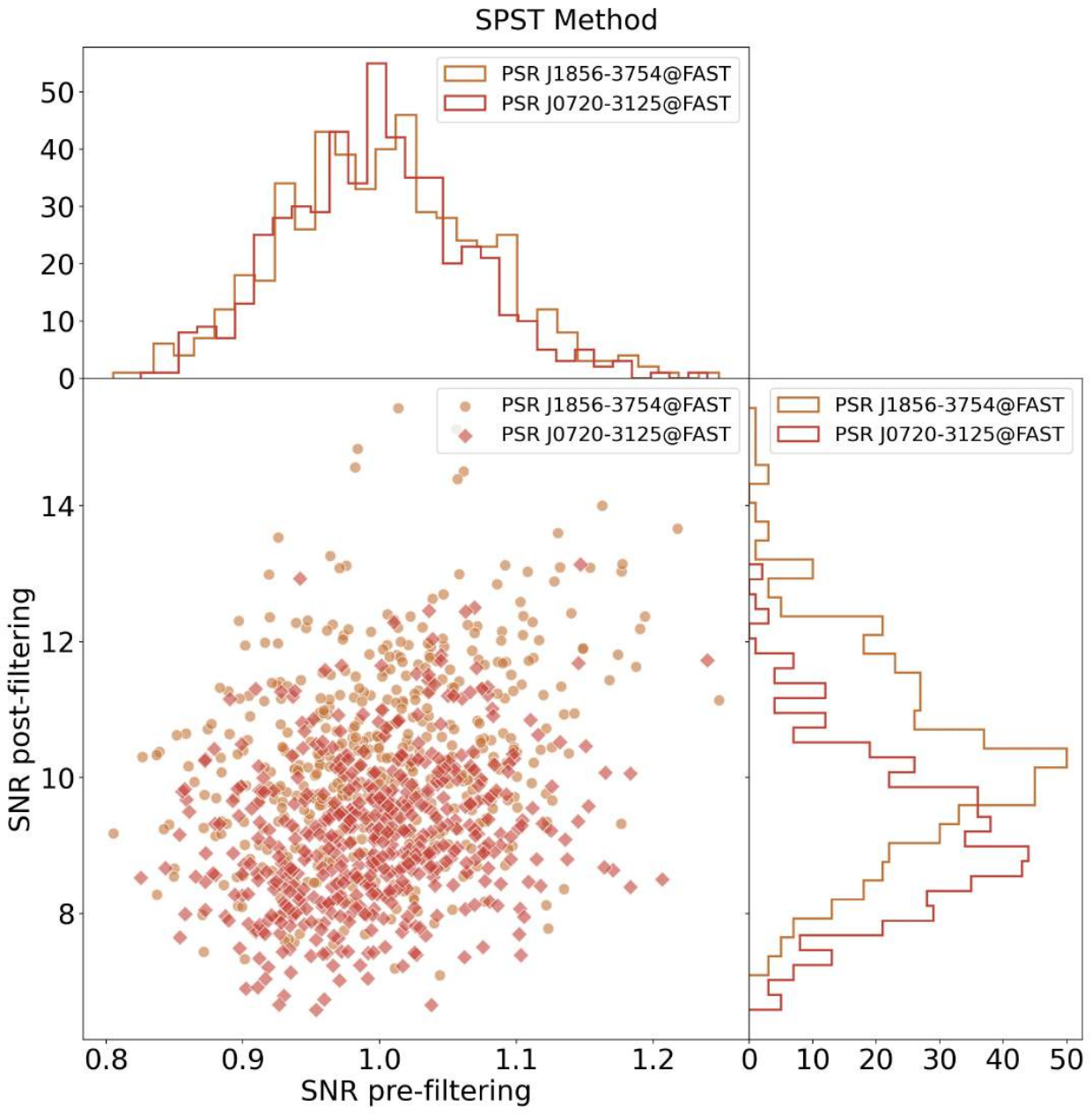}
	\includegraphics[width = 0.4\textwidth]{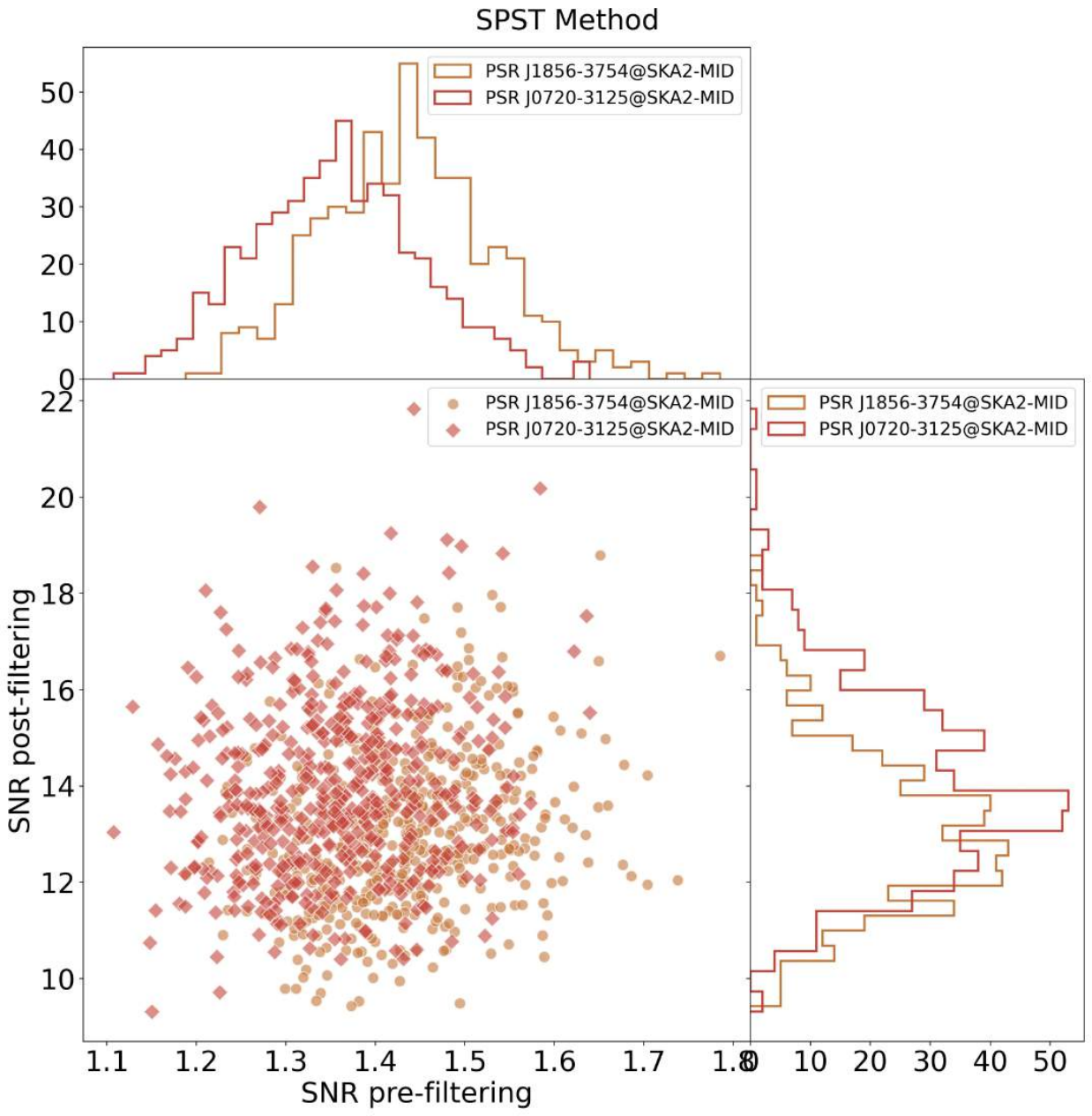}
	\caption{Comparison of the S/R pre- and post-filtering. The panel shows the S/R results of the SPST method, where the data for the left panel are from simulated observations of FAST and the data for the right panel are from simulated observations of SKA2-MID.}
	\label{fig:SNR-500-1}
\end{figure*}

\begin{figure*}
	\centering
	\includegraphics[width = 0.4\textwidth]{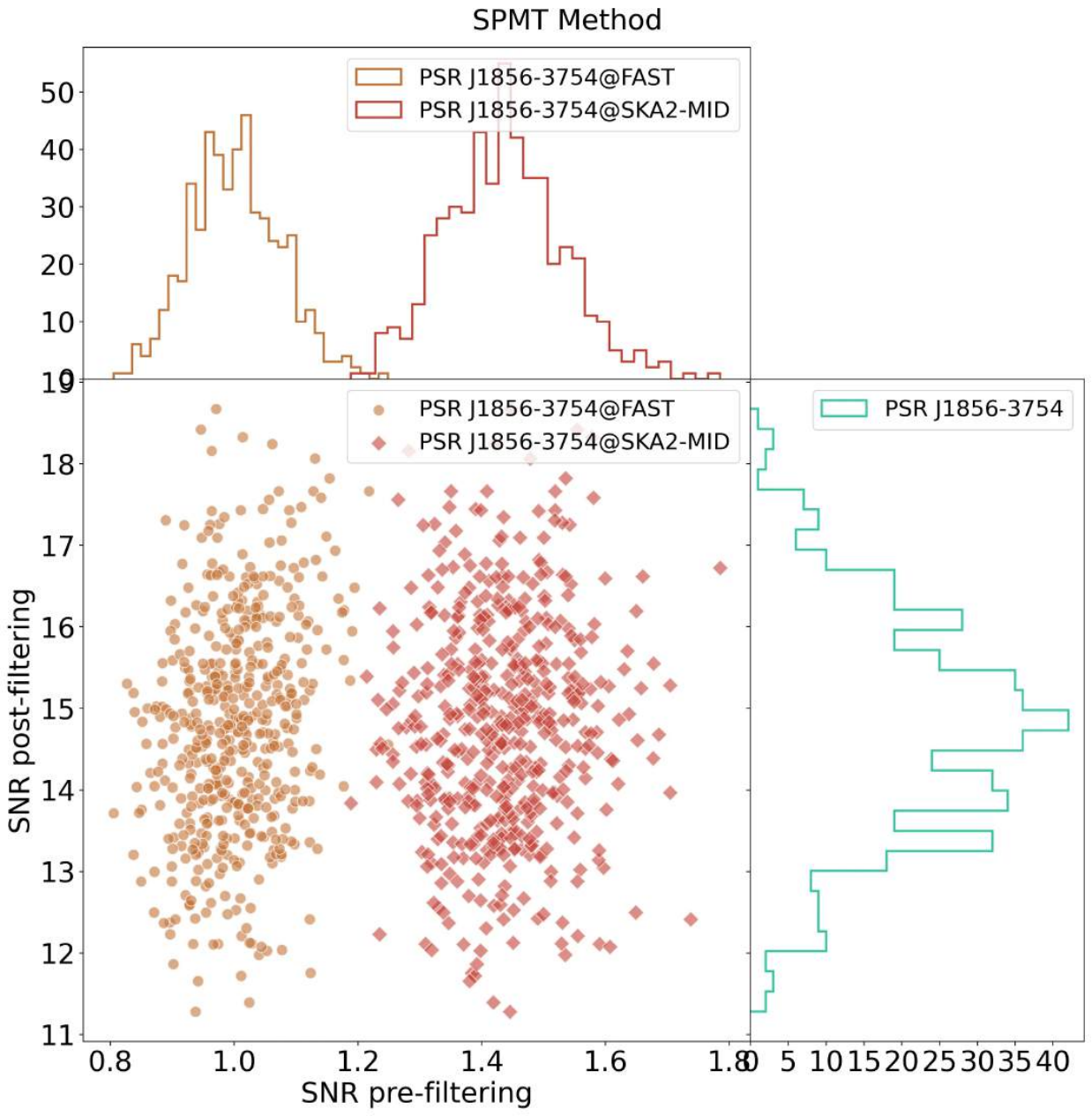}
	\includegraphics[width = 0.4\textwidth]{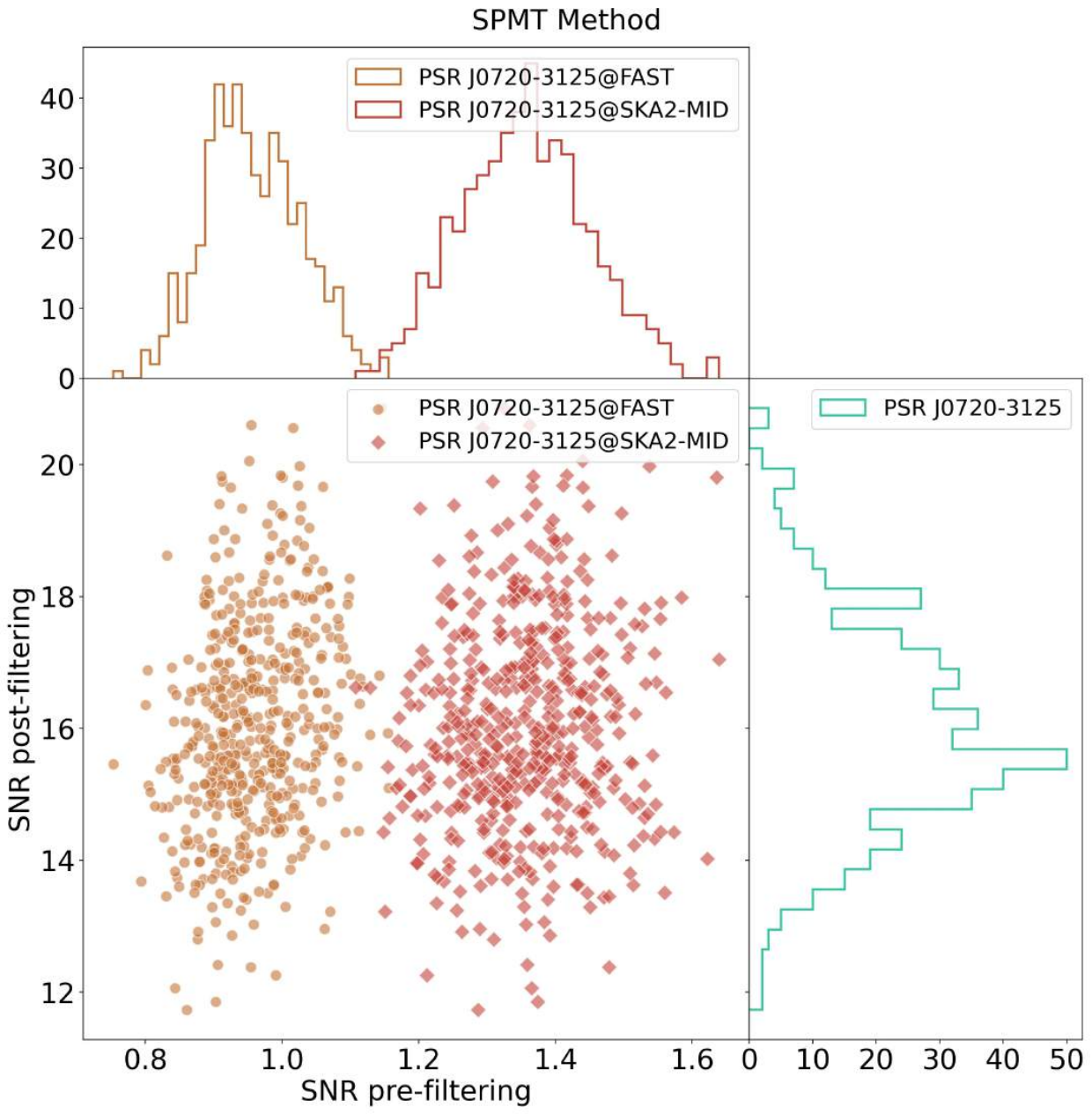}
	\caption{Comparison of the S/R pre- and post-filtering. The panel shows the S/R results of the SPMT method, showing pulsars PSR J1856-3754 and PSR J0720-3125 in order from left to right.}
	\label{fig:SNR-500-2}
\end{figure*}

\begin{figure*}
	\centering
	\includegraphics[width = 0.4\textwidth]{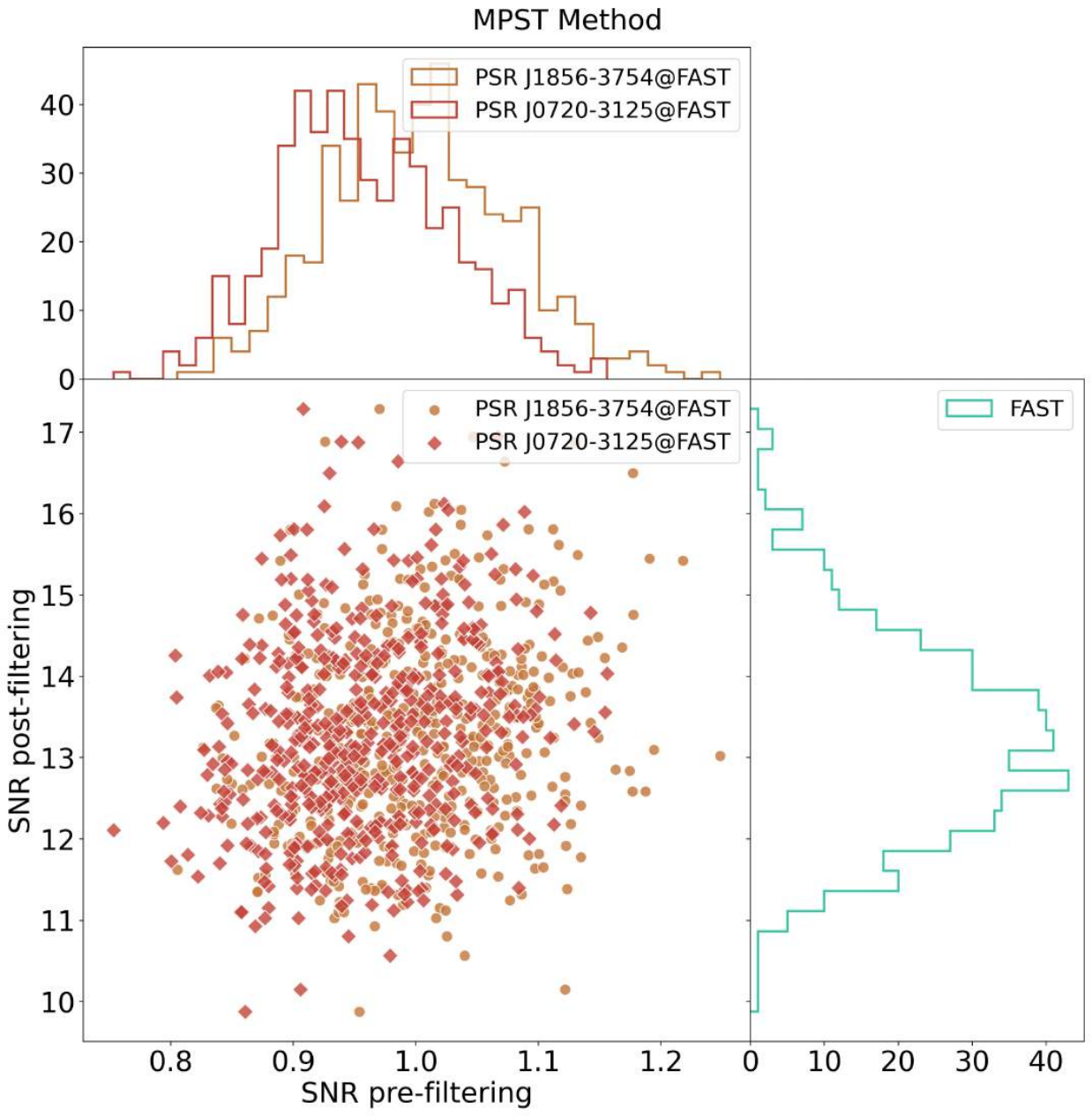}
	\includegraphics[width = 0.4\textwidth]{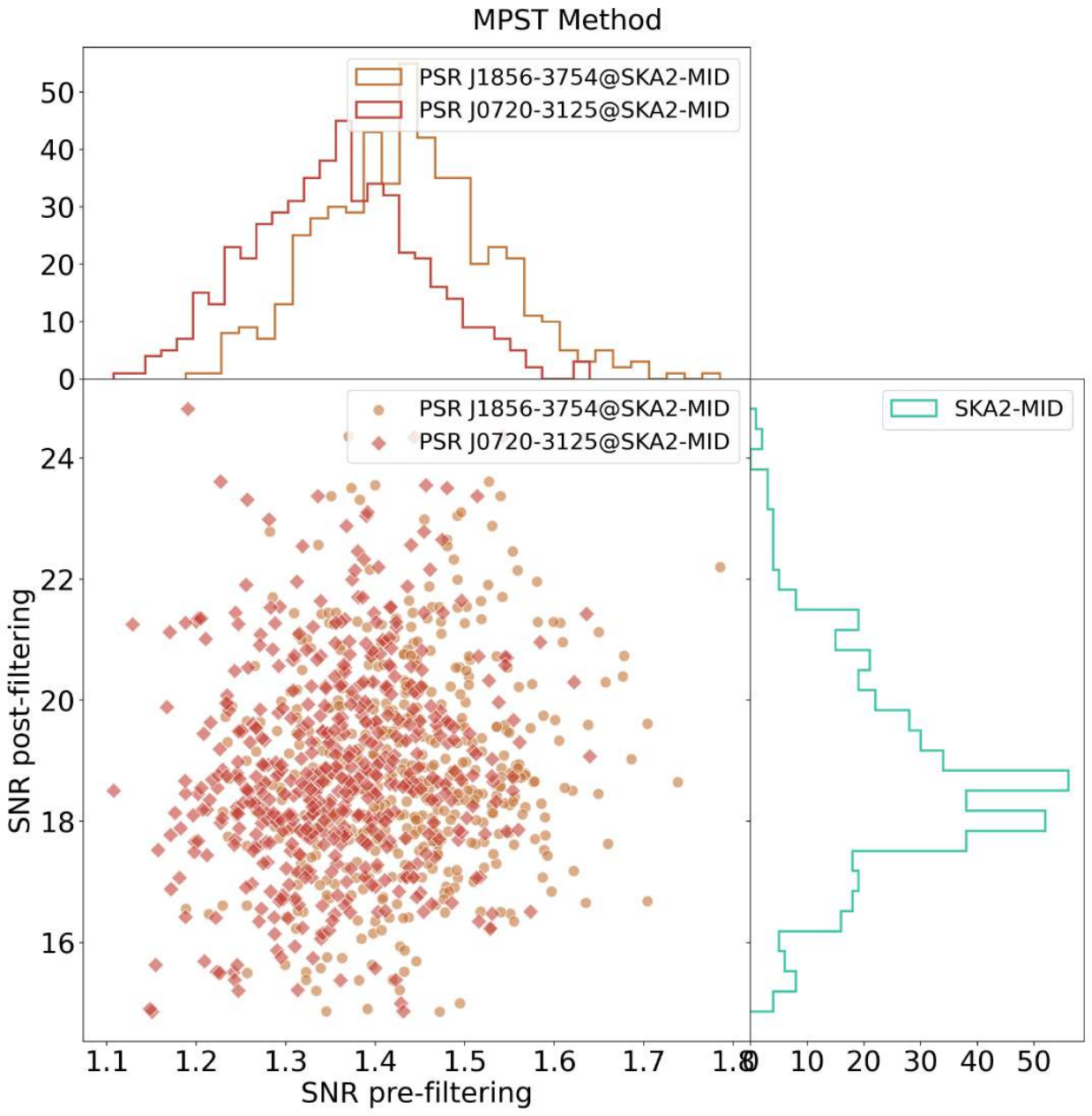}
	\caption{Comparison of the S/R pre- and post-filtering. The third shows the S/R results of the MPST method, showing simulated data from FAST and SKA2-MID in order from left to right.}
	\label{fig:SNR-500-3}
\end{figure*}

\begin{figure*}
	\centering
	\includegraphics[width = 0.4\textwidth]{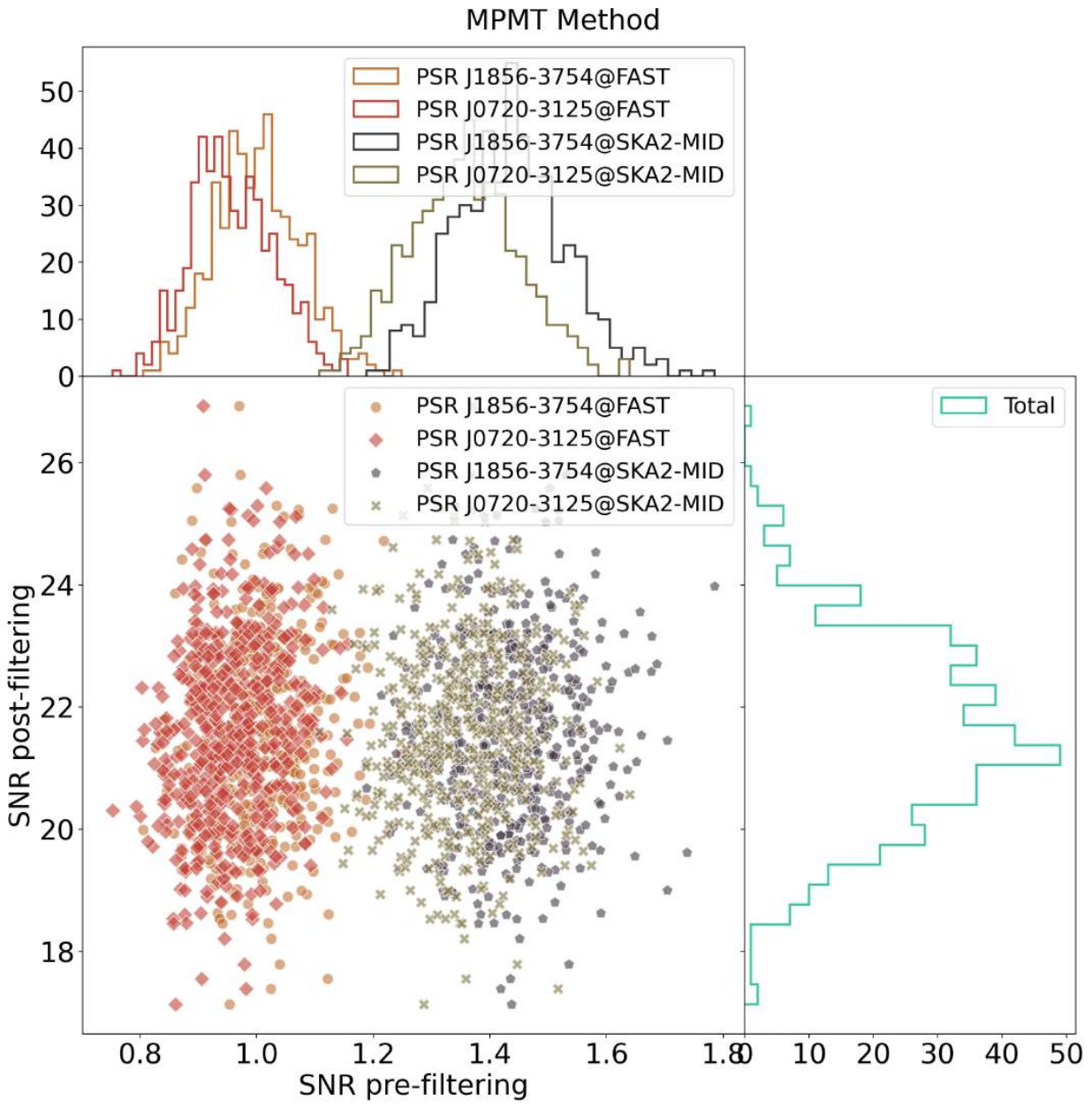}
	\caption{Comparison of the S/R pre- and post-filtering. The panel shows the S/R results of the MPMT method, from which it can be seen that this method improves the S/R the best.}
	\label{fig:SNR-500-4}
\end{figure*}

\section{Comparison of Graviton-Photon Conversion Approaches}
\label{app:comparison}
This appendix presents a structured overview of representative methodologies used to search for high‑frequency gravitational waves via graviton‑to‑photon conversion.
Table~\ref{tab:comparison} summarizes the key environments, frequency ranges, observables, and limiting factors considered in the literature. It also details the principal modelling assumptions that affect the conversion probability $P_{g\rightarrow\gamma}$—including the choice between geometric‑optics/WKB and plane‑wave limits, the treatment of dispersion and effective photon mass, the handling of coherence, and the prescriptions for resonant layers. By offering a concise side‑by‑side comparison, the table complements the discussion in Section~\ref{sec:discussion} and clarifies both the diversity of approaches found in the literature and the specific choices adopted in this work.	
\begin{table*}\label{tab:comparison}
	\begin{threeparttable}
		\scriptsize
		\setlength{\tabcolsep}{3pt}
		\renewcommand{\arraystretch}{1.15}
		
		\begin{tabular*}{\textwidth}{
				@{\extracolsep{\fill}}
				p{0.10\textwidth}
				p{0.18\textwidth}
				p{0.09\textwidth}
				p{0.18\textwidth}
				p{0.17\textwidth}
				p{0.22\textwidth}
				@{}
			}
			\hline
			Reference &
			Environment / method &
			Frequency focus &
			Main observable &
			$P_{g\rightarrow\gamma}$ modelling highlights \tnote{a}&
			Main limiting factors / what differs from this work \\
			\hline
			
			\citet{Domcke:2020yzq} &
			Large-scale cosmic magnetic fields; sky-averaged radio constraints &
			MHz--GHz &
			Distortion/excess in sky-averaged radio spectrum (EDGES/ARCADE~2 recast as bounds on $h_c$) &
			Domain / stochastic-field treatment; decoherence handled statistically (density-matrix style); large-scale dispersion modelling &
			Reach mainly controlled by assumed cosmic magnetic-field strength and foreground modelling; not a dedicated targeted campaign on individual magnetospheres. \tnote{b} \\
			
			\hline
			\citet{Ito:2023nkq} &
			Earth/Galaxy/IGM magnetic fields; ``photon-flux'' bounds &
			Very wide (reported across $10^6$--$10^{36}\,\mathrm{Hz}$) &
			Require converted photon flux not to exceed observed photon flux from telescopes &
			Plane-wave/domain picture across multiple media; effective photon-mass modelling differs by environment (Earth/Galaxy/IGM) &
			Uses existing photon observations to set broad-band limits; our work instead develops a radio-band observing strategy and radiometer-limited projections with FAST/SKA2-MID. \tnote{a} \\
			
			\hline
			\citet{Dandoy:2024oqg} &
			Galactic neutron-star population; non-resonant conversion &
			$10^8$--$10^{25}$\,Hz (reported) &
			Converted photon flux from NS population compared with Galactic emission &
			Population-based modelling; simplified magnetosphere ingredients; typically non-resonant treatment over wide bands &
			Population-based limits relying on population modelling and multi-band data coverage; our work targets the radio band with a multi-target/multi-instrument observing strategy. \tnote{b} \\
			
			\hline
			\citet{2024PhRvD.110j3003M} &
			Neutron-star magnetospheres; resonant conversion (birefringence/plasma) &
			IR/X-ray frequency &
			Strain limits from X-ray/IR flux measurements of specific NS targets &
			Resonant-layer emphasis (effective photon mass $\rightarrow 0$); birefringence/plasma dispersion treated in the high-$f$ regime; resonance size prescription matters &
			Resonant conversion yields strong limits at much higher frequencies; our focus is GHz radio observations with FAST/SKA2-MID and strategy-level sensitivity projections. \tnote{b} \\
			
			\hline
			This work &
			Pulsar magnetospheres in the radio band; strategy-driven (MPMT) &
			MHz--GHz radio band &
			Radio telescope sensitivity forecasts; multi-target/multi-instrument cross-validation &
			Geometric-optics/WKB LOS evaluation through time-dependent magnetospheric snapshots; phase and LOS sampling for operational averaging &
			Improvement mainly from deep integration, instrument sensitivity, and strategy-level systematics control; BBN used as benchmark for representative early-Universe SGWB scenarios. \tnote{c} \\
			
			\hline
		\end{tabular*}
		\begin{tablenotes}
			\footnotesize
			\item[a] Modelling highlights summarize for the main assumptions relevant to computing $P_{g\rightarrow\gamma}$ including: wave-propagation regime (geometric-optics/WKB vs.\ plane-wave/domain treatments), effective photon-mass/dispersion modelling (plasma and, where relevant, additional terms), treatment of phase mismatch and propagation effects, and whether the calculation focuses on resonant layers (and how their effective size is prescribed).
			\item[b] These entries derive upper limits by reinterpreting existing electromagnetic observations (e.g., sky-averaged spectra or source fluxes). In contrast, our work presents dedicated sensitivity projections for future radio observations with specified instruments and integration times.
			\item[c] In our MPMT strategy, FAST provides deep-integration flux thresholds, while SKA2-MID delivers an independent, high-sensitivity dataset with complementary systematics, enabling robust cross-validation.
		\end{tablenotes}
	\end{threeparttable}
\end{table*}

\bibliography{ROMVGW}{}

\begin{thebibliography}{}
\expandafter\ifx\csname natexlab\endcsname\relax\def\natexlab#1{#1}\fi
\providecommand{\url}[1]{\href{#1}{#1}}
\providecommand{\dodoi}[1]{doi:~\href{http://doi.org/#1}{\nolinkurl{#1}}}
\providecommand{\doeprint}[1]{\href{http://ascl.net/#1}{\nolinkurl{http://ascl.net/#1}}}
\providecommand{\doarXiv}[1]{\href{https://arxiv.org/abs/#1}{\nolinkurl{https://arxiv.org/abs/#1}}}

\bibitem[{B. {Abbott} {et~al.}(2005){Abbott}, {Abbott}, {Adhikari}, {Agresti},
  {Ajith}, {Allen}, {Allen}, {Amin}, {Anderson}, {Anderson}, {Araya},
  {Armandula}, {Ashley}, {Aulbert}, {Babak}, {Balasubramanian}, {Ballmer},
  {Barish}, {Barker}, {Barker}, {Barton}, {Bayer}, {Belczynski}, {Betzwieser},
  {Bhawal}, {Bilenko}, {Billingsley}, {Black}, {Blackburn}, {Blackburn},
  {Bland}, {Bogue}, {Bork}, {Bose}, {Brady}, {Braginsky}, {Brau}, {Brown},
  {Buonanno}, {Busby}, {Butler}, {Cadonati}, {Cagnoli}, {Camp}, {Cannizzo},
  {Cannon}, {Cardenas}, {Carter}, {Casey}, {Charlton}, {Chatterji}, {Chen},
  {Chin}, {Christensen}, {Cokelaer}, {Colacino}, {Coldwell}, {Cook}, {Corbitt},
  {Coyne}, {Creighton}, {Creighton}, {Dalrymple}, {D'Ambrosio}, {Danzmann},
  {Davies}, {Debra}, {Dergachev}, {Desai}, {Desalvo}, {Dhurandar}, {D{\'\i}az},
  {di Credico}, {Drever}, {Dupuis}, {Ehrens}, {Etzel}, {Evans}, {Evans},
  {Fairhurst}, {Finn}, {Franzen}, {Frey}, {Fritschel}, {Frolov}, {Fyffe},
  {Ganezer}, {Garofoli}, {Gholami}, {Giaime}, {Goda}, {Goggin}, {Gonz{\'a}lez},
  {Gray}, {Gretarsson}, {Grimmett}, {Grote}, {Grunewald}, {Guenther},
  {Gustafson}, {Hamilton}, {Hanna}, {Hanson}, {Hardham}, {Harry}, {Heefner},
  {Heng}, {Hewitson}, {Hindman}, {Hoang}, {Hough}, {Hua}, {Ito}, {Itoh},
  {Ivanov}, {Johnson}, {Johnson}, {Jones}, {Jones}, {Jones}, {Kalogera},
  {Katsavounidis}, {Kawabe}, {Kawamura}, {Kells}, {Khan}, {Kim}, {King},
  {Klimenko}, {Koranda}, {Kozak}, {Krishnan}, {Landry}, {Lantz}, {Lazzarini},
  {Lei}, {Leonor}, {Libbrecht}, {Lindquist}, {Liu}, {Lormand}, {Lubi{\'n}ski},
  {L{\"u}ck}, {Luna}, {Machenschalk}, {Macinnis}, {Mageswaran}, {Mailand},
  {Malec}, {Mandic}, {Marka}, {Maros}, {Mason}, {Matone}, {Mavalvala},
  {McCarthy}, {McClelland}, {McHugh}, {McNabb}, {Melissinos}, {Mendell},
  {Mercer}, {Meshkov}, {Messaritaki}, {Messenger}, {Mikhailov}, {Mitra},
  {Mitrofanov}, {Mitselmakher}, {Mittleman}, {Miyakawa}, {Mohanty}, {Moreno},
  {Mossavi}, {Mueller}, {Mukherjee}, {Myers}, {Myers}, {Nash}, {Nocera},
  {Noel}, {O'Reilly}, {O'Shaughnessy}, {Ottaway}, {Overmier}, {Owen}, {Pan},
  {Papa}, {Parameshwaraiah}, {Parameswariah}, {Pedraza}, {Penn}, {Pitkin},
  {Prix}, {Quetschke}, {Raab}, {Radkins}, {Rahkola}, {Rakhmanov}, \&
  {Rawlins}}]{2005PhRvL..95v1101A}
{Abbott}, B., {Abbott}, R., {Adhikari}, R., {et~al.} 2005,
  \bibinfo{title}{{Upper Limits on a Stochastic Background of Gravitational
  Waves},} \prl, 95, 221101, \dodoi{10.1103/PhysRevLett.95.221101}

\bibitem[{B.~P. Abbott {et~al.}(2016)Abbott {et~al.}}]{LIGOScientific:2016aoc}
Abbott, B.~P., {et~al.} 2016, \bibinfo{title}{{Observation of Gravitational
  Waves from a Binary Black Hole Merger},} Phys. Rev. Lett., 116, 061102,
  \dodoi{10.1103/PhysRevLett.116.061102}

\bibitem[{B.~P. Abbott {et~al.}(2017)Abbott {et~al.}}]{LIGOScientific:2017vwq}
Abbott, B.~P., {et~al.} 2017, \bibinfo{title}{{GW170817: Observation of
  Gravitational Waves from a Binary Neutron Star Inspiral},} Phys. Rev. Lett.,
  119, 161101, \dodoi{10.1103/PhysRevLett.119.161101}

\bibitem[{R. {Abbott} {et~al.}(2023){Abbott}, {Abbott}, {Acernese}, {Ackley},
  {Adams}, {Adhikari}, {Adhikari}, {Adya}, {Affeldt}, {Agarwal}, {Agathos},
  {Agatsuma}, {Aggarwal}, {Aguiar}, {Aiello}, {Ain}, {Ajith}, {Akcay},
  {Akutsu}, {Albanesi}, {Allocca}, {Altin}, {Amato}, {Anand}, {Anand},
  {Ananyeva}, {Anderson}, {Anderson}, {Ando}, {Andrade}, {Andres},
  {Andri{\'c}}, {Angelova}, {Ansoldi}, {Antelis}, {Antier}, {Appert}, {Arai},
  {Arai}, {Arai}, {Araki}, {Araya}, {Araya}, {Areeda}, {Ar{\`e}ne}, {Aritomi},
  {Arnaud}, {Arogeti}, {Aronson}, {Arun}, {Asada}, {Asali}, {Ashton}, {Aso},
  {Assiduo}, {Aston}, {Astone}, {Aubin}, {Austin}, {Babak}, {Badaracco},
  {Bader}, {Badger}, {Bae}, {Bae}, {Baer}, {Bagnasco}, {Bai}, {Baiotti},
  {Baird}, {Bajpai}, {Ball}, {Ballardin}, {Ballmer}, {Balsamo}, {Baltus},
  {Banagiri}, {Bankar}, {Barayoga}, {Barbieri}, {Barish}, {Barker}, {Barneo},
  {Barone}, {Barr}, {Barsotti}, {Barsuglia}, {Barta}, {Bartlett}, {Barton},
  {Bartos}, {Bassiri}, {Basti}, {Bawaj}, {Bayley}, {Baylor}, {Bazzan},
  {B{\'e}csy}, {Bedakihale}, {Bejger}, {Belahcene}, {Benedetto}, {Beniwal},
  {Bennett}, {Bentley}, {Benyaala}, {Bergamin}, {Berger}, {Bernuzzi}, {Berry},
  {Bersanetti}, {Bertolini}, {Betzwieser}, {Beveridge}, {Bhandare}, {Bhardwaj},
  {Bhattacharjee}, {Bhaumik}, {Bilenko}, {Billingsley}, {Bini}, {Birney},
  {Birnholtz}, {Biscans}, {Bischi}, {Biscoveanu}, {Bisht}, {Biswas}, {Bitossi},
  {Bizouard}, {Blackburn}, {Blair}, {Blair}, {Blair}, {Bobba}, {Bode}, {Boer},
  {Bogaert}, {Boldrini}, {Bonavena}, {Bondu}, {Bonilla}, {Bonnand}, {Booker},
  {Boom}, {Bork}, {Boschi}, {Bose}, {Bose}, {Bossilkov}, {Boudart},
  {Bouffanais}, {Bozzi}, {Bradaschia}, {Brady}, {Bramley}, {Branch},
  {Branchesi}, {Brandt}, {Brau}, {Breschi}, {Briant}, {Briggs}, {Brillet},
  {Brinkmann}, {Brockill}, {Brooks}, {Brooks}, {Brown}, {Brunett}, {Bruno},
  {Bruntz}, {Bryant}, {Bulik}, {Bulten}, {Buonanno}, {Buscicchio}, {Buskulic},
  {Buy}, {Byer}, {Davies}, {Cadonati}, {Cagnoli}, {Cahillane}, {Bustillo},
  {Callaghan}, {Callister}, {Calloni}, {Cameron}, {Camp}, {Canepa},
  {Canevarolo}, {Cannavacciuolo}, {Cannon}, {Cao}, {Cao}, {Capocasa}, {Capote},
  {Carapella}, \& {Carbognani}}]{2023PhRvX..13d1039A}
{Abbott}, R., {Abbott}, T.~D., {Acernese}, F., {et~al.} 2023,
  \bibinfo{title}{{GWTC-3: Compact Binary Coalescences Observed by LIGO and
  Virgo during the Second Part of the Third Observing Run},} Physical Review X,
  13, 041039, \dodoi{10.1103/PhysRevX.13.041039}

\bibitem[{S.~L. Adler(1971)Adler}]{Adler:1971wn}
Adler, S.~L. 1971, \bibinfo{title}{{Photon splitting and photon dispersion in a
  strong magnetic field},} Annals Phys., 67, 599,
  \dodoi{10.1016/0003-4916(71)90154-0}

\bibitem[{N. Aggarwal {et~al.}(2021)Aggarwal {et~al.}}]{Aggarwal:2020olq}
Aggarwal, N., {et~al.} 2021, \bibinfo{title}{{Challenges and opportunities of
  gravitational-wave searches at MHz to GHz frequencies},} Living Rev. Rel.,
  24, 4, \dodoi{10.1007/s41114-021-00032-5}

\bibitem[{A. {Agrawal} \& J. {Domke}(2024){Agrawal} \&
  {Domke}}]{2024arXiv240519747A}
{Agrawal}, A., \& {Domke}, J. 2024, \bibinfo{title}{{Understanding and
  mitigating difficulties in posterior predictive evaluation},} arXiv e-prints,
  arXiv:2405.19747, \dodoi{10.48550/arXiv.2405.19747}

\bibitem[{F.~A. {Aharonian} {et~al.}(2012){Aharonian}, {Bogovalov}, \&
  {Khangulyan}}]{2012Natur.482..507A}
{Aharonian}, F.~A., {Bogovalov}, S.~V., \& {Khangulyan}, D. 2012,
  \bibinfo{title}{{Abrupt acceleration of a `cold' ultrarelativistic wind from
  the Crab pulsar},} \nat, 482, 507, \dodoi{10.1038/nature10793}

\bibitem[{T. Akiba {et~al.}(2019)Akiba, Sano, Yanase, Ohta, \&
  Koyama}]{optuna_2019}
Akiba, T., Sano, S., Yanase, T., Ohta, T., \& Koyama, M. 2019, in Proceedings
  of the 25th {ACM} {SIGKDD} International Conference on Knowledge Discovery
  and Data Mining

\bibitem[{K. An(1933)An}]{an1933sulla}
An, K. 1933, \bibinfo{title}{Sulla determinazione empirica di una legge
  didistribuzione,} Giorn Dell'inst Ital Degli Att, 4, 89

\bibitem[{K.~N. {Ananda} {et~al.}(2007){Ananda}, {Clarkson}, \&
  {Wands}}]{2007PhRvD..75l3518A}
{Ananda}, K.~N., {Clarkson}, C., \& {Wands}, D. 2007,
  \bibinfo{title}{{Cosmological gravitational wave background from primordial
  density perturbations},} \prd, 75, 123518, \dodoi{10.1103/PhysRevD.75.123518}

\bibitem[{J. Anandan \& R.~Y. Chiao(1982)Anandan \& Chiao}]{Anandan:1982is}
Anandan, J., \& Chiao, R.~Y. 1982, \bibinfo{title}{{GRAVITATIONAL RADIATION
  ANTENNAS USING THE SAGNAC EFFECT},} Gen. Rel. Grav., 14, 515,
  \dodoi{10.1007/BF00756213}

\bibitem[{D. Andriot \& G. Lucena~G\'omez(2017)Andriot \&
  Lucena~G\'omez}]{Andriot:2017oaz}
Andriot, D., \& Lucena~G\'omez, G. 2017, \bibinfo{title}{{Signatures of extra
  dimensions in gravitational waves},} JCAP, 06, 048,
  \dodoi{10.1088/1475-7516/2017/06/048}

\bibitem[{J. Antoniadis {et~al.}(2023)Antoniadis {et~al.}}]{EPTA:2023fyk}
Antoniadis, J., {et~al.} 2023, \bibinfo{title}{{The second data release from
  the European Pulsar Timing Array - III. Search for gravitational wave
  signals},} Astron. Astrophys., 678, A50, \dodoi{10.1051/0004-6361/202346844}

\bibitem[{A. Arvanitaki \& A.~A. Geraci(2013)Arvanitaki \&
  Geraci}]{Arvanitaki:2012cn}
Arvanitaki, A., \& Geraci, A.~A. 2013, \bibinfo{title}{{Detecting
  high-frequency gravitational waves with optically-levitated sensors},} Phys.
  Rev. Lett., 110, 071105, \dodoi{10.1103/PhysRevLett.110.071105}

\bibitem[{Z. Arzoumanian {et~al.}(2023)Arzoumanian {et~al.}}]{NANOGrav:2023bts}
Arzoumanian, Z., {et~al.} 2023, \bibinfo{title}{{The NANOGrav 12.5 yr Data Set:
  Bayesian Limits on Gravitational Waves from Individual Supermassive Black
  Hole Binaries},} Astrophys. J. Lett., 951, L28,
  \dodoi{10.3847/2041-8213/acdbc7}

\bibitem[{P. {Auclair} {et~al.}(2020){Auclair}, {Blanco-Pillado}, {Figueroa},
  {Jenkins}, {Lewicki}, {Sakellariadou}, {Sanidas}, {Sousa}, {Steer},
  {Wachter}, {Kuroyanagi}, \& {LISA Cosmology Working
  Group}}]{2020JCAP...04..034A}
{Auclair}, P., {Blanco-Pillado}, J.~J., {Figueroa}, D.~G., {et~al.} 2020,
  \bibinfo{title}{{Probing the gravitational wave background from cosmic
  strings with LISA},} \jcap, 2020, 034, \dodoi{10.1088/1475-7516/2020/04/034}

\bibitem[{D. {Baker} {et~al.}(2022){Baker}, {Brisken}, {van Kerkwijk}, {Main},
  {Pen}, {Sprenger}, \& {Wucknitz}}]{2022MNRAS.510.4573B}
{Baker}, D., {Brisken}, W., {van Kerkwijk}, M.~H., {et~al.} 2022,
  \bibinfo{title}{{Interstellar interferometry: precise curvature measurement
  from pulsar secondary spectra},} \mnras, 510, 4573,
  \dodoi{10.1093/mnras/stab3599}

\bibitem[{R. Ballantini {et~al.}(2003)Ballantini, Bernard, Chiaveri,
  Chincarini, Gemme, Losito, Parodi, \& Picasso}]{Ballantini:2003nt}
Ballantini, R., Bernard, P., Chiaveri, E., {et~al.} 2003, \bibinfo{title}{{A
  detector of high frequency gravitational waves based on coupled microwave
  cavities},} Class. Quant. Grav., 20, 3505,
  \dodoi{10.1088/0264-9381/20/15/316}

\bibitem[{N. Barnaby \& M. Peloso(2011)Barnaby \& Peloso}]{Barnaby:2010vf}
Barnaby, N., \& Peloso, M. 2011, \bibinfo{title}{{Large Nongaussianity in Axion
  Inflation},} Phys. Rev. Lett., 106, 181301,
  \dodoi{10.1103/PhysRevLett.106.181301}

\bibitem[{D. {Baumann} {et~al.}(2007){Baumann}, {Steinhardt}, {Takahashi}, \&
  {Ichiki}}]{2007PhRvD..76h4019B}
{Baumann}, D., {Steinhardt}, P., {Takahashi}, K., \& {Ichiki}, K. 2007,
  \bibinfo{title}{{Gravitational wave spectrum induced by primordial scalar
  perturbations},} \prd, 76, 084019, \dodoi{10.1103/PhysRevD.76.084019}

\bibitem[{A. Bauswein {et~al.}(2019)Bauswein, Bastian, Blaschke, Chatziioannou,
  Clark, Fischer, \& Oertel}]{Bauswein:2018bma}
Bauswein, A., Bastian, N.-U.~F., Blaschke, D.~B., {et~al.} 2019,
  \bibinfo{title}{{Identifying a first-order phase transition in neutron star
  mergers through gravitational waves},} Phys. Rev. Lett., 122, 061102,
  \dodoi{10.1103/PhysRevLett.122.061102}

\bibitem[{A. Berlin {et~al.}(2022)Berlin, Blas, Tito~D'Agnolo, Ellis, Harnik,
  Kahn, \& Sch\"utte-Engel}]{Berlin:2021txa}
Berlin, A., Blas, D., Tito~D'Agnolo, R., {et~al.} 2022,
  \bibinfo{title}{{Detecting high-frequency gravitational waves with microwave
  cavities},} Phys. Rev. D, 105, 116011, \dodoi{10.1103/PhysRevD.105.116011}

\bibitem[{P. Bernard {et~al.}(2001)Bernard, Gemme, Parodi, \&
  Picasso}]{Bernard:2001kp}
Bernard, P., Gemme, G., Parodi, R., \& Picasso, E. 2001, \bibinfo{title}{{A
  Detector of small harmonic displacements based on two coupled microwave
  cavities},} Rev. Sci. Instrum., 72, 2428, \dodoi{10.1063/1.1366636}

\bibitem[{N.~D.~R. {Bhat} {et~al.}(2004){Bhat}, {Cordes}, {Camilo}, {Nice}, \&
  {Lorimer}}]{2004ApJ...605..759B}
{Bhat}, N.~D.~R., {Cordes}, J.~M., {Camilo}, F., {Nice}, D.~J., \& {Lorimer},
  D.~R. 2004, \bibinfo{title}{{Multifrequency Observations of Radio Pulse
  Broadening and Constraints on Interstellar Electron Density Microstructure},}
  \apj, 605, 759, \dodoi{10.1086/382680}

\bibitem[{S. Bird {et~al.}(2016)Bird, Cholis, Mu\~noz, Ali-Ha\"\i{}moud,
  Kamionkowski, Kovetz, Raccanelli, \& Riess}]{Bird:2016dcv}
Bird, S., Cholis, I., Mu\~noz, J.~B., {et~al.} 2016, \bibinfo{title}{{Did LIGO
  detect dark matter?},} Phys. Rev. Lett., 116, 201301,
  \dodoi{10.1103/PhysRevLett.116.201301}

\bibitem[{C. Bishop(2006)Bishop}]{bishop2006pattern}
Bishop, C. 2006, Pattern Recognition and Machine Learning (Springer).
\newblock
  \url{https://www.microsoft.com/en-us/research/publication/pattern-recognition-machine-learning/}

\bibitem[{G.~S. Bisnovatyi-Kogan \& V.~N. Rudenko(2004)Bisnovatyi-Kogan \&
  Rudenko}]{Bisnovatyi-Kogan:2004cdg}
Bisnovatyi-Kogan, G.~S., \& Rudenko, V.~N. 2004, \bibinfo{title}{{Very high
  frequency gravitational wave background in the universe},} Class. Quant.
  Grav., 21, 3347, \dodoi{10.1088/0264-9381/21/14/001}

\bibitem[{J.~J. {Blanco-Pillado} {et~al.}(2014){Blanco-Pillado}, {Olum}, \&
  {Shlaer}}]{2014PhRvD..89b3512B}
{Blanco-Pillado}, J.~J., {Olum}, K.~D., \& {Shlaer}, B. 2014,
  \bibinfo{title}{{Number of cosmic string loops},} \prd, 89, 023512,
  \dodoi{10.1103/PhysRevD.89.023512}

\bibitem[{A. {Bransgrove} {et~al.}(2023){Bransgrove}, {Beloborodov}, \&
  {Levin}}]{2023ApJ...958L...9B}
{Bransgrove}, A., {Beloborodov}, A.~M., \& {Levin}, Y. 2023,
  \bibinfo{title}{{Radio Emission and Electric Gaps in Pulsar Magnetospheres},}
  \apjl, 958, L9, \dodoi{10.3847/2041-8213/ad0556}

\bibitem[{R. {Braun} {et~al.}(2019){Braun} {et~al.}}]{2019arXiv191212699B}
{Braun}, R., {et~al.} 2019, \bibinfo{title}{{Anticipated Performance of the
  Square Kilometre Array -- Phase 1 (SKA1)},} arXiv e-prints, arXiv:1912.12699,
  \dodoi{10.48550/arXiv.1912.12699}

\bibitem[{G. Breit \& J.~A. Wheeler(1934)Breit \& Wheeler}]{Breit:1934zz}
Breit, G., \& Wheeler, J.~A. 1934, \bibinfo{title}{{Collision of two light
  quanta},} Phys. Rev., 46, 1087, \dodoi{10.1103/PhysRev.46.1087}

\bibitem[{O.~P. Buneman \& D.~A. Dunn(1965)Buneman \&
  Dunn}]{Buneman1965COMPUTEREI}
Buneman, O.~P., \& Dunn, D.~A. 1965.
\newblock \url{https://api.semanticscholar.org/CorpusID:117165226}

\bibitem[{S. {Burke-Spolaor} \& M. {Bailes}(2010){Burke-Spolaor} \&
  {Bailes}}]{2010MNRAS.402..855B}
{Burke-Spolaor}, S., \& {Bailes}, M. 2010, \bibinfo{title}{{The millisecond
  radio sky: transients from a blind single-pulse search},} \mnras, 402, 855,
  \dodoi{10.1111/j.1365-2966.2009.15965.x}

\bibitem[{B.~J. {Burn}(1966){Burn}}]{1966MNRAS.133...67B}
{Burn}, B.~J. 1966, \bibinfo{title}{{On the depolarization of discrete radio
  sources by Faraday dispersion},} \mnras, 133, 67,
  \dodoi{10.1093/mnras/133.1.67}

\bibitem[{R.-g. Cai {et~al.}(2019)Cai, Pi, \& Sasaki}]{Cai:2018dig}
Cai, R.-g., Pi, S., \& Sasaki, M. 2019, \bibinfo{title}{{Gravitational Waves
  Induced by non-Gaussian Scalar Perturbations},} Phys. Rev. Lett., 122,
  201101, \dodoi{10.1103/PhysRevLett.122.201101}

\bibitem[{A. {Camero} {et~al.}(2014){Camero}, , {et~al.}}]{2014MNRAS.438.3291C}
{Camero}, A., , {et~al.} 2014, \bibinfo{title}{{Quiescent state and outburst
  evolution of SGR 0501+4516},} \mnras, 438, 3291,
  \dodoi{10.1093/mnras/stt2432}

\bibitem[{F. {Camilo} {et~al.}(2007){Camilo}, {Ransom}, {Halpern}, \&
  {Reynolds}}]{2007ApJ...666L..93C}
{Camilo}, F., {Ransom}, S.~M., {Halpern}, J.~P., \& {Reynolds}, J. 2007,
  \bibinfo{title}{{1E 1547.0-5408: A Radio-emitting Magnetar with a Rotation
  Period of 2 Seconds},} \apjl, 666, L93, \dodoi{10.1086/521826}

\bibitem[{C. {Caprini} \& D.~G. {Figueroa}(2018){Caprini} \&
  {Figueroa}}]{2018CQGra..35p3001C}
{Caprini}, C., \& {Figueroa}, D.~G. 2018, \bibinfo{title}{{Cosmological
  backgrounds of gravitational waves},} Classical and Quantum Gravity, 35,
  163001, \dodoi{10.1088/1361-6382/aac608}

\bibitem[{ {CASA Team} {et~al.}(2022){CASA Team}, {Bean},
  {et~al.}}]{2022PASP..134k4501C}
{CASA Team}, {Bean}, B., {et~al.} 2022, \bibinfo{title}{{CASA, the Common
  Astronomy Software Applications for Radio Astronomy},} \pasp, 134, 114501,
  \dodoi{10.1088/1538-3873/ac9642}

\bibitem[{N. {Chkheidze}(2009){Chkheidze}}]{2009A&A...500..861C}
{Chkheidze}, N. 2009, \bibinfo{title}{{The emission polarization of RX
  J1856.5-3754},} \aap, 500, 861, \dodoi{10.1051/0004-6361/200911648}

\bibitem[{I.~V. {Chugunov} {et~al.}(1975){Chugunov}, {Eidman}, \&
  {Suvorov}}]{1975Ap&SS..32L...7C}
{Chugunov}, I.~V., {Eidman}, V.~I., \& {Suvorov}, E.~V. 1975,
  \bibinfo{title}{{The Motion of Charged Particles in a strong Electromagnetic
  Field and Curvature Radiation},} \apss, 32, L7, \dodoi{10.1007/BF00646233}

\bibitem[{T.~R. {Clifton} \& A.~G. {Lyne}(1986){Clifton} \&
  {Lyne}}]{1986Natur.320...43C}
{Clifton}, T.~R., \& {Lyne}, A.~G. 1986, \bibinfo{title}{{High-radio-frequency
  survey for young and millisecond pulsars},} \nat, 320, 43,
  \dodoi{10.1038/320043a0}

\bibitem[{W.~A. {Coles} {et~al.}(2010){Coles}, {Rickett}, {Gao}, {Hobbs}, \&
  {Verbiest}}]{2010ApJ...717.1206C}
{Coles}, W.~A., {Rickett}, B.~J., {Gao}, J.~J., {Hobbs}, G., \& {Verbiest},
  J.~P.~W. 2010, \bibinfo{title}{{Scattering of Pulsar Radio Emission by the
  Interstellar Plasma},} \apj, 717, 1206, \dodoi{10.1088/0004-637X/717/2/1206}

\bibitem[{J.~J. {Condon} \& S.~M. {Ransom}(2016){Condon} \&
  {Ransom}}]{2016era..book.....C}
{Condon}, J.~J., \& {Ransom}, S.~M. 2016, {Essential Radio Astronomy}

\bibitem[{Y. {Cong} {et~al.}(2021){Cong} {et~al.}}]{2021ApJ...914..128C}
{Cong}, Y., {et~al.} 2021, \bibinfo{title}{{An Ultralong-wavelength Sky Model
  with Absorption Effect},} \apj, 914, 128, \dodoi{10.3847/1538-4357/abf55c}

\bibitem[{I. {Contopoulos} {et~al.}(1999){Contopoulos}, {Kazanas}, \&
  {Fendt}}]{1999ApJ...511..351C}
{Contopoulos}, I., {Kazanas}, D., \& {Fendt}, C. 1999, \bibinfo{title}{{The
  Axisymmetric Pulsar Magnetosphere},} \apj, 511, 351, \dodoi{10.1086/306652}

\bibitem[{J.~M. Cordes \& T.~J.~W. Lazio(2002)Cordes \& Lazio}]{Cordes:2002wz}
Cordes, J.~M., \& Lazio, T. J.~W. 2002, \bibinfo{title}{{NE2001. 1. A New model
  for the galactic distribution of free electrons and its fluctuations},}
  \doarXiv{astro-ph/0207156}

\bibitem[{J.~M. Cordes \& T.~J.~W. Lazio(2003)Cordes \& Lazio}]{Cordes:2003ik}
Cordes, J.~M., \& Lazio, T. J.~W. 2003, \bibinfo{title}{{NE2001. 2. Using radio
  propagation data to construct a model for the galactic distribution of free
  electrons},} \doarXiv{astro-ph/0301598}

\bibitem[{B. {Crinquand} {et~al.}(2019){Crinquand}, {Cerutti}, \&
  {Dubus}}]{2019A&A...622A.161C}
{Crinquand}, B., {Cerutti}, B., \& {Dubus}, G. 2019, \bibinfo{title}{{Kinetic
  modeling of the electromagnetic precursor from an axisymmetric binary pulsar
  coalescence},} \aap, 622, A161, \dodoi{10.1051/0004-6361/201834610}

\bibitem[{A.~M. Cruise(2000)Cruise}]{Cruise:2000za}
Cruise, A.~M. 2000, \bibinfo{title}{{An electromagnetic detector for
  very-high-frequency gravitational waves},} Class. Quant. Grav., 17, 2525,
  \dodoi{10.1088/0264-9381/17/13/305}

\bibitem[{C. {Cutler} \& {\'E}.~E. {Flanagan}(1994){Cutler} \&
  {Flanagan}}]{1994PhRvD..49.2658C}
{Cutler}, C., \& {Flanagan}, {\'E}.~E. 1994, \bibinfo{title}{{Gravitational
  waves from merging compact binaries: How accurately can one extract the
  binary's parameters from the inspiral waveform\textbackslash?},} \prd, 49,
  2658, \dodoi{10.1103/PhysRevD.49.2658}

\bibitem[{R. D'Agostino \& E.~S. Pearson(1973)D'Agostino \&
  Pearson}]{272b2fa8-f3b3-371d-b34e-a4c6938458a1}
D'Agostino, R., \& Pearson, E.~S. 1973, \bibinfo{title}{Tests for Departure
  from Normality. Empirical Results for the Distributions of b2 and √ b1,}
  Biometrika, 60, 613.
\newblock \url{http://www.jstor.org/stable/2335012}

\bibitem[{R.~B. D'Agostino \& A. Belanger(1990)D'Agostino \&
  Belanger}]{25848ac6-50af-379a-93d8-ad15b9244648}
D'Agostino, R.~B., \& Belanger, A. 1990, \bibinfo{title}{A Suggestion for Using
  Powerful and Informative Tests of Normality,} The American Statistician, 44,
  316.
\newblock \url{http://www.jstor.org/stable/2684359}

\bibitem[{T. {Damour} \& A. {Vilenkin}(2000){Damour} \&
  {Vilenkin}}]{2000PhRvL..85.3761D}
{Damour}, T., \& {Vilenkin}, A. 2000, \bibinfo{title}{{Gravitational Wave
  Bursts from Cosmic Strings},} \prl, 85, 3761,
  \dodoi{10.1103/PhysRevLett.85.3761}

\bibitem[{V. Dandoy {et~al.}(2024)Dandoy, Bert\'olez-Mart\'\i{}nez, \&
  Costa}]{Dandoy:2024oqg}
Dandoy, V., Bert\'olez-Mart\'\i{}nez, T., \& Costa, F. 2024,
  \bibinfo{title}{{High Frequency Gravitational Wave Bounds from Galactic
  Neutron Stars},} \doarXiv{2402.14092}

\bibitem[{J.~G. {Davies} {et~al.}(1972){Davies}, {Lyne}, \&
  {Seiradakis}}]{1972Natur.240..229D}
{Davies}, J.~G., {Lyne}, A.~G., \& {Seiradakis}, J.~H. 1972,
  \bibinfo{title}{{Pulsar Associated with the Supernova Remnant IC 443},} \nat,
  240, 229, \dodoi{10.1038/240229a0}

\bibitem[{J.~M. Dawson(1983)Dawson}]{Dawson:1983zz}
Dawson, J.~M. 1983, \bibinfo{title}{{Particle simulation of plasmas},} Rev.
  Mod. Phys., 55, 403, \dodoi{10.1103/RevModPhys.55.403}

\bibitem[{D. {De Grandis} {et~al.}(2022){De Grandis}, {Rigoselli},
  {Mereghetti}, {Younes}, {Pizzochero}, {Taverna}, {Tiengo}, {Turolla}, \&
  {Zane}}]{2022MNRAS.516.4932D}
{De Grandis}, D., {Rigoselli}, M., {Mereghetti}, S., {et~al.} 2022,
  \bibinfo{title}{{Two decades of X-ray observations of the isolated neutron
  star RX J1856.5 - 3754: detection of thermal and non-thermal hard X-rays and
  refined spin-down measurement},} \mnras, 516, 4932,
  \dodoi{10.1093/mnras/stac2587}

\bibitem[{J. {Derouillat} {et~al.}(2018){Derouillat}, {Beck}, {P{\'e}rez},
  {Vinci}, {Chiaramello}, {Grassi}, {Fl{\'e}}, {Bouchard}, {Plotnikov},
  {Aunai}, {Dargent}, {Riconda}, \& {Grech}}]{2018CoPhC.222..351D}
{Derouillat}, J., {Beck}, A., {P{\'e}rez}, F., {et~al.} 2018,
  \bibinfo{title}{{SMILEI : A collaborative, open-source, multi-purpose
  particle-in-cell code for plasma simulation},} Computer Physics
  Communications, 222, 351, \dodoi{10.1016/j.cpc.2017.09.024}

\bibitem[{D.~A. Dickey \& W.~A. Fuller(1981)Dickey \&
  Fuller}]{Dickey1981LIKELIHOODRS}
Dickey, D.~A., \& Fuller, W.~A. 1981, \bibinfo{title}{LIKELIHOOD RATIO
  STATISTICS FOR AUTOREGRESSIVE TIME SERIES WITH A UNIT ROOT,} Econometrica,
  49, 1057.
\newblock \url{https://api.semanticscholar.org/CorpusID:154769359}

\bibitem[{W. Dittrich \& H. Gies(2000)Dittrich \& Gies}]{Dittrich:2000zu}
Dittrich, W., \& Gies, H. 2000, {Probing the quantum vacuum. Perturbative
  effective action approach in quantum electrodynamics and its application},
  Vol. 166, \dodoi{10.1007/3-540-45585-X}

\bibitem[{W. Dittrich \& M. Reuter(1985)Dittrich \& Reuter}]{Dittrich:1985yb}
Dittrich, W., \& Reuter, M. 1985, {Effective Lagrangians in Quantum
  Electrodynamics}, Vol. 220

\bibitem[{V. Domcke(2023)Domcke}]{Domcke:2023qle}
Domcke, V. 2023, in {57th Rencontres de Moriond on Electroweak Interactions and
  Unified Theories}.
\newblock \doarXiv{2306.04496}

\bibitem[{V. Domcke \& C. Garcia-Cely(2021)Domcke \&
  Garcia-Cely}]{Domcke:2020yzq}
Domcke, V., \& Garcia-Cely, C. 2021, \bibinfo{title}{{Potential of radio
  telescopes as high-frequency gravitational wave detectors},} Phys. Rev.
  Lett., 126, 021104, \dodoi{10.1103/PhysRevLett.126.021104}

\bibitem[{V. {Doroshenko} {et~al.}(2018){Doroshenko}, {Suleimanov}, \&
  {Santangelo}}]{2018A&A...618A..76D}
{Doroshenko}, V., {Suleimanov}, V., \& {Santangelo}, A. 2018,
  \bibinfo{title}{{CXOU J160103.1-513353: another central compact object with a
  carbon atmosphere?},} \aap, 618, A76, \dodoi{10.1051/0004-6361/201833271}

\bibitem[{R. Duclous {et~al.}(2011)Duclous, Kirk, \& Bell}]{Duclous:2010zb}
Duclous, R., Kirk, J.~G., \& Bell, A.~R. 2011, \bibinfo{title}{{Monte Carlo
  calculations of pair production in high-intensity laser-plasma
  interactions},} Plasma Phys. Control. Fusion, 53, 015009,
  \dodoi{10.1088/0741-3335/53/1/015009}

\bibitem[{R. Durrer \& A. Neronov(2013)Durrer \& Neronov}]{Durrer:2013pga}
Durrer, R., \& Neronov, A. 2013, \bibinfo{title}{{Cosmological Magnetic Fields:
  Their Generation, Evolution and Observation},} Astron. Astrophys. Rev., 21,
  62, \dodoi{10.1007/s00159-013-0062-7}

\bibitem[{R. {Easther} \& E.~A. {Lim}(2006){Easther} \&
  {Lim}}]{2006JCAP...04..010E}
{Easther}, R., \& {Lim}, E.~A. 2006, \bibinfo{title}{{Stochastic gravitational
  wave production after inflation},} \jcap, 2006, 010,
  \dodoi{10.1088/1475-7516/2006/04/010}

\bibitem[{C. {Eckart}(1948){Eckart}}]{1948RvMP...20..399E}
{Eckart}, C. 1948, \bibinfo{title}{{The Approximate Solution of One-Dimensional
  Wave Equations},} Reviews of Modern Physics, 20, 399,
  \dodoi{10.1103/RevModPhys.20.399}

\bibitem[{G.~G. {Fahlman} \& P.~C. {Gregory}(1981){Fahlman} \&
  {Gregory}}]{1981Natur.293..202F}
{Fahlman}, G.~G., \& {Gregory}, P.~C. 1981, \bibinfo{title}{{An X-ray pulsar in
  SNR G109.1-1.0},} \nat, 293, 202, \dodoi{10.1038/293202a0}

\bibitem[{Y. Feng {et~al.}(2022)Feng {et~al.}}]{Feng:2022ill}
Feng, Y., {et~al.} 2022, \bibinfo{title}{{Frequency-dependent polarization of
  repeating fast radio bursts\textemdash{}implications for their origin},}
  Science, 375, abl7759, \dodoi{10.1126/science.abl7759}

\bibitem[{M.~A. {Fine} {et~al.}(2023){Fine}, {Van Eck}, \&
  {Pratley}}]{2023MNRAS.520.4822F}
{Fine}, M.~A., {Van Eck}, C.~L., \& {Pratley}, L. 2023,
  \bibinfo{title}{{Correcting bandwidth depolarization by extreme Faraday
  rotation},} \mnras, 520, 4822, \dodoi{10.1093/mnras/stad423}

\bibitem[{B.~M. {Gaensler} {et~al.}(2005){Gaensler}, {Kouveliotou}, {Gelfand},
  {Taylor}, {Eichler}, {Wijers}, {Granot}, {Ramirez-Ruiz}, {Lyubarsky},
  {Hunstead}, {Campbell-Wilson}, {van der Horst}, {McLaughlin}, {Fender},
  {Garrett}, {Newton-McGee}, {Palmer}, {Gehrels}, \&
  {Woods}}]{2005Natur.434.1104G}
{Gaensler}, B.~M., {Kouveliotou}, C., {Gelfand}, J.~D., {et~al.} 2005,
  \bibinfo{title}{{An expanding radio nebula produced by a giant flare from the
  magnetar SGR 1806-20},} \nat, 434, 1104, \dodoi{10.1038/nature03498}

\bibitem[{S. {Galliou} {et~al.}(2013){Galliou}, {Goryachev}, {Bourquin},
  {Abb{\'e}}, {Aubry}, \& {Tobar}}]{2013NatSR...3.2132G}
{Galliou}, S., {Goryachev}, M., {Bourquin}, R., {et~al.} 2013,
  \bibinfo{title}{{Extremely Low Loss Phonon-Trapping Cryogenic Acoustic
  Cavities for Future Physical Experiments},} Scientific Reports, 3, 2132,
  \dodoi{10.1038/srep02132}

\bibitem[{J. {Garc{\'\i}a-Bellido} \& D.~G.
  {Figueroa}(2007){Garc{\'\i}a-Bellido} \& {Figueroa}}]{2007PhRvL..98f1302G}
{Garc{\'\i}a-Bellido}, J., \& {Figueroa}, D.~G. 2007,
  \bibinfo{title}{{Stochastic Background of Gravitational Waves from Hybrid
  Preheating},} \prl, 98, 061302, \dodoi{10.1103/PhysRevLett.98.061302}

\bibitem[{J. {Garc{\'\i}a-Bellido} {et~al.}(2008){Garc{\'\i}a-Bellido},
  {Figueroa}, \& {Sastre}}]{2008PhRvD..77d3517G}
{Garc{\'\i}a-Bellido}, J., {Figueroa}, D.~G., \& {Sastre}, A. 2008,
  \bibinfo{title}{{Gravitational wave background from reheating after hybrid
  inflation},} \prd, 77, 043517, \dodoi{10.1103/PhysRevD.77.043517}

\bibitem[{M. Gasperini \& G. Veneziano(2016)Gasperini \&
  Veneziano}]{Gasperini:2007vw}
Gasperini, M., \& Veneziano, G. 2016, \bibinfo{title}{{String Theory and
  Pre-big bang Cosmology},} Nuovo Cim. C, 38, 160,
  \dodoi{10.1393/ncc/i2015-15160-8}

\bibitem[{M.~E. Gertsenshtein(1962)Gertsenshtein}]{Gertsenshtein:1962}
Gertsenshtein, M.~E. 1962, \bibinfo{title}{{Wave Resonance of Light and
  Gravitational Waves},} Sov. Phys. JETP, 14, 84

\bibitem[{F. Giordano \& P. Coretto(2017)Giordano \& Coretto}]{Giordano2017AMC}
Giordano, F., \& Coretto, P. 2017, \bibinfo{title}{A Monte Carlo subsampling
  method for estimating the distribution of signal-to-noise ratio statistics in
  nonparametric time series regression models,} Statistical Methods \&
  Applications, 29, 483 .
\newblock \url{https://api.semanticscholar.org/CorpusID:202948915}

\bibitem[{M. {Giovannini}(1999){Giovannini}}]{1999PhRvD..60l3511G}
{Giovannini}, M. 1999, \bibinfo{title}{{Production and detection of relic
  gravitons in quintessential inflationary models},} \prd, 60, 123511,
  \dodoi{10.1103/PhysRevD.60.123511}

\bibitem[{E. {Gogus} {et~al.}(2008){Gogus}, {Woods}, \&
  {Kouveliotou}}]{2008GCN..8118....1G}
{Gogus}, E., {Woods}, P., \& {Kouveliotou}, C. 2008, \bibinfo{title}{{Discovery
  of the spin period of the new soft gamma repeater Sgr 0501+4516.},} GRB
  Coordinates Network, 8118, 1

\bibitem[{P. {Goldreich} \& W.~H. {Julian}(1969){Goldreich} \&
  {Julian}}]{1969ApJ...157..869G}
{Goldreich}, P., \& {Julian}, W.~H. 1969, \bibinfo{title}{{Pulsar
  Electrodynamics},} \apj, 157, 869, \dodoi{10.1086/150119}

\bibitem[{ {H.~E.~S.~S. Collaboration} {et~al.}(2023){H.~E.~S.~S.
  Collaboration}, {Aharonian}, {Ait Benkhali}, {Aschersleben}, {Ashkar},
  {Backes}, {Barbosa Martins}, {Batzofin}, {Becherini}, {Berge},
  {Bernl{\"o}hr}, {Bi}, {B{\"o}ttcher}, {Boisson}, {Bolmont}, {de Bony de
  Lavergne}, {Borowska}, {Bradascio}, {Breuhaus}, {Brose}, {Brun}, {Bruno},
  {Bulik}, {Burger-Scheidlin}, {Bylund}, {Cangemi}, {Caroff}, {Casanova},
  {Celic}, {Cerruti}, {Chand}, {Chandra}, {Chen}, {Chibueze}, {Cotter},
  {Damascene Mbarubucyeye}, {Djannati-Ata{\"\i}}, {Dmytriiev}, {Egberts},
  {Ernenwein}, {Feijen}, {Fiasson}, {Fichet de Clairfontaine}, {Fontaine},
  {F{\"u}{\ss}ling}, {Funk}, {Gabici}, {Gallant}, {Ghafourizadeh}, {Giavitto},
  {Giunti}, {Glawion}, {Glicenstein}, {Goswami}, {Grolleron}, {Grondin},
  {Haerer}, {Haupt}, {Hinton}, {Hofmann}, {Holch}, {Holler}, {Horns}, {Huang},
  {Jamrozy}, {Jankowsky}, {Joshi}, {Jung-Richardt}, {Kasai}, {Katarzy{\'n}ski},
  {Kh{\'e}lifi}, {Klepser}, {Klu{\v{z}}niak}, {Komin}, {Kosack}, {Kostunin},
  {Lang}, {Le Stum}, {Lemi{\`e}re}, {Lemoine-Goumard}, {Lenain}, {Leuschner},
  {Lohse}, {Luashvili}, {Lypova}, {Mackey}, {Malyshev}, {Malyshev}, {Marandon},
  {Marchegiani}, {Marcowith}, {Marinos}, {Mart{\'\i}-Devesa}, {Marx}, {Maurin},
  {Meyer}, {Mitchell}, {Moderski}, {Mohrmann}, {Montanari}, {Moulin}, {Muller},
  {Murach}, {Nakashima}, {de Naurois}, {Niemiec}, {Noel}, {O'Brien}, {Ohm},
  {Olivera-Nieto}, {de Ona Wilhelmi}, {Ostrowski}, {Panny}, {Panter},
  {Parsons}, {Peron}, {Pita}, {Prokhorov}, {Prokoph}, {P{\"u}hlhofer}, {Punch},
  {Quirrenbach}, {Reichherzer}, {Reimer}, {Reimer}, {Renaud}, {Rieger},
  {Rowell}, {Rudak}, {Ruiz-Velasco}, {Sahakian}, {Sailer}, {Salzmann},
  {Sanchez}, {Santangelo}, {Sasaki}, {Sch{\"u}ssler}, {Schwanke}, {Shapopi},
  {Sinha}, {Sol}, {Specovius}, {Spencer}, {Spir-Jacob}, {Stawarz}, {Steenkamp},
  {Steinmassl}, {Steppa}, {Sushch}, {Suzuki}, {Takahashi}, {Tanaka},
  {Tavernier}, {Terrier}, {Thorpe-Morgan}, {Tluczykont}, {Tsirou}, {Tsuji},
  {van Eldik}, {Vecchi}, {Veh}, {Venter}, {Vink}, {Wagner}, {Werner}, {White},
  {Wierzcholska}, {Wun Wong}, {Yassin}, {Zacharias}, {Zargaryan}, {Zdziarski},
  {Zech}, {Zhu}, {Zouari}, {{\.Z}ywucka}, {Zanin}, {Kerr}, {Johnston},
  {Shannon}, \& {Smith}}]{2023NatAs...7.1341H}
{H.~E.~S.~S. Collaboration}, {Aharonian}, F., {Ait Benkhali}, F., {et~al.}
  2023, \bibinfo{title}{{Discovery of a radiation component from the Vela
  pulsar reaching 20 teraelectronvolts},} Nature Astronomy, 7, 1341,
  \dodoi{10.1038/s41550-023-02052-3}

\bibitem[{F. {Haberl} {et~al.}(1997){Haberl}, {Motch}, {Buckley}, {Zickgraf},
  \& {Pietsch}}]{1997AA...326..662H}
{Haberl}, F., {Motch}, C., {Buckley}, D.~A.~H., {Zickgraf}, F.~J., \&
  {Pietsch}, W. 1997, \bibinfo{title}{{RXJ0720.4-3125: strong evidence for an
  isolated pulsating neutron star.},} \aap, 326, 662

\bibitem[{H. {Hakobyan} {et~al.}(2023){Hakobyan}, {Philippov}, \&
  {Spitkovsky}}]{2023ApJ...943..105H}
{Hakobyan}, H., {Philippov}, A., \& {Spitkovsky}, A. 2023,
  \bibinfo{title}{{Magnetic Energy Dissipation and {\ensuremath{\gamma}}-Ray
  Emission in Energetic Pulsars},} \apj, 943, 105,
  \dodoi{10.3847/1538-4357/acab05}

\bibitem[{V. {Hambaryan} {et~al.}(2017){Hambaryan}, {Suleimanov}, {Haberl},
  {Schwope}, {Neuh{\"a}user}, {Hohle}, \& {Werner}}]{2017A&A...601A.108H}
{Hambaryan}, V., {Suleimanov}, V., {Haberl}, F., {et~al.} 2017,
  \bibinfo{title}{{The compactness of the isolated neutron star RX
  J0720.4-3125},} \aap, 601, A108, \dodoi{10.1051/0004-6361/201630368}

\bibitem[{M. Haverkorn {et~al.}(2004)Haverkorn, Katgert, \&
  de~Bruyn}]{Haverkorn:2004kx}
Haverkorn, M., Katgert, P., \& de~Bruyn, A.~G. 2004,
  \bibinfo{title}{{Properties of the warm magnetized ISM, as inferred from WSRT
  polarimetric imaging},} Astron. Astrophys., 427, 169,
  \dodoi{10.1051/0004-6361:200400042}

\bibitem[{W. Heisenberg \& H. Euler(1936)Heisenberg \&
  Euler}]{Heisenberg:1936nmg}
Heisenberg, W., \& Euler, H. 1936, \bibinfo{title}{{Consequences of Dirac's
  theory of positrons},} Z. Phys., 98, 714, \dodoi{10.1007/BF01343663}

\bibitem[{N. Herman {et~al.}(2023)Herman, Lehoucq, \&
  F\'{u}zfa}]{Herman:2022fau}
Herman, N., Lehoucq, L., \& F\'{u}zfa, A. 2023,
  \bibinfo{title}{{Electromagnetic antennas for the resonant detection of the
  stochastic gravitational wave background},} Phys. Rev. D, 108, 124009,
  \dodoi{10.1103/PhysRevD.108.124009}

\bibitem[{W.~C.~G. {Ho} \& C.~O. {Heinke}(2009){Ho} \&
  {Heinke}}]{2009Natur.462...71H}
{Ho}, W. C.~G., \& {Heinke}, C.~O. 2009, \bibinfo{title}{{A neutron star with a
  carbon atmosphere in the Cassiopeia A supernova remnant},} \nat, 462, 71,
  \dodoi{10.1038/nature08525}

\bibitem[{C.~J. {Hogan}(1986){Hogan}}]{1986MNRAS.218..629H}
{Hogan}, C.~J. 1986, \bibinfo{title}{{Gravitational radiation from cosmological
  phase transitions},} \mnras, 218, 629, \dodoi{10.1093/mnras/218.4.629}

\bibitem[{E.~W. {Hones} \& J.~E. {Bergeson}(1965){Hones} \&
  {Bergeson}}]{1965JGR....70.4951H}
{Hones}, Jr., E.~W., \& {Bergeson}, J.~E. 1965, \bibinfo{title}{{Electric Field
  Generated by a Rotating Magnetized Sphere},} \jgr, 70, 4951,
  \dodoi{10.1029/JZ070i019p04951}

\bibitem[{W. Hong(2025)Hong}]{whzencode}
Hong, W. 2025, \bibinfo{title}{High Sensitivity Methodologies to Detect Radio
  Band Gravitational Waves Simulation Code,} Zenodo,
  \dodoi{10.5281/zenodo.15680536}

\bibitem[{W. Hong {et~al.}(2025)Hong, Tao, He, \& Zhang}]{Hong:2024ofh}
Hong, W., Tao, Z.-Z., He, P., \& Zhang, T.-J. 2025,
  \bibinfo{title}{{Theoretical Radio Signals from Radio-band Gravitational
  Waves Converted from the Neutron Star Magnetic Field},} Astrophys. J., 990,
  156, \dodoi{10.3847/1538-4357/adf19a}

\bibitem[{A. Hook {et~al.}(2018)Hook, Kahn, Safdi, \& Sun}]{Hook:2018iia}
Hook, A., Kahn, Y., Safdi, B.~R., \& Sun, Z. 2018, \bibinfo{title}{{Radio
  Signals from Axion Dark Matter Conversion in Neutron Star Magnetospheres},}
  Phys. Rev. Lett., 121, 241102, \dodoi{10.1103/PhysRevLett.121.241102}

\bibitem[{R. {Hu} \& A.~M. {Beloborodov}(2022){Hu} \&
  {Beloborodov}}]{2022ApJ...939...42H}
{Hu}, R., \& {Beloborodov}, A.~M. 2022, \bibinfo{title}{{Axisymmetric Pulsar
  Magnetosphere Revisited},} \apj, 939, 42, \dodoi{10.3847/1538-4357/ac961d}

\bibitem[{A.~I. {Ibrahim} {et~al.}(2004){Ibrahim}
  {et~al.}}]{2004ApJ...609L..21I}
{Ibrahim}, A.~I., {et~al.} 2004, \bibinfo{title}{{Discovery of a Transient
  Magnetar: XTE J1810-197},} \apjl, 609, L21, \dodoi{10.1086/422636}

\bibitem[{Y.~P. {Ilyasov}(1971){Ilyasov}}]{1971R&QE...14..425I}
{Ilyasov}, Y.~P. 1971, \bibinfo{title}{{The influence of the ``confusion''
  effect on the sensitivity of large radio telescopes in observing discrete
  sources},} Radiophysics and Quantum Electronics, 14, 425,
  \dodoi{10.1007/BF01030727}

\bibitem[{G.~L. {Israel} {et~al.}(1994){Israel}, {Mereghetti}, \&
  {Stella}}]{1994ApJ...433L..25I}
{Israel}, G.~L., {Mereghetti}, S., \& {Stella}, L. 1994, \bibinfo{title}{{The
  Discovery of 8.7 Second Pulsations from the Ultrasoft X-Ray Source 4U
  0142+61},} \apjl, 433, L25, \dodoi{10.1086/187539}

\bibitem[{G.~L. {Israel} {et~al.}(2016){Israel} {et~al.}}]{2016MNRAS.457.3448I}
{Israel}, G.~L., {et~al.} 2016, \bibinfo{title}{{The discovery, monitoring and
  environment of SGR J1935+2154},} \mnras, 457, 3448,
  \dodoi{10.1093/mnras/stw008}

\bibitem[{A. Ito {et~al.}(2020)Ito, Ikeda, Miuchi, \& Soda}]{Ito:2019wcb}
Ito, A., Ikeda, T., Miuchi, K., \& Soda, J. 2020, \bibinfo{title}{{Probing GHz
  gravitational waves with graviton\textendash{}magnon resonance},} Eur. Phys.
  J. C, 80, 179, \dodoi{10.1140/epjc/s10052-020-7735-y}

\bibitem[{A. Ito {et~al.}(2024{\natexlab{a}})Ito, Kohri, \&
  Nakayama}]{Ito:2023nkq}
Ito, A., Kohri, K., \& Nakayama, K. 2024{\natexlab{a}},
  \bibinfo{title}{{Gravitational Wave Search through Electromagnetic
  Telescopes},} PTEP, 2024, 023E03, \dodoi{10.1093/ptep/ptae004}

\bibitem[{A. Ito {et~al.}(2024{\natexlab{b}})Ito, Kohri, \&
  Nakayama}]{Ito:2023fcr}
Ito, A., Kohri, K., \& Nakayama, K. 2024{\natexlab{b}},
  \bibinfo{title}{{Probing high frequency gravitational waves with pulsars},}
  Phys. Rev. D, 109, 063026, \dodoi{10.1103/PhysRevD.109.063026}

\bibitem[{C. Itzykson \& J.~B. Zuber(1980)Itzykson \& Zuber}]{Itzykson:1980rh}
Itzykson, C., \& Zuber, J.~B. 1980, {Quantum Field Theory}, International
  Series In Pure and Applied Physics (New York: McGraw-Hill)

\bibitem[{K. Jedamzik \& A. Saveliev(2019)Jedamzik \&
  Saveliev}]{Jedamzik:2018itu}
Jedamzik, K., \& Saveliev, A. 2019, \bibinfo{title}{{Stringent Limit on
  Primordial Magnetic Fields from the Cosmic Microwave Background Radiation},}
  Phys. Rev. Lett., 123, 021301, \dodoi{10.1103/PhysRevLett.123.021301}

\bibitem[{P. Jiang {et~al.}(2019)Jiang {et~al.}}]{Jiang:2019rnj}
Jiang, P., {et~al.} 2019, \bibinfo{title}{{Commissioning progress of the
  FAST},} Sci. China Phys. Mech. Astron., 62, 959502,
  \dodoi{10.1007/s11433-018-9376-1}

\bibitem[{P. {Jiang} {et~al.}(2020){Jiang} {et~al.}}]{2020RAA....20...64J}
{Jiang}, P., {et~al.} 2020, \bibinfo{title}{{The fundamental performance of
  FAST with 19-beam receiver at L band},} Research in Astronomy and
  Astrophysics, 20, 064, \dodoi{10.1088/1674-4527/20/5/64}

\bibitem[{M. {Kamionkowski} {et~al.}(1994){Kamionkowski}, {Kosowsky}, \&
  {Turner}}]{1994PhRvD..49.2837K}
{Kamionkowski}, M., {Kosowsky}, A., \& {Turner}, M.~S. 1994,
  \bibinfo{title}{{Gravitational radiation from first-order phase
  transitions},} \prd, 49, 2837, \dodoi{10.1103/PhysRevD.49.2837}

\bibitem[{D.~L. {Kaplan} {et~al.}(2003){Kaplan}, {van Kerkwijk}, {Marshall},
  {Jacoby}, {Kulkarni}, \& {Frail}}]{2003ApJ...590.1008K}
{Kaplan}, D.~L., {van Kerkwijk}, M.~H., {Marshall}, H.~L., {et~al.} 2003,
  \bibinfo{title}{{The Nearby Neutron Star RX J0720.4-3125 from Radio to
  X-Rays},} \apj, 590, 1008, \dodoi{10.1086/375052}

\bibitem[{S. {Khlebnikov} \& I. {Tkachev}(1997){Khlebnikov} \&
  {Tkachev}}]{1997PhRvD..56..653K}
{Khlebnikov}, S., \& {Tkachev}, I. 1997, \bibinfo{title}{{Relic gravitational
  waves produced after preheating},} \prd, 56, 653,
  \dodoi{10.1103/PhysRevD.56.653}

\bibitem[{J. {Kijak} \& J. {Gil}(1998){Kijak} \& {Gil}}]{1998MNRAS.299..855K}
{Kijak}, J., \& {Gil}, J. 1998, \bibinfo{title}{{Radio emission regions in
  pulsars},} \mnras, 299, 855, \dodoi{10.1046/j.1365-8711.1998.01832.x}

\bibitem[{J. {Kijak} \& J. {Gil}(2002){Kijak} \& {Gil}}]{2002A&A...392..189K}
{Kijak}, J., \& {Gil}, J. 2002, \bibinfo{title}{{Structure of pulsar beams:
  Conal versus patchy},} \aap, 392, 189, \dodoi{10.1051/0004-6361:20020946}

\bibitem[{D.~A. Kirzhnits(1972)Kirzhnits}]{Kirzhnits:1972iw}
Kirzhnits, D.~A. 1972, \bibinfo{title}{{Weinberg model in the hot universe},}
  JETP Lett., 15, 529

\bibitem[{L. Kofman {et~al.}(1994)Kofman, Linde, \&
  Starobinsky}]{Kofman:1994rk}
Kofman, L., Linde, A.~D., \& Starobinsky, A.~A. 1994,
  \bibinfo{title}{{Reheating after inflation},} Phys. Rev. Lett., 73, 3195,
  \dodoi{10.1103/PhysRevLett.73.3195}

\bibitem[{S.~S. {Komissarov}(2006){Komissarov}}]{2006MNRAS.367...19K}
{Komissarov}, S.~S. 2006, \bibinfo{title}{{Simulations of the axisymmetric
  magnetospheres of neutron stars},} \mnras, 367, 19,
  \dodoi{10.1111/j.1365-2966.2005.09932.x}

\bibitem[{A. {Kosowsky} {et~al.}(1992){Kosowsky}, {Turner}, \&
  {Watkins}}]{1992PhRvD..45.4514K}
{Kosowsky}, A., {Turner}, M.~S., \& {Watkins}, R. 1992,
  \bibinfo{title}{{Gravitational radiation from colliding vacuum bubbles},}
  \prd, 45, 4514, \dodoi{10.1103/PhysRevD.45.4514}

\bibitem[{C. Kouveliotou {et~al.}(1998)Kouveliotou
  {et~al.}}]{Kouveliotou:1998ze}
Kouveliotou, C., {et~al.} 1998, \bibinfo{title}{{An X-ray pulsar with a
  superstrong magnetic field in the soft gamma-ray repeater SGR 1806-20.},}
  Nature, 393, 235, \dodoi{10.1038/30410}

\bibitem[{D. Kwiatkowski {et~al.}(1992)Kwiatkowski, Phillips, Schmidt, \&
  Shin}]{KWIATKOWSKI1992159}
Kwiatkowski, D., Phillips, P.~C., Schmidt, P., \& Shin, Y. 1992,
  \bibinfo{title}{Testing the null hypothesis of stationarity against the
  alternative of a unit root: How sure are we that economic time series have a
  unit root?} Journal of Econometrics, 54, 159,
  \dodoi{https://doi.org/10.1016/0304-4076(92)90104-Y}

\bibitem[{D. Lai(2001)Lai}]{Lai:2000at}
Lai, D. 2001, \bibinfo{title}{{Matter in strong magnetic fields},} Rev. Mod.
  Phys., 73, 629, \dodoi{10.1103/RevModPhys.73.629}

\bibitem[{L.~D. {Landau} \& E.~M. {Lifshitz}(1960){Landau} \&
  {Lifshitz}}]{1960ecm..book.....L}
{Landau}, L.~D., \& {Lifshitz}, E.~M. 1960, {Electrodynamics of continuous
  media}

\bibitem[{L.~D. {Landau} \& E.~M. {Lifshitz}(1971){Landau} \&
  {Lifshitz}}]{1971ctf..book.....L}
{Landau}, L.~D., \& {Lifshitz}, E.~M. 1971, {The classical theory of fields}

\bibitem[{M.~I. Large {et~al.}(1968)Large {et~al.}}]{Large:1968mi}
Large, M.~I., {et~al.} 1968, \bibinfo{title}{{Pulsar Search at the Molonglo
  Radio Observatory.},} Nature, 220, 753, \dodoi{10.1038/220753a0}

\bibitem[{A. {Lecacheux} {et~al.}(2004){Lecacheux}, {Konovalenko}, \&
  {Rucker}}]{2004P&SS...52.1357L}
{Lecacheux}, A., {Konovalenko}, A.~A., \& {Rucker}, H.~O. 2004,
  \bibinfo{title}{{Using large radio telescopes at decametre wavelengths},}
  \planss, 52, 1357, \dodoi{10.1016/j.pss.2004.09.006}

\bibitem[{A. {Levinson} {et~al.}(2005){Levinson}, {Melrose}, {Judge}, \&
  {Luo}}]{2005ApJ...631..456L}
{Levinson}, A., {Melrose}, D., {Judge}, A., \& {Luo}, Q. 2005,
  \bibinfo{title}{{Large-Amplitude, Pair-creating Oscillations in Pulsar and
  Black Hole Magnetospheres},} \apj, 631, 456, \dodoi{10.1086/432498}

\bibitem[{F. Li {et~al.}(2008)Li, Baker, Fang, Stephenson, \& Chen}]{Li:2008qr}
Li, F., Baker, Jr., R. M.~L., Fang, Z., Stephenson, G.~V., \& Chen, Z. 2008,
  \bibinfo{title}{{Perturbative Photon Fluxes Generated by High-Frequency
  Gravitational Waves and Their Physical Effects},} Eur. Phys. J. C, 56, 407,
  \dodoi{10.1140/epjc/s10052-008-0656-9}

\bibitem[{F. Li {et~al.}(2009)Li, Yang, Fang, Baker, Stephenson, \&
  Wen}]{Li:2009zzy}
Li, F., Yang, N., Fang, Z., {et~al.} 2009, \bibinfo{title}{{Signal Photon Flux
  and Background Noise in a Coupling Electromagnetic Detecting System for High
  Frequency Gravitational Waves},} Phys. Rev. D, 80, 064013,
  \dodoi{10.1103/PhysRevD.80.064013}

\bibitem[{F.-Y. Li {et~al.}(2000)Li, Tang, Luo, \& Li}]{Li:2000du}
Li, F.-Y., Tang, M.-X., Luo, J., \& Li, Y.-C. 2000,
  \bibinfo{title}{{Electrodynamical response of a high-energy photon flux to a
  gravitational wave},} Phys. Rev. D, 62, 044018,
  \dodoi{10.1103/PhysRevD.62.044018}

\bibitem[{F.-Y. Li {et~al.}(2003)Li, Tang, \& Shi}]{Li:2003tv}
Li, F.-Y., Tang, M.-X., \& Shi, D.-P. 2003, \bibinfo{title}{{Electromagnetic
  response of a Gaussian beam to high frequency relic gravitational waves in
  quintessential inflationary models},} Phys. Rev. D, 67, 104008,
  \dodoi{10.1103/PhysRevD.67.104008}

\bibitem[{F.-Y. Li {et~al.}(2013)Li, Wen, \& Fang}]{Li:2013fna}
Li, F.-Y., Wen, H., \& Fang, Z.-Y. 2013, \bibinfo{title}{{High-frequency
  gravitational waves having large spectral densities and their electromagnetic
  response},} Chin. Phys. B, 22, 120402, \dodoi{10.1088/1674-1056/22/12/120402}

\bibitem[{F.-Y. Li \& N. Yang(2004)Li \& Yang}]{Li:2004df}
Li, F.-Y., \& Yang, N. 2004, \bibinfo{title}{{Resonant interaction between a
  weak gravitational wave and a microwave beam in the double polarized states
  through a static magnetic field},} Chin. Phys. Lett., 21, 2113,
  \dodoi{10.1088/0256-307X/21/11/011}

\bibitem[{F.-Y. Li {et~al.}(2023)Li, Yu, Li, Wei, \& Jiang}]{Li:2023tzw}
Li, F.-Y., Yu, H., Li, J., Wei, L.-F., \& Jiang, Q.-Q. 2023,
  \bibinfo{title}{{Electromagnetic counterparts of high-frequency gravitational
  waves in a rotating laboratory frame system and their detection},} Phys. Rev.
  D, 108, 065014, \dodoi{10.1103/PhysRevD.108.065014}

\bibitem[{J. Li {et~al.}(2011)Li, Lin, Li, \& Zhong}]{Li:2011zzl}
Li, J., Lin, K., Li, F., \& Zhong, Y. 2011, \bibinfo{title}{{The signal photon
  flux, background photons and shot noise in electromagnetic response of
  high-frequency relic gravitational waves},} Gen. Rel. Grav., 43, 2209,
  \dodoi{10.1007/s10714-011-1176-8}

\bibitem[{J. {Li} {et~al.}(2012){Li}, {Spitkovsky}, \&
  {Tchekhovskoy}}]{2012ApJ...746...60L}
{Li}, J., {Spitkovsky}, A., \& {Tchekhovskoy}, A. 2012,
  \bibinfo{title}{{Resistive Solutions for Pulsar Magnetospheres},} \apj, 746,
  60, \dodoi{10.1088/0004-637X/746/1/60}

\bibitem[{J. Li {et~al.}(2016)Li, Zhang, Lin, \& Wen}]{Li:2014bma}
Li, J., Zhang, L., Lin, K., \& Wen, H. 2016, \bibinfo{title}{{Resonance of
  Gaussian electromagnetic field to the high frequency gravitational waves},}
  Int. J. Theor. Phys., 55, 3506, \dodoi{10.1007/s10773-016-2977-z}

\bibitem[{M. {Lobet} {et~al.}(2016){Lobet}, {d'Humi{\`e}res}, {Grech}, {Ruyer},
  {Davoine}, \& {Gremillet}}]{2016JPhCS.688a2058L}
{Lobet}, M., {d'Humi{\`e}res}, E., {Grech}, M., {et~al.} 2016, in Journal of
  Physics Conference Series, Vol. 688, Journal of Physics Conference Series
  (IOP), 012058, \dodoi{10.1088/1742-6596/688/1/012058}

\bibitem[{D.~R. {Lorimer} \& M. {Kramer}(2004){Lorimer} \&
  {Kramer}}]{2004hpa..book.....L}
{Lorimer}, D.~R., \& {Kramer}, M. 2004, {Handbook of Pulsar Astronomy}, Vol.~4

\bibitem[{A.~G. {Lyne} \& B.~J. {Rickett}(1968){Lyne} \&
  {Rickett}}]{1968Natur.218..326L}
{Lyne}, A.~G., \& {Rickett}, B.~J. 1968, \bibinfo{title}{{Measurements of the
  Pulse Shape and Spectra of the Pulsating Radio Sources},} \nat, 218, 326,
  \dodoi{10.1038/218326a0}

\bibitem[{M. {Maggiore}(2007){Maggiore}}]{2007gwte.book.....M}
{Maggiore}, M. 2007, {Gravitational Waves: Volume 1: Theory and Experiments},
  \dodoi{10.1093/acprof:oso/9780198570745.001.0001}

\bibitem[{R.~N. {Manchester} {et~al.}(1978){Manchester}, ,
  {et~al.}}]{1978MNRAS.185..409M}
{Manchester}, R.~N., , {et~al.} 1978, \bibinfo{title}{{The second Molonglo
  pulsar survey - discovery of 155 pulsars.},} \mnras, 185, 409,
  \dodoi{10.1093/mnras/185.2.409}

\bibitem[{S.~A. {Mao} {et~al.}(2017){Mao}, {Carilli}, {Gaensler}, {Wucknitz},
  {Keeton}, {Basu}, {Beck}, {Kronberg}, \& {Zweibel}}]{2017NatAs...1..621M}
{Mao}, S.~A., {Carilli}, C., {Gaensler}, B.~M., {et~al.} 2017,
  \bibinfo{title}{{Detection of microgauss coherent magnetic fields in a galaxy
  five billion years ago},} Nature Astronomy, 1, 621,
  \dodoi{10.1038/s41550-017-0218-x}

\bibitem[{J.~I. {McDonald} \& S.~A.~R. {Ellis}(2024){McDonald} \&
  {Ellis}}]{2024PhRvD.110j3003M}
{McDonald}, J.~I., \& {Ellis}, S. A.~R. 2024, \bibinfo{title}{{Resonant
  conversion of gravitational waves in neutron star magnetospheres},} \prd,
  110, 103003, \dodoi{10.1103/PhysRevD.110.103003}

\bibitem[{D.~B. Melrose \& R.~J. Stoneham(1976)Melrose \&
  Stoneham}]{Melrose:1976dr}
Melrose, D.~B., \& Stoneham, R.~J. 1976, \bibinfo{title}{{Vacuum Polarization
  and Photon Propagation in a Magnetic Field},} Nuovo Cim. A, 32, 435,
  \dodoi{10.1007/BF02730208}

\bibitem[{C.~W. {Misner} {et~al.}(1973){Misner}, {Thorne}, \&
  {Wheeler}}]{1973grav.book.....M}
{Misner}, C.~W., {Thorne}, K.~S., \& {Wheeler}, J.~A. 1973, {Gravitation}

\bibitem[{K. {Mori} {et~al.}(2013){Mori}, {Gotthelf}, {Zhang}, {An},
  {Baganoff}, {Barri{\`e}re}, {Beloborodov}, {Boggs}, {Christensen}, {Craig},
  {Dufour}, {Grefenstette}, {Hailey}, {Harrison}, {Hong}, {Kaspi}, {Kennea},
  {Madsen}, {Markwardt}, {Nynka}, {Stern}, {Tomsick}, \&
  {Zhang}}]{2013ApJ...770L..23M}
{Mori}, K., {Gotthelf}, E.~V., {Zhang}, S., {et~al.} 2013,
  \bibinfo{title}{{NuSTAR Discovery of a 3.76 s Transient Magnetar Near
  Sagittarius A*},} \apjl, 770, L23, \dodoi{10.1088/2041-8205/770/2/L23}

\bibitem[{A. Neronov \& I. Vovk(2010)Neronov \& Vovk}]{Neronov:2010gir}
Neronov, A., \& Vovk, I. 2010, \bibinfo{title}{{Evidence for strong
  extragalactic magnetic fields from Fermi observations of TeV blazars},}
  Science, 328, 73, \dodoi{10.1126/science.1184192}

\bibitem[{C. Pankow {et~al.}(2018)Pankow {et~al.}}]{Pankow:2018qpo}
Pankow, C., {et~al.} 2018, \bibinfo{title}{{Mitigation of the instrumental
  noise transient in gravitational-wave data surrounding GW170817},} Phys. Rev.
  D, 98, 084016, \dodoi{10.1103/PhysRevD.98.084016}

\bibitem[{A.~A. Penzias \& R.~W. Wilson(1965)Penzias \&
  Wilson}]{Penzias:1965wn}
Penzias, A.~A., \& Wilson, R.~W. 1965, \bibinfo{title}{{A Measurement of excess
  antenna temperature at 4080-Mc/s},} Astrophys. J., 142, 419,
  \dodoi{10.1086/148307}

\bibitem[{A. Philippov \& M. Kramer(2022)Philippov \& Kramer}]{annurev}
Philippov, A., \& Kramer, M. 2022, \bibinfo{title}{Pulsar Magnetospheres and
  Their Radiation,} Annual Review of Astronomy and Astrophysics, 60, 495,
  \dodoi{https://doi.org/10.1146/annurev-astro-052920-112338}

\bibitem[{A. {Philippov} {et~al.}(2020){Philippov}, {Timokhin}, \&
  {Spitkovsky}}]{2020PhRvL.124x5101P}
{Philippov}, A., {Timokhin}, A., \& {Spitkovsky}, A. 2020,
  \bibinfo{title}{{Origin of Pulsar Radio Emission},} \prl, 124, 245101,
  \dodoi{10.1103/PhysRevLett.124.245101}

\bibitem[{P.~C.~B. PHILLIPS \& P. PERRON(1988)PHILLIPS \&
  PERRON}]{10.1093/biomet/75.2.335}
PHILLIPS, P. C.~B., \& PERRON, P. 1988, \bibinfo{title}{Testing for a unit root
  in time series regression,} Biometrika, 75, 335,
  \dodoi{10.1093/biomet/75.2.335}

\bibitem[{ {Planck Collaboration} {et~al.}(2020){Planck Collaboration},
  {Aghanim}, {et~al.}}]{2020AA...641A...6P}
{Planck Collaboration}, {Aghanim}, N., {et~al.} 2020, \bibinfo{title}{{Planck
  2018 results. VI. Cosmological parameters},} \aap, 641, A6,
  \dodoi{10.1051/0004-6361/201833910}

\bibitem[{M.~S. Pshirkov {et~al.}(2016)Pshirkov, Tinyakov, \&
  Urban}]{Pshirkov:2015tua}
Pshirkov, M.~S., Tinyakov, P.~G., \& Urban, F.~R. 2016, \bibinfo{title}{{New
  limits on extragalactic magnetic fields from rotation measures},} Phys. Rev.
  Lett., 116, 191302, \dodoi{10.1103/PhysRevLett.116.191302}

\bibitem[{G. {Raffelt} \& L. {Stodolsky}(1988){Raffelt} \&
  {Stodolsky}}]{1988PhRvD..37.1237R}
{Raffelt}, G., \& {Stodolsky}, L. 1988, \bibinfo{title}{{Mixing of the photon
  with low-mass particles},} \prd, 37, 1237, \dodoi{10.1103/PhysRevD.37.1237}

\bibitem[{N. {Rea} {et~al.}(2013){Rea} {et~al.}}]{2013ApJ...770...65R}
{Rea}, N., {et~al.} 2013, \bibinfo{title}{{The Outburst Decay of the Low
  Magnetic Field Magnetar SGR 0418+5729},} \apj, 770, 65,
  \dodoi{10.1088/0004-637X/770/1/65}

\bibitem[{D.~J. {Reardon} \& W.~A. {Coles}(2023){Reardon} \&
  {Coles}}]{2023MNRAS.521.6392R}
{Reardon}, D.~J., \& {Coles}, W.~A. 2023, \bibinfo{title}{{Determining electron
  column density fluctuations in a dominant scattering region using pulsar
  scintillation},} \mnras, 521, 6392, \dodoi{10.1093/mnras/stad962}

\bibitem[{D.~J. {Reardon} {et~al.}(2020){Reardon}, {Coles}, {Bailes}, {Bhat},
  {Dai}, {Hobbs}, {Kerr}, {Manchester}, {Os{\l}owski}, {Parthasarathy},
  {Russell}, {Shannon}, {Spiewak}, {Toomey}, {Tuntsov}, {van Straten},
  {Walker}, {Wang}, {Zhang}, \& {Zhu}}]{2020ApJ...904..104R}
{Reardon}, D.~J., {Coles}, W.~A., {Bailes}, M., {et~al.} 2020,
  \bibinfo{title}{{Precision Orbital Dynamics from Interstellar Scintillation
  Arcs for PSR J0437-4715},} \apj, 904, 104, \dodoi{10.3847/1538-4357/abbd40}

\bibitem[{D.~J. Reardon {et~al.}(2023)Reardon {et~al.}}]{Reardon:2023gzh}
Reardon, D.~J., {et~al.} 2023, \bibinfo{title}{{Search for an Isotropic
  Gravitational-wave Background with the Parkes Pulsar Timing Array},}
  Astrophys. J. Lett., 951, L6, \dodoi{10.3847/2041-8213/acdd02}

\bibitem[{B.~J. {Rickett}(1990){Rickett}}]{1990ARA&A..28..561R}
{Rickett}, B.~J. 1990, \bibinfo{title}{{Radio propagation through the turbulent
  interstellar plasma.},} \araa, 28, 561,
  \dodoi{10.1146/annurev.aa.28.090190.003021}

\bibitem[{J.~D. {Romano} \& N.~J. {Cornish}(2017){Romano} \&
  {Cornish}}]{2017LRR....20....2R}
{Romano}, J.~D., \& {Cornish}, N.~J. 2017, \bibinfo{title}{{Detection methods
  for stochastic gravitational-wave backgrounds: a unified treatment},} Living
  Reviews in Relativity, 20, 2, \dodoi{10.1007/s41114-017-0004-1}

\bibitem[{R. Roshan \& G. White(2025)Roshan \& White}]{Roshan:2024qnv}
Roshan, R., \& White, G. 2025, \bibinfo{title}{{Using gravitational waves to
  see the first second of the Universe},} Rev. Mod. Phys., 97, 015001,
  \dodoi{10.1103/RevModPhys.97.015001}

\bibitem[{S.~E. SAID \& D.~A. DICKEY(1984)SAID \&
  DICKEY}]{10.1093/biomet/71.3.599}
SAID, S.~E., \& DICKEY, D.~A. 1984, \bibinfo{title}{Testing for unit roots in
  autoregressive-moving average models of unknown order,} Biometrika, 71, 599,
  \dodoi{10.1093/biomet/71.3.599}

\bibitem[{S. Sanidas {et~al.}(2019)Sanidas {et~al.}}]{Sanidas:2019stw}
Sanidas, S., {et~al.} 2019, \bibinfo{title}{{The LOFAR Tied-Array All-Sky
  Survey (LOTAAS): Survey overview and initial pulsar discoveries},} Astron.
  Astrophys., 626, A104, \dodoi{10.1051/0004-6361/201935609}

\bibitem[{M. Sasaki {et~al.}(2016)Sasaki, Suyama, Tanaka, \&
  Yokoyama}]{Sasaki:2016jop}
Sasaki, M., Suyama, T., Tanaka, T., \& Yokoyama, S. 2016,
  \bibinfo{title}{{Primordial Black Hole Scenario for the Gravitational-Wave
  Event GW150914},} Phys. Rev. Lett., 117, 061101,
  \dodoi{10.1103/PhysRevLett.117.061101}

\bibitem[{B.~S. {Sathyaprakash} \& B.~F. {Schutz}(2009){Sathyaprakash} \&
  {Schutz}}]{2009LRR....12....2S}
{Sathyaprakash}, B.~S., \& {Schutz}, B.~F. 2009, \bibinfo{title}{{Physics,
  Astrophysics and Cosmology with Gravitational Waves},} Living Reviews in
  Relativity, 12, 2, \dodoi{10.12942/lrr-2009-2}

\bibitem[{P.~A.~G. {Scheuer}(1968){Scheuer}}]{1968Natur.218..920S}
{Scheuer}, P.~A.~G. 1968, \bibinfo{title}{{Amplitude Variations in Pulsed Radio
  Sources},} \nat, 218, 920, \dodoi{10.1038/218920a0}

\bibitem[{J.~S. Schwinger(1951)Schwinger}]{Schwinger:1951nm}
Schwinger, J.~S. 1951, \bibinfo{title}{{On gauge invariance and vacuum
  polarization},} Phys. Rev., 82, 664, \dodoi{10.1103/PhysRev.82.664}

\bibitem[{A. {Seta} \& C. {Federrath}(2021){Seta} \&
  {Federrath}}]{2021MNRAS.502.2220S}
{Seta}, A., \& {Federrath}, C. 2021, \bibinfo{title}{{Magnetic fields in the
  Milky Way from pulsar observations: effect of the correlation between thermal
  electrons and magnetic fields},} \mnras, 502, 2220,
  \dodoi{10.1093/mnras/stab128}

\bibitem[{Y. Shao {et~al.}(2023)Shao, Xu, Wang, Yang, Li, Zhang, \&
  Chen}]{Shao:2023agv}
Shao, Y., Xu, Y., Wang, Y., {et~al.} 2023, \bibinfo{title}{{The 21-cm forest as
  a simultaneous probe of dark matter and cosmic heating history},} Nature
  Astron., 7, 1116, \dodoi{10.1038/s41550-023-02024-7}

\bibitem[{T. {Shenar} {et~al.}(2023){Shenar}, {Wade}, {Marchant}, {Bagnulo},
  {Bodensteiner}, {Bowman}, {Gilkis}, {Langer}, {Nicolas-Chen{\'e}},
  {Oskinova}, {Van Reeth}, {Sana}, {St-Louis}, {de Oliveira}, {Todt}, \&
  {Toonen}}]{2023Sci...381..761S}
{Shenar}, T., {Wade}, G.~A., {Marchant}, P., {et~al.} 2023, \bibinfo{title}{{A
  massive helium star with a sufficiently strong magnetic field to form a
  magnetar},} Science, 381, 761, \dodoi{10.1126/science.ade3293}

\bibitem[{X. {Siemens} {et~al.}(2007){Siemens}, {Mandic}, \&
  {Creighton}}]{2007PhRvL..98k1101S}
{Siemens}, X., {Mandic}, V., \& {Creighton}, J. 2007,
  \bibinfo{title}{{Gravitational-Wave Stochastic Background from Cosmic
  Strings},} \prl, 98, 111101, \dodoi{10.1103/PhysRevLett.98.111101}

\bibitem[{N. Smirnov(1948)Smirnov}]{smirnov1948table}
Smirnov, N. 1948, \bibinfo{title}{Table for estimating the goodness of fit of
  empirical distributions,} The annals of mathematical statistics, 19, 279

\bibitem[{N.~V. Smirnov(1939)Smirnov}]{smirnov1939estimate}
Smirnov, N.~V. 1939, \bibinfo{title}{Estimate of deviation between empirical
  distribution functions in two independent samples,} Bulletin Moscow
  University, 2, 3

\bibitem[{T.~L. {Smith} {et~al.}(2006){Smith}, {Pierpaoli}, \&
  {Kamionkowski}}]{2006PhRvL..97b1301S}
{Smith}, T.~L., {Pierpaoli}, E., \& {Kamionkowski}, M. 2006,
  \bibinfo{title}{{New Cosmic Microwave Background Constraint to Primordial
  Gravitational Waves},} \prl, 97, 021301,
  \dodoi{10.1103/PhysRevLett.97.021301}

\bibitem[{J. {Snoek} {et~al.}(2015){Snoek}, {Rippel}, {Swersky}, {Kiros},
  {Satish}, {Sundaram}, {Patwary}, {Prabhat}, \& {Adams}}]{2015arXiv150205700S}
{Snoek}, J., {Rippel}, O., {Swersky}, K., {et~al.} 2015,
  \bibinfo{title}{{Scalable Bayesian Optimization Using Deep Neural Networks},}
  arXiv e-prints, arXiv:1502.05700, \dodoi{10.48550/arXiv.1502.05700}

\bibitem[{A. {Soudais} {et~al.}(2024){Soudais}, {Cerutti}, \&
  {Contopoulos}}]{2024A&A...690A.170S}
{Soudais}, A., {Cerutti}, B., \& {Contopoulos}, I. 2024,
  \bibinfo{title}{{Scaling up global kinetic models of pulsar magnetospheres
  using a hybrid force-free-PIC numerical approach},} \aap, 690, A170,
  \dodoi{10.1051/0004-6361/202450238}

\bibitem[{T. {Sprenger} {et~al.}(2021){Sprenger}, {Wucknitz}, {Main}, {Baker},
  \& {Brisken}}]{2021MNRAS.500.1114S}
{Sprenger}, T., {Wucknitz}, O., {Main}, R., {Baker}, D., \& {Brisken}, W. 2021,
  \bibinfo{title}{{The {\ensuremath{\theta}}-{\ensuremath{\theta}} diagram:
  transforming pulsar scintillation spectra to coordinates on highly
  anisotropic interstellar scattering screens},} \mnras, 500, 1114,
  \dodoi{10.1093/mnras/staa3353}

\bibitem[{D.~H. {Staelin} \& E.~C. {Reifenstein}(1968){Staelin} \&
  {Reifenstein}}]{1968Sci...162.1481S}
{Staelin}, D.~H., \& {Reifenstein}, III, E.~C. 1968, \bibinfo{title}{{Pulsating
  Radio Sources near the Crab Nebula},} Science, 162, 1481,
  \dodoi{10.1126/science.162.3861.1481}

\bibitem[{F. Staff(2021)Staff}]{fastref}
Staff, F. 2021, \bibinfo{title}{{FAST official website},},
  \url{https://fast.bao.ac.cn/}

\bibitem[{S. Staff(2019)Staff}]{skaref}
Staff, S. 2019, \bibinfo{title}{{SKA official website},},
  \url{https://www.skao.int/}

\bibitem[{G.~V. Stephenson(2009)Stephenson}]{Stephenson:2009zz}
Stephenson, G.~V. 2009, \bibinfo{title}{{The standard quantum limit for the
  Li-Baker HFGW detector},} AIP Conf. Proc., 1103, 542,
  \dodoi{10.1063/1.3115563}

\bibitem[{K. Takahashi {et~al.}(2013)Takahashi, Mori, Ichiki, Inoue, \&
  Takami}]{Takahashi:2013lba}
Takahashi, K., Mori, M., Ichiki, K., Inoue, S., \& Takami, H. 2013,
  \bibinfo{title}{{Lower Bounds on Magnetic Fields in Intergalactic Voids from
  Long-term GeV-TeV Light Curves of the Blazar Mrk 421},} Astrophys. J. Lett.,
  771, L42, \dodoi{10.1088/2041-8205/771/2/L42}

\bibitem[{F. Tavecchio {et~al.}(2010)Tavecchio, Ghisellini, Foschini, Bonnoli,
  Ghirlanda, \& Coppi}]{Tavecchio:2010mk}
Tavecchio, F., Ghisellini, G., Foschini, L., {et~al.} 2010,
  \bibinfo{title}{{The intergalactic magnetic field constrained by Fermi/LAT
  observations of the TeV blazar 1ES 0229+200},} Mon. Not. Roy. Astron. Soc.,
  406, L70, \dodoi{10.1111/j.1745-3933.2010.00884.x}

\bibitem[{ {The LIGO Scientific Collaboration} {et~al.}(2025){The LIGO
  Scientific Collaboration}, {the Virgo Collaboration}, {the KAGRA
  Collaboration}, {Abac}, {Abouelfettouh}, {Acernese}, {Ackley}, {Adamcewicz},
  {Adhicary}, {Adhikari}, {Adhikari}, {Adhikari}, {Adkins}, {Afroz}, {Agapito},
  {Agarwal}, {Agathos}, {Aggarwal}, {Aggarwal}, {Aguiar}, {Ahrend}, {Aiello},
  {Ain}, {Ajith}, {Akutsu}, {Albanesi}, {Ali}, {Al-Kershi}, {All{\'e}n{\'e}},
  {Allocca}, {Al-Shammari}, {Altin}, {Alvarez-Lopez}, {Amar}, {Amarasinghe},
  {Amato}, {Amicucci}, {Amra}, {Ananyeva}, {Anderson}, {Anderson}, {Andia},
  {Ando}, {Andr{\'e}s-Carcasona}, {Andri{\'c}}, {Anglin}, {Ansoldi}, {Antelis},
  {Antier}, {Aoumi}, {Appavuravther}, {Appert}, {Apple}, {Arai}, {Araya},
  {Araya}, {Arca Sedda}, {Areeda}, {Aritomi}, {Armato}, {Armstrong}, {Arnaud},
  {Arogeti}, {Aronson}, {Arun}, {Ashton}, {Aso}, {Asprea}, {Assiduo}, {Assis de
  Souza Melo}, {Aston}, {Astone}, {Attadio}, {Aubin}, {AultONeal}, {Avallone},
  {Avila}, {Babak}, {Badger}, {Bae}, {Bagnasco}, {Baiotti}, {Bajpai}, {Baka},
  {Baker}, {Baker}, {Baker}, {Baldi}, {Baldicchi}, {Ball}, {Ballardin},
  {Ballmer}, {Banagiri}, {Banerjee}, {Bankar}, {Baptiste}, {Baral}, {Baratti},
  {Barayoga}, {Barish}, {Barker}, {Barman}, {Barneo}, {Barone}, {Barr},
  {Barsotti}, {Barsuglia}, {Barta}, {Bartoletti}, {Barton}, {Bartos},
  {Basalaev}, {Bassiri}, {Basti}, {Bawaj}, {Baxi}, {Bayley}, {Baylor},
  {Baynard}, {Bazzan}, {Bedakihale}, {Beirnaert}, {Bejger}, {Belardinelli},
  {Bell}, {Bellie}, {Bellizzi}, {Benoit}, {Bentara}, {Bentley}, {Ben Yaala},
  {Bera}, {Bergamin}, {Berger}, {Bernuzzi}, {Beroiz}, {Berry}, {Bersanetti},
  {Bertheas}, {Bertolini}, {Betzwieser}, {Beveridge}, {Bevilacqua}, {Bevins},
  {Bhandare}, {Bhatt}, {Bhattacharjee}, {Bhattacharyya}, {Bhaumik},
  {Biancalana}, {Bianchi}, {Bilenko}, {Billingsley}, {Binetti}, {Bini}, {Binu},
  {Biot}, {Birnholtz}, {Biscoveanu}, {Bisht}, {Bitossi}, {Bizouard}, {Blaber},
  {Blackburn}, {Blagg}, {Blair}, {Blair}, {Bode}, {Boettner}, {Boileau},
  {Boldrini}, {Bolingbroke}, {Bolliand}, {Bonavena}, {Bondarescu}, {Bondu},
  {Bonilla}, {Bonilla}, {Bonino}, {Bonnand}, {Borchers}, {Borhanian}, {Boschi},
  {Bose}, {Bossilkov}, {Bothra}, {Boudon}, {Bourg}, {Boyle}, {Bozzi},
  {Bradaschia}, {Brady}, {Branch}, {Branchesi}, {Braun}, {Briant}, {Brillet},
  {Brinkmann}, {Brockill}, \& {Brockmueller}}]{2025arXiv250818082T}
{The LIGO Scientific Collaboration}, {the Virgo Collaboration}, {the KAGRA
  Collaboration}, {et~al.} 2025, \bibinfo{title}{{GWTC-4.0: Updating the
  Gravitational-Wave Transient Catalog with Observations from the First Part of
  the Fourth LIGO-Virgo-KAGRA Observing Run},} arXiv e-prints,
  arXiv:2508.18082, \dodoi{10.48550/arXiv.2508.18082}

\bibitem[{A.~R. {Thompson} {et~al.}(2017){Thompson}, {Moran}, \&
  {Swenson}}]{2017isra.book.....T}
{Thompson}, A.~R., {Moran}, J.~M., \& {Swenson}, Jr., G.~W. 2017,
  {Interferometry and Synthesis in Radio Astronomy, 3rd Edition},
  \dodoi{10.1007/978-3-319-44431-4}

\bibitem[{K.~S. {Thorne}(1980){Thorne}}]{1980RvMP...52..299T}
{Thorne}, K.~S. 1980, \bibinfo{title}{{Multipole expansions of gravitational
  radiation},} Reviews of Modern Physics, 52, 299,
  \dodoi{10.1103/RevModPhys.52.299}

\bibitem[{A. {Tiengo} \& S. {Mereghetti}(2007){Tiengo} \&
  {Mereghetti}}]{2007ApJ...657L.101T}
{Tiengo}, A., \& {Mereghetti}, S. 2007, \bibinfo{title}{{XMM-Newton Discovery
  of 7 s Pulsations in the Isolated Neutron Star RX J1856.5-3754},} \apjl, 657,
  L101, \dodoi{10.1086/513143}

\bibitem[{M.-l. Tong {et~al.}(2008)Tong, Zhang, \& Li}]{Tong:2008rz}
Tong, M.-l., Zhang, Y., \& Li, F.-Y. 2008, \bibinfo{title}{{Using polarized
  maser to detect high-frequency relic gravitational waves},} Phys. Rev. D, 78,
  024041, \dodoi{10.1103/PhysRevD.78.024041}

\bibitem[{W.-y. Tsai(1974)Tsai}]{Tsai:1974ap}
Tsai, W.-y. 1974, \bibinfo{title}{{Vacuum Polarization in Homogeneous Magnetic
  Fields},} Phys. Rev. D, 10, 2699, \dodoi{10.1103/PhysRevD.10.2699}

\bibitem[{W.-y. Tsai \& T. Erber(1974)Tsai \& Erber}]{Tsai:1974fa}
Tsai, W.-y., \& Erber, T. 1974, \bibinfo{title}{{Photon Pair Creation in
  Intense Magnetic Fields},} Phys. Rev. D, 10, 492,
  \dodoi{10.1103/PhysRevD.10.492}

\bibitem[{W.-y. Tsai \& T. Erber(1975)Tsai \& Erber}]{Tsai:1975iz}
Tsai, W.-y., \& Erber, T. 1975, \bibinfo{title}{{The Propagation of Photons in
  Homogeneous Magnetic Fields: Index of Refraction},} Phys. Rev. D, 12, 1132,
  \dodoi{10.1103/PhysRevD.12.1132}

\bibitem[{M. Unger \& G.~R. Farrar(2024)Unger \& Farrar}]{Unger:2023lob}
Unger, M., \& Farrar, G.~R. 2024, \bibinfo{title}{{The Coherent Magnetic Field
  of the Milky Way},} Astrophys. J., 970, 95, \dodoi{10.3847/1538-4357/ad4a54}

\bibitem[{L.~F. Urrutia(1978)Urrutia}]{Urrutia:1977xb}
Urrutia, L.~F. 1978, \bibinfo{title}{{Vacuum Polarization in Parallel
  Homogeneous Electric and Magnetic Fields},} Phys. Rev. D, 17, 1977,
  \dodoi{10.1103/PhysRevD.17.1977}

\bibitem[{B. {Uyaniker} \& T.~L. {Landecker}(2002){Uyaniker} \&
  {Landecker}}]{2002ApJ...575..225U}
{Uyaniker}, B., \& {Landecker}, T.~L. 2002, \bibinfo{title}{{A Highly Ordered
  Faraday Rotation Structure in the Interstellar Medium},} \apj, 575, 225,
  \dodoi{10.1086/341266}

\bibitem[{T. {Vachaspati} \& A. {Vilenkin}(1985){Vachaspati} \&
  {Vilenkin}}]{1985PhRvD..31.3052V}
{Vachaspati}, T., \& {Vilenkin}, A. 1985, \bibinfo{title}{{Gravitational
  radiation from cosmic strings},} \prd, 31, 3052,
  \dodoi{10.1103/PhysRevD.31.3052}

\bibitem[{ {VERITAS Collaboration} {et~al.}(2011){VERITAS Collaboration},
  {Aliu}, {Arlen}, {Aune}, {Beilicke}, {Benbow}, {Bouvier}, {Bradbury},
  {Buckley}, {Bugaev}, {Byrum}, {Cannon}, {Cesarini}, {Christiansen}, {Ciupik},
  {Collins-Hughes}, {Connolly}, {Cui}, {Dickherber}, {Duke}, {Errando},
  {Falcone}, {Finley}, {Finnegan}, {Fortson}, {Furniss}, {Galante}, {Gall},
  {Gibbs}, {Gillanders}, {Godambe}, {Griffin}, {Grube}, {Guenette}, {Gyuk},
  {Hanna}, {Holder}, {Huan}, {Hughes}, {Hui}, {Humensky}, {Imran}, {Kaaret},
  {Karlsson}, {Kertzman}, {Kieda}, {Krawczynski}, {Krennrich}, {Lang},
  {Lyutikov}, {Madhavan}, {Maier}, {Majumdar}, {McArthur}, {McCann},
  {McCutcheon}, {Moriarty}, {Mukherjee}, {Nu{\~n}ez}, {Ong}, {Orr}, {Otte},
  {Park}, {Perkins}, {Pizlo}, {Pohl}, {Prokoph}, {Quinn}, {Ragan}, {Reyes},
  {Reynolds}, {Roache}, {Rose}, {Ruppel}, {Saxon}, {Schroedter}, {Sembroski},
  {{\c{S}}ent{\"u}rk}, {Smith}, {Staszak}, {Te{\v{s}}i{\'c}}, {Theiling},
  {Thibadeau}, {Tsurusaki}, {Tyler}, {Varlotta}, {Vassiliev}, {Vincent},
  {Vivier}, {Wakely}, {Ward}, {Weekes}, {Weinstein}, {Weisgarber}, {Williams},
  \& {Zitzer}}]{2011Sci...334...69V}
{VERITAS Collaboration}, {Aliu}, E., {Arlen}, T., {et~al.} 2011,
  \bibinfo{title}{{Detection of Pulsed Gamma Rays Above 100 GeV from the Crab
  Pulsar},} Science, 334, 69, \dodoi{10.1126/science.1208192}

\bibitem[{V.~V. {Vitkevich}(1957){Vitkevich}}]{1957AZh....34..349V}
{Vitkevich}, V.~V. 1957, \bibinfo{title}{{Radio Stars and the Methods of Their
  Investigation.},} \azh, 34, 349

\bibitem[{J. {Wang} {et~al.}(2025){Wang}, {Li}, {Pan}, {Deng}, {Liu}, {Yang},
  {Hu}, {Wang}, {Zhang}, \& {Chen}}]{2025arXiv250111872W}
{Wang}, J., {Li}, Y., {Pan}, H., {et~al.} 2025, \bibinfo{title}{{FAST drift
  scan survey for HI intensity mapping: simulation on Bayesian-stacking-based
  HI mass function estimation},} arXiv e-prints, arXiv:2501.11872,
  \dodoi{10.48550/arXiv.2501.11872}

\bibitem[{G. {Wentzel}(1926){Wentzel}}]{1926ZPhy...38..518W}
{Wentzel}, G. 1926, \bibinfo{title}{{Eine Verallgemeinerung der
  Quantenbedingungen f{\"u}r die Zwecke der Wellenmechanik},} Zeitschrift fur
  Physik, 38, 518, \dodoi{10.1007/BF01397171}

\bibitem[{E. {Witten}(1984){Witten}}]{1984PhRvD..30..272W}
{Witten}, E. 1984, \bibinfo{title}{{Cosmic separation of phases},} \prd, 30,
  272, \dodoi{10.1103/PhysRevD.30.272}

\bibitem[{K. Yee(1966)Yee}]{Kane:1138693}
Yee, K. 1966, \bibinfo{title}{Numerical solution of initial boundary value
  problems involving maxwell's equations in isotropic media,} IEEE Transactions
  on Antennas and Propagation, 14, 302, \dodoi{10.1109/TAP.1966.1138693}

\bibitem[{H.-R. Yu {et~al.}(2014)Yu, Zhang, \& Pen}]{Yu:2013bia}
Yu, H.-R., Zhang, T.-J., \& Pen, U.-L. 2014, \bibinfo{title}{{Method for Direct
  Measurement of Cosmic Acceleration by 21-cm Absorption Systems},} Phys. Rev.
  Lett., 113, 041303, \dodoi{10.1103/PhysRevLett.113.041303}

\bibitem[{B. Zhang(2023)Zhang}]{Zhang:2022uzl}
Zhang, B. 2023, \bibinfo{title}{{The physics of fast radio bursts},} Rev. Mod.
  Phys., 95, 035005, \dodoi{10.1103/RevModPhys.95.035005}

\end{thebibliography}
\bibliographystyle{aasjournalv7}

\end{document}